\documentclass[prl,aps,amssymb,twocolumn,nofootinbib,superscriptaddress]{revtex4-1}
\usepackage{amsmath}
\usepackage{amssymb}
\usepackage{amsthm}
\usepackage{amsfonts}
\usepackage{listings}
\usepackage{enumerate}
\usepackage{latexsym}
\usepackage{psfrag}

\usepackage{bm}
\usepackage{graphicx}
\usepackage[caption=false,position=top]{subfig}

\usepackage{color}
\usepackage[normalem]{ulem}

\begin{document}
\author{Barry Bradlyn}
\thanks{These authors contributed equally to the preparation of this work.}
\affiliation{Princeton Center for Theoretical Science, Princeton University, Princeton, New Jersey 08544, USA}
\author{Jennifer Cano}
\thanks{These authors contributed equally to the preparation of this work.}
\affiliation{Princeton Center for Theoretical Science, Princeton University, Princeton, New Jersey 08544, USA}
\author{Zhijun Wang}
\thanks{These authors contributed equally to the preparation of this work.}
\affiliation{Department of Physics, Princeton University, Princeton, New Jersey 08544, USA}
\author{M.~G. Vergniory}
\affiliation{Donostia International Physics Center, P. Manuel de Lardizabal 4, 20018 Donostia-San Sebasti\'{a}n, Spain}
\author{C. Felser}
\affiliation{Max Planck Institute for Chemical Physics of Solids, 01187 Dresden, Germany}
\author{R.~J. Cava}
\affiliation{Department of Chemistry, Princeton University, Princeton, New Jersey 08544, USA}
\author{B. Andrei Bernevig}
\affiliation{Department of Physics, Princeton University, Princeton, New Jersey 08544, USA}
\date{March 9, 2016}
\title{Beyond Dirac and Weyl fermions: Unconventional quasiparticles in conventional crystals}
\begin{abstract}
In quantum field theory, we learn that fermions come in three varieties: Majorana, Weyl, and Dirac. Here we show that in solid state systems this classification is incomplete and find several additional types of crystal symmetry-protected free fermionic excitations . We exhaustively classify linear and quadratic 3-, 6- and 8- band crossings stabilized by space group symmetries in solid state systems with spin-orbit coupling and time-reversal symmetry. Several distinct types of fermions arise, differentiated by their degeneracies at and along high symmetry points, lines, and surfaces. Some notable consequences of these fermions are the presence of Fermi arcs in non-Weyl systems and the existence of Dirac lines. Ab-initio calculations identify a number of materials that realize these exotic fermions close to the Fermi level.
\end{abstract}
\maketitle
\paragraph{Introduction}
Condensed matter systems have recently become a fertile ground for the discovery of fermionic particles and phenomena predicted in high-energy physics, but which lack experimental observation. Starting with graphene and its Dirac fermions\cite{CastroNeto09}, continuing to Majorana fermions in superconducting heterostructures\cite{Read00,Fu08,Lutchyn10,Oreg10,Mourik12,NadjPerge14}, and, most recently, with the discovery of Weyl\cite{Wan11,Weng15,Huang15,Xu15,Lv15,Xu15a,Lv15a,Hirschberger16,Felser2016} and Dirac\cite{Young12,Wang12,Liu14,Liu14a,Steinberg14,Xiong2015} semimetals, solid-state physics has proven to be an efficient simulacrum of relativistic free fermions. There is, however, a fundamental difference between electrons in a solid and those at high energy: for relativistic fermions, the constraints imposed by Poincar\'{e} symmetry greatly limit the types of particles that may occur. The situation in condensed matter physics is less constrained; only certain subgroups of Poincar\'{e} symmetry -- the $230$ space groups that exist in 3d lattices -- need be respected. There is the potential, then, to find free fermionic excitations in solid state systems that have \emph{no high energy analogues}.

Here, we find and classify these exotic fermions, propose experiments to demonstrate their topological character, and { point out a large number of different classes of candidate materials where these fermions appear close to the Fermi level.} We consider materials with time reversal (TR) symmetry and spin-orbit coupling.
We find that three of the space groups host half-integer angular momentum fermionic excitations with \emph{three-fold degeneracies}, stabilized by non-symmorphic symmetries. 
The existence of three-fold (and higher) degeneracies has long been known from a band theory perspective\cite{Cracknell,Faddeyev2014,Kovalev1965,Miller1967,Casher1969}. { Our purpose is to elucidate their topological nature.}
We show that all of the three-fold degeneracies either carry a Chern number $\pm 2$ or sit at the critical point separating the two Chern numbers.
In two closely related space groups, the combination of TR and inversion results in six-fold degeneracies that consist of a three-fold degeneracy and its time-reversed partner, a ``six-fold Dirac point.'' There are two other types of six-fold degeneracies, both of which are distinct from free spin-$5/2$ particles.
We also discuss eightfold degeneracies, which were recently introduced in Ref.~\cite{Wieder2015}. Here, we prove that there exists a finer classification of the $8-$fold degenerate fermions.
{ Finally, we show that for the 3-fold, as well as for a class of 4-fold degeneracies discussed in a future manuscript, the low energy Hamiltonian is of the form $\mathbf{k}\cdot\mathbf{S}$, where $\mathbf{S}$ is the vector of spin-1 or -3/2 matrices. This provides the first natural generalization of the recently discovered Weyl fermion Hamiltonian, $\mathbf{k}\cdot \sigma$\cite{Wan11,Weng15,Huang15,Xu15,Lv15,Xu15a,Lv15a,Hirschberger16,Felser2016}.}

We enumerate all possible $3$-, $6$- and $8$- fold degenerate fermions in the subsequent sections.
We include a symmetry analysis for each degeneracy type; an exhaustive search of the 230 space groups\cite{Cracknell} guarantees that the list is complete.
For each new fermion, we investigate the low-energy $\mathbf{k}\cdot \mathbf{p}$ Hamiltonian allowed by the space group symmetries\cite{Cracknell,Bilbao1,Bilbao2,Bilbao3}. 
This determines which degeneracies carry non-trivial Berry curvature or host exotic symmetry-enforced degeneracies along high-symmetry lines or planes in the Brillouin zone (BZ)\cite{Chiu14,Bian15,Chan15}. We derive the role these features play in transport and surface properties. 
Finally, using \emph{ab-initio} techniques, we predict existing material realizations for each of our new fermions, close to the Fermi level.
We conclude with a discussion of the experimental outlook for synthesizing new fermions.

\paragraph{Space groups with 3-, 6-, and 8-band crossings}
The guiding principle of our classification is to find irreducible representations (irreps) of the (little) group of lattice symmetries at high-symmetry points in the BZ for each of the $230$ space groups (SGs); the dimension of these representations corresponds to the number of bands that meet at the high-symmetry point, and is one of the characteristics of the fermion type.
Since we are interested in fermions with spin-orbit coupling, we consider only the double-valued representations; TR symmetry is an antiunitary that squares to $-1$.
Table \ref{table1} summarizes the results of our search.
All the space groups include non-symmorphic generators, { and all representations are projective;} these are in fact necessary ingredients for the 3-, 6- and 8-d irreps, as elaborated upon in the Supplementary Material.

We find that space groups $199, 214,$ and $220$ may host a three-dimensional representation at the $P$ point in the BZ (the high-symmetry points are defined in the Supplementary Material). These space groups have a body-centered cubic Bravais lattice, and the $P$ point is a TR non-invariant point at a corner of the BZ (that is, $P\neq-P$). All three of these systems host a complementary $3$-fold degeneracy at $-P$ due to TR symmetry; Kramers's theorem requires this to be the case. SG 214 is unique in that the $3$-fold degeneracy at $-P$ persists even if time reversal symmetry is broken, as the $P$ and $-P$ points are related by a two-fold screw rotation in the full symmetry group.

In the presence of TR symmetry, six space groups can host $6$-fold degeneracies. In all cases, these arise as $3$-fold degeneracies which are doubled by the presence of TR symmetry. Four of these -- SGs $198,205,212,$ and $213$ -- correspond to simple-cubic Bravais lattice, and the $6$-fold degeneracy occurs at the TR invariant $R$ point at the corner of the BZ. The other two $6$-fold degeneracies occur in SGs $206$ and $230$ at the $P$ point. Although this point is not TR invariant, these SGs are inversion symmetric, and hence all degeneracies are doubled.

Finally, we find, in agreement with previous work\cite{Wieder2015}, that seven SGs may host $8$-fold degeneracies. { However, as shown below, the resulting fermions fall into distinct classes}. Two of these, SGs $130$ and $135$ have a tetragonal Bravais lattice; these are special in that they \emph{require} $8$-fold degeneracies at the time-reversal invariant $A$ point. In addition, SGs $222,223$ and $230$ may host $8$-fold degeneracies. SGs $222$ and $223$ are simple-cubic, and an $8$-fold fermion can occur at the $R$ point in the BZ; for SG $230$, it occurs at the time-reversal invariant $H$ point.

There are two more SGs that can host $8$-fold degeneracies, SG $218$ and SG $220$. These differ from the others in that they lack inversion symmetry. Energy bands away from high symmetry points need no longer come in pairs. SG $218$ has a simple cubic Bravais lattice, and an $8$-fold degeneracy may occur at the $R$ point. In SG $220$ the degeneracy may occur at the $H$ point.

\begin{table}[t]
\begin{tabular}{l | c | c | c | l  }
SG& La & $k$ & d & Generators \\
  \hline			
198 & cP & R & 6  & $\lbrace C_{3,111}^-|010\rbrace, \lbrace C_{2x}|\frac{1}{2}\frac{3}{2}0\rbrace, \lbrace C_{2y}|0\frac{3}{2}\frac{1}{2}\rbrace$ \\
199 & cB & P & 3  & $\lbrace C_{3,111}^-|101\rbrace, \lbrace C_{2x}|\bar{\frac{1}{2}}\frac{1}{2}0\rbrace, \lbrace C_{2y}|0\frac{1}{2}\bar{\frac{1}{2}}\rbrace$ \\
205 & cP & R & 6 & $\lbrace C_{3,111}^-|010\rbrace, \lbrace C_{2x}|\frac{1}{2}\frac{3}{2}0\rbrace, \lbrace C_{2y}|0\frac{3}{2}\frac{1}{2}\rbrace,\lbrace I|000\rbrace$ \\
206 & cB & P & 6 & $\lbrace C_{3,111}^-|101\rbrace, \lbrace C_{2x}|\bar{\frac{1}{2}}\frac{1}{2}0\rbrace, \lbrace C_{2y}|0\frac{1}{2}\bar{\frac{1}{2}}\rbrace$\\
212 & cP & R & 6 & $ \lbrace C_{2x}|\frac{1}{2}\frac{1}{2}0\rbrace,\!  \lbrace C_{2y}|0\frac{1}{2}\frac{1}{2}\rbrace,\! \lbrace C_{3,111}^-|000\rbrace,\! \lbrace C_{2,1\bar{1}0}|\frac{1}{4}\frac{1}{4}\frac{1}{4}\rbrace$\\
213 & cP & R & 6 & $ \lbrace C_{2x}|\frac{1}{2}\frac{1}{2}0\rbrace,\!  \lbrace C_{2y}|0\frac{1}{2}\frac{1}{2}\rbrace,\!\lbrace C_{3,111}^-|000\rbrace, \!\lbrace C_{2,1\bar{1}0}|\frac{3}{4}\frac{3}{4}\frac{3}{4}\rbrace$\\
214 & cB & P & 3 & $\lbrace C_{3,111}^-|101\rbrace, \lbrace C_{2x}|\bar{\frac{1}{2}}\frac{1}{2}0\rbrace, \lbrace C_{2y}|0\frac{1}{2}\bar{\frac{1}{2}}\rbrace$\\
220 & cB & P & 3 & $\lbrace C_{3,\bar{1}\bar{1}1}|0\frac{1}{2}\frac{1}{2}\rbrace,\! \lbrace C_{2y}|0\frac{1}{2}\frac{1}{2}\rbrace,\! \lbrace C_{2x}|\frac{3}{2}\frac{3}{2}0\rbrace,\! \lbrace IC_{4x}^-|\frac{1}{2}11\rbrace$ \\
230 & cB & P & 6 & $\lbrace C_{3,\bar{1}\bar{1}1}|0\frac{1}{2}\frac{1}{2}\rbrace, \! \lbrace C_{2y}|0\frac{1}{2}\frac{1}{2}\rbrace,\! \lbrace C_{2x}|\frac{3}{2}\frac{3}{2}0\rbrace,\! \lbrace IC_{4x}^-|\frac{1}{2}11\rbrace$ \\
\hline
130 & tP & A & 8 & $\lbrace C_{4z}|000\rbrace, \lbrace \sigma_{\bar{x}y} | 00\frac{1}{2}\rbrace, \lbrace I | \frac{1}{2}\frac{1}{2}\frac{1}{2}\rbrace $\\
135 & tP & A & 8 & $\lbrace C_{4z}|\frac{1}{2}\frac{1}{2}\frac{1}{2}\rbrace, \lbrace \sigma_{\bar{x}y} | 00\frac{1}{2}\rbrace, \lbrace I | 000\rbrace $\\
218 & cP & R & 8 & $\lbrace C_{2x}|001\rbrace, \lbrace C_{2y}|000\rbrace, \lbrace C_{3,111}^-|001\rbrace, \lbrace \sigma_{\bar{x}y}|\frac{1}{2}\frac{1}{2}\frac{1}{2}\rbrace$ \\
220 & cB & H & 8 & $\lbrace C_{2x}|\frac{1}{2}\frac{1}{2}0\rbrace, \lbrace C_{2y}|0\frac{1}{2}\frac{3}{2}\rbrace, \lbrace C_{3,111}^-|001\rbrace, \lbrace \sigma_{\bar{x}y}|\frac{1}{2}\frac{1}{2}\frac{1}{2}\rbrace$\\
222 & cP & R & 8 & $\lbrace C_{4z}^-|000\rbrace, \lbrace C_{2x} | 000\rbrace, \lbrace C_{3,111}^-|010\rbrace,\lbrace I | \frac{1}{2}\frac{1}{2}\frac{1}{2}\rbrace $\\
223 & cP & R & 8 &  $\lbrace C_{4z}^-|\frac{1}{2}\frac{1}{2}\frac{1}{2}\rbrace, \lbrace C_{2x} | 000\rbrace, \lbrace C_{3,111}^-|010\rbrace, \lbrace I | 000\rbrace $\\
230 & cB & H & 8 & $\lbrace C_{4z}| 0\frac{1}{2}0\rbrace, \lbrace C_{2y} | 1\frac{1}{2}\frac{1}{2}\rbrace, \lbrace C_{3,111}|111\rbrace,\lbrace I | 000\rbrace $ \\
\end{tabular}
\caption{Summary of all new fermion types in solid state systems. La indicates the type of lattice (cP is cubic primitive, cB is cubic body-centered, and tP is tetragonal primitive), d indicates the maximum degeneracy at the relevant $k$ point in the presence of time reversal symmetry. Group generators are defined in the Supplementary Material.}
\label{table1}
\end{table}

\begin{figure*}
\centering
\subfloat[SGs 199 and 214\label{fig:kdotp199}]{%
  \includegraphics[height=1.5in]{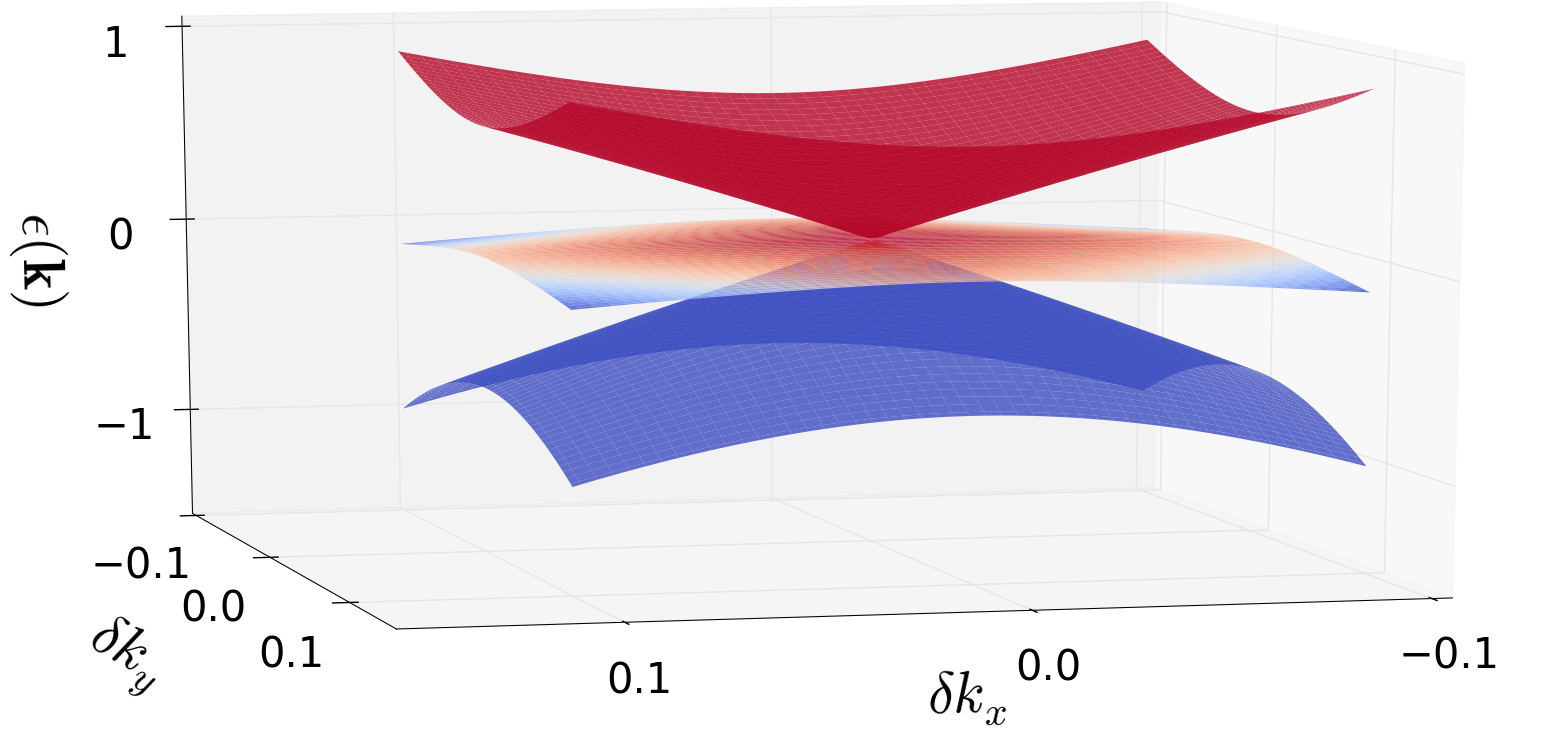}%
}\quad\quad
\subfloat[SG 220\label{fig:kdotp220}]{%
  \includegraphics[height=1.7in]{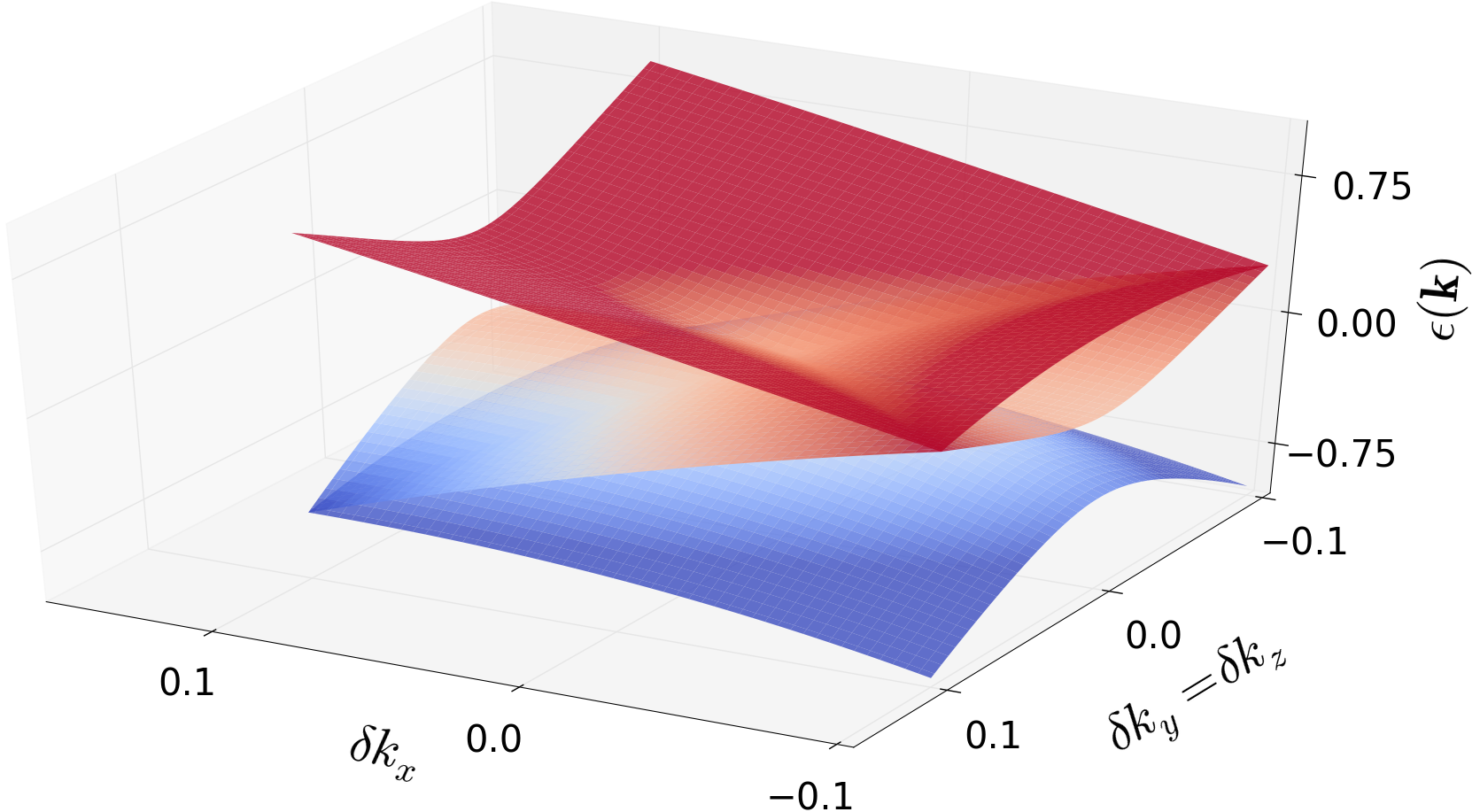}%
}
\caption{Energy dispersion near a three-fold degeneracy at the $P$ point in (a) SGs 199 and 214 and (b) SG 220. In the latter case, pairs of bands remain degenerate in energy along the high-symmetry lines $|\delta k_x|=|\delta k_y|=|\delta k_z|$.}
\label{fig:kdotp3}
\end{figure*}

\begin{figure*}
\centering
\subfloat[SG 205\label{fig:kdotp205}]{\includegraphics[height=1.05in]{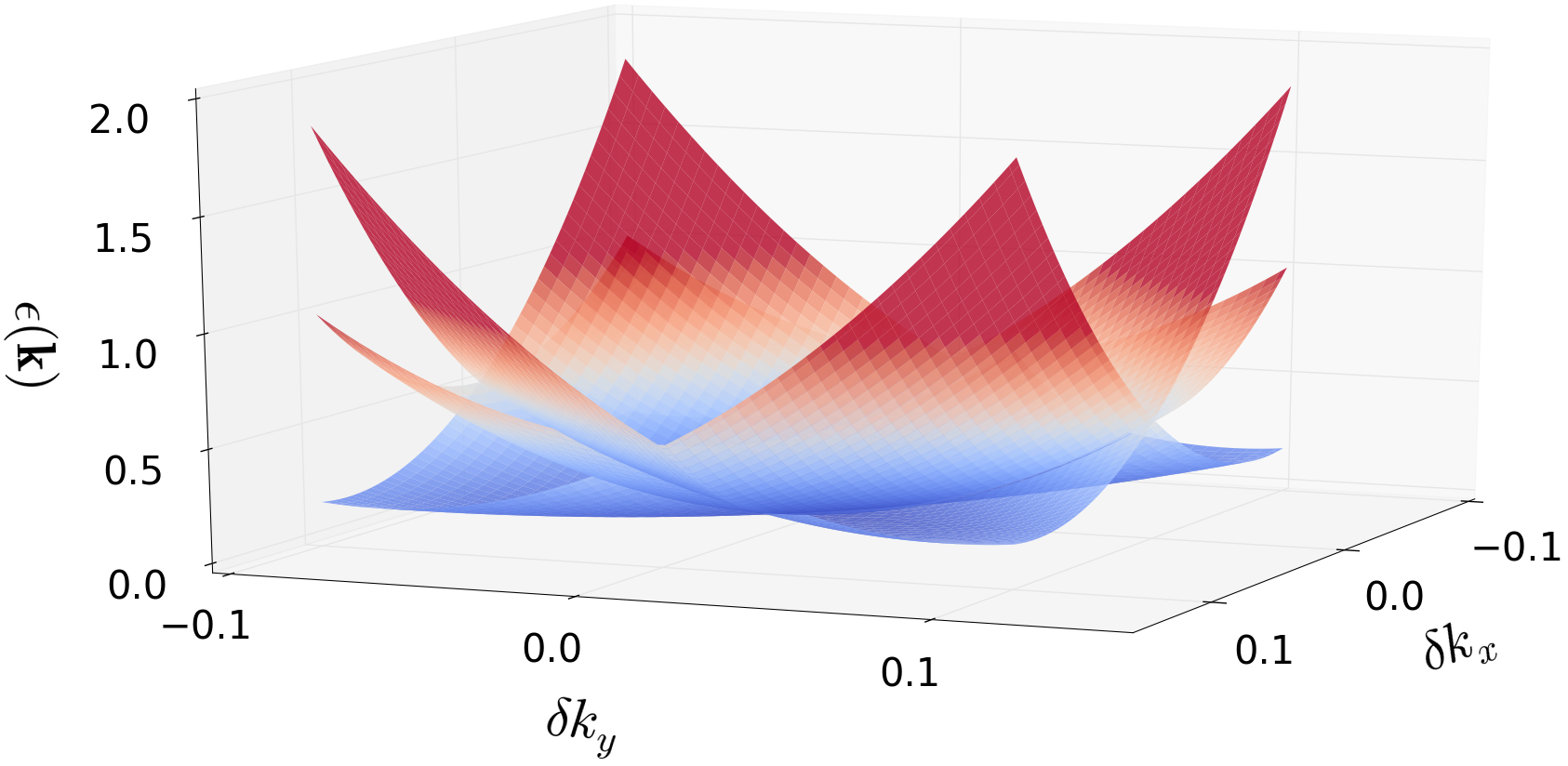}}\quad
\subfloat[SGs 206 and 230\label{fig:kdotp206}]{ \includegraphics[height=1.05in]{sg199-test.png}}\quad
\subfloat[SGs 198, 212 and 213\label{fig:kdotp198}]{\includegraphics[height=1.05in]{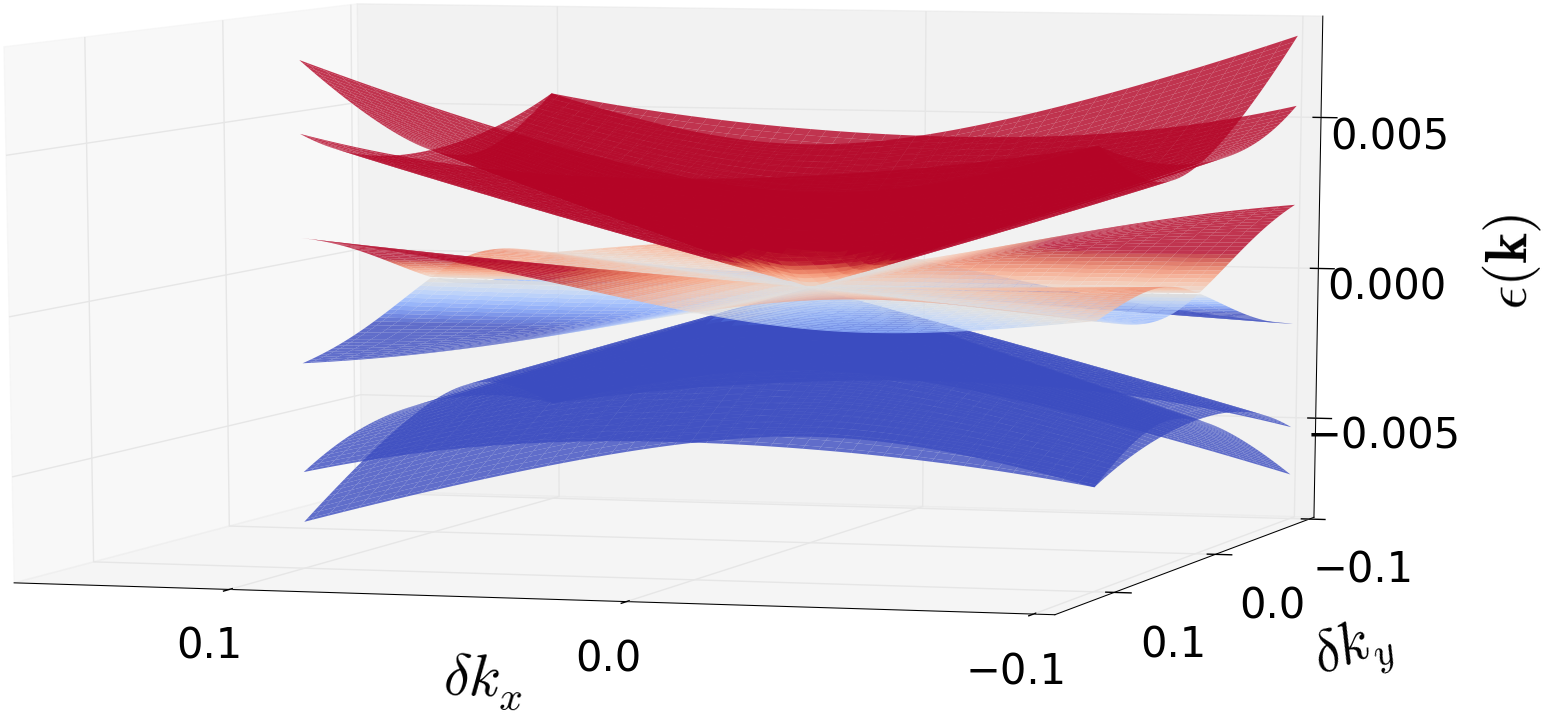}} 
\caption{Energy dispersion near a six-fold degeneracy in (a) SG 205, (b) SG 206 and 230, and (c) SGs 198, 212, and 213. In SGs $198, 212,$ and $213$ bands become degenerate in pairs along the faces $\delta k_i=0$ of the BZ. In SGs $205, 206$ and $230$, all bands are two-fold degenerate due to inversion symmetry.}
\label{fig:kdotp6}
\end{figure*}

\begin{figure*}
\centering
\subfloat[SGs 130 and 135\label{fig:kdotp130}]{%
  \includegraphics[height=1.15in]{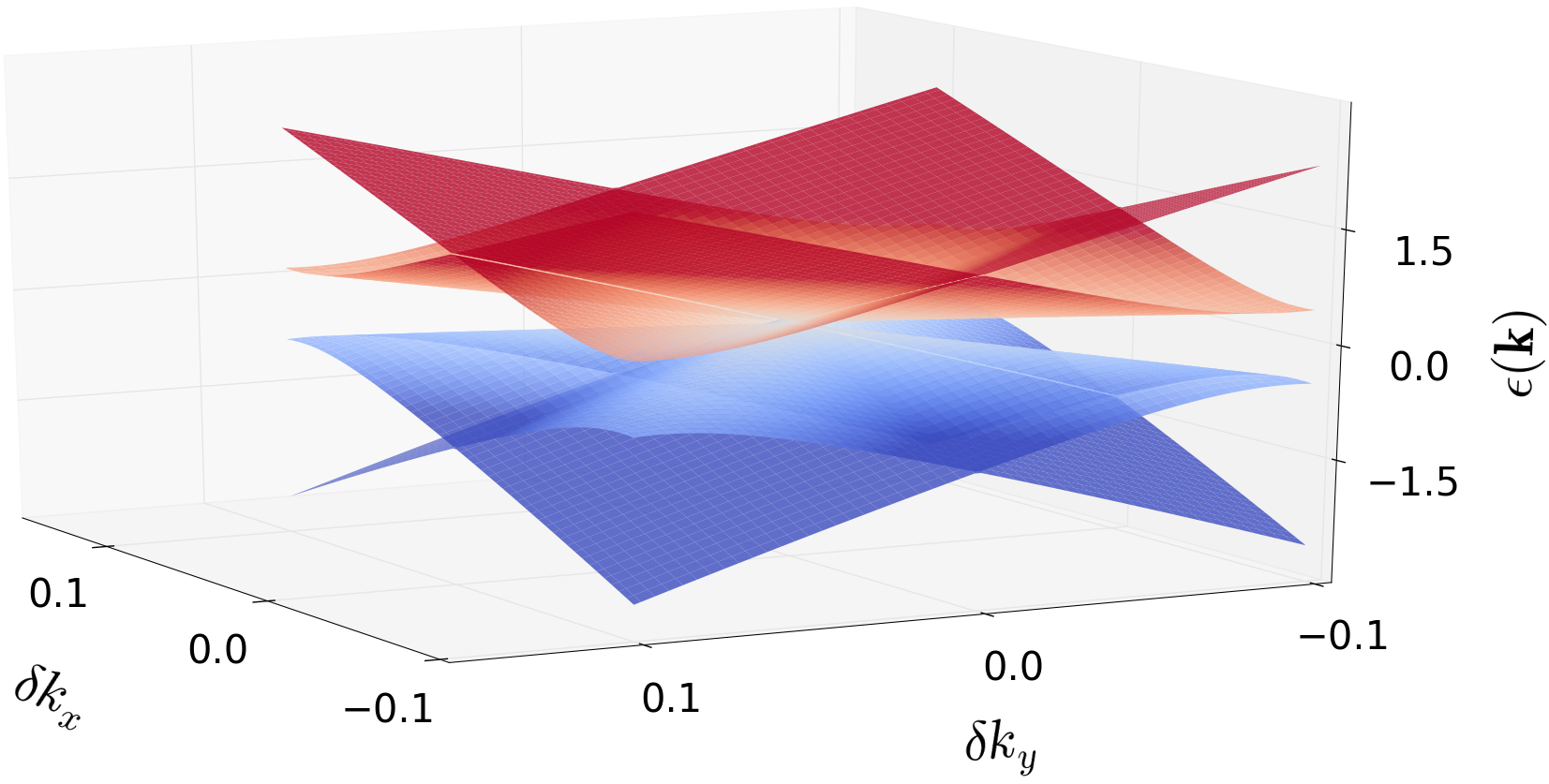}%
}\quad 
\subfloat[SGs 222, 223 and 230\label{fig:kdotp222}]{%
  \includegraphics[height=1.15in]{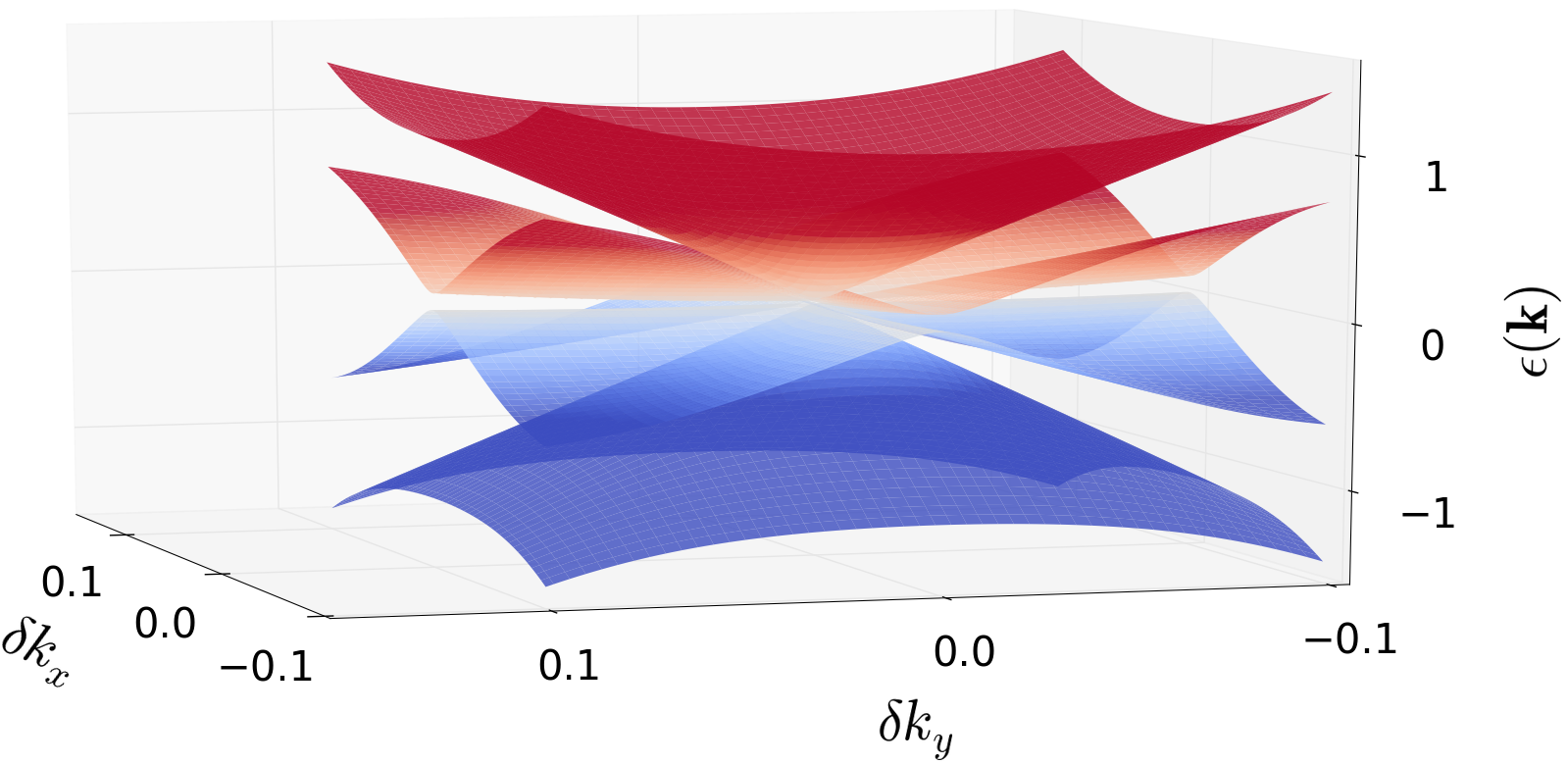}%
	}\quad
\subfloat[SGs 218 and 220\label{fig:kdotp218}]{%
  \includegraphics[height=1.15in]{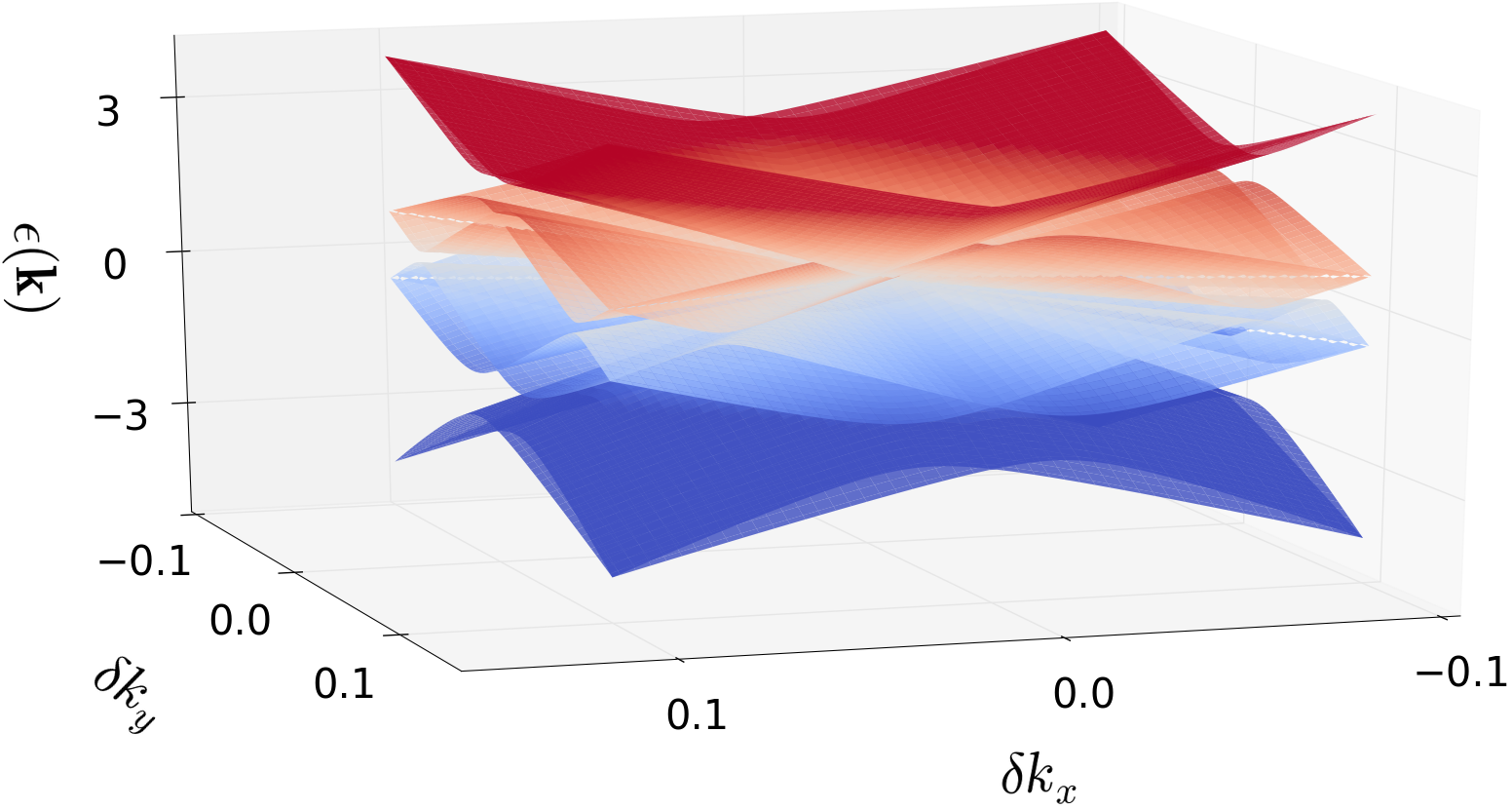}%
}
\caption{Energy dispersion near an eight-fold degeneracy in (a) SGs 130 and 135, (b) SGs 222, 223 and 230, and (c) SGs 218 and 220. (a) and (b) show pairwise degeneracy due to inversion symmetry. In addition, in (a), two degenerate bands form four-fold degenerate line nodes along the edges of the BZ. In (c) the eight-fold degeneracy splits into four non-degenerate and two doubly degenerate pairs of bands along the high symmetry $|\delta k_x|=|\delta k_y|=|\delta k_z|$ lines.}
\label{fig:kdotp8}
\end{figure*}

\paragraph{Low energy effective models}
For each of the band crossings in Table \ref{table1}, we compute a low-energy expansion of the most general Hamiltonian consistent with the symmetries of the little group near the degeneracy point, $\mathbf{k}_0$, in terms of $\delta\mathbf{k}\equiv\mathbf{k}-\mathbf{k}_0$.
%We do this by writing down the most general matrix function of $\delta\mathbf{k}\equiv\mathbf{k}-\mathbf{k}_0$ that is invariant under all the symmetries of the little group of $\mathbf{k}_0$, taking into account how these symmetries act on $\delta\mathbf{k}$. 
Full details of the constructions are in the Supplementary Material. 
Representative plots of the band dispersion along high symmetry lines are shown in Figs. \ref{fig:kdotp3}--\ref{fig:kdotp8}, where inessential higher-order terms have been added for the sake of clarity. 
%\footnote{While in the discussion below we retain only terms of the lowest order needed for a topological analysis, In these plots we show the spectrum with enough powers of $\delta\mathbf{k}$ retained to remove all non-essential degeneracies. We comment on this in cases where it is relevant.}

We begin by analyzing the threefold degeneracy points. The linearized $\mathbf{k}\cdot\mathbf{p}$ Hamiltonian for SGs 199 and 214 take the form
\begin{equation}
H_{199}(\phi,\delta\mathbf{k})= \left(\begin{array}{ccc}
0 & e^{i\phi}\delta k_x & e^{-i\phi}\delta k_y \\
e^{-i\phi}\delta k_x & 0 & e^{i\phi}\delta k_z \\
e^{i\phi}\delta k_y & e^{-i\phi}\delta k_z & 0
\end{array}\right),
\end{equation}
where $\phi$ is a real parameter; without loss of generality we set the zero of energy at zero throughout and omit an overall energy scale.
The bands are non-degenerate away from $\delta \mathbf{k}=0$, unless $\phi=n\pi/3$ for integer $n$, in which case bands become degenerate along the lines $|\delta k_x|=|\delta k_y|=|\delta k_z|$. While the locations of these degeneracies in $(\delta\mathbf{k},\phi)$ change in the presence of higher order terms, they identify two topologically distinct phases. First, for $\pi/3<\phi<2\pi/3$, the $\delta\mathbf{k}\neq 0$ Hamiltonian is adiabatically connected to the one with $\phi=\pi/2$ for sufficiently small $|\delta\mathbf{k}|>0$. At this value of $\phi$, the Hamiltonian takes the form
\begin{equation}
H_{199}(\pi/2,\delta\mathbf{k})=\delta\mathbf{k}\cdot\mathbf{S},
\end{equation}
where the matrices $S_i$ are the generators of the rotation group $SO(3)$ in the spin-1 representation. This shows that our three-fold fermion is a \emph{fermionic} spin-1 generalization of an ordinary Weyl fermion. The translation phases of the non-symmorphic little-group symmetries have effectively converted a half-integer spin representation into an integer-spin representation.
 The three bands $\psi_\pm,\psi_0$ of the Hamiltonian have energies $\epsilon_{\pm}=\pm|\delta\mathbf{k}|,\epsilon_0=0$. Furthermore, the Chern numbers of each of these bands evaluated over any closed surface enclosing the degeneracy point are $\nu_\pm=\mp 2$ and $\nu_0=0$. These Berry fluxes characterize the entire phase $\pi/3<\phi<2\pi/3$. 

%We perform a similar analysis for the phase $0<\phi<\pi/3$ by exactly computing the Berry curvature when $\phi = \pi/6$ and find the bands of positive and negative flux flipped in order in energy; the same is true when $2\pi/3<\phi<\pi$. The remainder of the phase diagram is identical to that with $0<\phi<\pi$ after the replacement $H_{199}\rightarrow-H_{199}$. 
At $\phi=n\pi/3$, the $\nu=0$ band becomes degenerate with \emph{both} the bands $\psi_\pm$ at different points in momentum space; these degeneracies transport Berry curvature between $\psi_+$ and $\psi_-$. { The formation of line nodes at the transition is an artifact of linearization:} when higher order terms are included in the Hamiltonian, the line nodes break up into sets of four single Weyl nodes, which carry Berry curvature away from the degeneracy point. The properties of all the phases for the other values of $\phi$ can be derived from those for $\pi/3<\phi<2\pi/3$; all regions feature bands with Chern number $\pm 2$ (see Supplementary Material). 
Thus, this three-band crossing has the topological character of a double Weyl point\cite{Fang12}, but the dispersion of a single Weyl point; this behavior is facilitated by the trivial ($\nu = 0$) band passing through the gapless point.
The energy spectrum is pictured in Fig.~\ref{fig:kdotp199}.

Having identified a fermion with a spin-$1$ $\mathbf{k}\cdot\mathbf{S}$ Hamiltonian, it is natural to ask if there exist similar particles for higher values of angular momentum $j$. { Our classification of new fermions rules out the possibility of $j\geq 2$, since these would either have degeneracy greater than eight, which we have ruled out via an exhaustive search, or would have appeared on our list.} In the Supplementary Material, we present a full classification of $j=3/2$ fermions, which we find can be stabilized by \emph{either} { symmorphic or non-symmorphic symmetries;} we find seven space groups which can host this excitation. Three of these overlap with groups mentioned earlier: SGs 212 and 213 can host spin-$3/2$ fermions at the $\Gamma$ point and SG 214 can host a spin-$3/2$ fermion at the $\Gamma$ and $H$ points.

The $3$-fold degeneracy in SG $220$ is distinct from that in SGs 199 and 214. The linear-order $\mathbf{k}\cdot\mathbf{p}$ reads
\begin{equation}
H_{220}=H_{199}(0,(\delta k_y,\delta k_x,-\delta k_z))
% H_{220}=a_1\left(\begin{array}{ccc}
% 0 & \delta k_y & \delta k_x \\
% \delta k_y & 0 & -\delta k_z \\
% \delta k_x & -\delta k_z & 0
% \end{array}\right).
\end{equation}
This 3-fold degeneracy sits at a critical point in the phase diagram for SG 199 described in the previous paragraph.
Consequently, pairs of two bands are degenerate along the lines $|\delta k_x|=|\delta k_y|=|\delta k_z|$ (Fig.~\ref{fig:kdotp220}). { Mirror and $3$-fold rotation symmetry dictate that -- unlike for SGs 199 and 214 discussed above -- these line nodes persist to $\emph{all}$ orders in the $\mathbf{k}\cdot\mathbf{p}$ expansion, as proved in the Supplementary Material.} The line nodes are characterized by the holonomy of the wavefunction (i.e. Berry phase), around any loop encircling the line, given by $w=-1$.

Next, we consider the $6$-fold band degeneracies. We start with SGs 205, 206 and 230, in which TR symmetry, $\mathcal{T}$, times inversion, $I$, forces all bands to be two-fold degenerate, as shown in Figs~\ref{fig:kdotp205} and \ref{fig:kdotp206}.
In SGs $206$ and $230$, the $\mathbf{k}\cdot\mathbf{p}$ Hamiltonian can be written as
\begin{equation} H_{206} = H_{199}\oplus H_{199}^*
\end{equation}
%For space group $206$, the $\mathbf{k}\cdot\mathbf{p}$ Hamiltonian depends on two complex parameters $b=b_1+ib_2$ and $c=c_1+ic_2$. However, there exists a $\delta\mathbf{k}$-independent unitary transformation $U$ such that
%\begin{equation}
%U^\dag H_{206}U=\left.\left(H_{199}\oplus H_{199}^*\right)\right|_{a=b_1+i\sqrt{b_2^2+|c|^2}}
%\end{equation}
%The Hamiltonian for SG $230$ obeys an analogous relation, except for this space group $b_2=0$ identically by symmetry. 
Due to $\mathcal{T}I$, there is no 
%{$U(1)$ topological number (Berry flux) associated with these degeneracies. On the other hand, the eigenvalues of $SU(2)$ Wilson loop operators\cite{Aris1,Aris2,Hourglass} come in complex conjugate pairs, which wind twice (in opposite directions) as the Wilson loop is moved from the top to the bottom of a sphere encircling the degeneracy point.}
abelian Berry curvature (i.e. Chern number) associated with these degeneracies. Instead, we can consider the non-Abelian $SU(2)$-valued holonomy (Wilson loop\cite{Aris1,Aris-new}) of the wavefunction in each twofold degenerate pair of bands. Evaluated along $C_2$ symmetric loops, the eigenvalues of the $SU(2)$ holonomy matrix wind twice -- and in opposite directions -- around the unit circle as a function of position in the Brillouin zone (see Supplementary Material.). Note, that, as gauge-invariant quantities, these eigenvalues are in-principle measurable\cite{wilsonloopmsr,wilsonloopexp}, and hence their winding provides a meaningful topological classification.

Unlike the previous cases, SG 205 contains inversion symmetry in the little group of the $R$ point. This forces the effective Hamiltonian to be quadratic in $\delta \mathbf{k}$. However, it is still related to $H_{199}$ by,
\begin{align}
%H_{205}(\delta\mathbf{k})&=U(H_{199}(k')\oplus H_{199}^*(k'))U^\dag, \nonumber \\
H_{205}(\delta\mathbf{k})&=H_{199}(\delta k')\oplus H_{199}^*(\delta k')+F(\delta\mathbf{k})\oplus F(\delta\mathbf{k}), \nonumber \\
\end{align}
where $F(\delta\mathbf{k})$ is a diagonal matrix whose entries are $E_1\delta k_x^2+E_2\delta k_y^2+E_3\delta k_z^2$, and { all cyclic permutations of the $\delta k_i$'s.} Due to its quadratic coordinate dependence, $H_{205}(\delta\mathbf{k})$ has only bands of zero net Berry flux, and Wilson loop eigenvalues do not wind.

We conclude our analysis of the $3$- and $6$-fold fermions, with SGs $198$, $212$, and $213$. Unlike the other $6$-band systems, these lack inversion symmetry, and so host six bands with distinct energies. The linearized $\mathbf{k}\cdot\mathbf{p}$ Hamiltonians may be written as,
\begin{align}
H_{198}(\delta\mathbf{k})&=\left(\begin{array}{cc}
H_{199}(\phi,\delta\mathbf{k}) &  bH_{199}(0,\delta\mathbf{k}) \\
b^* H_{199}(0,\delta\mathbf{k}) & -H^*_{199}(\phi,\delta\mathbf{k})
\end{array}\right), \\
H_{212,213}(\delta\mathbf{k})&=\left(\begin{array}{cc}
H_{199}(\pi/2,\delta\mathbf{k'}) & b H_{199}(0,\delta\mathbf{k'}) \\
b^* H_{199}(0,\delta\mathbf{k'}) & -H^*_{199}(\pi/2,\delta\mathbf{k'})
\end{array}\right), \nonumber
\end{align}
where $\delta\mathbf{k}'=(\delta k_z,\delta k_x,-\delta k_y)$ and $b$ is an arbitrary parameter. The six eigenstates of these Hamiltonians have distinct energies except along the faces of the BZ, where the spectrum degenerates into pairs related by the composition of a non-symmorphic $C_2$ rotation and time reversal; this degeneracy is shown in Fig.~\ref{fig:kdotp198}. Since this symmetry is antiunitary and squares to $-1$, these degeneracies are stable to higher order terms in $\mathbf{k}\cdot\mathbf{p}$.

Next, we examine the $8$-fold fermions. In SGs $130$ and $135$, $\mathcal{T}I$ symmetry mandates doubly degenerate bands. Close to the $A$ point, the linearized $\mathbf{k}\cdot\mathbf{p}$ Hamiltonian reads\begin{align}
H_{130}&=H_{135}=\delta k_z(a\sigma_2\sigma_3\sigma_3+b\sigma_2\sigma_3\sigma_2+c\sigma_2\sigma_3\sigma_1)\\
&+\delta k_x(-d\sigma_1\sigma_3\sigma_0+e\sigma_1\sigma_2\sigma_3+f\sigma_1\sigma_2\sigma_2+g\sigma_1\sigma_2\sigma_1)\nonumber \\
&+\delta k_y(d\sigma_3\sigma_3\sigma_0+e\sigma_3\sigma_2\sigma_3+f\sigma_3\sigma_2\sigma_2+g\sigma_3\sigma_2\sigma_1)\nonumber
\end{align}
where $a,b,\dots,g$ are real-valued parameters. This Hamiltonian has \emph{fourfold} degenerate line nodes along lines $\delta k_i=\delta k_j=0$ with $i\neq j; i,j\in\{x,y,z\}$ which follow the BZ edges, as shown in Fig.~\ref{fig:kdotp130}. This is seen by noting that the matrices multiplying any given $\delta k_i$ are part of a Clifford algebra. These lines are generally protected by composites of time reversal and non-symmorphic mirror symmetry.
%(the linearized Hamiltonian also has another set of line nodes in the $|\delta k_x|=|\delta k_y|$ planes, but these are artifacts of the linearization). 
Due to $\mathcal{T}I$ symmetry, abelian Berry phase of these line nodes vanishes. However, they can be characterized by the two ($-1$) eigenvalues of the $SU(2)$ Wilson loop encircling them.

A similar story holds for SGs $222,223$ and $230$, with
\begin{align}
H_{222}&=H_{223}=\delta k_z(a\sigma_3\sigma_1\sigma_3+b\sigma_1\sigma_1\sigma_1+c\sigma_1\sigma_1\sigma_2)\nonumber \\
&-\delta k_x(\frac{a}{2}\sigma_1\sigma_1\sigma_3+\frac{a\sqrt{3}}{2}\sigma_1\sigma_2\sigma_0+b\sigma_3\sigma_1\sigma_1+c\sigma_3\sigma_1\sigma_2) \nonumber \\
&+\delta k_y(\frac{a}{2}\sigma_2\sigma_1\sigma_0-\frac{a\sqrt{3}}{2}\sigma_2\sigma_2\sigma_3+b\sigma_0\sigma_1\sigma_2-c\sigma_0\sigma_1\sigma_1),
\end{align}
and a similar expression for $H_{230}$ after a permutation of the $\delta k$'s. Besides the $\mathcal{T}I$ double degeneracy of all bands, there are no additional degeneracies, as shown in Fig.~\ref{fig:kdotp222}.

Finally, we examine the $8$-fold degeneracy in SGs $218$ and $220$. Because both of these cases lack $\mathcal{T}I$, they have eight non-degenerate bands away from the high-symmetry point. However, there is a degeneracy along high-symmetry lines emanating from it. Along lines $|\delta k_x|=|\delta k_y|=|\delta k_z|$, the $8$-fold degeneracy splits into four singly degenerate bands and two pairs of doubly degenerate bands. In addition, along lines where two of the $\delta k_i$ are zero, and along lines where $\delta k_i=\delta k_j, \delta k_k=0$, there are four pairs of doubly degenerate bands. Unlike SGs $198,212$ and $213$ above, however, there are no additional degeneracies along high-symmetry planes. The spectrum is shown in Fig.~\ref{fig:kdotp218}. { The $\mathbf{k}\cdot\mathbf{p}$ Hamiltonian is given in the Supplementary Material.}

\begin{figure}
\includegraphics[width=.4\textwidth]{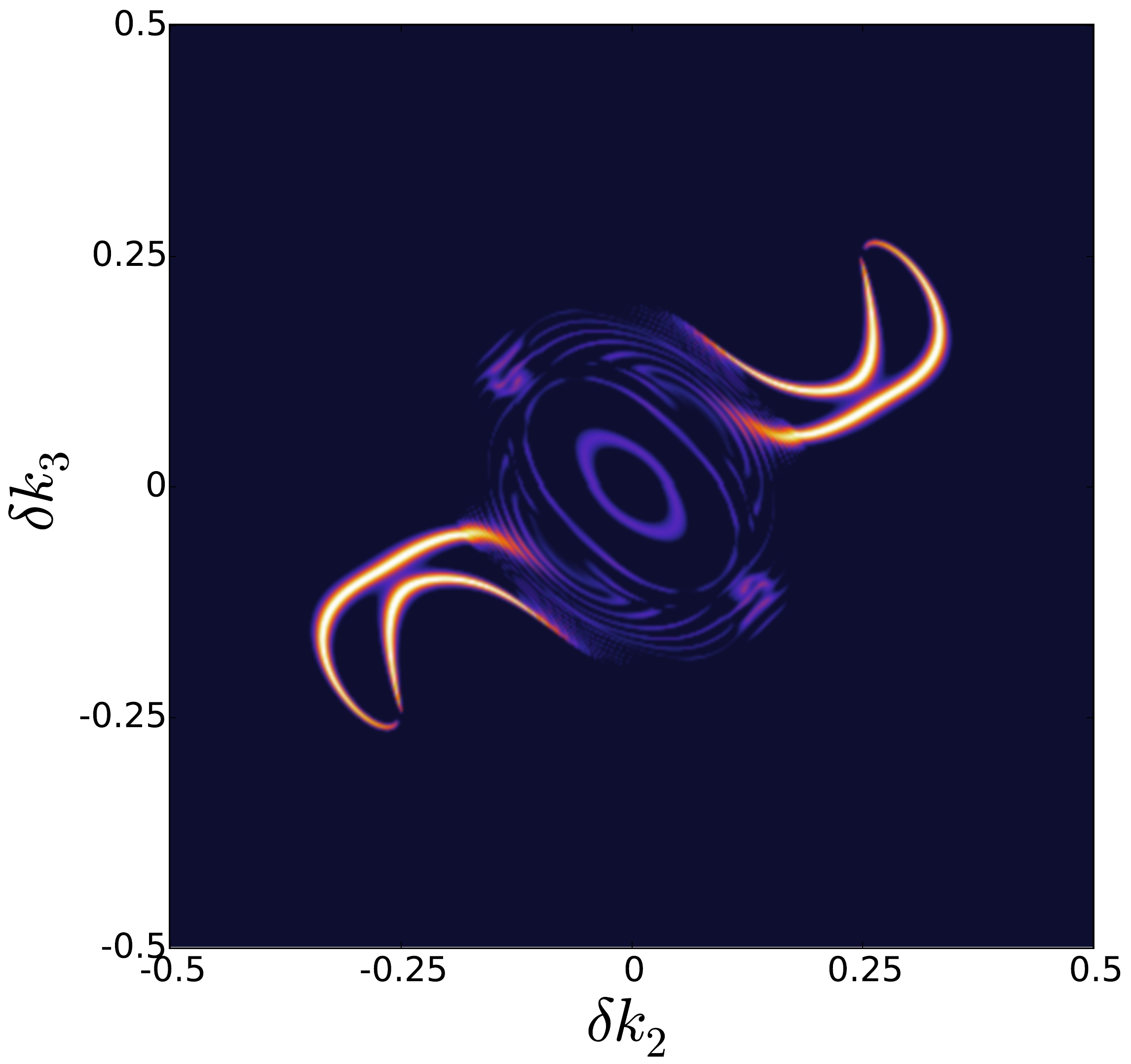}
\begin{picture}(0,0)
\put(-175,143){\includegraphics[height=1.5cm]{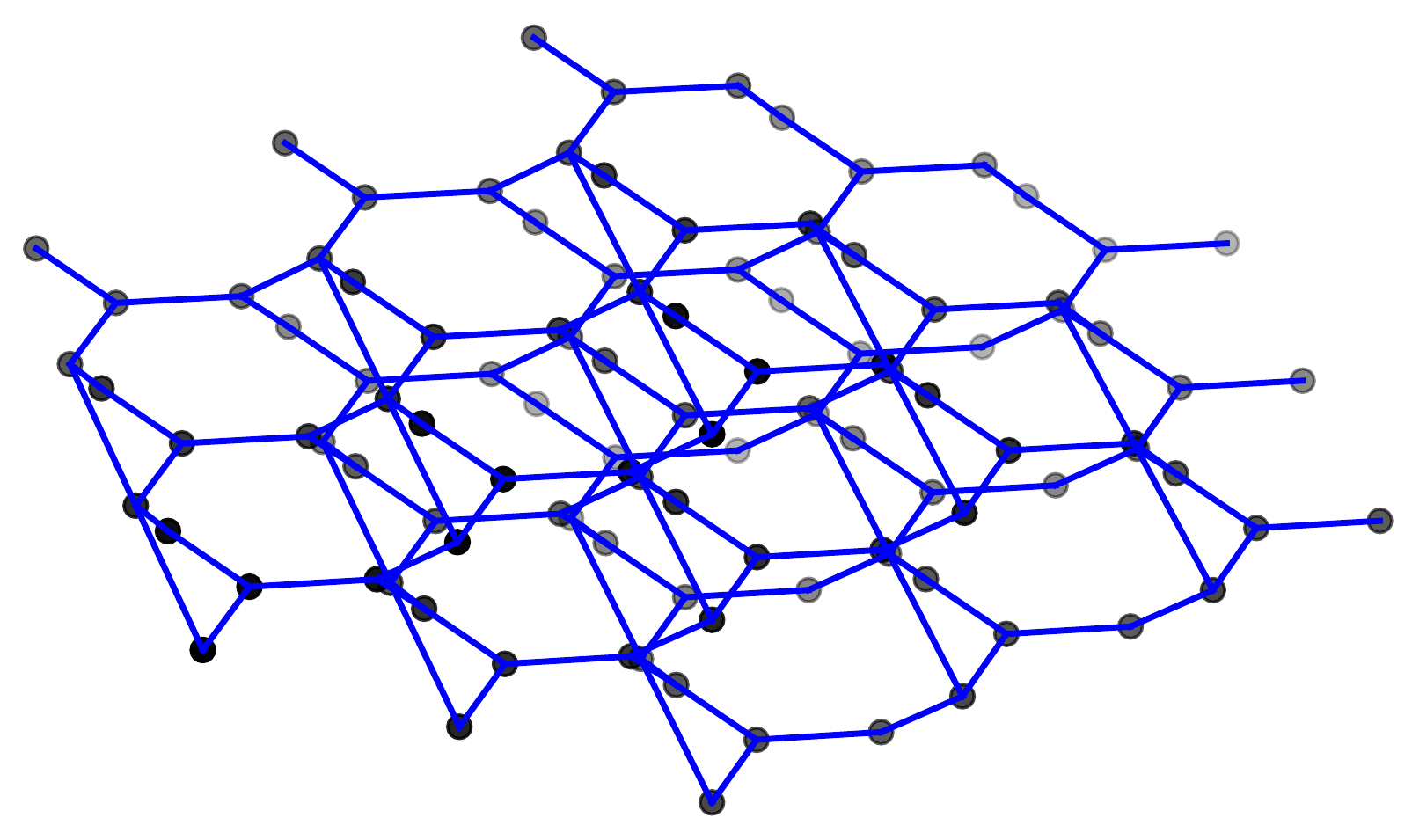}}
\end{picture}
\caption{Tight binding surface states for SG 214, showing the surface density of states for a surface in the $\bar{1}{1}{1}$ direction, calculated using the open-source PythTB package\cite{PythTB}. The $x$ and $y$ axes correspond to multiples of the reciprocal lattice vectors $g_2=2\pi(1,0,1)$ and $g_3=2\pi(1,1,0)$ respectively. There are Fermi arcs emanating from the points $\pm(0.25,0.25)$ which correspond to the surface projection of the $P$ and $-P$ points. Inset shows the atoms in nine unit cells with lines to indicate the nonzero hopping amplitudes; each unit cell consists of four atoms with three $p$ orbitals per atom. Only $p$ orbitals with intersite spin-orbit coupling are included. }\label{fig:arcs}
\end{figure}

\paragraph{Experimental signatures}
We now consider how to experimentally detect the topological character of the new fermions.
%The new fermions exhibit novel experimental signatures.
We start with the $3-$fold degeneracy in SGs 199 and 214. Because the degeneracy at the $P$ point carries net Berry flux $|\nu|=2$, the surface spectrum will host two Fermi arcs that emerge from the surface projection of the $P$ point\cite{Wan11},
similar to those that appear from double Weyl points\cite{Fang12}.
In the presence of TR, an additional $3-$fold degeneracy exists at the $-P$ point at the same energy; its surface projection will be the origin for two more Fermi arcs. { Furthermore, since the monopole charge is invariant under the action of time reversal, materials in this symmetry group will \emph{always} exhibit Fermi arcs. Whether these arcs are masked by other spurious Fermi pockets is a quantitative detail; the arcs will nevertheless always exist.} These four Fermi arcs must terminate on the surface projection of four Weyl points (or two double Weyl points), which \emph{must} exist elsewhere in the BZ. These can be identified with the Weyl points that drive the topological phase transition between Chern number $\nu=\pm 2$: at the phase transition, two Weyl fermions emerge from the threefold degeneracy to carry away the Fermi arcs of the $\nu=+2$ phase, while two other Weyl points emerge as the endpoints of the Fermi arcs for the $\nu=-2$ phase. 
We have verified this with a toy tight-binding model for SG 214. Fig.~\ref{fig:arcs} shows the surface density of states for a surface in the $\bar{1}11$ direction in the first surface BZ. Two pairs of Fermi arcs are visible, emanating from the surface projections of the $P$ and $-P$ points.
Breaking time-reversal symmetry with an external Zeeman field will split each threefold degeneracy into a number of Weyl points, as detailed in the Supplementary Material; generically these will be a mix of type-I and type-II Weyl fermions\cite{Soluyanov2015}. While in SG 199 this will destroy the exact degeneracy between the $P$ and $-P$ points, the degeneracy will persist in SG 214 for perturbations invariant under $C_{2,1\bar{1}0}$.

%can split the degeneracy between the Fermi arcs at the $P$ and $-P$ point in SG 199, but not in SG 214, where a $C_2$ rotation relates these points.

\begin{figure}
\includegraphics[width=.45\textwidth]{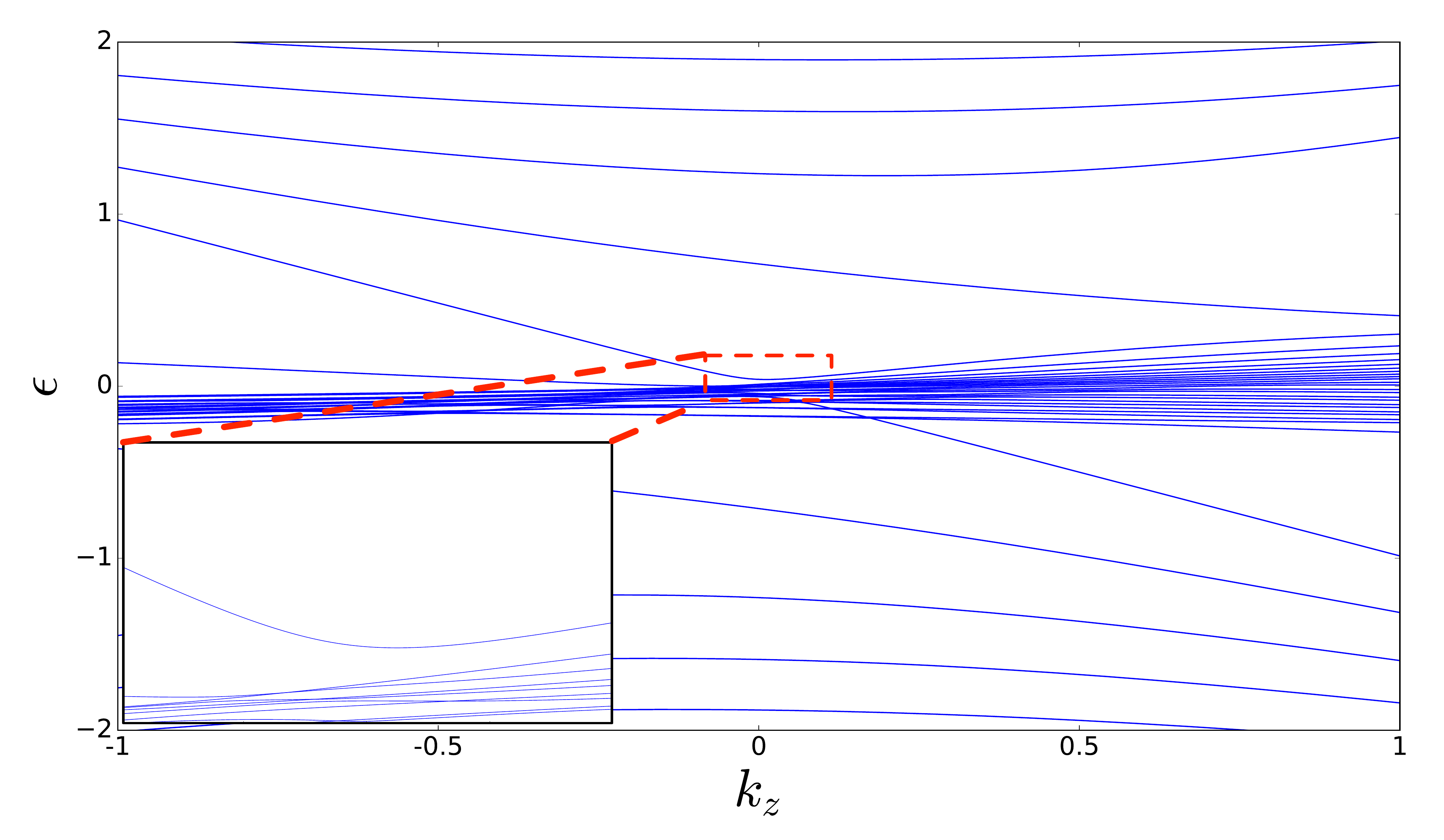}
\caption{Landau Level spectrum of the 3-fold degenerate fermion as a function of $\delta k_z$ with magnetic field along the $z$ direction. The two chiral modes can be clearly seen. The inset shows that in fact both chiral have a spectral flow that goes through $k_z\rightarrow\pm\infty$.}\label{fig:3dLLspec}
\end{figure}

In addition to Fermi arc surface states, the threefold fermions will exhibit anomalous negative magnetoresistance and a chiral anomaly distinct from that of either a single or double Weyl point.
%While a full analysis is forthcoming, 
For weak magnetic fields, semiclassical considerations\cite{Son2013} suggest that the magnetoresistance in SGs 199 and 214 match that of a double Weyl point\cite{Fang12}, although the density of states corresponds to a linear dispersion. At large magnetic fields, the Landau level spectrum for the 3-fold fermions displays two chiral modes; unlike the case of an ordinary (single or double) Weyl point, the spectral flow of these chiral modes does not pass through zero momentum, but instead flows to $k_z\rightarrow\pm\infty$. We show the numerically computed Landau level spectrum in Fig.~\ref{fig:3dLLspec}; the analytical derivation of the spectrum at the exactly solvable $\phi=\pi/2$ point can be found in the Supplementary Material.
%Due to the nonvanishing Berry curvature, the 3-fold fermions cannot be gapped by small crystal-symmetry breaking perturbations, but can be split into a pair of single Weyl points.

Since the rest of our new fermions do not host net Berry curvature, there is no guarantee of topologically protected surface states in the strictest sense. However, the non-trivial Berry phase associated with the line node in SG 220 implies the presence of a ``Fermi drum '' surface state\cite{BurkovHookBalents2011,Chan15}, although this will not be robust to breaking of the crystal symmetry in the bulk. Also note that, despite the name, these states need not be flat -- or even nearly flat -- in energy. Similarly, the non-abelian Berry phase associated to the Dirac lines in SG 130 and 135 suggests the existence of \emph{pairs} of drumhead states. { Finally, in the presence of an external magnetic field (or strain perturbation), these Dirac lines may be split to yield \emph{any} of the usual gapless topological phases: Weyl, Dirac, and line node semimetals. This opens up the possibility of tuning topological Dirac semimetals in these materials with a Zeeman field, similar to the recent progress made with field-created Weyl semimetals in half-Heusler materials\cite{Hirschberger16,Felser2016}. Finally, all of our new fermions are also detectable via ARPES (see, for example similar experimental results in Ref.~[\onlinecite{ZrZnSi}]) and through quantum oscillation experiments.}

\paragraph{Material realizations}

\begin{figure*}\centering
\subfloat[Ag$_3$Se$_2$Au (SG 214)\label{fig:dft214-2}]{
	\includegraphics[width=.3\textwidth]{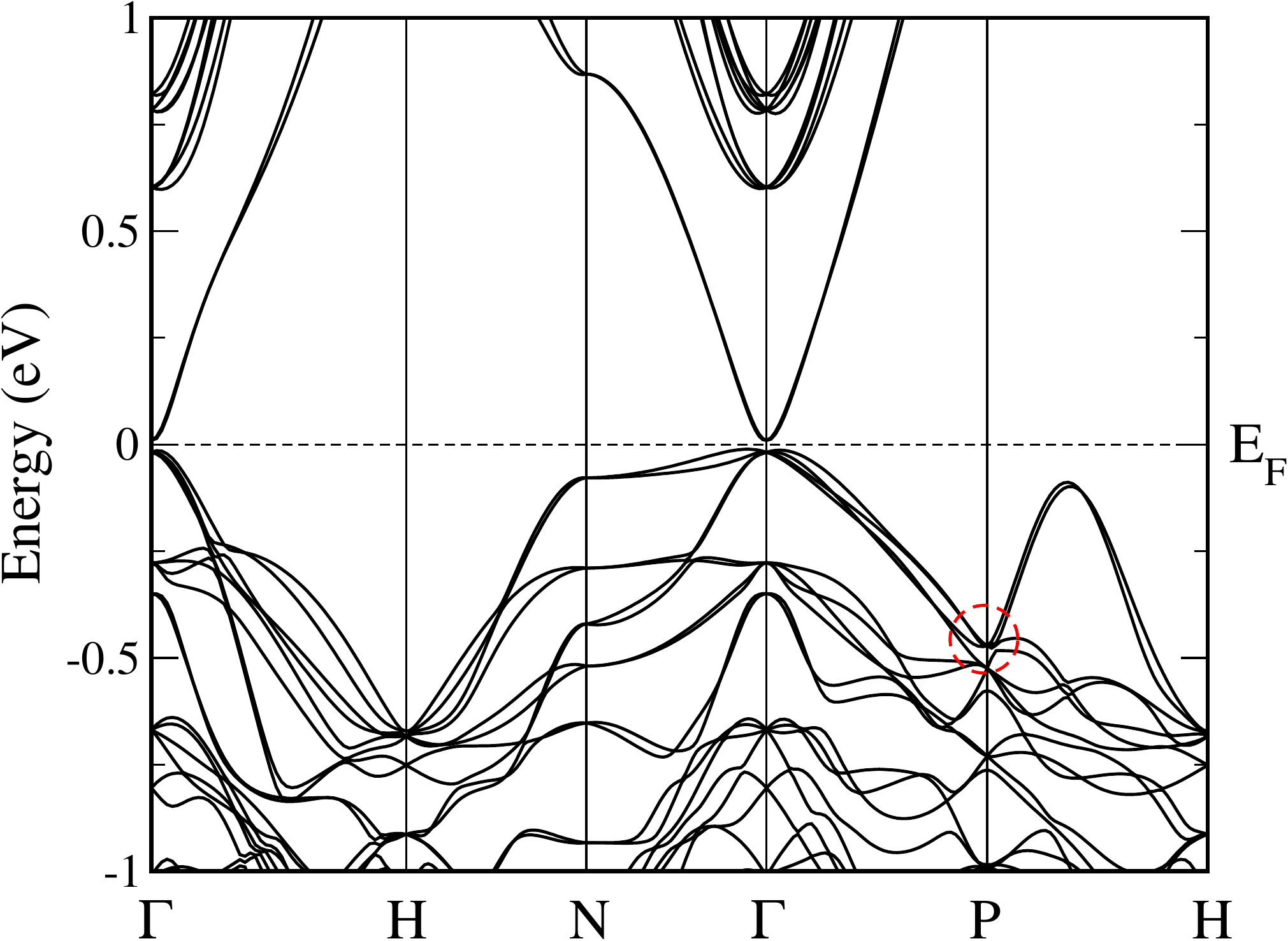}
	}\quad
\subfloat[Pd$_3$Bi$_2$S$_2$ (SG 199)\label{fig:dft1992}]{
	\includegraphics[width=.3\textwidth]{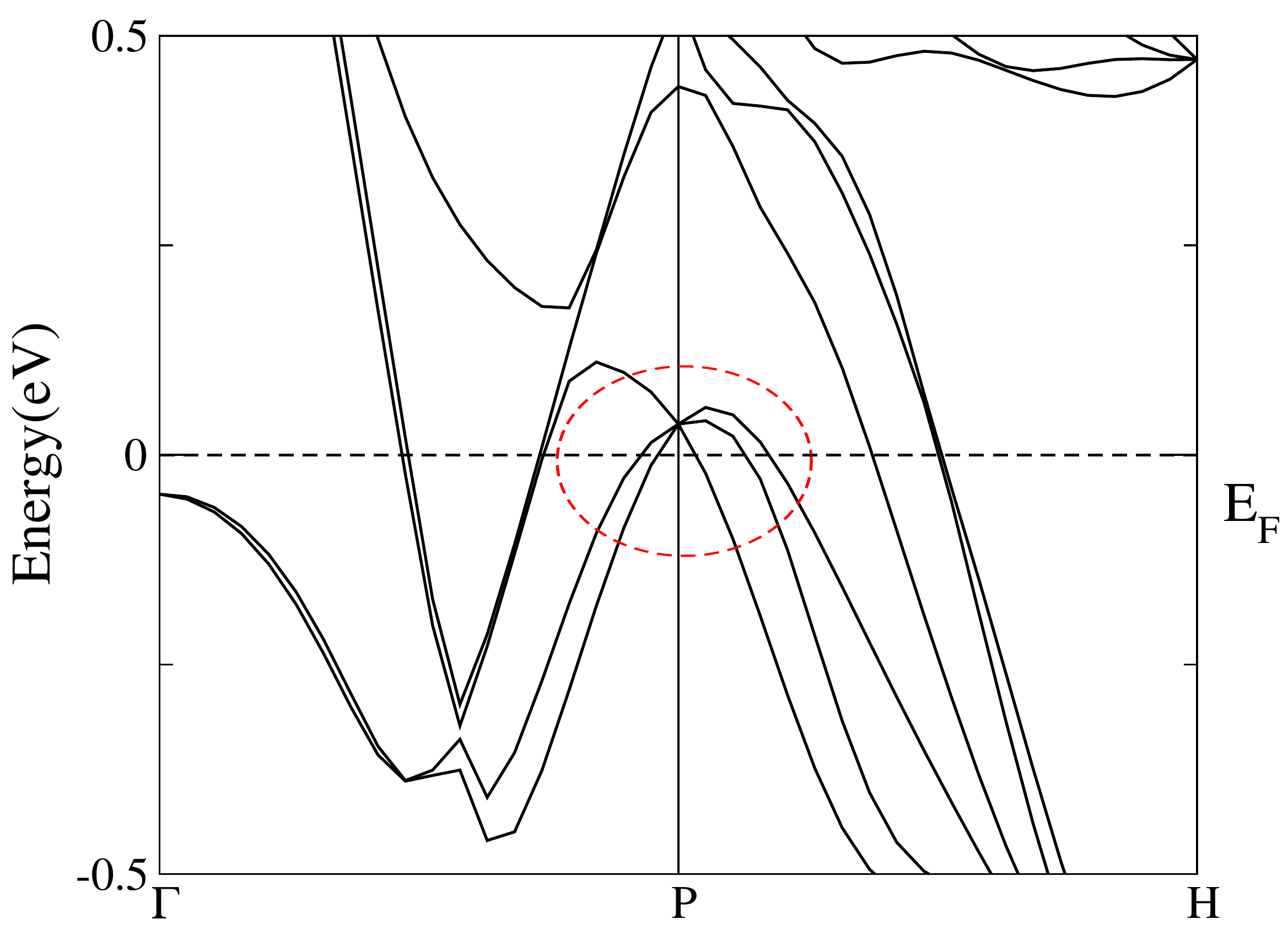}
}
\caption{Materials exhibiting 3-fold fermions near the Fermi level.}
\label{fig:dft3}
\end{figure*}

\begin{figure*}\centering
\subfloat[Ba$_4$Bi$_3$ (SG 220)\label{fig:dft220-1}]{%
  \includegraphics[width=.3\textwidth]{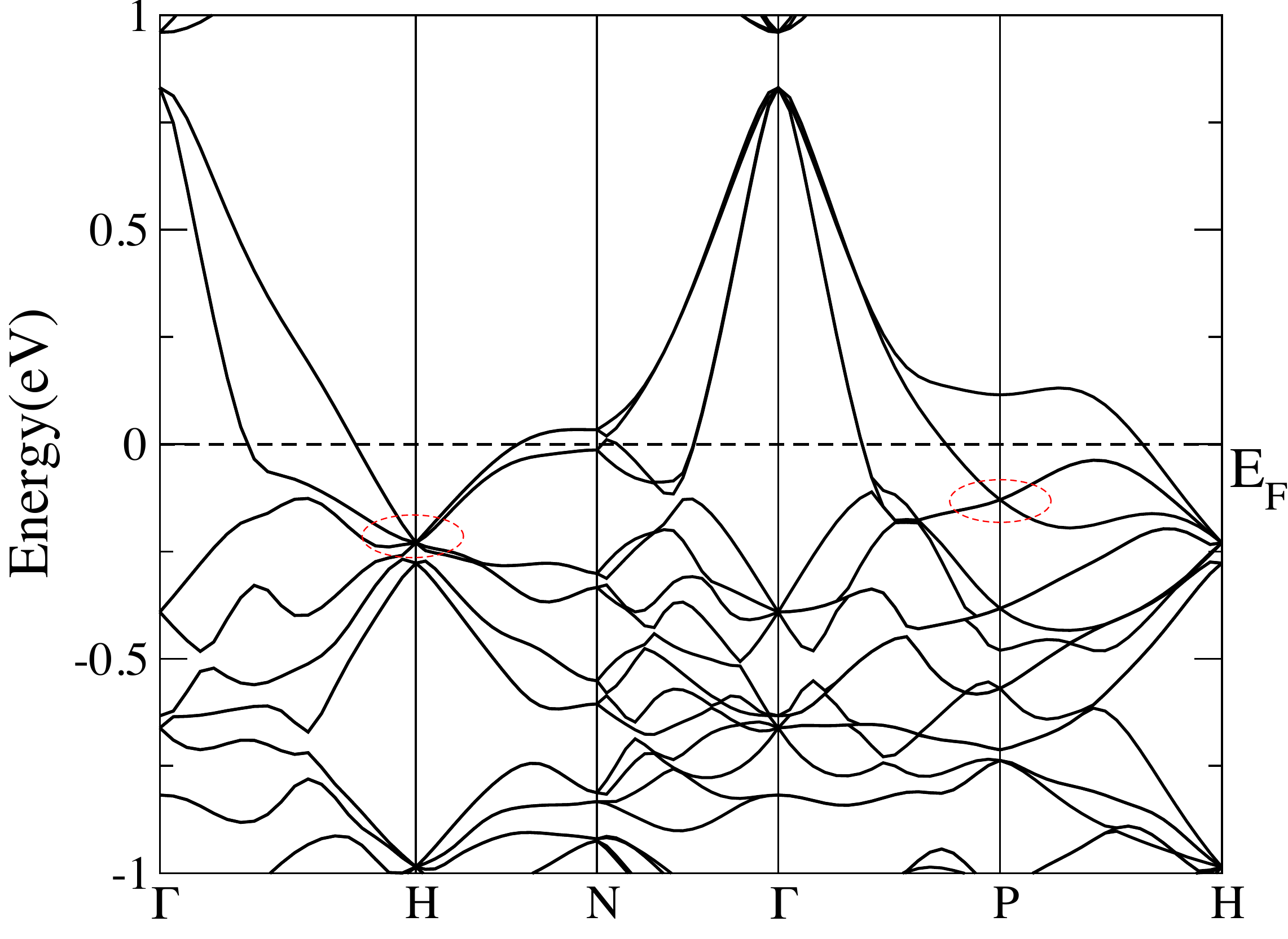}%
}\quad 
\subfloat[La$_4$Bi$_3$ (SG 220)\label{fig:dft220-2}]{%
  \includegraphics[width=.3\textwidth]{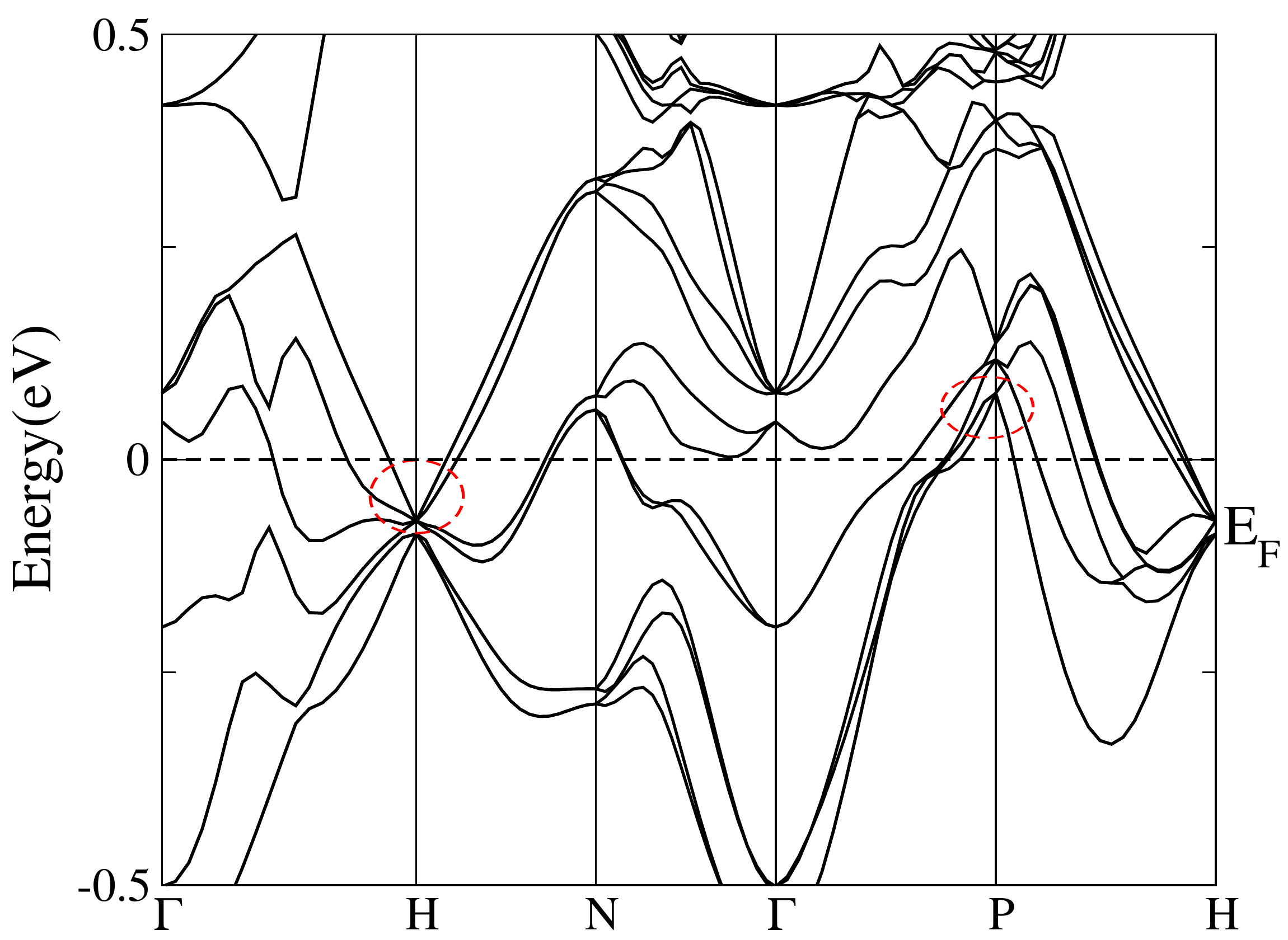}%
	}
\caption{Compounds in SG 220 display 3- and 8-fold fermions near the Fermi level at the $P$ and $H$ points, respectively.}
\label{fig:dft220}
\end{figure*}

\begin{figure*}
\subfloat[MgPt (SG 198)\label{fig:dft198-3}]{%
  \includegraphics[width=.3\textwidth]{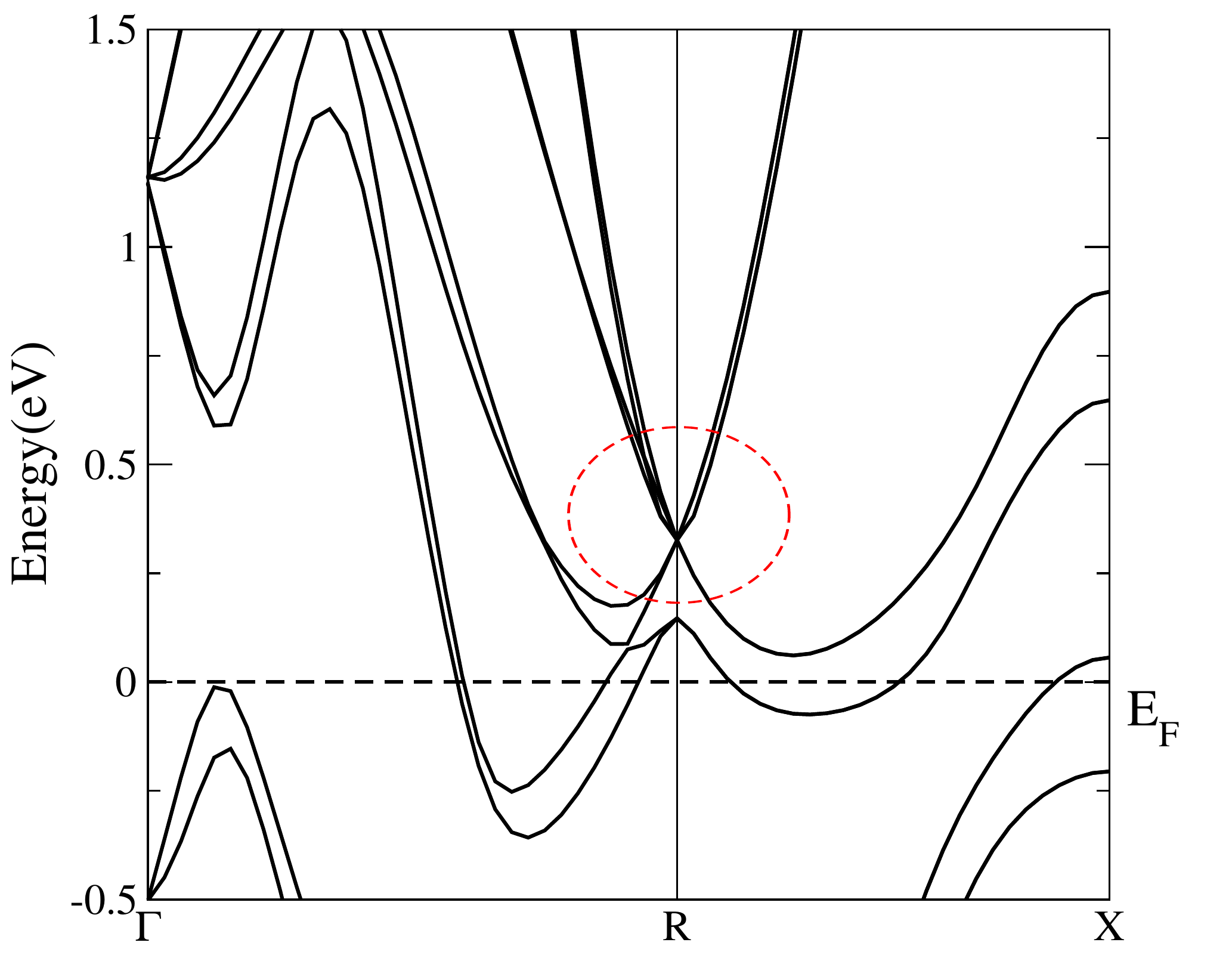}%
}\quad 
\subfloat[AsPdS (SG 198)\label{fig:dft198-1}]{%
  \includegraphics[width=.3\textwidth]{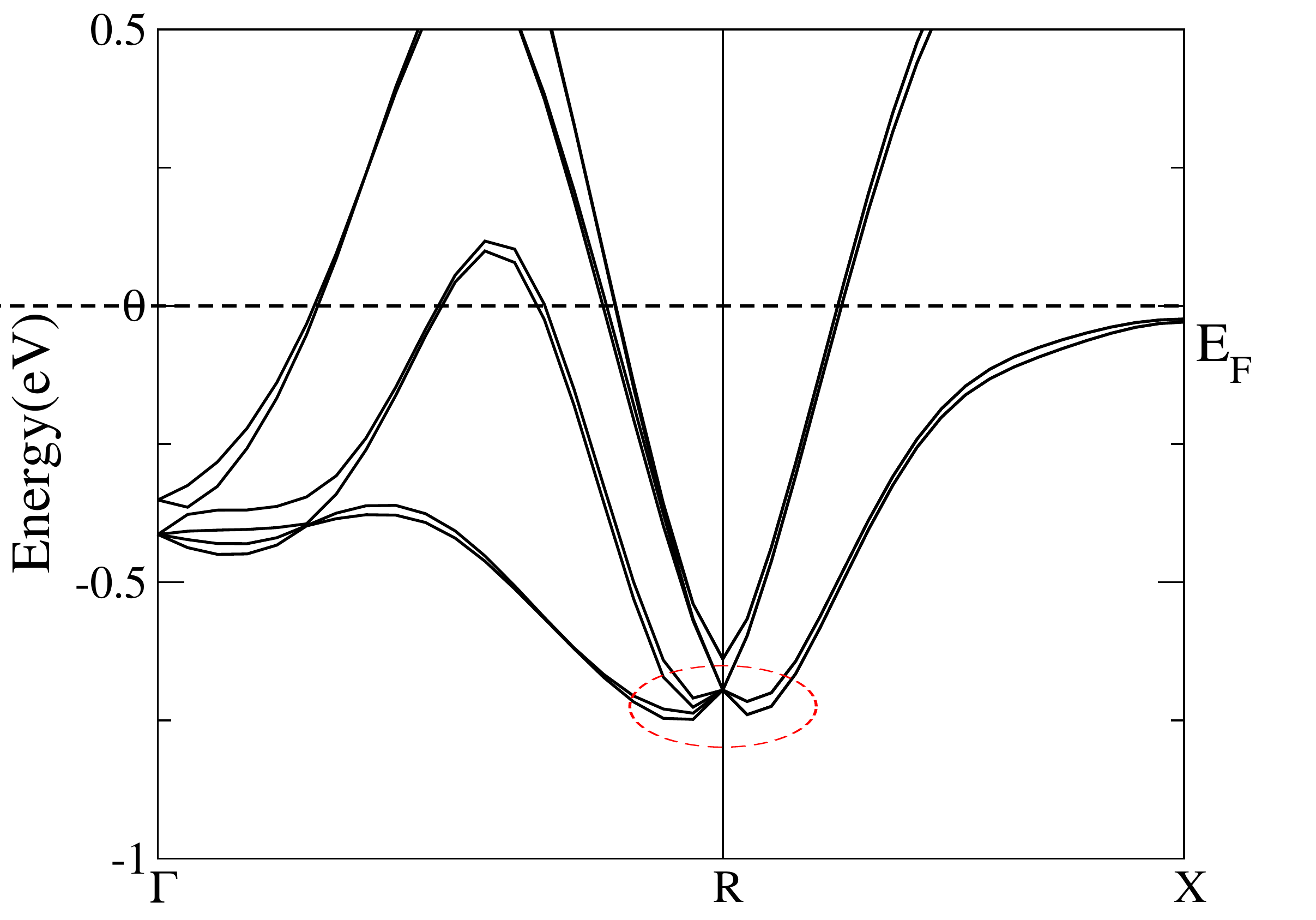}%
}\quad 
\subfloat[K$_3$BiTe$_3$ (SG 198)\label{fig:dft198-2}]{%
  \includegraphics[width=.3\textwidth]{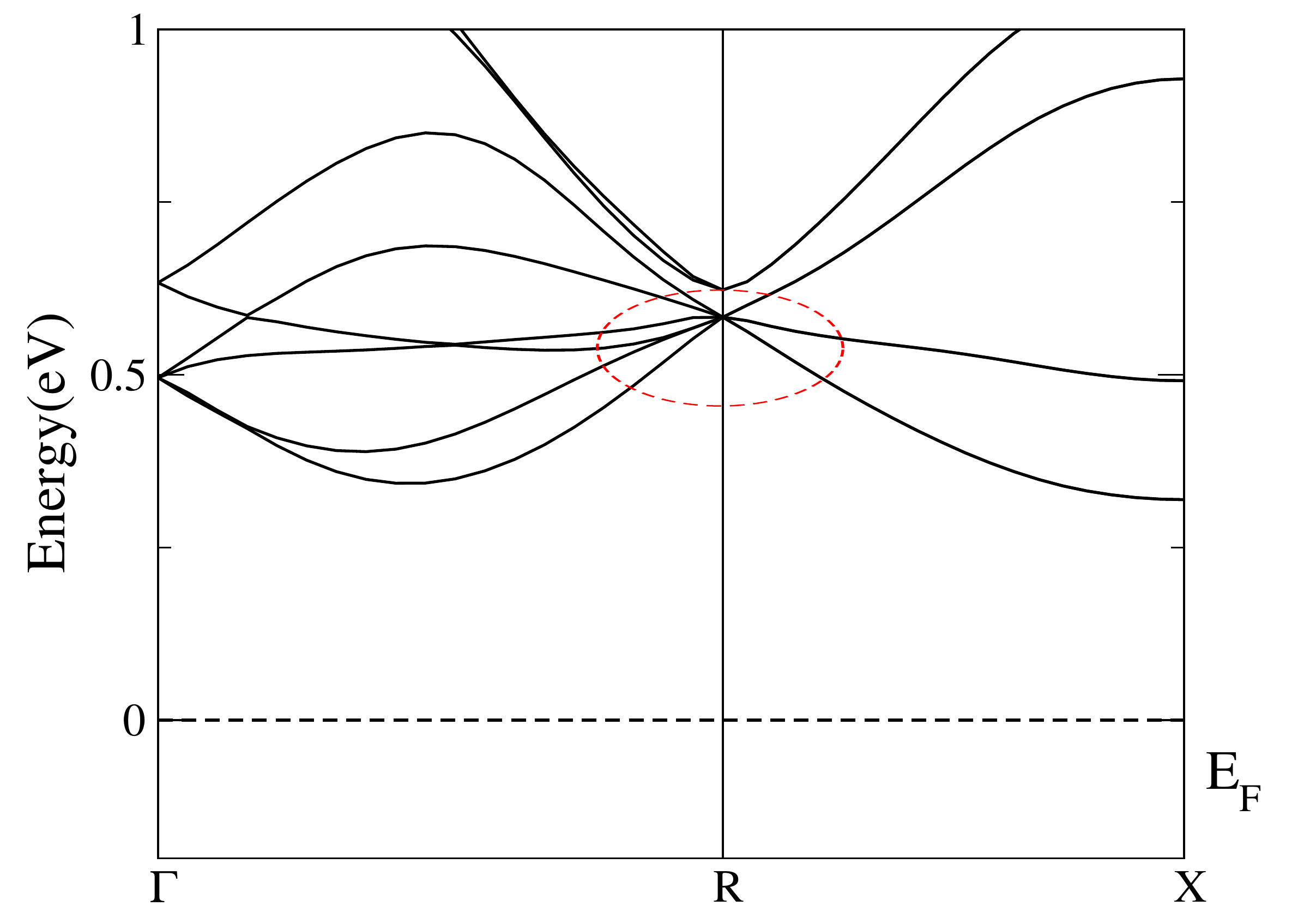}%
	}\quad
\subfloat[Li$_2$Pd$_3$B (SG 212)\label{fig:dft212}]{
	\includegraphics[width=.3\textwidth]{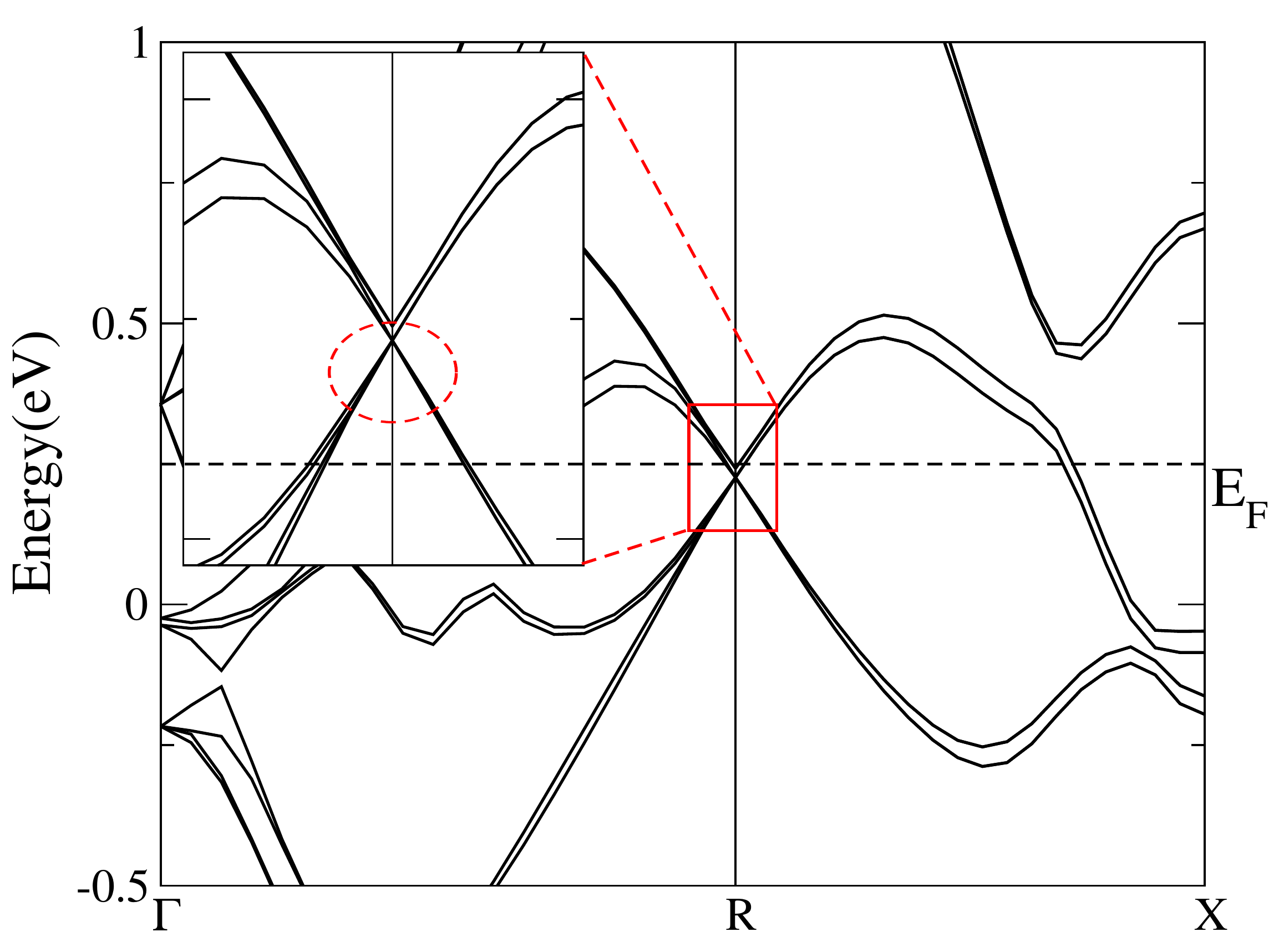}%%% this is actually SG 212
}\quad
\subfloat[Mg$_3$Ru$_2$ (SG 213)\label{fig:dft213-3}]{
	\includegraphics[width=.3\textwidth]{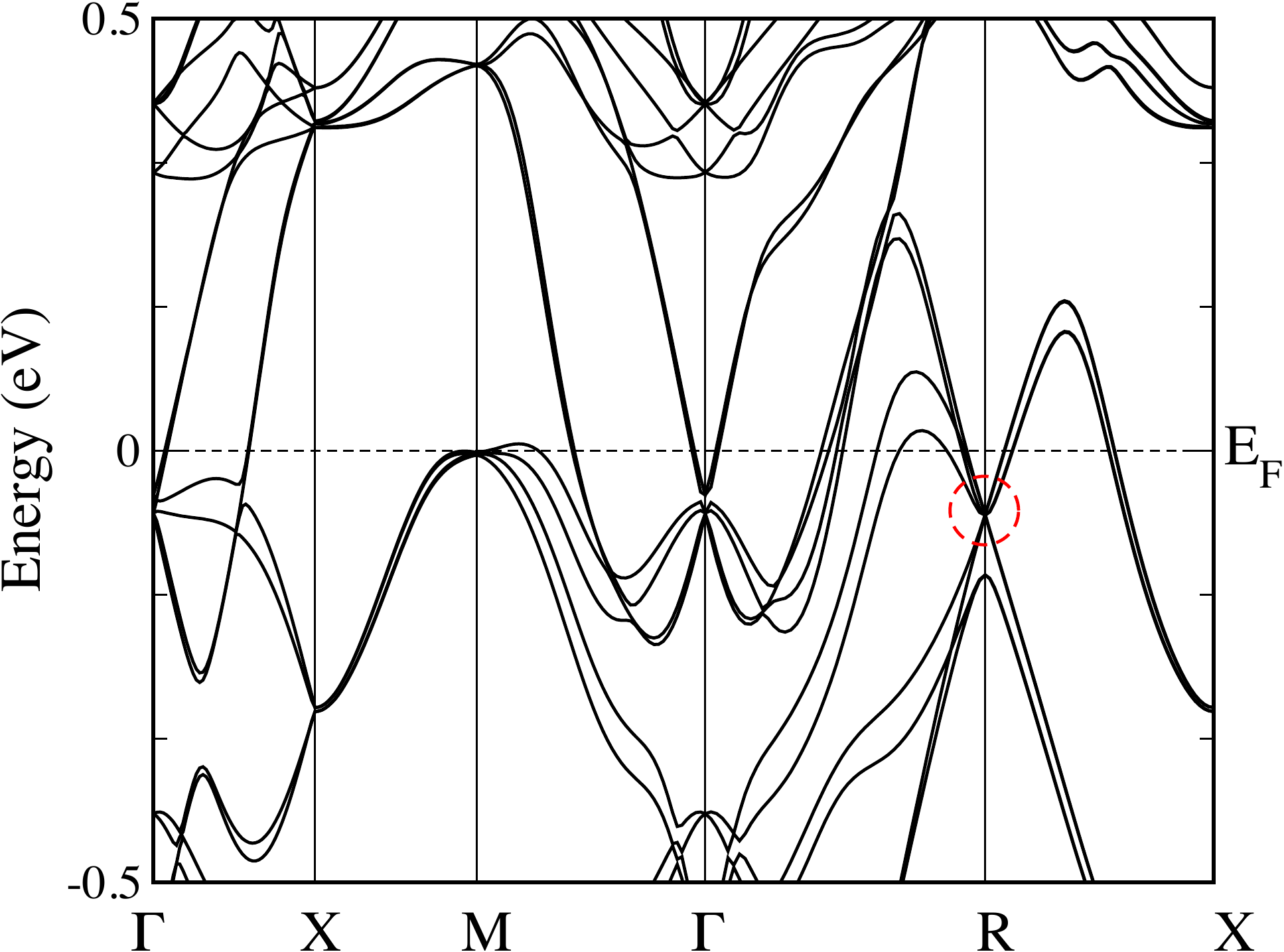}
}\quad
\subfloat[PdSb$_2$ (SG 205)\label{fig:dft205}]{%
  \includegraphics[width=.3\textwidth]{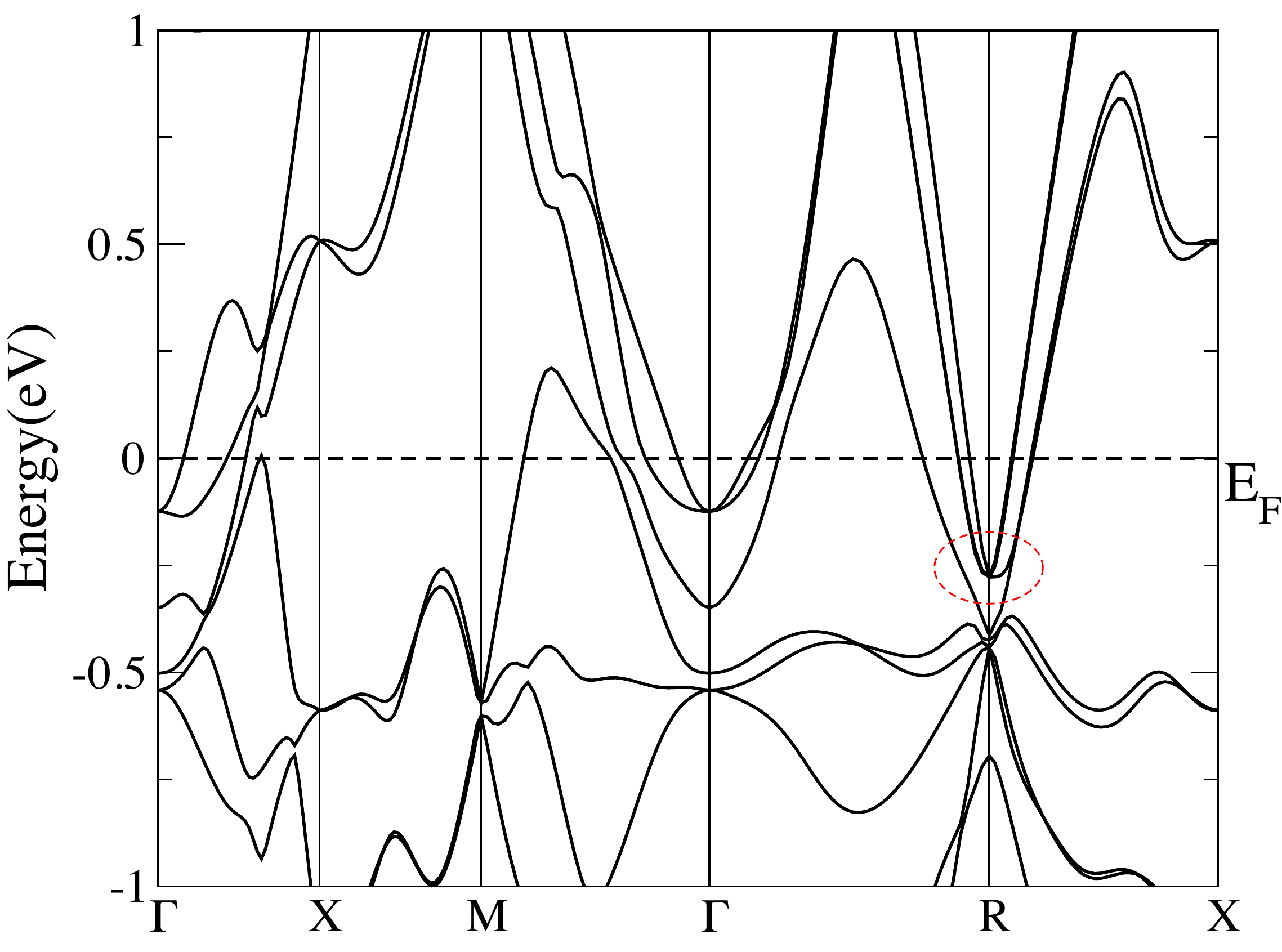}%
}
\caption{6-fold fermions at the $R$ point in (a,b,c) SG 198, (d) SG 212, (e) SG 213, and (f) SG 205.}
\label{fig:dft6}
\end{figure*}

\begin{figure*}
\subfloat[CuBi$_2$O$_4$ (SG 130)\label{fig:dft130}]{
	\includegraphics[width=.3\textwidth]{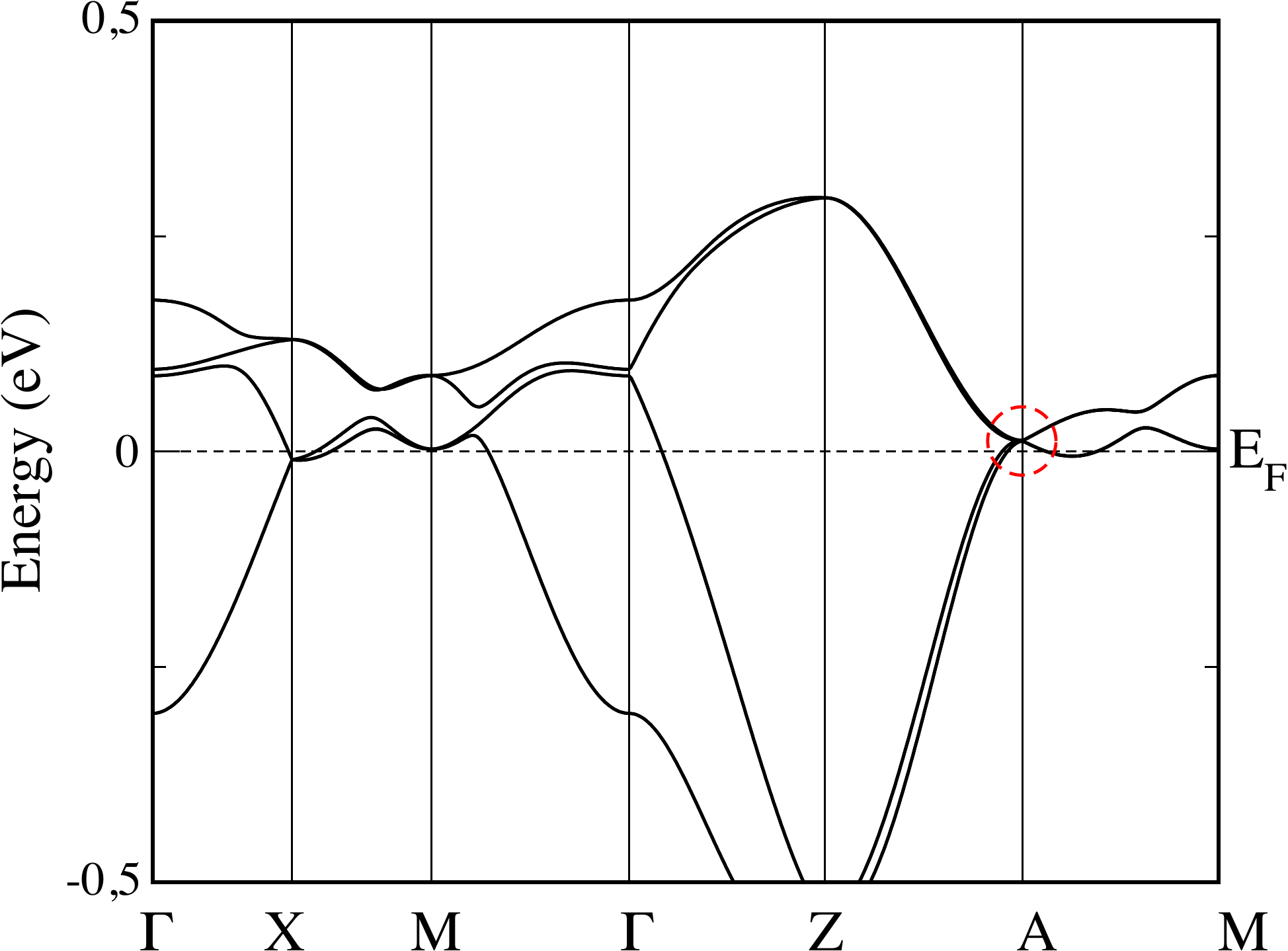}
}\quad
\subfloat[PdBi$_2$O$_4$ (SG 130)\label{fig:dft130-1}]{
	\includegraphics[width=.3\textwidth]{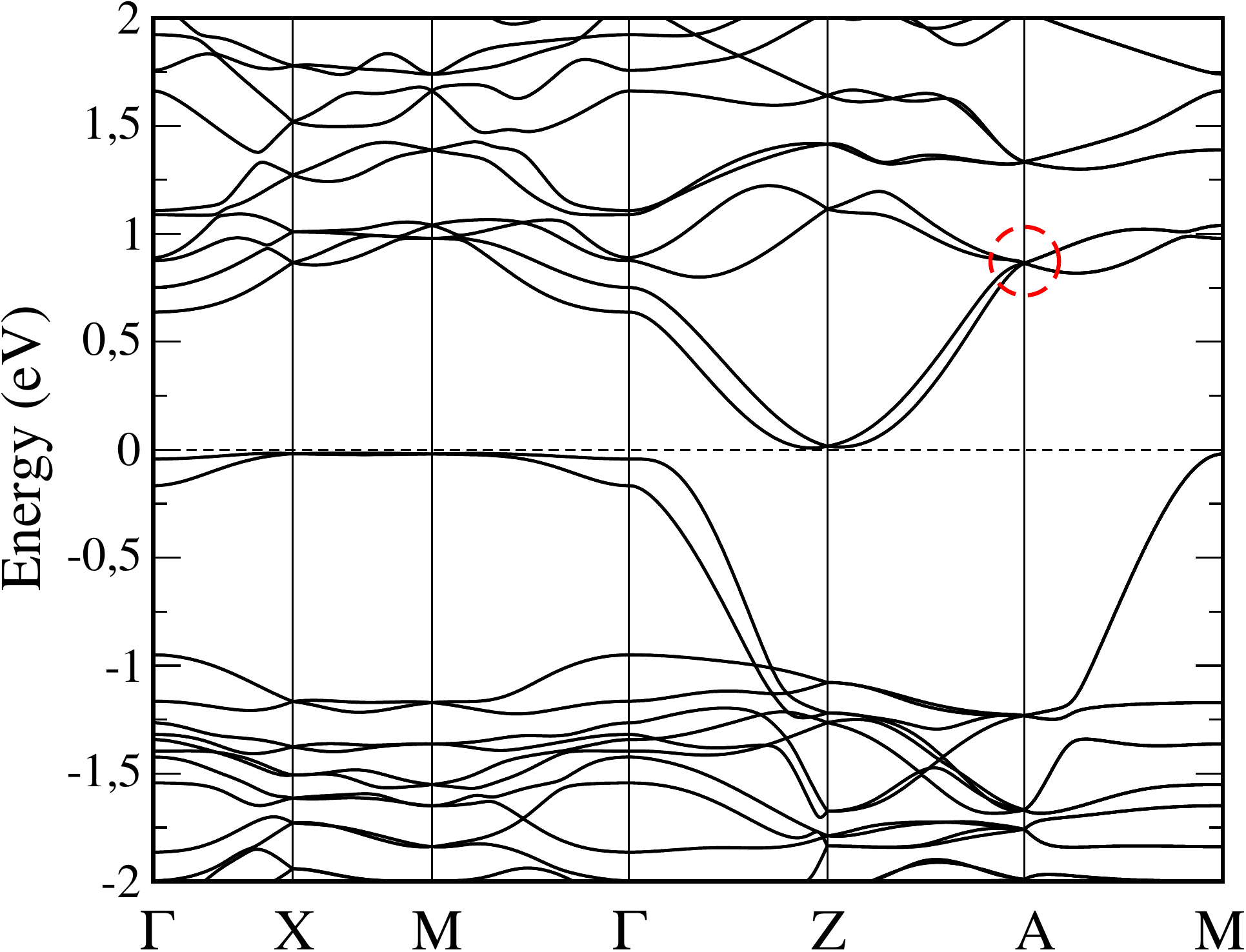}
}\quad
\subfloat[PdS (SG 135)\label{fig:dft135-2}]{%
  \includegraphics[width=.3\textwidth]{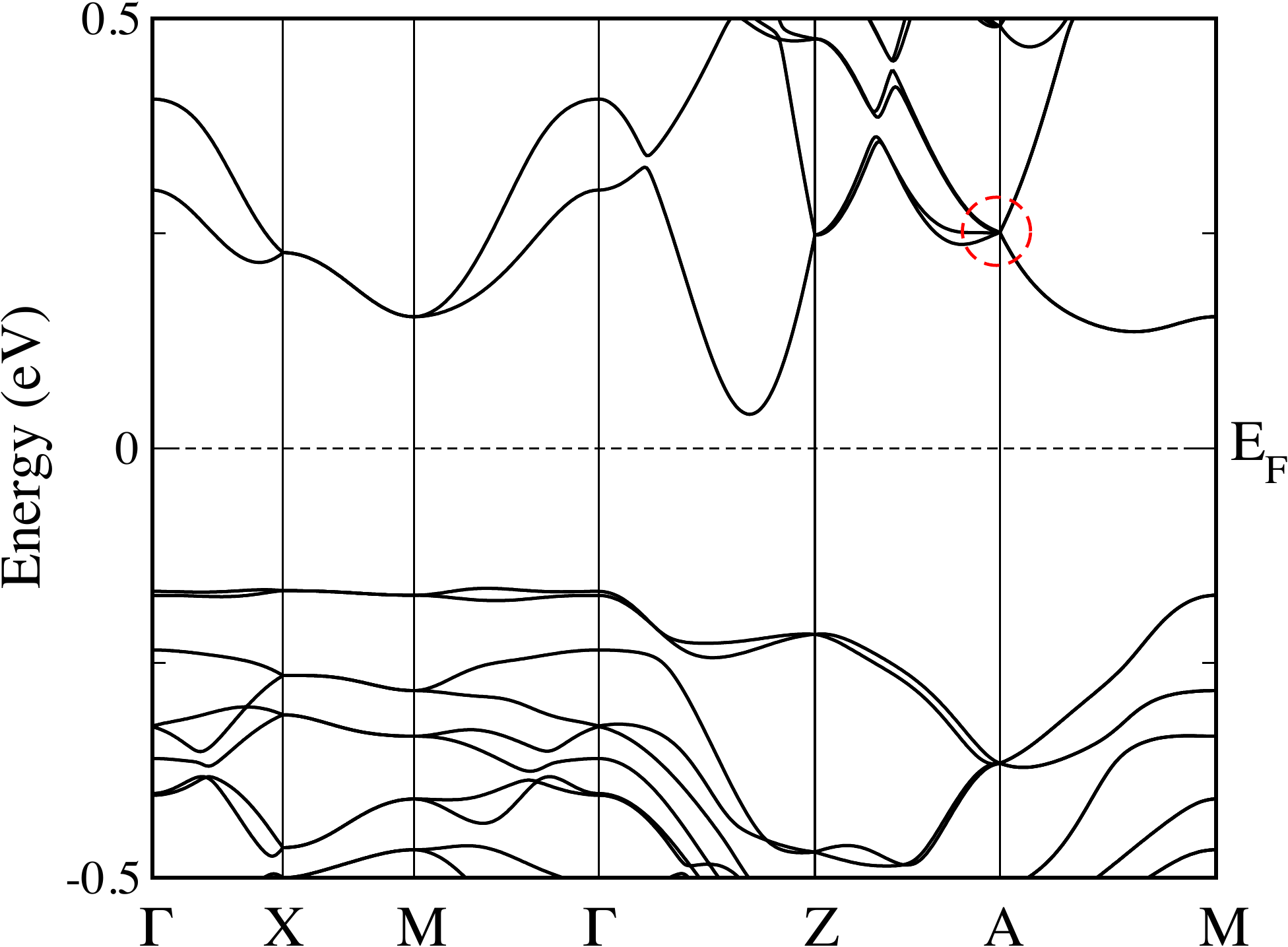}%
}\quad
\subfloat[CsSn (SG 218)\label{fig:dft218out}]{%
  \includegraphics[width=.3\textwidth]{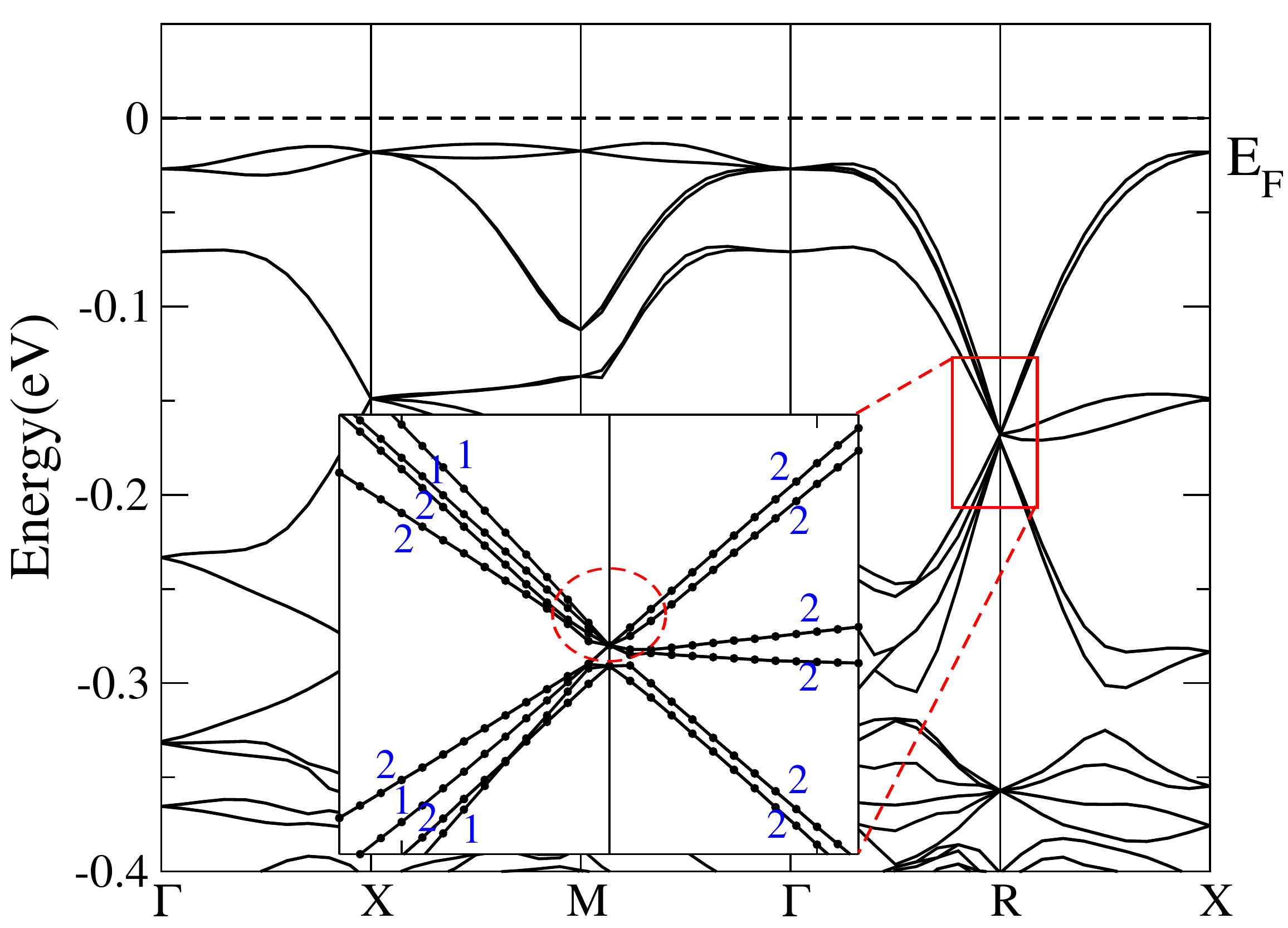}%
}%\quad 
%\subfloat[CsSn (SG 218) \label{fig:dft218in}]{%
  %\includegraphics[width=.3\textwidth]{218ABzoom.png}%}
\caption{8-fold fermions at the $A$ point in (a) CuBi$_2$O$_4$, (b) PdBi$_2$O$_4$, and (c) PdS, and at the $R$ point in (d) CsSn.}
\label{fig:dft8}
\end{figure*}

\begin{figure*}
\subfloat[Ta$_3$Sb (SG 223)\label{fig:dft223-1}]{%
  \includegraphics[width=.3\textwidth]{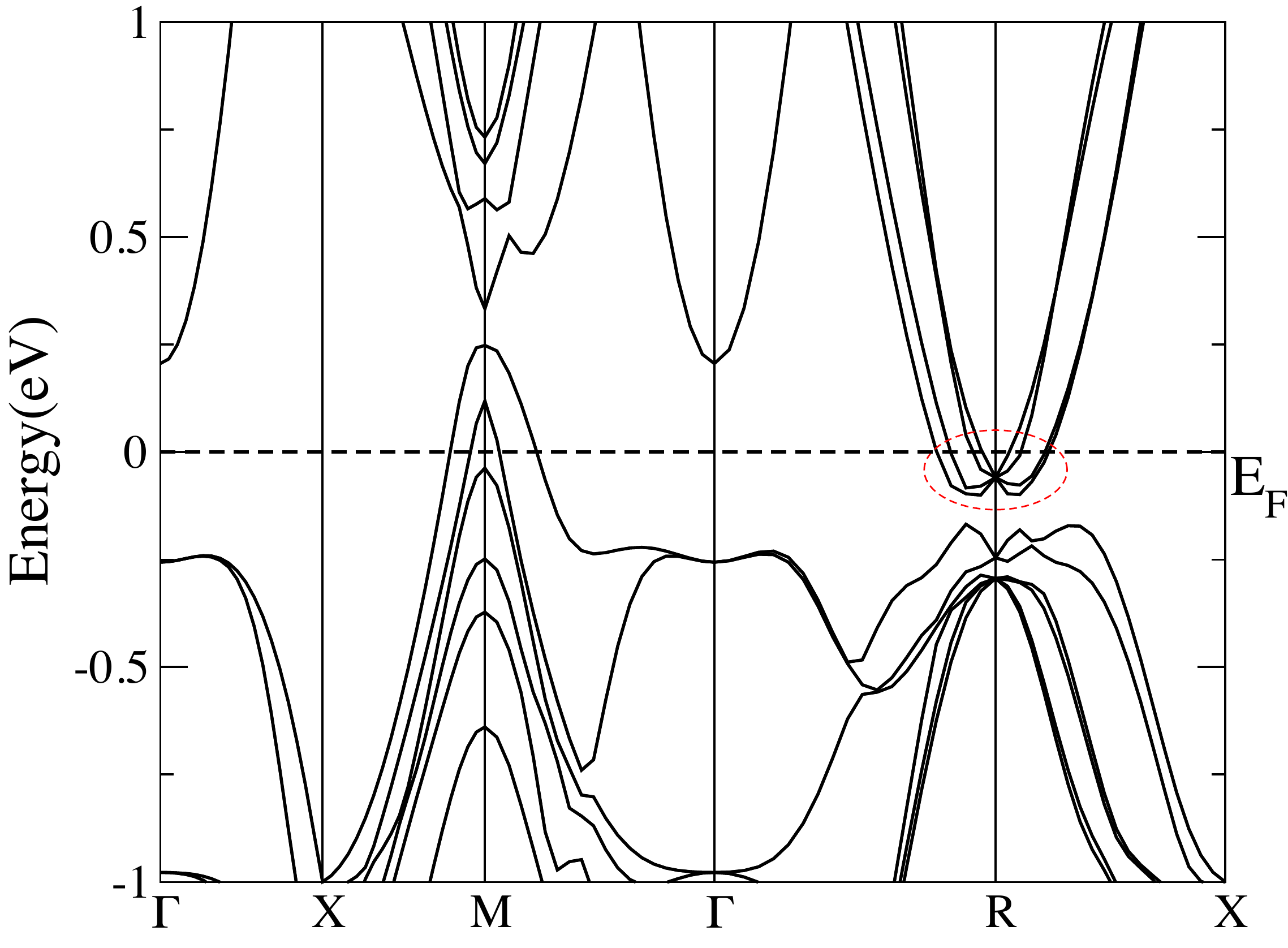}%
}\quad 
\subfloat[LaPd$_3$S$_4$ (SG 223)\label{fig:dft223-2}]{%
  \includegraphics[width=.3\textwidth]{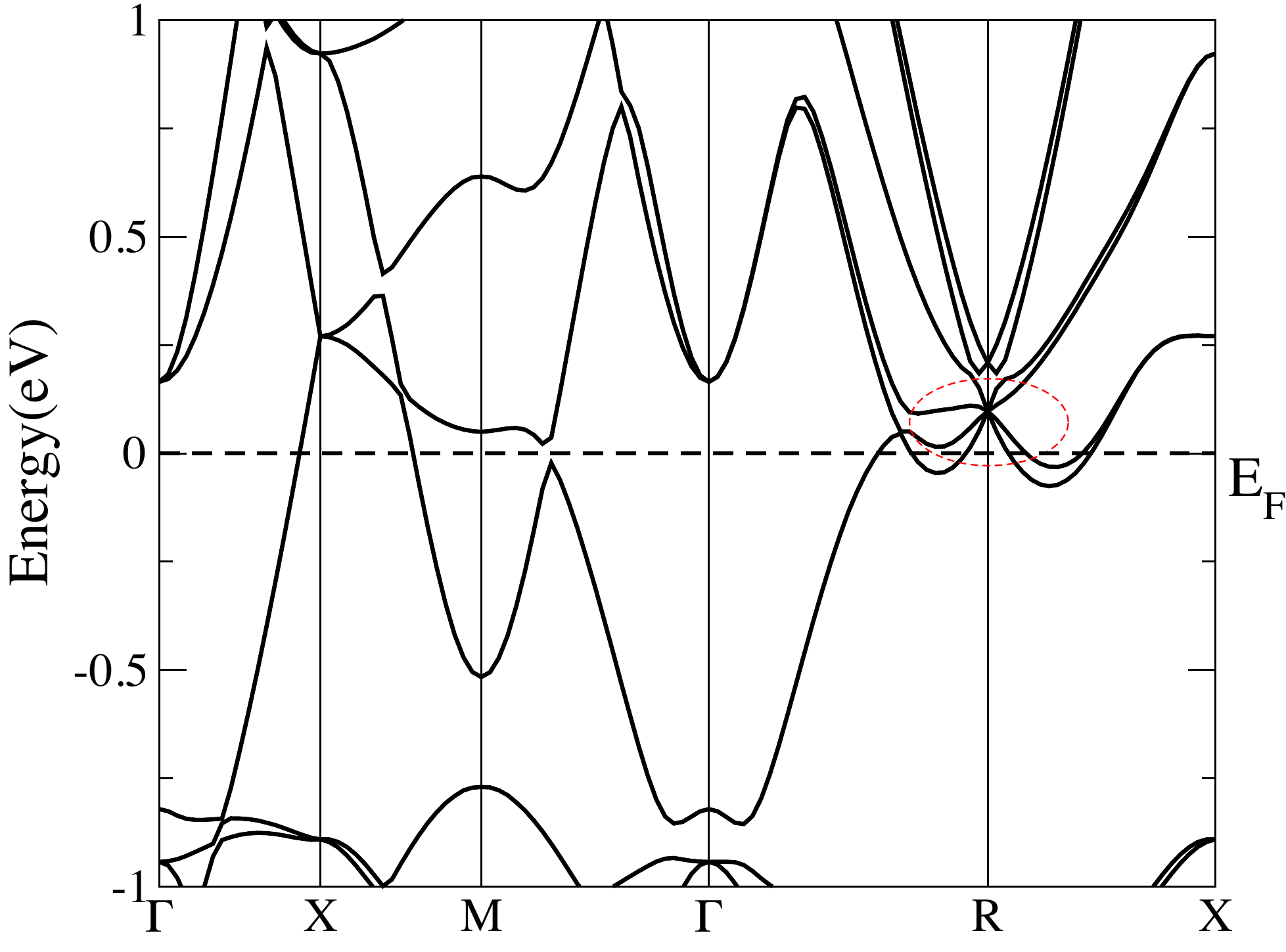}%
}\quad
\subfloat[Nb$_3$Bi (SG 223)\label{fig:dft223-3}]{%
  \includegraphics[width=.3\textwidth]{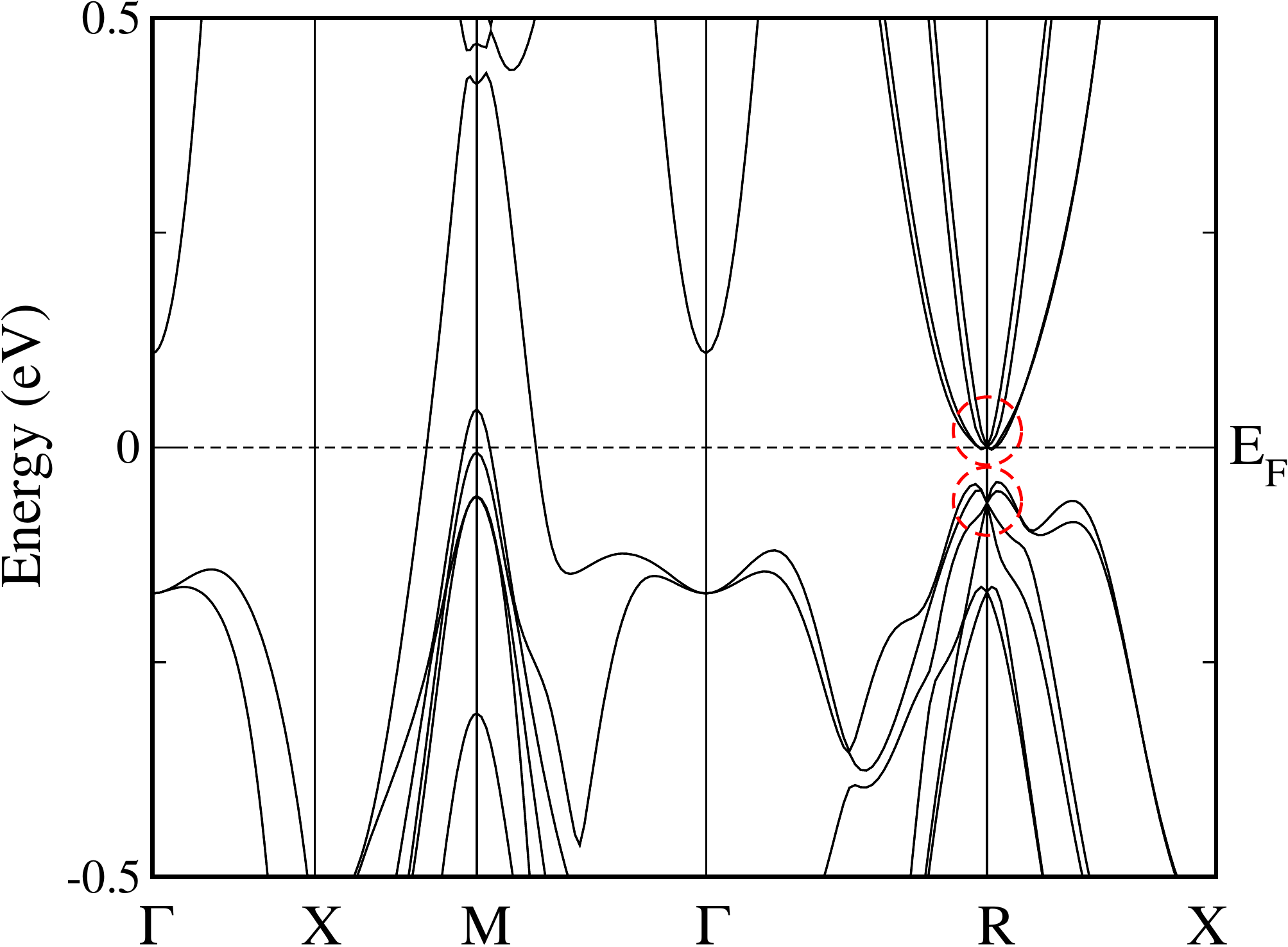}%
}
\caption{8-fold fermions at the $R$ point in (a) Ta$_3$Sb, (b) LaPd$_3$S$_4$ and (c) Nb$_3$Bi.}
\label{fig:dft9}
\end{figure*}

We propose candidate materials\cite{Jain13,Karlsruhe16} that realize each of the new types of fermions near the Fermi level.
In the Supplementary Material, we provide many more examples that require doping to bring the Fermi level to the band crossing, { but which, in the cases where the fermions are below the Fermi level, are still observable in ARPES experiments.}
We have computed the band structure of each candidate -- shown in Figs~\ref{fig:dft3}--\ref{fig:dft9} -- to confirm that the desired band crossings exist and are relatively close to the Fermi level. We performed electronic structure calculations within density-functional theory (DFT) as implemented in the Vienna \textit{ab initio} simulation package\cite{kresse1993}, and use the core-electron projector-augmented-wave basis in the generalized-gradient method\cite{perdew1996}. Spin-orbital coupling (SOC) is accounted for self-consistently. Unless otherwise noted, \emph{all} materials we report have been synthesized as single crystals.

We begin with an exotic 3-band fermion is SG 199, in Pd$_3$Bi$_2$S$_2$, which exists in single crystal form\cite{Weihrich07}. The band crossing at the $P$ point is only .1eV above the Fermi level, and its position could be further tuned by doping. The band structure is shown in Fig.~\ref{fig:dft1992}.

Next, we consider the exotic 3-band fermion in SG 214 in Ag$_3$Se$_2$Au, which can be grown as a single crystal\cite{Bindi04}. Fig.~\ref{fig:dft214-2} shows that although the spin-$1$ Weyl point is located $0.5$eV below the Fermi level, there are fourfold degeneracies at the $\Gamma$ and $H$ points located only $.02$eV below the Fermi level and there are no other bands in the vicinity: 
{ this remarkable material exhibits a $\mathbf{k}\cdot\mathbf{S}$ type spin-$3/2$ Hamiltonian close to the $\Gamma$ and $H$ points.}

Space groups 220 and 230 can host new fermions at both the $P$ and $H$ points.
In space group 220, we find 3-fold and 8-fold fermions in Ba$_4$Bi$_3$\cite{Li03} and La$_4$Bi$_3$\cite{Hulliger77}, shown in Figs.~\ref{fig:dft220-1} and \ref{fig:dft220-2}. In the latter case, the band crossings are less than .1eV from the Fermi level. These materials are parts of the families of compounds A$_4$Pn$_3$ and R$_4$Pn$_3$ (A = Ca, Sr, Ba, Eu; R = rare-earth element, i.e. La, Ce; Pn (pnictogen) = As, Sb, Bi), which are also potential candidates.

MgPt\cite{StadelmaierHardy1961} is a near ideal example of a 6-band fermion in SG 198; Fig.~\ref{fig:dft198-3} shows the band crossing is about .3eV above the Fermi level and isolated from other bands. More examples can be found in the families of PdAsS\cite{Hulliger63} and K$_3$BiTe$_3$\cite{Eisenmann91}, as shown in Figs~\ref{fig:dft198-1} and \ref{fig:dft198-2}. These band crossings are about .7eV below and .5eV above the Fermi level, respectively. Similar fermions can be found closer to the Fermi level in the compounds Li$_2$Pd$_3$B\cite{Eibenstein97} (SG 212) and Mg$_3$Ru$_2$\cite{Pottgen08}, shown in Figs~\ref{fig:dft212} and \ref{fig:dft213-3}.

The quadratic 6-band fermions in SGs 205 can be found in PdSb$_2$\cite{Furuseth67}, as shown in Fig.~\ref{fig:dft205}, as well as in the similar compounds FeS$_2$ and PtP$_2$. 

The 8-band fermions required to exist in SG 130 sit almost exactly at the Fermi level in CuBi$_2$O$_4$ and are isolated from all other bands, as shown in Fig.~\ref{fig:dft130}. { This material has a filling of $180=8*22+4$ electrons per unit cell, and hence it represents the first realizable example of a filling enforced semimetal\cite{Po2015}.}
CuBi$_2$O$_4$ exists in powder form, { and growth of single crystals is currently being attempted.} The material is magnetic below 50K, so experiments will need to be carried out above the Ne{\`e}l temperature, if not for interaction effects, which appear to make this material insulating\cite{Sharma2016}.
There are other bismuth oxides in this space group, which are also required to exhibit 8-fold fermions; Fig.~\ref{fig:dft130-1} shows PdBi$_2$O$_4$\cite{Conflant80}, which exists in single crystal form, and two predicted compounds are shown in the Supplementary Material.

8-band fermions are also required to exist in SG 135. One example is shown in Fig.~\ref{fig:dft135-2} in PdS, sitting .25eV above the Fermi level. This material is naturally insulating, but could be potentially doped. It has been observed in polycrystalline form\cite{Kliche92}.

The 8-band fermions predicted to occur in SG 218 exist in CsSn\cite{Hoch02} and CsSi\cite{Schnering05}, and more generally, in the class $AB$ for A=K, Rb, Cs and B=Si, Ge, Sn; the band structure of CsSn shows its unique splitting into four two-fold degenerate bands in the $k_x=k_z$ direction away from the $R$ point in Fig.~\ref{fig:dft218out}.
%\textbf{Moving along the $k_x=k_y=k_z$ direction should show the splitting into two 2-fold and four single degenerate bands, although the splitting between two of the single-degenerate bands is not resolvable. POTENTIAL CONFLICT BTWN BAND STRUCTURE AND KP.} 
There is a similar 8-band fermion at the $H$ point in SG 220, which is shown in Fig.~\ref{fig:dft220} for Ba$_4$Bi$_3$\cite{Li03} and La$_4$Bi$_3$\cite{Hohnke66}.

The 8-band fermions predicted to occur in SG 223 are exhibited in the candidates X$_3$Y, where X is either Nb or Ta and Y is any group A-IV or A-V element in the beta-tungsten structure A15, as well as in the family MPd$_3$S$_4$, where M is any rare-earth metal. The band structures for Ta$_3$Sb (powder)\cite{Furuseth65a} and LaPd$_3$S$_4$\cite{Keszler85} show the 8-band crossing within nearly .1eV of Fermi level, as shown in Figs~\ref{fig:dft223-1} and \ref{fig:dft223-2}. Fig.~\ref{fig:dft223-3} shows Nb$_3$Bi (powder)\cite{Killpatrick64}, which has two 8-fold fermions within .1eV of the Fermi level. { An exhaustive database search is currently underway for all filling-enforced semimetals with new fermions close to the Fermi level.}

\paragraph{Outlook}
In this letter we have analyzed all possible exotic fermion types that can occur in spin-orbit coupled crystals with time reversal symmetry going beyond the Majorana-Weyl-Dirac classification. By virtue of their band topology, these fermions can play host to novel surface states, magnetotransport properties, and ARPES signatures. { Growth of many of the material candidates mentioned above, including AsPdS, La$_3$PbI$_3$, La$_4$Bi$_3$, LaPd$_3$S$_4$ and Ta$_3$Sb is currently underway, and should yield fruitful results in ARPES and magnetotransport experiments.}

As we have emphasized throughout, non-symmorphic crystal symmetries were essential for stabilizing these fermions -- it is the presence of half-lattice translations that allow spin-$1/2$ electrons to transform under integer spin representations of the rotation group, yielding $3-$fold and $6-$fold degeneracies. The new fermions also provide an explicit realization of the non-symmorphic insulating filling bounds derived recently in Refs.~\cite{Watanabe2015,Po2015}, as follows. Due to the presence of TR symmetry, we know that our threefold fermions must occur in connection with an additional non-degenerate band, so that all bands have Kramers partners at TR invariant momenta; this suggests that the minimal band connectivity in these space groups is four, consistent with the filling bounds. Similarly, since the $6-$fold fermions arise from doubling a three-fold degeneracy, { each of which comes along with an aforementioned additional non-degenerate band.} We thus expect -- and indeed confirm -- that the minimal insulating filling is $8$ in these cases. Finally, as noted in Ref.~\cite{Wieder2015}, the $8-$fold degenerate fermions saturate the filling bound for those SGs.

Looking ahead, there are several open questions which deserve future attention. First, gapping these degeneracies by breaking the symmetries that protect them can lead to novel symmetry-protected topological phases, with new classes of $2d$ gapless surface modes. { Furthermore, our symmetry analysis can be extended to crystals with magnetic order, and hence with interactions.}
This requires an investigation of representations of the $1191$ remaining magnetic space groups, which we are currently undertaking.%\cite{future-us}. 

\paragraph{Acknowledgements}
The authors thank Aris Alexandradinata, Titus Neupert, Alexey Soluyanov, and Ali Yazdani for helpful discussions. MGV acknowledges the Fellow Gipuzkoa Program  through FEDER “Una Manera de hacer Europa” and the FIS2013-48286-C2-1-P national project of the Spanish MINECO. BAB acknowledges the support of the ARO MURI on topological insulators, grant W911-NF-12-1-0461, ONR-N00014-11-1-0635, NSF CAREER DMR-0952428, NSF-MRSEC DMR-1005438, the Packard Foundation and a Keck grant. RJC similarly acknowledges the support of the ARO MURI and the NSF MRSEC grants.

\bibliography{newfermions}

\end{document}

\begingroup
\let\clearpage\relax
\author{Barry Bradlyn}
\affiliation{Princeton Center for Theoretical Science, Princeton University, Princeton, New Jersey 08544, USA}
\author{Jennifer Cano}
\affiliation{Princeton Center for Theoretical Science, Princeton University, Princeton, New Jersey 08544, USA}
\author{Zhijun Wang}
\affiliation{Department of Physics, Princeton University, Princeton, New Jersey 08544, USA}
\author{M.~G. Vergniory}
\affiliation{Donostia International Physics Center, P. Manuel de Lardizabal 4, 20018 Donostia-San Sebasti\'{a}n, Spain}
\author{C. Felser}
\affiliation{Max Planck Institute for Chemical Physics of Solids, 01187 Dresden, Germany}
\author{R.~J. Cava}
\affiliation{Department of Chemistry, Princeton University, Princeton, New Jersey 08544, USA}
\author{B. Andrei Bernevig}
\affiliation{Department of Physics, Princeton University, Princeton, New Jersey 08544, USA}
\date{\today}
\title{Supplementary Material for Beyond Dirac and Weyl fermions: Unconventional quasiparticles in conventional crystals}
\maketitle

\onecolumngrid

%\section{Symmetry analysis of 3d and 6d Irreps}
\section{Symmetry analysis of 3d and 6d Irreps}
\subsection{Notation}\label{sec:notation}

A Bravais lattice in three dimensions has three basis vectors, indicated by $\mathbf{t}_i, i=1,2,3$.
Reciprocal space lattice vectors are indicated by $\mathbf{g}_i$, where $\mathbf{g}_i\cdot \mathbf{t}_j=2\pi\delta_{ij}$. 
The new fermions described in the main text occur only in the primitive cubic, body-centered cubic, and primitive tetragonal lattices; their lattice and reciprocal lattice vectors are shown in Table~\ref{tab:vectors}. Diagrams of the first Brillouin zone for each of these Bravais lattices are shown in Figure \ref{bzfig}, with high symmetry points labelled.

\begin{table}[b]
\begin{tabular}{ccc}
Bravais lattice & Lattice vectors & Reciprocal lattice vectors\\
\hline
Primitive cubic & $(a,0,0),(0,a,0),(0,0,a)$ & $\frac{2\pi}{a}(1,0,0),\frac{2\pi}{a}(0,1,0),\frac{2\pi}{a}(0,0,1)$\\
Body-centered cubic & $\frac{a}{2}(-1,1,1),\frac{a}{2}(1,-1,1),\frac{a}{2}(1,1,-1)$ & $\frac{2\pi}{a}(0,1,1),\frac{2\pi}{a}(1,0,1),\frac{2\pi}{a}(1,1,0)$\\
Primitive tetragonal & $(a,0,0),(0,a,0),(0,0,c)$ & $\frac{2\pi}{a}(1,0,0),\frac{2\pi}{a}(0,1,0),\frac{2\pi}{c}(0,0,1)$
\end{tabular}
\caption{Lattice and reciprocal lattice vectors}\label{tab:vectors}
\end{table}

We indicate non-symmorphic symmetry operations using Seitz notation, i.e., a point group operation $\mathcal{O}$ followed by a translation $\mathbf{v}=v_i\mathbf{t}_i$ is indicated by $\{\mathcal{O}|\mathbf{v}\}$ or, component-wise, $\{\mathcal{O}|v_1v_2v_3\}$.
The rules for combining operations is as follows:
\begin{equation} \lbrace \mathcal{O}_2 |\mathbf{v}_2 \rbrace \lbrace \mathcal{O}_1 | \mathbf{v}_1 \rbrace = \lbrace \mathcal{O}_2\mathcal{O}_1 | \mathbf{v}_2 + R_2\mathbf{v}_1 \rbrace.
\label{eq:product} \end{equation}
We thus have the following useful relations:
\begin{equation} \lbrace \mathcal{O}|\mathbf{v}\rbrace^{-1} = \lbrace \mathcal{O}^{-1} | -\mathcal{O}^{-1}\mathbf{v}\rbrace,
\label{eq:inverse}\end{equation}
\begin{equation} \lbrace \mathcal{O}|\mathbf{v}\rbrace= \lbrace E|\mathbf{v}\rbrace \lbrace \mathcal{O}|0\rbrace = \lbrace \mathcal{O}|0\rbrace \lbrace E|\mathcal{O}^{-1}\mathbf{v}\rbrace.
\end{equation}
We will always use $E$ for the identity operator and $I$ for inversion. We frequently use $\mathcal{R}$ to indicate a $2\pi$ rotation; since we are interested in spin-1/2 particles, this operator is always represented by $-\mathbb{I}$. 
We use $C_{2x}, C_{2y}, C_{2z}$ to indicate 2-fold rotations about the $x$-, $y$- or $z$- hat axes; otherwise, we use $C_{n,n_xn_yn_z}$ to indicate an $n$-fold rotation about the $n_x\hat{x}+n_y\hat{y}+n_z\hat{z}$ axis.
Similarly, $\sigma_{x},\sigma_y,\sigma_z$ indicate mirror operations through the planes perpendicular to the indicated axis and $\sigma_{n_xn_yn_z}$ indicates a mirror operation through the plane perpendicular to the $n_x\hat{x}+n_y\hat{y}+n_z\hat{z}$ direction.
We also encounter four-fold roto-inversions; we define $S_{4x}\equiv IC_{4x}^{-1}$ and similarly for $y$ and $z$.
Pure translations are indicated by $\{E|\mathbf{t}\}$.
Irreducible representation (irrep) of the group of translations are labeled by reciprocal space vectors; in the irrep labeled by $\mathbf{k}$, an integer translation $\mathbf{t}\equiv n_i\mathbf{t}_i$ is represented by the phase $e^{-i\mathbf{k}\cdot \mathbf{t}}$.

\begin{figure}[b]
\subfloat[\label{fig:tetrabz}]{\includegraphics[width=.3\textwidth]{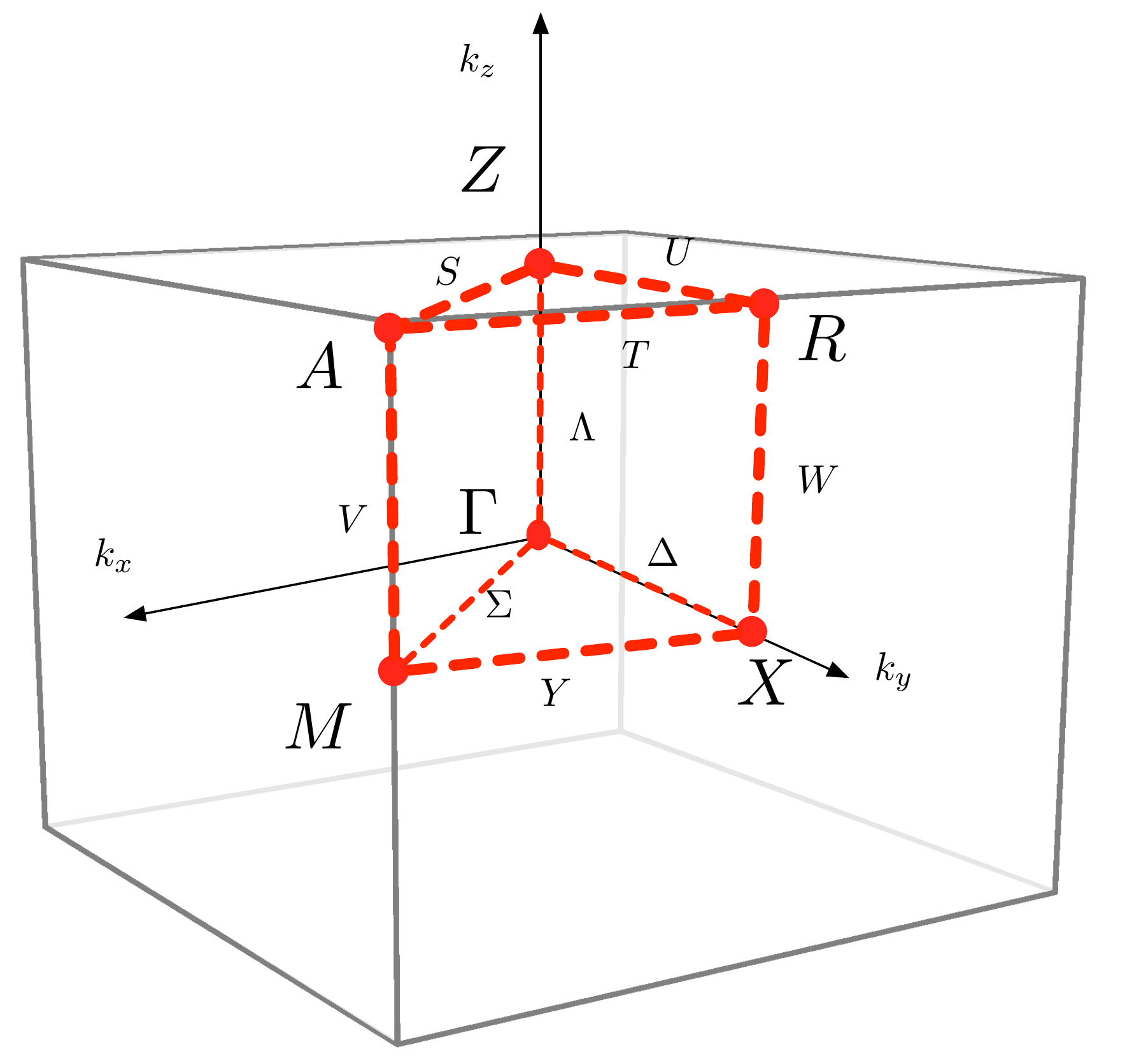}}
\subfloat[\label{fig:scbz}]{\includegraphics[width=.3\textwidth]{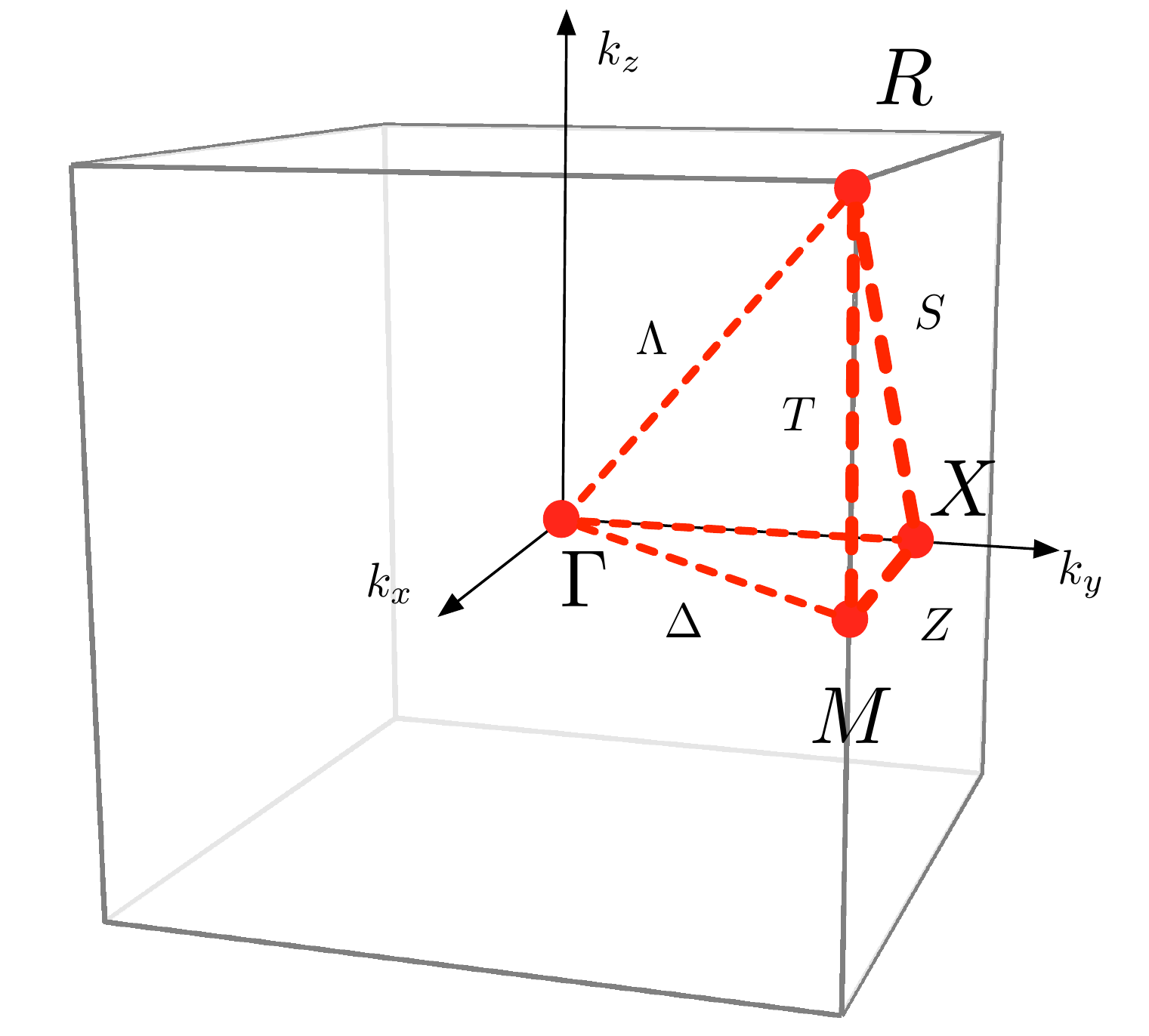}}
\subfloat[\label{fig:bccbz}]{\includegraphics[width=.3\textwidth]{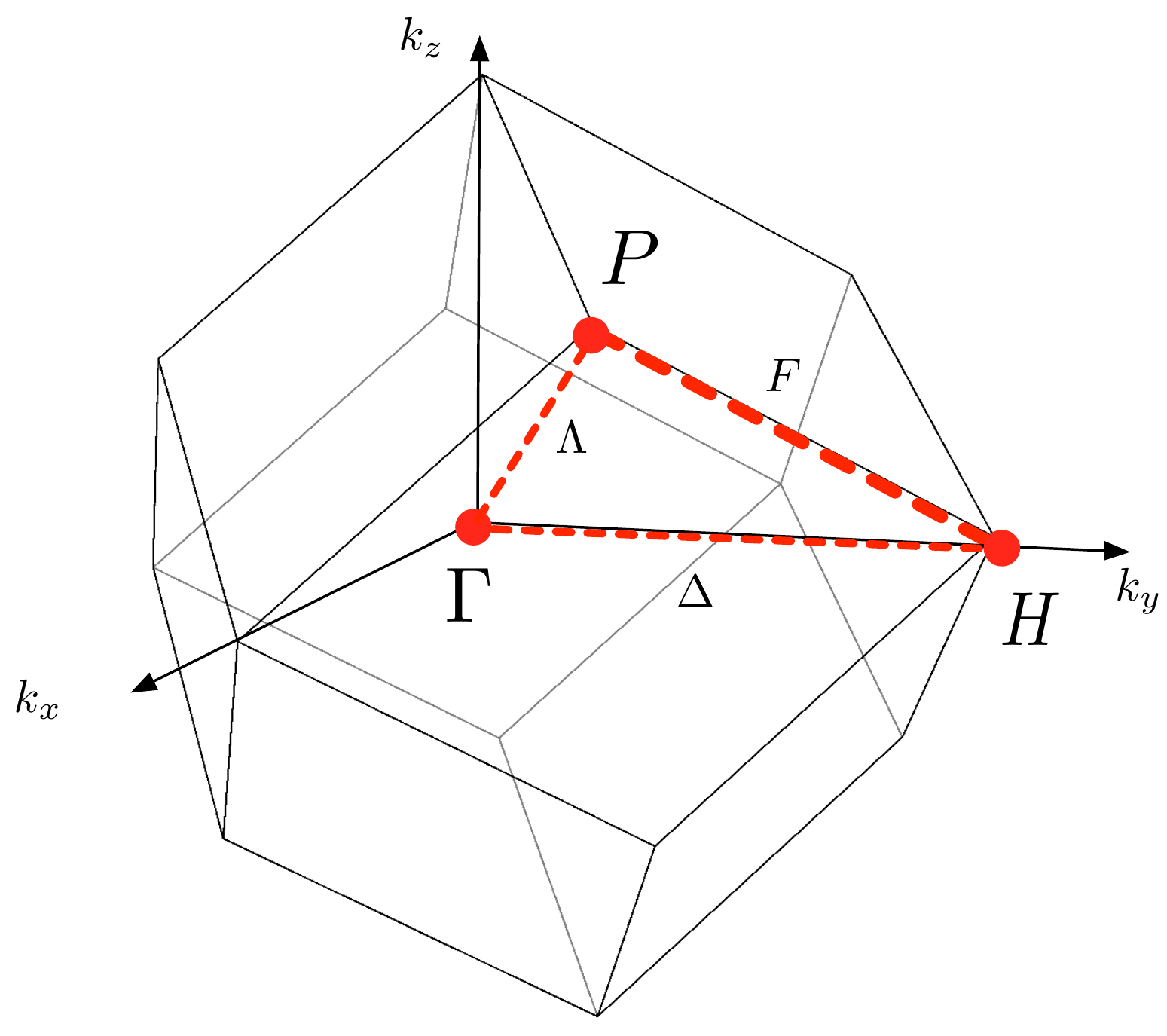}}
\caption{(a) The Brillouin zone of a tetragonal Bravais lattice. (b) The Brillouin zone of a simple cubic Bravais lattice. (c) The Brillouin zone of a body-centered cubic Bravais lattice. In cases (a) and (b) all of the labelled high-symmetry points are time-reversal invariant. In case (c) only the $P$ point is not time-reversal invariant. (A particular point, $\mathbf{k}$, is (not) time-reversal invariant if $2\mathbf{k}$ is (not) an integer reciprocal lattice vector.)}
\label{bzfig}
\end{figure}

The little group $G^{\mathbf{k}_0}$ of a point $\mathbf{k}_0$ in reciprocal space is the set of all space group operations $\{\mathcal{O}|\mathbf{v} \}$ such that $\mathcal{O}\mathbf{k}_0 =\mathbf{k}_0+n_i\mathbf{g}_i$, i.e., the set of all space group operations whose `symmorphic part' leaves $\mathbf{k}$ invariant up to an integer reciprocal lattice vector; later we will consider the effect of including time reversal in this definition.
%A $d$-dimensional irrep indicates that the $d$ bands that transform like the irrep at the $\mathbf{k}$ point will be degenerate at this point. 
If a $d$-dimensional irrep exists, a generic Hamiltonian which respects the space group symmetries can display a $d$-dimensional degeneracy at $\mathbf{k}_0$.
However, if multiple irreps with dimensions $d_1, d_2, ...$ exist, not all will necessarily be realized in a given material.
Furthremore, notice that two-fold degeneracies (Weyl fermions) can exist without protection by a space group symmetry.

If a $d$-band crossing exists, and if the Fermi level is near the crossing, then these bands constitute the low-energy dispersion relation of a fermion with $d$ components. Here we are exploring fermions beyond the Weyl and Dirac paradigm. These new fermions consist of 3-, 6- and 8-band crossings in the presence of time reversal symmetry; the 6- and 8-band crossings emerge from 3- and 4-band crossings without time reversal symmetry. In addition, we show that the degeneracy of the band crossing does not fully characterize the behavior of the fermion. Degeneracies on high-symmetry lines and planes provide a further, finer grading, which completes our fermion classification.

In the main text, we discussed the particular space groups that can support 3-, 6-, or 8-band crossings at prescribed points in the Brillouin zone. 
In the appendices, we will address each space group mentioned in the main text and show that the little group at the prescribed point has an irrep of the correct degeneracy to support the band crossings\cite{Cracknell,Faddeyev2014,Kovalev1965,Miller1967,Casher1969}.
In particular, we will first find the little group without time reversal symmetry, where it will display either a 3- or 4-band crossings, and then show that the irrep either remains 3-dimensional or doubles in size to 6- or 8-dimensional (since we are interested in 3-, 6-, and 8-band crossings, we do not include in our search 4d irreps that remain the same size in the presence of time reversal.)

\subsection{Space groups with 3d irreps}

In this appendix, we prove a sufficient condition for the existence of a 3d irrep given three group generators.
We show that it is satisfied by the generators of the little group at the point $\mathbf{k}_0=(\pi,\pi,\pi)$ for the space groups 198, 199, 205, 206, 212, 213, 214, 220 and 230. These are all cubic lattices\cite{Cracknell}; hence, symmetry operators that leave the $(\pi,\pi,\pi)$ point invariant (up to a reciprocal lattice vector) include a 3-fold rotation about the $\hat{x}+\hat{y}+\hat{z}$ axis, a 2-fold rotation about the $\hat{x}$ or $\hat{y}$ axes, or any combination of these elements. These operators, potentially combined with non-symmorphic translations, generate the little group at the $(\pi,\pi,\pi)$ point. In some cases, additional generators, which do not change the size of the 3d irrep, are also present.

We then consider the presence of time reversal symmetry. We show that in the presence of time reversal symmetry, space groups 198, 205, 206, 212, 213 and 230 can host 6-fold degeneracies at the $(\pi,\pi,\pi)$ point, while space groups 199, 214 and 220 can host 3-fold degeneracies at this point.

\subsection{Sufficient condition for a 3d irrep}\label{sec:condition}

Consider the case when the little group at a particular high-symmetry point has three generators, $\mathcal{G}_1, \mathcal{G}_2,$ and $\mathcal{G}_3$, with matrix representations
$G_1,G_2,$ and $G_3$, perhaps up to an overall phase (i.e, the representation of $\mathcal{G}_i$ is $e^{i\theta_i}G_i$). Then if the matrices satisfy
\begin{equation}G_1^2=G_2^2=G_3^3=1,\quad [G_1,G_2]=0,\quad G_1G_3=G_3G_2,\quad G_2G_3=G_3G_1G_2, \label{eq:condition}\end{equation}
there exists a 3d irrep. Before proving this, we comment that Eq.~(\ref{eq:condition}) places tight constraints on the operators: clearly $\mathcal{G}_3$ is a three-fold rotation or screw and $\mathcal{G}_1$ and $\mathcal{G}_2$ are either two-fold rotations or screws (they could also be mirrors or glides without any change to the logic). The defining axes of $\mathcal{G}_1$ and $\mathcal{G}_2$ must be perpendicular, otherwise $G_1G_2G_1^{-1}G_2^{-1}$ would not be proportional to the identity. Now assume that $\mathcal{G}_1$ and $\mathcal{G}_2$ are both symmorphic: then $G_1$ and $G_2$ anti-commute, regardless of putative phases $\theta_{1,2}$; hence, the assumption must be wrong, and at least one of $\mathcal{G}_1$ or $\mathcal{G}_2$ must be non-symmorphic. Thus, non-symmorphicity plays a crucial role in obtaining 3-fold fermions.

In addition, since $\mathcal{G}_1$ and $\mathcal{G}_2$ are 2-fold rotations or screws about orthogonal axes -- which implies $\mathcal{G}_1\mathcal{G}_2$ must be a 2-fold rotation or screw about their mutually orthogonal axis -- then the last two equalities in Eq.~(\ref{eq:condition}) require $\mathcal{G}_3$ to have equal components (up to a sign) along the axes of $\mathcal{G}_1,\mathcal{G}_2$ and $\mathcal{G}_1\mathcal{G}_2$. Consequently, Eq.~(\ref{eq:condition}) can only hold in a cubic lattice.

We now show that Eq.~(\ref{eq:condition}) requires a 3d irrep.
We show this by considering an eigenstate of the Hamiltonian, $\psi$, which is a simultaneous eigenstate of $G_1$ and $G_2$ with respective eigenvalues $\lambda_1$ and $\lambda_2$. Eq.~(\ref{eq:condition}) shows that $G_3\psi$ and $G_3^2\psi$ are also a simultaneous eigenstates of $G_1$ and $G_2$: 
\begin{align}G_1G_3\psi &= G_3G_2\psi = \lambda_2G_3\psi,\nonumber\\
G_2G_3\psi &= G_3G_1G_2\psi=\lambda_1\lambda_2G_3\psi, \nonumber\\
G_1G_3^2\psi &= G_3G_2G_3\psi=G_3^2G_1G_2\psi = \lambda_1\lambda_2G_3^2\psi,\nonumber\\
G_2G_3^2\psi &= G_3G_1G_2G_3\psi= G_3^2G_1\psi=\lambda_1G_3^2\psi.
\end{align}
As long as either $\lambda_1\neq 1$ or $\lambda_2\neq 1$, $\psi, G_3\psi$ and $G_3^2\psi$ all have distinct pairs of eigenvalues under $G_1$ and $G_2$, which means that they are three distinct states, and transform as a 3d irrep of $G_3, G_1$ and $G_2$. 
(The case $\lambda_1=\lambda_2=1$ corresponds to the trivial representation, where all $G_i\propto\mathbb{I}$).

If additional generators are present, we show on a case-by-case basis that they are consistent with the 3d irrep.

\subsection{Space groups 198 and 205}

Space groups 198 and 205 are primitive cubic lattices. 
%The basis vectors in Cartesian coordinates are $\hat{x},\hat{y},\hat{z}$ and the reciprocal lattice vectors are $2\pi \hat{x},2\pi\hat{y},2\pi\hat{z}$.
Recall that we are interested in irreps at the point $R=(\pi,\pi,\pi)$ in momentum space.
The point group that leaves this point invariant is generated by a 3-fold rotation $C_{3,111}^{-1}$ along the $111$ axis, $C_{2x}$ and $C_{2y}$ (and, in space group 205, inversion). Including their non-symmorphic parts, the generators are:
\begin{equation} \{C_{3,111}^{-1}|010\}, \{C_{2x}|\frac{1}{2}\frac{3}{2}0\}, \{C_{2y}|0\frac{3}{2}\frac{1}{2}\}. 
\label{eq:gens198}\end{equation}
Using the group operations defined in Sec~\ref{sec:notation}, these generators satisfy:
\begin{align}
\{C_{3,111}^{-1}|010\}^3 &= \{\mathcal{R}|111\}, \nonumber\\
\{C_{2x}|\frac{1}{2}\frac{3}{2}0\}^2 &= \{\mathcal{R}|100\}, \nonumber\\
\{C_{2y}|0\frac{3}{2}\frac{1}{2}\}^2 &= \{\mathcal{R}|030\},
\label{eq:rels198}
\end{align}
as well as:
\begin{gather}
\{C_{2x}|\frac{1}{2}\frac{3}{2}0\}\{C_{2y}|0\frac{3}{2}\frac{1}{2}\}=\{C_{2x}C_{2y}|\frac{1}{2}0\bar{\frac{1}{2}}\}=\{\mathcal{R}C_{2y}C_{2x}|\frac{1}{2}0\bar{\frac{1}{2}}\}=\{C_{2y}|0\frac{3}{2}\frac{1}{2}\}\{\mathcal{R}C_{2x}|\bar{\frac{1}{2}}\bar{\frac{3}{2}}1\}=\{C_{2y}|0\frac{3}{2}\frac{1}{2}\}\{C_{2x}|\frac{1}{2}\frac{3}{2}0\}\{\mathcal{R}|\bar{1}3\bar{1}\},\nonumber\\
\{C_{3,111}^{-1}|010\}^{-1}\{C_{2x}|\frac{1}{2}\frac{3}{2}0\}\{C_{3,111}^{-1}|010\}=\{C_{2y}|0\frac{1}{2}\bar{\frac{1}{2}} \}=\{C_{2y}|0\frac{3}{2}\frac{1}{2}\}\{E|0\bar{1}1\},\nonumber\\
\{C_{3,111}^{-1}|010\}^{-1}\{C_{2y}|0\frac{3}{2}\frac{1}{2}\}\{C_{3,111}^{-1}|010\}=\{C_{2z}|\frac{1}{2}0\frac{3}{2}\} = \{C_{2x}|\frac{1}{2}\frac{3}{2}0\}\{C_{2y}|0\frac{3}{2}\frac{1}{2}\}\{E|002\}.
\label{eq:rels198-2}
\end{gather}
We seek a matrix representation for these abstract group elements.
Let $\{C_{3,111}^{-1}|010\},\{C_{2x}|\frac{1}{2}\frac{3}{2}0\}$ and $\{C_{2y}|0\frac{3}{2}\frac{1}{2}\}$ be represented by some to-be-determined matrices $G_3,G_1$ and $G_2$.
As described in Sec~\ref{sec:notation}, integer lattice translations $\{E|\mathbf{v}\}$ are represented by $e^{-i\mathbf{k}\cdot \mathbf{v}}=e^{-i\pi(v_x+v_y+v_z)}$, where the equality follows from evaluating at the point $\mathbf{k}=(\pi,\pi,\pi)$. Similarly, $\{\mathcal{R}|\mathbf{v}\}$ is represented by $-e^{-i\pi(v_x+v_y+v_z)}$. Consequently, Eqs.~(\ref{eq:rels198}) and (\ref{eq:rels198-2}) can be translated to the following relations between the matrix representations:
\begin{equation}
G_3^3=G_1^2=G_2^2=1,\quad G_1G_2=G_2G_1,\quad  G_3^{-1}G_1G_3=G_2,\quad G_3^{-1}G_2G_3=G_1G_2,
\end{equation}
which are exactly Eq.~(\ref{eq:condition}).
Thus, we immediately see from Sec~\ref{sec:condition} that these space groups have a 3d irrep.

As mentioned above, the inversion operator is also in the little group at the $R$ point for space group 205.
In general, it satisfies:
\begin{equation} \{I|000\}^{-1}\{\mathcal{O}|v\}\{I|000\}=\{\mathcal{O}|v\}\{E|-2\mathcal{O}^{-1}v\}.
\end{equation}
Thus, $\{I|000\}$ commutes with all the generators (\ref{eq:gens198}).
Hence, it is consistent to take the representation of $\{I|000\}$ to be $\pm\mathbb{I}$ without changing the other matrix representations in the 3d irrep.

\subsection{Space groups 199, 206 and 214}\label{sec:199}

Space groups 199, 206 and 214 are body-centered cubics. 
Again, we are interested in the little group at the $(\pi,\pi,\pi)$ point (the $P$ point).
As for space groups 198 and 205, the point group that leaves this point invariant is generated by a 3-fold rotation along the $111$ axis, $C_{2x}$ and $C_{2y}$. Including their non-symmorphic parts, the generators are:
\begin{equation} \{C_{3,111}^{-1}|101\}, \{C_{2x}|\bar{\frac{1}{2}}\frac{1}{2}0\}, \{C_{2y}|0\frac{1}{2}\bar{\frac{1}{2}}\}.
\end{equation}
These operators are very similar to those in (\ref{eq:gens198}); thus, we do not write out the equivalent relations to Eq.~(\ref{eq:rels198}) and (\ref{eq:rels198-2}), but merely say that if the matrix representations of $\{C_{3,111}^{-1}|101\}, \{C_{2x}|\bar{\frac{1}{2}}\frac{1}{2}0\}$ and $ \{C_{2y}|0\frac{1}{2}\bar{\frac{1}{2}}\}$ are called $G_3, G_1$ and $G_2$, respectively, then Eq.~(\ref{eq:condition}) is also obeyed for these space groups, guaranteeing that they have a 3d irrep.

\subsection{Space groups 212 and 213}

Space groups 212 and 213 are primitive cubics. The $R$ point $(\pi,\pi,\pi)$ is left invariant by $C_{3,111}$, $C_{2x}$ and $C_{2y}$, as in the previous cases, as well as $C_{2,1\bar{1}0}$.
The particular generators are
\begin{equation} \{C_{3,111}^{-1}|000\},\{C_{2x}|\frac{1}{2}\frac{1}{2}0\}, \{C_{2y}|0\frac{1}{2}\frac{1}{2}\}\text{ and }\begin{array}{ll} \lbrace C_{2,1\bar{1}0}|\frac{1}{4}\frac{1}{4}\frac{1}{4}\rbrace & \text{(SG 212)}\\
\lbrace C_{2,1\bar{1}0}|\frac{3}{4}\frac{3}{4}\frac{3}{4}\rbrace & \text{(SG 213)}\
\end{array}.
\label{eq:gens212}\end{equation}
Using the operations in Sec~\ref{sec:notation}, we compute:
\begin{align}
\{C_{3,111}^{-1}|000\}^3&=\{\mathcal{R}|000\},\nonumber\\
\{C_{2x}|\frac{1}{2}\frac{1}{2}0\}^2&=\{\mathcal{R}|100\},\nonumber\\
\{C_{2y}|0\frac{1}{2}\frac{1}{2}\}^2&=\{\mathcal{R}|010\},\nonumber\\
\{C_{2,1\bar{1}0}|\frac{1}{4}\frac{1}{4}\frac{1}{4}\}^2 &= \{C_{2,1\bar{1}0}|\frac{3}{4}\frac{3}{4}\frac{3}{4}\}^2 = \{\mathcal{R}|000\}.
\label{eq:genrels212}
\end{align}
as well as the relations:
\begin{align}
\{C_{3,111}^{-1}|000\}\{C_{2x}|\frac{1}{2}\frac{1}{2}0\}\{C_{3,111}^{-1}|000\}^{-1}&=\{C_{2x}|\frac{1}{2}\frac{1}{2}0\}\{C_{2y}|0\frac{1}{2}\frac{1}{2}\}\{E|001\},\nonumber\\
\{C_{3,111}^{-1}|000\}\{C_{2y}|0\frac{1}{2}\frac{1}{2}\}\{C_{3,111}^{-1}|000\}^{-1}&=\{C_{2x}|\frac{1}{2}\frac{1}{2}0\},\nonumber\\
\{C_{2x}|\frac{1}{2}\frac{1}{2}0\}\{C_{2y}|0\frac{1}{2}\frac{1}{2}\}&=\{C_{2y}|0\frac{1}{2}\frac{1}{2}\}\{C_{2x}|\frac{1}{2}\frac{1}{2}0\}\{\mathcal{R}|\bar{1}1\bar{1}\},\nonumber\\
\{C_{2,1\bar{1}0}|vvv\}\{C_{2x}|\frac{1}{2}\frac{1}{2}0\}\{C_{2,1\bar{1}0}|vvv\}^{-1}&=\begin{cases}\{C_{2y}|0\frac{1}{2}\frac{1}{2}\}\{\mathcal{R}|0\bar{1}0\}&\text{SG 212}\\\{C_{2y}|0\frac{1}{2}\frac{1}{2}\}\{\mathcal{R}|\bar{1}\bar{1}\bar{1}\} &\text{SG 213}\end{cases},\nonumber\\
\{C_{2,1\bar{1}0}|vvv\}\{C_{3,111}^{-1}|000\}\{C_{2,1\bar{1}0}|vvv\}^{-1}&= \{C_{3,111}^{-1}|000\}^2\{\mathcal{R}|000\},
\label{eq:genrels212-2}
\end{align}
where we have used $v$ to be $\frac{1}{4}$ for space group 212 and $\frac{3}{4}$ for space group 213. 
Now let $-G_3^{-1}, -G_1, -G_2$ and $-iG_4$ be the respective matrix representations for 
$\{C_{3,111}^{-1}|000\}, \{C_{2x}|\frac{1}{2}\frac{1}{2}0\}, \{C_{2y}|0\frac{1}{2}\frac{1}{2}\}$ and $\{C_{2,1\bar{1}0}|\frac{1}{4}\frac{1}{4}\frac{1}{4}\} (\{C_{2,1\bar{1}0}|\frac{3}{4}\frac{3}{4}\frac{3}{4}\})$ in space group 212(213).
The phases are for convenience. Eqs.~(\ref{eq:genrels212}) and (\ref{eq:genrels212-2}) translate to the matrix equations:
\begin{equation}G_3^3=G_1^2=G_2^2=G_4^2=1\label{eq:squares212}\end{equation}
and
\begin{equation}G_3^{-1}G_2G_3=G_2G_1, \quad G_3^{-1}G_1G_3=G_2,\quad  [G_1,G_2]=0, \quad G_4G_2G_4^{-1}=G_1,\quad G_4G_3^{-1}G_4^{-1}=G_3^{-2}.
\end{equation}
The last line is trivially equivalent to:
\begin{equation}G_2G_3=G_3G_2G_1, \quad G_1G_3=G_3G_2,\quad  [G_1,G_2]=0, \quad G_4G_2=G_1G_4,\quad G_4G_3^2G_4^{-1}=G_3.\label{eq:cond212}
\end{equation}
Eqs.~(\ref{eq:squares212}) and (\ref{eq:cond212}) are exactly Eq.~(\ref{eq:condition}) for $G_3,G_1,G_2$. It remains to show that there is a matrix $G_4$ compatible with the 3d irrep that satisfies Eqs.~(\ref{eq:squares212}) and (\ref{eq:cond212}). To do this, we follow the argument in Sec~\ref{sec:condition}: consider an eigenstate $\psi$ of $G_1$ and $G_2$ with eigenvalues $\lambda_1$ and $\lambda_2$, which we write here as a pair $(\lambda_1,\lambda_2)$. It was shown in Sec~\ref{sec:condition} that $\psi, G_3\psi$ and $G_3^2\psi$ have eigenvalues $(\lambda_1,\lambda_2), (\lambda_2,\lambda_1\lambda_2), (\lambda_1\lambda_2,\lambda_1)$ and transform as a 3d irrep as long as $\lambda_1\neq 1$ or $\lambda_2\neq 1$.
From Eq.~(\ref{eq:cond212}), we compute the eigenvalues of $G_4\psi$:
\begin{align}
G_1G_4\psi &= G_4G_2\psi = \lambda_2G_4\psi,\nonumber\\
G_2G_4\psi &= (G_4G_2)^{-1}\psi=(G_1G_4)^{-1}\psi = G_4G_1\psi=\lambda_1G_4\psi,
\end{align}
where we have used, following Eq.~(\ref{eq:squares212}) $G_{2,4}=G_{2,4}^{-1}$.
Thus, $G_4\psi$ has the swapped eigenvalues $(\lambda_2,\lambda_1)$.
Since the case $\lambda_1=\lambda_2=1$ only has 1d irreps, a 3d irrep requires $\lambda_1\neq 1$ or $\lambda_2\neq 1$. Without loss of generality, we take $\lambda_1=1, \lambda_2=-1$, which yields two 3d irreps:
\begin{equation}
G_1=\begin{pmatrix}1&0&0\\0&-1&0\\0&0&-1\end{pmatrix},G_2=\begin{pmatrix}-1&0&0\\0&-1&0\\0&0&1\end{pmatrix},G_3=\begin{pmatrix}0&0&1\\1&0&0\\0&1&0\end{pmatrix},G_4=\pm \begin{pmatrix}0&0&1\\0&1&0\\1&0&0\end{pmatrix}. \label{eq:212rep}
\end{equation}
The other choices of $\lambda_1$ and $\lambda_2$ yield representations in which the trace of each matrix is unchanged; hence, the representations are equivalent.

One might consider a larger-dimensional representation, i.e., where $G_4\psi$ is orthogonal to $G_3^2\psi$, but shares the same eigenvalues. Then $G_4\psi, G_3G_4\psi$ and $G_3^2G_4\psi$ form a 3d irrep under $G_1, G_2$ and $G_3$. However, the 6d irrep consisting of $\psi, G_3\psi, G_3^2\psi, G_4\psi, G_3(G_4\psi)$ and $G_3^2(G_4\psi)$ is reducible: the states $\psi + G_3G_4\psi, G_3\psi + G_3^2G_4\psi$ and $G_3^2\psi+G_4\psi$ comprise an invariant subset, as do $\psi - G_3G_4\psi, G_3\psi - G_3^2G_4\psi$ and $G_3^2\psi-G_4\psi$.

\subsection{Space groups 220 and 230}\label{sec:220}
Space groups 220 and 230 are body-centered cubics. As in the previous cases, the $P$ point, $(\pi,\pi,\pi)$, is invariant under $C_{3,111}$, $C_{2x}$ and $C_{2y}$. In space groups 220 and 230, there is an additional non-symmorphic 4-fold roto-inversion, $S_{4x}\equiv IC_{4x}^{-1}$, which also leaves the $P$ point invariant. Including their non-symmorphic parts, the generators are:
\begin{equation}\{C_{3,111}|000\},\{C_{2x}|\frac{3}{2}\frac{3}{2}0\},\{C_{2y}|0\frac{1}{2}\frac{1}{2}\},\{S_{4x}|\frac{1}{2}11\}.
\end{equation}
They satisfy
\begin{align} \{C_{3,111}|000\}^3&=\{C_{2x}|\frac{3}{2}\frac{3}{2}0\}^2=\{C_{2y}|0\frac{1}{2}\frac{1}{2}\}^2=\{S_{4x}|\frac{1}{2}11\}^4=\{\mathcal{R}|000\},
\label{eq:rels220}
\end{align}
as well as
\begin{align}
\{S_{4x}|\frac{1}{2}11\}^2&=\{C_{2x}|\frac{3}{2}\frac{3}{2}0\}\{\mathcal{R}|110\}, \nonumber\\
\{C_{3,111}|000\}^{-1}\{C_{2y}|0\frac{1}{2}\frac{1}{2}\}\{C_{3,111}|000\}&= \{C_{2x}|\frac{3}{2}\frac{3}{2}0\}\{E|110\},\nonumber\\
\{C_{3,111}|000\}^{-1}\{C_{2x}|\frac{3}{2}\frac{3}{2}0\}\{C_{3,111}|000\}&=\{C_{2x}|\frac{3}{2}\frac{3}{2}0\}\{C_{2y}|0\frac{1}{2}\frac{1}{2}\}\{E|\bar{3}\bar{1}\bar{1}\} ,\nonumber\\
\{S_{4x}|\frac{1}{2}11\}\{C_{3,111}|000\}\{S_{4x}|\frac{1}{2}11\}^{-1}&=\{C_{2x}|\frac{3}{2}\frac{3}{2}0\}\{C_{3,111}|000\}^2\{\mathcal{R}|001\},\nonumber\\
\{S_{4x}|\frac{1}{2}11\}\{C_{2y}|0\frac{1}{2}\frac{1}{2}\}\{S_{4x}|\frac{1}{2}11\}^{-1}&=\{C_{2y}|0\frac{1}{2}\frac{1}{2}\}\{C_{2x}|\frac{3}{2}\frac{3}{2}0\}\{E|113\},\nonumber\\
\{C_{2x}|\frac{3}{2}\frac{3}{2}0\}\{C_{2y}|0\frac{1}{2}\frac{1}{2}\}&=\{C_{2y}|0\frac{1}{2}\frac{1}{2}\}\{C_{2x}|\frac{3}{2}\frac{3}{2}0\}\{\mathcal{R}|330\}. 
\label{eq:rels220-2}
\end{align}
Now let $-G_3^{-1},iG_1, -iG_2$ and $e^{i\pi/4}G_4$ be the matrix representatives for $\{C_{3,111}|000\},\{C_{2x}|\frac{3}{2}\frac{3}{2}0\},\{C_{2y}|0\frac{1}{2}\frac{1}{2}\}$ and $\{S_{4x}|\frac{1}{2}11\}$, respectively. Again, we have chosen phases for convenience; they represent a unitary transformation of the operators.
Then Eqs.~(\ref{eq:rels220}) and (\ref{eq:rels220-2}) give us the matrix relations between $G_3,G_1,G_2,G_4$:
\begin{equation}G_3^3=G_1^2=G_2^2=G_4^4=1\label{eq:squares220}\end{equation}
and
\begin{equation} G_3G_2=G_1G_3,\quad G_3G_1=G_1G_2G_3,\quad [G_1,G_2]=0,\quad G_4^2=G_1,\quad  G_4G_3^2G_4^{-1}=G_1G_3,\quad G_4G_2G_4^{-1}=G_2G_1 .\label{eq:cond220}
%\tilde{G_4}^2&=\tilde{G_1} \nonumber\\
%\tilde{G_3}^{-1}\tilde{G_2}\tilde{G_3}&=-\tilde{G_1}\nonumber\\
%\tilde{G_3}^{-1}\tilde{G_1}\tilde{G_3}&=i\tilde{G_1}\tilde{G_2}\nonumber\\
%\tilde{G_4}\tilde{G_3}\tilde{G_4}^{-1}&=-i\tilde{G_1}\tilde{G_3}^2\nonumber\\
%\tilde{G_4}\tilde{G_2}\tilde{G_4}^{-1}&=-i\tilde{G_2}\tilde{G_1}\nonumber\\
%\tilde{G_1}\tilde{G_2}&=\tilde{G_2}\tilde{G_1}
\end{equation}
The equalities between $G_3, G_1$ and $G_2$ are exactly the condition in Eq.~(\ref{eq:condition}) for a 3d irrep. Furthermore, the equalities involving $G_4$ are compatible with the 3d irrep; one can see this by arguments similar to those in the previous sections: if $\psi$ is an eigenstate of $G_1$ and $G_2$ with eigenvalues $(\lambda_1,\lambda_2)$, then $G_3\psi$ is an eigenstate with values $(\lambda_2,\lambda_1\lambda_2)$ and $G_3^2\psi$ an eigenstate with values $(\lambda_1\lambda_2,\lambda_1)$. From Eq.~(\ref{eq:cond220}), $G_4\psi$ is an eigenstate with eigenvalues $(\lambda_1,\lambda_1\lambda_2)$. Thus, taking  $\lambda_1=1,\lambda_2=-1$, we construct the two 3d irreps:
\begin{equation}
G_1=\begin{pmatrix}1&0&0\\0&-1&0\\0&0&-1\end{pmatrix},G_2=\begin{pmatrix}-1&0&0\\0&-1&0\\0&0&1\end{pmatrix},G_3=\begin{pmatrix}0&0&1\\1&0&0\\0&1&0\end{pmatrix},G_4= \pm \begin{pmatrix} 1&0&0\\0&0&-1\\0&1&0\end{pmatrix},
\label{eq:irrep220}
\end{equation}
which satisfy Eqs.~(\ref{eq:squares220}) and (\ref{eq:cond220}). Hence, Eq.~(\ref{eq:irrep220}) gives the two inequivalent 3d irreps for these space groups. As in the previous section, the other choices of $\lambda_{1,2}$ yield equivalent irreps. Furthermore, as explained at the end of the previous section in detail, if, for example, one constructed a representation where $G_4\psi$ was orthogonal to $\psi$, but shared the same eigenvalues, this representation would be reducible.

\subsection{Time reversal symmetry}\label{sec:TR3}

We consider the primitive cubic and body-centered cubic lattices separately: in the former, the action of time reversal leaves the $(\pi,\pi,\pi)$ point invariant, up to an integer reciprocal lattice translation, while in the latter case, it does not. To understand this, we refer to reciprocal lattice vectors in Table~\ref{tab:vectors}. In the primitive cubic case, $(\pi,\pi,\pi) = \frac{1}{2}(\mathbf{g}_1+\mathbf{g}_2+\mathbf{g}_3)$ (in units $a=1$). Under time reversal, $(\pi,\pi,\pi) \rightarrow \mathcal{T}(\pi,\pi,\pi) = -\frac{1}{2}(\mathbf{g}_1+\mathbf{g}_2+\mathbf{g}_3)$. Evidently, $(\pi,\pi,\pi)$ and $\mathcal{T}(\pi,\pi,\pi)$ differ by the integer lattice translation $\mathbf{g}_1+\mathbf{g}_2+\mathbf{g}_3$. On the other hand, in the body-centered cubic case, $(\pi,\pi,\pi) = \frac{1}{4}(\mathbf{g}_1+\mathbf{g}_2+\mathbf{g}_3)$ and $\mathcal{T}(\pi,\pi,\pi)=-\frac{1}{4}(\mathbf{g}_1+\mathbf{g}_2+\mathbf{g}_3)$. Hence, $(\pi,\pi,\pi)$ and $\mathcal{T}(\pi,\pi,\pi)$ differ by a half-integer reciprocal lattice translation $\frac{1}{2}(\mathbf{g}_1+\mathbf{g}_2+\mathbf{g}_3)$. 

In all cases, time reversal commutes with all space group symmetries. 
In addition, $\mathcal{T}^2=-1$, since we deal exclusively with spin-orbit coupled fermions. 
Consequently, by Kramer's theorem, $\mathcal{T}\psi$ is always orthogonal to $\psi$.
%Consequently, $\mathcal{T}\psi$ is always orthogonal to $\psi$: if we write $\mathcal{T}\psi = \alpha \psi + \beta \psi_\perp$, for some $\psi_\perp$ orthogonal to $\psi$ and $|\alpha|^2+|\beta|^2=1$, then $\mathcal{T}^2\psi = -\psi = |\alpha|^2\psi + \alpha^*\beta\psi_\perp + \beta^*\mathcal{T}\psi_\perp$. It follows that $\beta \neq 0$ and $\mathcal{T}\psi_\perp= \frac{1}{\beta^*}\left((-1-|\alpha|^2)\psi - \alpha^*\beta\psi_\perp\right)$. Since $\mathcal{T}^2=-1$ is unitary, $\langle\mathcal{T}\psi_\perp|\mathcal{T}\psi_\perp\rangle=1= (1+|\alpha|^2)^2/|\beta|^2+|\alpha|^2$; this is only satisfied if $\alpha=0$.

In the following, we will expand upon Sec~\ref{sec:condition}, using the following notation: let $\psi_1$ be a simultaneous eigenstate of $G_1$ and $G_2$ with eigenvalues $1$ and $-1$, which we write as a pair $(1,-1)$. Then $\psi_2 \equiv G_3\psi_1$ and $ \psi_3\equiv G_3^2\psi_1$ are also simultaneous eigenstates of $G_1$ and $G_2$ with eigenvalues $(-1,-1)$ and $(-1,1)$.

\subsubsection{Primitive cubics: space groups 198, 205, 212 and 213}

As mentioned above, in these space groups the $R$ point is invariant under time reversal symmetry. 
Now consider the eigenvalues of the states $\mathcal{T}\psi_i$. 
Since $[\mathcal{T},G_1]=[\mathcal{T},G_2]=0$ and the eigenvalues of $G_1$ and $G_2$ are real, $\mathcal{T}\psi_i$ has the same eigenvalues of $G_1$ and $G_2$ as $\psi_i$ and thus distinct from $\psi_{j\neq i}$. Hence $\langle \psi_{j\neq i}|\mathcal{T}\psi_i\rangle=0$. Furthermore, by Kramer's theorem, $\langle \psi_i|\mathcal{T}\psi_i\rangle =0$. Thus, the six states $\psi_i,\mathcal{T}\psi_i$ are orthonormal. Consequently, these space groups can generically host six-fold degeneracies at the $(\pi,\pi,\pi)$ point. 

\subsubsection{Body-centered cubics: space groups 199, 206, 214, 220 and 230}\label{subsec:199TR}

In these space groups the $P$ point is not invariant under time reversal symmetry. There are two possible cases:
\begin{enumerate}
\item If the space group contains an element $\mathcal{G}_0$ not in the little group of the $P$ point such that $\mathcal{G}_0(\pi,\pi,\pi)=-(\pi,\pi,\pi)$, then $\mathcal{T}\mathcal{G}_0$ leaves the $P$ point invariant, up to an integer reciprocal lattice translation. Then we must consider whether $\mathcal{T}\mathcal{G}_0$ permutes the states in the 3d irrep, in which case the 3-fold degeneracy is unchanged, or whether it takes them outside the 3d irrep, in which case the degeneracy doubles. In either cases, there is a distinct point in the Brillouin zone, $-P$, which has the same degeneracy as the $P$ point. 
%Below we consider this question for space groups 206, 214, and 230.
\item If the space group does not contain such an element, then the degeneracy at the $P$ point remains 3-fold and the three states are equal in energy to another 3-fold degeneracy at the $-P$ point. 
%This describes space groups 199 and 220.
\end{enumerate}

We now consider space groups 206, 214 and 230, which have an element $\mathcal{G}_0$ satisfying $\mathcal{G}_0(\pi,\pi,\pi) = -(\pi,\pi,\pi)$:
\begin{description}
\item[SG 206] In this group, $\mathcal{G}_0 = \{I|000\}$. We compute the relevant group relations:
\begin{align} \{I | 000\} \{C_{2x}|\bar{\frac{1}{2}}\frac{1}{2}0\} &=\{C_{2x}|\bar{\frac{1}{2}}\frac{1}{2}0\}\{I|000\}\{E|112\}, \nonumber\\
\{I | 000\}\{C_{2y}|0\frac{1}{2}\bar{\frac{1}{2}}\}&=\{C_{2y}|0\frac{1}{2}\bar{\frac{1}{2}}\}\{I|000\}\{E|211\},\nonumber\\
\{I|000\}^2&=\{E|000\}.
\label{eq:rels206T}
\end{align}
Referring to the matrix representations $G_3,G_1,G_2$ in Sec~\ref{sec:199}, and defining $G_0=U\mathcal{K}$ to be the matrix representation of $\mathcal{T}\{I|000\}$, where $\mathcal{K}$ is the anti-unitary complex conjugation operator, Eq.~(\ref{eq:rels206T}) translates to matrix relations:
\begin{equation} [G_0,G_1]=[G_0,G_2],\quad G_0^2=-1 .\end{equation}
Thus, the role of $G_0$ is identical to the role of time reversal in the primitive cubic case: it squares to $-1$ and leaves the eigenvalues of $G_1$ and $G_2$ invariant. Consequently, time reversal leads to a 6-fold degeneracy at the $P$ point. Furthermore, since the entire Brillouin zone is invariant under the combination $\mathcal{T}\mathcal{G}_0$, all bands in the spectrum are doubly degenerate.
\item[SG 214] Here $\mathcal{G}_0 = \{C_{2,1\bar{1}0}|\frac{1}{2}\frac{1}{2}\frac{1}{2}\}$. We compute the relevant group relations:
\begin{align}
\{C_{2,1\bar{1}0}|\frac{1}{2}\frac{1}{2}\frac{1}{2}\}\{C_{2x}|\bar{\frac{1}{2}}\frac{1}{2}0\} &= \{C_{2y}|0\frac{1}{2}\bar{\frac{1}{2}}\}\{C_{2,1\bar{1}0}|\frac{1}{2}\frac{1}{2}\frac{1}{2}\}\{\mathcal{R}|101\}, \nonumber\\
\{C_{2,1\bar{1}0}|\frac{1}{2}\frac{1}{2}\frac{1}{2}\}\{C_{2y}|0\frac{1}{2}\bar{\frac{1}{2}}\} &= \{C_{2x}|\bar{\frac{1}{2}}\frac{1}{2}0\}\{C_{2,1\bar{1}0}|\frac{1}{2}\frac{1}{2}\frac{1}{2}\}\{\mathcal{R}|011\}, \nonumber\\
\{C_{2,1\bar{1}0}|\frac{1}{2}\frac{1}{2}\frac{1}{2}\}^2&=\{\mathcal{R}|000\}. \label{eq:rels214T}
\end{align}
Referring to the matrix representations $G_3,G_1,G_2$ in Sec~\ref{sec:199}, and now defining $G_0=U\mathcal{K}$ to be the matrix representation of $\mathcal{T}\{C_{2,1\bar{1}0}|\frac{1}{2}\frac{1}{2}\frac{1}{2}\}$, Eq.~(\ref{eq:rels214T}) translates to matrix relations:
\begin{equation} G_0G_1=G_2G_0,\quad G_0G_2=G_1G_0,\quad G_0^2=1 .\label{eq:TR2-214}
\end{equation}
Thus, $G_0\psi_2$ has the same eigenvalues as $\psi_2$, while $G_0\psi_{1,3}$ has the same eigenvalues as $\psi_{3,1}$. Since $G_0^2=1$, Kramer's theorem does not apply, and $G_0$ permutes the group elements as follows: $G_0\psi_2=\psi_2, G_0\psi_{1,3}=\psi_{3,1}$. Thus, the 3 band crossing remains 3-fold degenerate when time reversal symmetry is present.
\item[SG 230] As in SG 206, $\mathcal{G}_0 = \{I|000\}$. The relevant group relations are:
\begin{align} 
\{I|000\}\{C_{2y}|0\frac{1}{2}\frac{1}{2} \}&=\{C_{2y}|0\frac{1}{2}\frac{1}{2} \} \{I|000\}\{E|0\bar{1}\bar{1}\}, \nonumber\\
\{I|000\}\{C_{2x}|\frac{3}{2}\frac{3}{2}0\} &= \{C_{2x}|\frac{3}{2}\frac{3}{2}0\}  \{I|000\}\{E|\bar{3}\bar{3}0\},\nonumber\\
\{I|000\}^2&=\{E|000\}. \label{eq:rels230T}
\end{align}
We define $G_0=U\mathcal{K}$ to be the matrix representation of $\mathcal{T}\{I|000\}$. Since time reversal commutes with all group elements, the definitions of $G_1$ and $G_2$ in Sec~\ref{sec:220}, combined with Eq.~(\ref{eq:rels230T}), yields matrix relations:
 \begin{equation} G_0(-iG_2)=(-1)(-iG_2)G_0,\quad G_0(iG_1)=(-1)(iG_1)G_0, \quad G_0^2=-1,\end{equation} 
 or, invoking the fact that $G_0$ is anti-unitary:
 \begin{equation} [G_0,G_2]=[G_0,G_1] =0, \quad G_0^2=-1.\end{equation}
Thus, the role of $G_0$ is identical to the role of time reversal for the primitive cubic lattices and hence leads to a 6-fold degeneracy at the $P$ point.
\end{description}
%\endgroup
%\section{Symmetry analysis of 8d irreps}
\section{Symmetry analysis of 8d irreps}
\subsection{Space groups that have 8-fold degeneracies with time reversal}

In this appendix we prove a sufficient condition for a 4d irrep given three group generators. This applies to the little group at the point $(\pi,\pi,\pi)$ for space groups 130, 135, 222, and 223 and the little group at the $(0,2\pi,0)$ point for space group 230 (since SG 230 is body-centered cubic, $(0,2\pi,0)\neq (0,0,0)$, as can be seen from Table~\ref{tab:vectors}). We then show how the presence of time reversal can cause an 8-fold degeneracy (and must cause such a degeneracy for space groups 130 and 135). Last, we consider the little groups of space group 218 at the $(\pi,\pi,\pi)$ point and space group 220 at the $(0,2\pi,0)$ point, which have 2d irreps in addition to a 4d irrep. We show that in the presence of time reversal symmetry, the 4d irrep leads to an 8-fold degeneracy.

Note that there are many other space groups which have 4-fold degeneracies, but no others that lead to an 8-fold degeneracy in the presence of time reversal, as proven by an exhaustive search of Bradley and Cracknell\cite{Cracknell}.

\subsection{Sufficient condition for a 4d irrep}\label{sec:condition4}
Consider a particular high-symmetry point that is left invariant under a 4-fold rotation, a 2-fold rotation, and inversion.
Then the little group is generated by a 4-fold rotation or screw, a 2-fold rotation or screw and either inversion or inversion followed by a translation.
Let $G_1, G_2$ and $G_3$ be the respective matrix representations of these operators, perhaps after a unitary transformation. Then if $G_1, G_2, G_3$ satisfy
\begin{equation} G_1^4=-1, \quad [G_2,G_3]=\{G_3,G_1\}=0, \quad G_2G_1=G_1^3G_2\label{eq:condition4}\end{equation} 
there exists a 4d irrep. This is evident by considering an eigenstate of $G_1$ satisfying $G_1\psi = \lambda\psi$. Then the states $G_2\psi, G_3\psi$ and $G_2G_3\psi$ are also eigenstates of $G_1$ with eigenvalues $\lambda^3=-\lambda^*, -\lambda$ and $-\lambda^3=\lambda^*$, respectively. Since $\lambda^4=-1$, these four eigenvalues are distinct. Thus, any irrep of the little group is at least 4d. 

(Notice that Eq.~(\ref{eq:condition4}) can be phrased entirely in terms of commuting and anti-commuting operators, namely:
\begin{equation} G_1^4=-1, [G_2,G_3]=\{G_3,G_1\}=\{G_2,G_1^2\}=[G_2,\frac{1}{\sqrt{2}}\left(G_1+G_1^3\right)]=\{G_2,\frac{1}{\sqrt{2}}\left(G_1-G_1^3\right)\}\end{equation}
Notice that $\frac{1}{\sqrt{2}}\left(G_1\pm G_1^3\right)$ are both unitary operators.
Now consider a simultaneous eigenstate $\varphi$ of $G_2$ and $G_3$, with eigenvalues $\lambda_2$ and $\lambda_3$, respectively. 
Then $\frac{1}{\sqrt{2}}(G_1\pm G_1^3)\varphi$ are also simultaneous eigenstates of $G_2$ and $G_3$, with eigenvalues $\pm \lambda_2$ and $-\lambda_3$ respectively. 
Furthermore, $G_1^2\varphi$ is an eigenstate with eigenvalues $-\lambda_2,\lambda_3$. Thus, the four states $\varphi, \frac{1}{\sqrt{2}}(G_1\pm G_1^3)\varphi$ and $G_1^2\varphi$ are all linearly independent. Hence, any irrep must be at least 4d.)

If, instead of (\ref{eq:condition4}), the following conditions are satisfied:
\begin{equation} G_1^4=-1, \quad [G_2,G_3]=\{G_3,G_1\}=0, \quad G_2G_1=-G_1^3G_2, \label{eq:condition4-2}\end{equation} 
there must also exist a 4d irrep, using the same logic.
In this case, if $G_1\psi = \lambda\psi$, the states $G_2\psi, G_3\psi$ and $G_2G_3\psi$ have eigenvalues $-\lambda^3=\lambda^*, -\lambda$ and $\lambda^3=-\lambda^*$. 

We now comment on one implication of Eq.~(\ref{eq:condition4}) and (\ref{eq:condition4-2}):
momentarily, assume that the generators are symmorphic. Then since inversion commutes with all rotations, $G_3$ commutes with $G_1$. Thus, the fact that $G_1$ and $G_3$ anticommute in Eq.~(\ref{eq:condition4}) indicates that the generators are \emph{not} symmorphic. Thus, as in the case of 3-fold degeneracies, non-symmorphic operators play a crucial role in generating nontrivial irreps.

\subsection{Role of time reversal}\label{sec:TR4}

Before showing how Sec~\ref{sec:condition4} applies to specific space groups, we consider the action of time reversal at time reversal invariant points. Let $T=U\mathcal{K}$ be the matrix representation of the time reversal operator, including the anti-unitary operator $\mathcal{K}$. We want to know the eigenvalues under $G_1$ of $T\psi_i$, where $\psi_1\equiv \psi, \psi_2\equiv G_2\psi, \psi_3\equiv G_3\psi, \psi_4\equiv G_2G_3\psi$ are eigenstates of $G_1$ with eigenvalues $\lambda_i$ which satisfy $\lambda_1=-\lambda_2^*=-\lambda_3=\lambda_4^*$, according to the previous section. Without loss of generality, take $G_2^2=G_3^2=1$; this can be accomplished by a unitary transformation that does not affect Eq.~(\ref{eq:condition4}).

First, consider the case where $[T,G_1]=0$ and $T(G_2G_3)=e^{i\phi}(G_2G_3)T$, where $e^{i\phi}=\pm 1$. The phase $e^{i\phi}$ is necessary because $G_2G_3$ need only be a representation of a space group operator up to a unitary transformation; time reversal commutes with all space group operations, but not with all unitary transformations (in particular, time reversal does not commute with multiplication by the phase $i$).
 Then $G_1T\psi_i=TG_1\psi_i=\lambda_i^*T\psi_i$. Thus, $T\psi_i$ has the same eigenvalue as $G_2G_3\psi_i$ and a distinct eigenvalue from the other states, i.e., $\langle G_2G_3\psi_{j\neq i} | T\psi_i\rangle=0$ (recall that all four eigenvalues $\lambda, \lambda^*, -\lambda$ and $-\lambda^*$ are distinct because $\lambda^4=-1$.). Suppose $T\psi_i=e^{i\theta_i}G_2G_3\psi_i$. It follows that
\begin{equation}
-\psi_i=T^2\psi_i=Te^{i\theta_i}G_2G_3\psi_i=e^{-i\theta_i}TG_2G_3\psi_i=e^{-i\theta_i}e^{i\phi}G_2G_3T\psi_i=e^{i\phi}(G_2G_3)^2\psi_i=e^{i\phi}\psi_i
\end{equation}
Clearly this is only consistent if $e^{i\phi}=-1$. Hence, we reach the following conclusion: if $[T,G_2G_3]=0$ then all the time-reversed partners $T\psi_i$ are orthogonal to all the $\psi_j$ and hence there must be an 8-band crossing in the presence of time reversal; if $\{T,G_2G_3\}=0$ then there is no necessary 8-band crossing, unless the little group has additional generators.

If instead $\{T,G_1\}=0$, the same analysis follows, with $G_2G_3$ replaced by $G_2$. 

\subsection{Space groups 130 and 135}

Space groups 130 and 135 are primitive tetragonal. We are interested in the little group at the $A$ point $(\pi,\pi,\pi)$, which is generated by $\{C_{4z}|000\}, \{C_{2x}|\frac{1}{2}\frac{1}{2}0\}$ and $\{I|\frac{1}{2}\frac{1}{2}\frac{1}{2}\}$ for space group 130 and $\{C_{4z}|\frac{1}{2}\frac{1}{2}\frac{1}{2}\}, \{C_{2x}|\frac{1}{2}\frac{1}{2}0\}$ and $\{I|000\}$ for space group 135.
Calling the three generators $\mathcal{G}_1, \mathcal{G}_2, \mathcal{G}_3$, they obey the following relations:
\begin{equation}
\begin{array}{rcc}
&\text{SG 130}& \text{SG 135}\\
\hline
\mathcal{G}_1^4=&\{\mathcal{R}|000\} & \{\mathcal{R}|002\} \\
\mathcal{G}_2^2=&\{\mathcal{R}|100\} & \{\mathcal{R}|100\} \\
\mathcal{G}_3^2=&\{E|000\} & \{E|000\}\\
\mathcal{G}_2^{-1}\mathcal{G}_3^{-1}\mathcal{G}_2\mathcal{G}_3=&\{E|\bar{1}0\bar{1}\} & \{E|\bar{1}10\}\\
\mathcal{G}_1^{-1}\mathcal{G}_3^{-1}\mathcal{G}_1\mathcal{G}_3=&\{E|0\bar{1}0\} & \{E|\bar{1}1\bar{1}\}\\
\mathcal{G}_2^{-1}\mathcal{G}_1\mathcal{G}_2\mathcal{G}_1=&\{E|\bar{1}00\} & \{E|0\bar{1}0\}
\end{array}
\label{eq:rels130}
\end{equation}
Notice that the non-symmorphic translations in Eq.~(\ref{eq:rels130}) are the same in both space groups when evaluated at the $(\pi,\pi,\pi)$ point. 
Thus, if we define $G_1$ to be the matrix representation of $\{C_{4z}|000\}(\{C_{4z}|\frac{1}{2}\frac{1}{2}\frac{1}{2}\})$, $G_2$ the representation of $\{C_{2x}|\frac{1}{2}\frac{1}{2}0\}$ and $G_3$ the representation of $\{I|\frac{1}{2}\frac{1}{2}\frac{1}{2}\}(\{I|000\})$ for space group 130(135), then Eq.~(\ref{eq:rels130}) gives us the matrix relations in Eq.~(\ref{eq:condition4}) along with $G_2^2=G_3^2=1$. Thus, we see that without time reversal, these space groups necessarily possess a 4d irrep.

Now consider the presence of time reversal symmetry: since time reversal commutes with all space group elements, $[T,G_3]=[T,G_1]=[T,G_4]=0$. As described in Sec~\ref{sec:TR4}, this implies an 8-fold degeneracy at the $A$ point.

\subsection{Space groups 222, 223 and 230}
\label{sec:222}

Space groups 222 and 223 are primitive cubic while space group 230 is body-centered cubic. In space groups 222 and 223, we consider the little group at the $R$ point, while in space group 230 we consider the $H$ point $(0,2\pi,0)$; both points are time reversal invariant. All three little groups have similar generators, which we abbreviate as $\mathcal{G}_1, \mathcal{G}_2, \mathcal{G}_3, \mathcal{G}_4$ for convenience:
\begin{equation}
\begin{array}{c|ccc}
& \text{SG 222}& \text{SG 223}& \text{SG 230}\\
\hline
\mathcal{G}_1 & \{C_{4z}^-|000\} & \{C_{4z}^-|\frac{1}{2}\frac{1}{2}\frac{1}{2}\} & \{C_{4z}^+|0\frac{1}{2}0\}\\
\mathcal{G}_2 & \{C_{2x}|000\} &  \{C_{2x}|000\}  &  \{\mathcal{R}C_{2y}|1\frac{1}{2}\frac{1}{2}\}\\ 
\mathcal{G}_3 & \{I|\frac{1}{2}\frac{1}{2}\frac{1}{2}\} & \{I|000\} & \{I|000\}\\
\mathcal{G}_4 & \{C_{3,111}^-|010\} & \{C_{3,111}^-|010\} & \{C_{3,111}^+|111\} 
\end{array}
\end{equation}
These obey the following relations in each of the space groups:
\begin{equation}
\begin{array}{rccc}
&\text{SG 222}& \text{SG 223}& \text{SG 230}\\
\hline
\mathcal{G}_1^4=&\{\mathcal{R}|000\} & \{\mathcal{R}|002\} & \{\mathcal{R}|000\} \\
\mathcal{G}_2^2=&\{\mathcal{R}|000\} & \{\mathcal{R}|000\} & \{\mathcal{R}|101\} \\
\mathcal{G}_3^2=&\{E|000\} & \{E|000\} & \{E|000\}\\
\mathcal{G}_4^3=&\{\mathcal{R}|111\} & \{\mathcal{R}|111\} & \{\mathcal{R}|333\} \\
\mathcal{G}_1\mathcal{G}_2^{-1}\mathcal{G}_1\mathcal{G}_2=&\{E|000\} & \{E|000\} & \{E|011\}\\
\mathcal{G}_4^{-2}\mathcal{G}_1^{-3}\mathcal{G}_4\mathcal{G}_1=&\{\mathcal{R}|\bar{1}00\} & \{\mathcal{R}|\bar{1}\bar{1}1\}  & \{\mathcal{R}|\bar{2}\bar{2}\bar{3}\}\\
\mathcal{G}_4^{-1}\mathcal{G}_1^{-2}\mathcal{G}_4\mathcal{G}_2=&\{\mathcal{R}|00\bar{2}\} & \{\mathcal{R}|\bar{1}10\} & \{\mathcal{R}|\bar{1}\bar{2}\bar{1}\}\\
\mathcal{G}_1^{-1}\mathcal{G}_3^{-1}\mathcal{G}_1\mathcal{G}_3= &\{E|100\} &\{E|1\bar{1}\bar{1}\} & \{E|001\} \\
\mathcal{G}_2^{-1}\mathcal{G}_3^{-1}\mathcal{G}_2\mathcal{G}_3= & \{E|0\bar{1}\bar{1}\} & \{E|000\} & \{E|01\bar{1}\} \\
\mathcal{G}_4^{-1}\mathcal{G}_3^{-1}\mathcal{G}_4\mathcal{G}_3= & \{E|00\bar{2}\} & \{E|00\bar{2}\} & \{E|\bar{2}\bar{2}\bar{2}\} \\
\end{array}
\label{eq:rels222}
\end{equation}
Let $G_1,G_2, G_3$ and $G_4$ be the matrix representatives of $\mathcal{G}_1, \mathcal{G}_2, \mathcal{G}_3$ and $\mathcal{G}_4$, respectively. Then Eq.~(\ref{eq:rels222}) translates to the following matrix relations:
\begin{equation} -G_1^4=-G_2^2=G_3^2=G_4^3=1, G_1G_2=-G_2G_1^3,G_4G_1 = G_1^3G_4^2,G_4G_2=-G_1^2G_4, \{G_1,G_3\}=[G_2,G_3]=[G_3,G_4]=0.
\label{eq:matrixrels222}
\end{equation}
The generators $G_{1,2,3}$ satisfy Eq.~(\ref{eq:condition4-2}). Hence, there must be a 4d irrep. Following Sec~\ref{sec:condition4}, we choose an eigenstate $\psi$ with $\lambda=e^{i\pi/4}$. This yields
\begin{equation}
G_1=e^{i\pi/4}\begin{pmatrix}1&0&0&0\\0&-i&0&0\\0&0&-1&0\\0&0&0&i\end{pmatrix},\quad G_2=\begin{pmatrix}0&-1&0&0\\1&0&0&0\\0&0&0&-1\\0&0&1&0\end{pmatrix},\quad G_3=\begin{pmatrix}0&0&1&0\\0&0&0&1\\1&0&0&0\\0&1&0&0\end{pmatrix}.
\label{eq:reps222}
\end{equation}
The other three choices of $\psi$ would yield equivalent representations. We then need to show that $G_4$ does not increase the size of the irrep; this is done by brute force: the only three choices of a $4\times 4$ matrix which satisfies Eq.~(\ref{eq:matrixrels222}) are
\begin{equation} G_4=\alpha\begin{pmatrix}\rho&0\\0&\rho\end{pmatrix},
\label{eq:reps222-2}
\end{equation}
where $\alpha^3=e^{-i\pi/4}/(2\sqrt{2})$ and $\rho=\begin{pmatrix}1&-i\\-1&-i\end{pmatrix}$;
It is evident that the three values of $\alpha$ yield three distinct irreps because they each give a different value for the trace of $G_4$.

We now consider the action of time reversal; let $T=U\mathcal{K}$ be the matrix representative of $\mathcal{T}$, where $\mathcal{K}$ is the complex conjugation operation. 
We first try to apply the argument of Sec~\ref{sec:TR4}: since $G_2^2=-1$, we need to apply a unitary transformation $G_2 \rightarrow G_2'\equiv  iG_2$ . Then $T(G_2'G_3)=T(iG_2G_3)=-iG_2G_3T=-(G_2'G_3)T$; hence, following the notation of Sec~\ref{sec:TR4}, $e^{i\phi}=-1$. Thus, this argument does not tell us that time reversal symmetry will double the 4d irrep.

Instead, we apply a different argument: assume that $\mathcal{T}\psi_i$ is a linear combination of the $\psi_i$. Since $\mathcal{T}\psi_i$ has the same eigenvalues as $G_2\psi_i$ (and a distinct eigenvalue from the other three states), this implies that $T$ takes a form where each $2\times 2$ block is proportional to the corresponding block of $G_2$:
\begin{equation} T=\left(\begin{array}{cc}e^{i\theta_1} i\sigma_y & 0\\ 0 &e^{i\theta_2} i\sigma_y\end{array}\right)\mathcal{K}, \end{equation}
where $\theta_1$ and $\theta_2$ are phases.
We now check whether $[T,G_4]=0$; since both are block diagonal, we compute each $2\times 2$ block of $TG_4T^{-1}G_4^{-1}$:
\begin{equation} (e^{i\theta_i}i\sigma_y)\mathcal{K}\alpha\rho(-e^{i\theta_i}i\sigma_y) \mathcal{K} \frac{1}{\alpha}\rho^{-1} =  (e^{i\theta_i}i\sigma_y)\alpha^*\rho^* (-e^{-i\theta_i}i\sigma_y)\frac{1}{\alpha}\rho^{-1}=\frac{\alpha^*}{\alpha}\sigma_y\rho^* \sigma_y\rho^{-1}=\frac{\alpha^*}{\alpha}(i\mathbb{I}).
\end{equation}
Thus, $[T,G_4]=0$ if and only if $\alpha^*/\alpha=-i$. Since $\alpha^3=e^{-i\pi/4}/(2\sqrt{2})$, $[T,G_4]=0$ if and only if $\alpha=8^{-1/6}e^{15\pi i/12}$.
Hence, our assumption that $\mathcal{T}\psi_i$ is a linear combination of the $\psi_i$ is only true for the irrep where $\alpha=8^{-1/6}e^{15\pi i/12}$. Thus, this irrep remains 4d in the presence of time reversal. On the other hand, for the two irreps with $\alpha =8^{-1/6}e^{-i\pi/12}$ and $8^{-1/6}e^{7\pi i/12}$, the set of  $\mathcal{T}\psi_i$ must be linearly independent of the $\psi_i$ and, consequently, time reversal doubles the dimension of these irreps, so that they are 8d.

\subsection{Space groups 218 and 220}
\label{sec:218}

Space group 218 is a primitive cubic while 220 is a body centered cubic, but their little groups at the $R$ and $H$ points, respectively, have identical generating relations. Unfortunately, they require separate consideration from the other space groups in this section because they are not required to have 4d irreps. The generators of the respective little groups are given by,
\begin{equation}
\begin{array}{c|cc}
& \text{SG 218}& \text{SG 220}\\
\hline
\mathcal{G}_1 & \{C_{2y}|000\} & \{C_{2y}|0\frac{1}{2}\frac{3}{2} \}\\
\mathcal{G}_2 & \{C_{3,111}^- | 001\} &  \{C_{3,111}^- | 001\} \\
\mathcal{G}_3 & \{C_{2x}|001\} & \{C_{2x}|\frac{1}{2}\frac{1}{2}0\}\\
\mathcal{G}_4 & \{\sigma_{\bar{x}y}|\frac{1}{2}\frac{1}{2}\frac{1}{2} \} &  \{\sigma_{\bar{x}y}|\frac{1}{2}\frac{1}{2}\frac{1}{2} \}
\end{array}
\end{equation}
which satisfy
\begin{equation}
\begin{array}{rcc}
&\text{SG 218}& \text{SG 220}\\
\hline 
\mathcal{G}_1^2= & \{\mathcal{R}|000\} & \{\mathcal{R}|101\}\\
\mathcal{G}_2^3= & \{\mathcal{R}|111\} & \{\mathcal{R}|111\}\\
\mathcal{G}_3^2= & \{\mathcal{R}|000\} & \{\mathcal{R}|000\}\\
\mathcal{G}_4^2= & \{\mathcal{R}|111\} & \{\mathcal{R}|111\}\\
\mathcal{G}_1^{-1}\mathcal{G}_3^{-1}\mathcal{G}_1\mathcal{G}_3= & \{\mathcal{R}|00\bar{2}\} & \{\mathcal{R}|\bar{3}\bar{2}\bar{1}\}\\
\mathcal{G}_2^{-1}\mathcal{G}_1^{-1}\mathcal{G}_3^{-1}\mathcal{G}_2\mathcal{G}_3= & \{E|\bar{1}0\bar{1}\} & \{E|\bar{2}\bar{3}\bar{1}\}\\
\mathcal{G}_2^{-1}\mathcal{G}_3^{-1}\mathcal{G}_2\mathcal{G}_1= & \{E|\bar{1}00\} & \{E|001\}\\
\mathcal{G}_4^{-1}\mathcal{G}_1^{-1}\mathcal{G}_4\mathcal{G}_3= & \{\mathcal{R}|0\bar{1}\bar{2}\} & \{\mathcal{R}|\bar{1}\bar{2}0\}\\
\mathcal{G}_4^{-1}\mathcal{G}_2^{-2}\mathcal{G}_4\mathcal{G}_2= & \{\mathcal{R}|0\bar{1}0\} &  \{\mathcal{R}|0\bar{1}0\}
\end{array}
\label{eq:rels218}
\end{equation}
Letting $G_1,G_2, G_3, G_4$ be the matrix representations of $\mathcal{G}_1,\mathcal{G}_2,\mathcal{G}_3,\mathcal{G}_4$, respectively, these relations translate into the matrix relations:
\begin{equation} -G_3^2=-G_1^2=G_2^3=G_4^2=1, \{G_1,G_3\}=0, G_2G_3=G_3G_1G_2, G_2G_1=-G_3G_2, G_4G_3=G_1G_4, G_4G_2=G_2^2G_4.
\label{eq:matrixrels218}
\end{equation}
Notice that $iG_1, iG_3$ and $iG_1G_3$ all anticommute and square to $1$; from this observation, we guess that a 2d irrep exists, where $iG_1$ and $iG_3$ are Pauli matrices. Solving for $G_2$ and $G_4$ in Eq.~(\ref{eq:matrixrels218}) yields two 2d irreps:
\begin{equation} G_1=\begin{pmatrix} -q & 0 \\ 0 & q \end{pmatrix}, G_2=\frac{1}{2}\begin{pmatrix} -1-q & 1+q \\ -1+q & -1+q \end{pmatrix}
, G_3=\begin{pmatrix} 0 & -1 \\ 1 & 0 \end{pmatrix}, G_4=\frac{1}{\sqrt{2}}\begin{pmatrix} 1 & -q \\ q & -1 \end{pmatrix}
\label{eq:2182d},\end{equation}
where $q=\pm i$.

However, we would like to find the 4d irrep.
We start by noting that the combination $(G_3-G_1+G_3G_1)$ commutes with $G_2$; furthermore, since $(G_3-G_1+G_3G_1)^2=3$, this combination of operators is nonzero.
Our strategy to find the 4d irrep is to consider simultaneous eigenstates of $G_2$ and 
\begin{equation}M\equiv \frac{1}{\sqrt{3}}(G_3-G_1+G_3G_1).\end{equation}
Using Eq.~(\ref{eq:rels218}),
\begin{equation}[M,G_2]=\{M,G_4\}=0,\quad M^2=-1.\end{equation}
Now define
\begin{equation} M_\pm = \frac{1}{\sqrt{3}}\left( G_3-e^{\pm 2\pi i/3}G_1+e^{\pm 4\pi i/3}G_3G_1\right).
\label{eq:Mpmdef}
\end{equation}
The $M_\pm$ act as raising/lowering operators for $G_2$:
\begin{equation} G_2M_\pm = e^{\pm 2\pi i/3}M_\pm G_2,\end{equation}
and satisfy
\begin{equation} MM_\pm = \pm iM_\pm,\quad M_\pm M_\mp=\pm iM-1,\quad M_\pm^2=0.
\label{eq:Mpm218}
\end{equation}

Now consider a state $\psi$ which is a simultaneous eigenstate of $M$ and $G_2$ with eigenvalues $m,\lambda$, where $-m^2=\lambda^3=1$. Then $G_4\psi$ is also a simultaneous eigenstate of $M$ and $G_2$ with eigenvalues $-m$ and $\lambda^2$, respectively. Since $G_4\psi$ and $\psi$ have distinct eigenvalues, they are orthogonal.
Now consider $M_\pm\psi$: if $\psi$ is not annihilated by $M_\pm$, this state is an eigenstate of $M$ and $G_2$ with eigenvalues $\pm i$ and $\lambda e^{\pm 2\pi i/3}$. Similarly, if $G_4\psi$ is not annihilated by $M_\pm$, then $M_\pm G_4\psi$ is an eigenstate of $M$ and $G_2$ with eigenvalues $\pm i$ and $\lambda^2e^{\pm 2\pi i/3}$.

In the case where $m=i, \lambda=1$, the states $\psi, M_-\psi, G_4\psi, M_+G_4\psi$ all have distinct pairs of eigenvalues under $M$ and $G_2$ (in particular, their pairs of eigenvalues are $(i,1),(-i,e^{-2\pi i/3}),(-i,1),(i,e^{2\pi i/3})$). One might worry that $\psi(G_4\psi)$ would be annihilated by $M_-(M_+)$, but the middle equality in Eq.~(\ref{eq:Mpm218}) proves that this is not the case, since $M_+(M_-\psi)=-2\psi\neq 0$ and $M_-(M_+G_4\psi)=-2G_4\psi\neq 0$. Thus, the states $\psi, M_-\psi, G_4\psi, M_+G_4\psi$ transform as a 4d irrep corresponding to the matrix representation:
\begin{equation} 
G_2=\begin{pmatrix} 1&0&0&0\\0&e^{-2\pi i/3}&0&0\\0&0&1&0\\0&0&0&e^{2\pi i/3} \end{pmatrix},
G_4=\begin{pmatrix}0&0&1&0\\0&0&0&-e^{2\pi i/3}\\1&0&0&0\\0&-e^{-2\pi i/3}&0&0\end{pmatrix},
M=\begin{pmatrix}i&0&0&0\\0&-i&0&0\\0&0&-i&0\\0&0&0&i\end{pmatrix},
\end{equation} and 
\begin{equation}
 M_+ = \begin{pmatrix}0&-2&0&0\\0&0&0&0\\0&0&0&0\\0&0&1&0\end{pmatrix}, 
 M_- = \begin{pmatrix}0&0&0&0\\1&0&0&0\\0&0&0&-2\\0&0&0&0\end{pmatrix}.\end{equation}
Inverting Eq.~(\ref{eq:Mpmdef}) yields:
\begin{equation}G_3=\frac{1}{\sqrt{3}}\left(
\begin{array}{cccc}
 i & -2 & 0 & 0 \\
 1 & -i & 0 & 0 \\
 0 & 0 & -i & -2 \\
 0 & 0 & 1 & i \\
\end{array}
\right), G_1=-\frac{1}{\sqrt{3}}\left(
\begin{array}{cccc}
 i & -2e^{4\pi i/3} & 0 & 0 \\
e^{2\pi i/3} & -i & 0 & 0 \\
 0 & 0 & -i & -2e^{2\pi i/3} \\
 0 & 0 & e^{4\pi i/3} & i \\
\end{array}
\right).
\end{equation}
%which indicate that 1) if $M\psi = \lambda\psi$, then $M_\pm \psi$, if it is non-zero, is an eigenstate of $M$ with eigenvalue $\pm i$, 2) $M_\pm$ annihilates eigenstates of $M$ with eigenvalues $\mp i$, and 3) a state can only be raised/lowered once. 
%Applying this information to a state $\psi$ which is a simultaneous eigenstate of $M$ and $G_2$ with eigenvalues $i$ and $1$, respectively, we see that $\psi, M_-\psi, G_4\psi$ and $M_+G_4\psi$ all have distinct pairs of eigenvalues under $M$ and $G_2$ (in particular, their pairs of eigenvalues are $(i,1),(-i,e^{-2\pi i/3}),(-i,1),(i,e^{2\pi i/3})$. Hence, they are linearly independent states that transform as a 4d irrep of the little group.
One can check that other choices of $(m,\lambda)$ would yield either the same 4d irrep or a reducible 2d irrep equivalent to those in Eq.~(\ref{eq:2182d}).

Now consider the role of time reversal: both the $R$ point in space group 218 and the $H$ point in space group 220 are time reversal invariant points (see Fig.~\ref{bzfig}). We first suppose that time reversal, when acting on a state in the 4d irrep, keeps it within the irrep. Then, since time reversal commutes with all group elements, it takes a state with eigenvalues $(m,\lambda)$ under $M$ and $G_2$ to a state with eigenvalues $(m^*,\lambda^*)$. Considering the eigenvalues in the previous paragraph, this shows that $\mathcal{T}\psi=e^{i\theta}G_4\psi$, where $\theta$ is a phase that can depend on $\psi$. Applying time reversal twice, $-\psi = \mathcal{T}^2\psi = G_4^2\psi =\psi$, yields a contradiction. Hence, it must be that $\mathcal{T}\psi, \mathcal{T}M_-\psi, \mathcal{T}G_4\psi$ and $\mathcal{T}M_+G_4\psi$ are orthogonal to $\psi, M_-\psi, G_4\psi$ and $M_+G_4\psi$; thus, the presence of time reversal yields an 8-band crossing at the relevant points in momentum space.
%\section{Constructing $\mathbf{k}\cdot\mathbf{p}$ Hamiltonians}
\section{Constructing $\mathbf{k}\cdot\mathbf{p}$ Hamiltonians}
The degeneracy of a fermion at a single point in $\mathbf{k}$ space does not fully determine the properties of that fermion. The dispersion, and any degeneracies in the dispersion away from the given point are also fundamental properties of the fermion. In the following sections, we will show how to derive these properties.
\subsection{General strategy}
Here we construct the most general low energy $\mathbf{k}\cdot\mathbf{p}$ Hamiltonians near the degeneracy points $\mathbf{k}_0$ in the space groups mentioned in the main text. Our strategy will be to find the most general matrix function $H(\delta\mathbf{k})$ that transforms covariantly under all of the $\mathcal{G}_i$ in the little group $G^{\mathbf{k}_0}$ of $\mathbf{k}_0$. Explicitly, we demand for a given representation $\Delta$ of $G^{\mathbf{k}_0}$ that
\begin{equation}
\Delta(\mathcal{G}_i)H(\delta\mathbf{k})\Delta(\mathcal{G}_i)^{-1}=H(\mathcal{G}_i\delta\mathbf{k}) \label{eq:Htrans}
\end{equation}
for all unitary elements of the little group. If present, the antiunitary elements of the little group are generated by products with one additional element $\mathcal{T}\mathcal{G}_0$
where $\mathcal{T}$ is the time-reversal operation, and $\mathcal{G}_0$ is an element of the \emph{full} space group that maps $\mathbf{k}_0$ to $-\mathbf{k}_0$ (this may be the identity element if $-\mathbf{k}_0\equiv \mathbf{k}_0$. The precise matrix representation -- and the dimension of said representation -- of $\mathcal{T}\mathcal{G}_0$ depends on whether $\mathcal{G}_0^2$ is plus or minus the identity. In either case, we demand that
\begin{align}
\Delta(\mathcal{T}\mathcal{G}_0)\Delta(\mathcal{G}_i)\Delta(\mathcal{T}\mathcal{G}_0)^{-1}&=\Delta^*(\mathcal{G}_0\mathcal{G}_i\mathcal{G}_0^{-1}), \label{eq:tr1} \\
\Delta(\mathcal{T}\mathcal{G}_0)H(\delta\mathbf{k})\Delta(\mathcal{T}\mathcal{G}_0)^{-1}&=H^*(-\mathcal{G}_0\delta\mathbf{k}). \label{eq:tr2}
\end{align}
Eq.~(\ref{eq:tr1}) follows because time reversal must be represented as an antiunitary operator that commutes with all space group elements. We use the convention for antiunitary elements $\mathcal{T}\mathcal{G}_0$ that $\Delta(\mathcal{T}\mathcal{G}_0)$ is a unitary matrix, and must be accompanied by the complex conjugation operator $K$ in order to obtain the representative $G_0$ of $\mathcal{T}\mathcal{G}_0$. Eq.~(\ref{eq:tr2}) is the statement that $H$ must transform covariantly under the action of $\mathcal{T}\mathcal{G}_0$. 

Our strategy will be as follows. For each of the space groups and high symmetry points discussed in the main text, we will construct a representation 
\begin{equation}
\{G_0,G_1,G_2,\dots\}=\{\Delta(\mathcal{T}\mathcal{G}_0)K,\Delta(\mathcal{G}_1),\Delta(\mathcal{G}_2),\dots\}
\end{equation}
of the little group of $\mathbf{k}_0$. We will then solve Eqs.~(\ref{eq:Htrans}) and (\ref{eq:tr2}) for the most general Hermitian matrix function of $\delta\mathbf{k}$ of the appropriate dimension consistent with the representation $\Delta$; the dimensions of the representations we consider are those found in the previous sections.

\subsection{Notation}
Before we proceed, it will be necessary to define Hermitian matrix bases in which to express our Hamiltonians. We will make use of the standard Pauli matrices
\begin{equation}
\sigma_0=\left(\begin{array}{cc}1&0\\0&1\end{array}\right),\sigma_1=\left(\begin{array}{cc}0&1\\1&0\end{array}\right),\sigma_2=\left(\begin{array}{cc}0&-i\\i&0\end{array}\right),\sigma_3=\left(\begin{array}{cc}1&0\\0&-1\end{array}\right). \label{Pauli}
\end{equation}
Additionally, for the three and sixfold degeneracies we will use the Gell-Mann matrices
\begin{align}
\lambda_0&=\left(\begin{array}{ccc}
1 & 0 & 0 \\
0 & 1 & 0 \\
0 & 0 & 1
\end{array}\right)
,\lambda_1=\left(\begin{array}{ccc}
0 & 1 & 0 \\
1 & 0 & 0 \\
0 & 0 & 0
\end{array}\right)
,\lambda_2=\left(\begin{array}{ccc}
0 & -i & 0 \\
i & 0 & 0 \\
0 & 0 & 0
\end{array}\right),\\
\lambda_3&=\left(\begin{array}{ccc}
1 & 0 & 0 \\
0 & -1 & 0 \\
0 & 0 & 0
\end{array}\right)
,\lambda_4=\left(\begin{array}{ccc}
0 & 0 & 1 \\
0 & 0 & 0 \\
1 & 0 & 0
\end{array}\right)
,\lambda_5=\left(\begin{array}{ccc}
0 & 0 & -i \\
0 & 0 & 0 \\
i & 0 & 0
\end{array}\right), \\
\lambda_6&=\left(\begin{array}{ccc}
0 & 0 & 0 \\
0 & 0 & 1 \\
0 & 1 & 0
\end{array}\right)
,\lambda_7=\left(\begin{array}{ccc}
0 & 0 & 0 \\
0 & 0 & -i \\
0 & i & 0
\end{array}\right)
,\lambda_8=\frac{1}{\sqrt{3}}\left(\begin{array}{ccc}
1 & 0 & 0 \\
0 & 1 & 0 \\
0 & 0 & -2
\end{array}\right).
\end{align}
These bases are convenient in that they satisfy the Hilbert-Schmidt orthonormality condition
\begin{align}
\frac{1}{2}\mathrm{tr}\left(\sigma_i^\dag\sigma_j\right)=\delta_{ij}, \mathrm{and}\;\;
\frac{1}{2}\mathrm{tr}\left(\lambda_\mu^\dag\lambda_\nu\right)=\delta_{\mu\nu}\;\;\mathrm{for} \;\;\mu,\nu>0.
\end{align}

\subsection{$\mathbf{k}\cdot\mathbf{p}$ Hamiltonians for Threefold degeneracies}
We start by analyzing the threefold degenerate space groups. These will also serve as the building blocks for the $6-$fold degeneracies.
\subsection{SG 199}\label{sec:199kdotp}
We start by considering the $P$ point in SG $199$. There are no antiunitary symmetries in $G^{\mathbf{k}_0}$ in this case. A minimal set of generators for the threefold representation can be chosen to be
\begin{align}
{G}_3\equiv \Delta(\{C_{31}^{-1}|101\})=\left(\begin{array}{ccc}
					0 & 0 & 1\\
					1 & 0 & 0\\
					0 & 1 & 0
					\end{array}\right),\;\;
{G}_1\equiv \Delta(\{C_{2x}|\bar{\frac{1}{2}}\frac{1}{2} 0\})=\left(\begin{array}{ccc}
					-1 & 0 & 0\\
					0 & -1 & 0\\
					0 & 0 & 1
					\end{array}\right). \label{eq:199irrep}
\end{align}
Note that $\mathcal{G}_2=\{C_{2y}|0\frac{1}{2}\bar{\frac{1}{2}}\}$ from Section \ref{sec:199} may be represented as $G_2=\Delta(\mathcal{G}_2)=G_3^{-1}G_1G_3^{-1}$. It is straightforward to verify that these matrices satisfy all the requirements Eq.~(\ref{eq:condition}) of an irreducible 3d representation. We now write
\begin{equation}
H_{199}(\delta\mathbf{k})=\sum_{i=0}^{8}f_i(\delta\mathbf{k})\lambda_i
\end{equation} 
and impose Eq.~(\ref{eq:Htrans}) as a constraint on the $f_i$. First, we find that $G_3$ and $G_1$ partition the Gell-Mann matrices into three classes
\begin{align}
\mathbf{Q}_1=\{\lambda_1,\lambda_4,\lambda_6\},\;\;
\mathbf{Q}_2=\{-\lambda_2,\lambda_5,-\lambda_7\},\;\;
\mathbf{Q}_3=\{\lambda_0,\lambda_3,\lambda_8\}.
\end{align}
In each of these classes the action of $G_3$ and $G_1$ by conjugation has a three dimensional representation $D_i$ satisfying,
\begin{align}
G_3(\mathbf{v}\cdot \mathbf{Q}_i)G_3^{-1}&=(D_i(G_3) \mathbf{v})\cdot \mathbf{Q}_i,\nonumber\\
G_1(\mathbf{v}\cdot \mathbf{Q}_i)G_1^{-1}&=(D_i(G_1) \mathbf{v})\cdot \mathbf{Q}_i
\end{align}
%\begin{align}
%P(a\lambda_1+b\lambda_4+c\lambda_6)P^{-1}&=D_1(P)\left(\begin{array}{c}a\\b\\c\end{array}\right),&& Q(a\lambda_1+b\lambda_4+c\lambda_6)Q^{-1}=D_1(Q)\left(\begin{array}{c}a\\b\\c\end{array}\right) \\
%P(-a\lambda_2+b\lambda_5-c\lambda_7)P^{-1}&=D_2(P)\left(\begin{array}{c}a\\b\\c\end{array}\right),&& Q(-a\lambda_2+b\lambda_5-c\lambda_7)Q^{-1}=D_2(Q)\left(\begin{array}{c}a\\b\\c\end{array}\right) \\
%P(a\lambda_0+b\lambda_3+c\lambda_8)P^{-1}&=D_3(P)\left(\begin{array}{c}a\\b\\c\end{array}\right),&& Q(a\lambda_0+b\lambda_3+c\lambda_8)Q^{-1}=D_3(Q)\left(\begin{array}{c}a\\b\\c\end{array}\right).
%\end{align}
for an arbitrary vector $\mathbf{v}$, where
\begin{align}
D_1(G_3)=D_2(G_3)&=\left(\begin{array}{ccc}0&1&0\\0&0&1\\1&0&0\end{array}\right), \;\;D_1(G_1)=D_2(G_1)=\left(\begin{array}{ccc}1&0&0\\0&-1&0\\0&0&-1\end{array}\right), \\
D_3(G_3)&=\left(\begin{array}{ccc}1&0&0\\0&-\frac{1}{2}&-\frac{\sqrt{3}}{2}\\0&\frac{\sqrt{3}}{2}&-\frac{1}{2} \end{array}\right), \;\;D_3(G_1)=\left(\begin{array}{ccc}1&0&0\\0&1&0\\0&0&1\end{array}\right).
\end{align}
We may now impose the constraints of symmetry in each class $\mathbf{Q}_i$ separately. For class $\mathbf{Q}_1$ we find
\begin{align}
f_1(\delta k_x,\delta k_y,\delta k_z)&=f_4(\delta k_y,\delta k_z,\delta k_x)=f_1(\delta k_x,-\delta k_y,-\delta k_z), \\
f_4(\delta k_x,\delta k_y,\delta k_z)&=f_6(\delta k_y,\delta k_z,\delta k_x)=-f_4(\delta k_x,-\delta k_y,-\delta k_z),\\
f_6(\delta k_x,\delta k_y,\delta k_z)&=f_1(\delta k_y,\delta k_z,\delta k_x)=-f_6(\delta k_x,-\delta k_y,-\delta k_z).
\end{align} These equations can be solved to any desired order in $\delta\mathbf{k}$. To linear order, we easily find
\begin{equation}
f_1=a_1\delta k_x, f_4=a_1\delta k_z, f_6=a_1\delta k_y.
\end{equation}
Carrying out this same procedure for classes $\mathbf{Q}_2$ and $\mathbf{Q}_3$, we find that to linear order:
\begin{equation}
H_{199}=E_0\lambda_0+\left(\begin{array}{ccc}
0 & a\delta k_x & a^*\delta k_y \\
a^*\delta k_x & 0 & a\delta k_z \\
a\delta k_y & a^*\delta k_z & 0
\end{array}\right), \label{eq:H199}
\end{equation}
where $a=a_1+ia_2$ is a complex parameter, and $E_0$ sets the zero of energy, which we now set to zero without loss of generality. For generic values of the parameter $a$, this Hamiltonian is fully gapped for all $\delta\mathbf{k}\neq 0$. To see this, we look at the characteristic polynomial
\begin{equation}
p_{199}=\det\left(H_{199}-\epsilon\lambda_0\right)=2\mathrm{Re}(a^3)\delta k_x\delta k_y\delta k_z+|a|^2|\delta\mathbf{k}|^2\epsilon-\epsilon^3. \label{eq:199charp}
\end{equation}
Away from $\delta\mathbf{k}=0$, the bands are non-degenerate, unless the characteristic polynomial factors as
\begin{equation}
p_{199}=-(\epsilon-\epsilon_*)^2(\epsilon+2\epsilon_*)
\end{equation}
for some $\epsilon_*$ (where the non-degenerate root is fixed at $\epsilon=-2\epsilon_*$, because $p_{199}$ contains no term of order $\epsilon^2$). Using this fact, and writing $a=|a|e^{i\phi}$, we find that the spectrum is degenerate only when
\begin{align}
|\delta k_x|=|\delta k_y|=|\delta k_z|,\\
\phi=\frac{n\pi}{3}, n\in\mathbb{Z}.
\end{align}
While the precise location of these gapless points in momentum \emph{and} parameter space depends on the fact that we truncated our Hamiltonian to linear order, the fact that they \emph{must} occur is universal, and they partition the space of Hamiltonians into distinct phases that differ in their Fermi surface topology.

First, we consider the case $0<\phi<\pi/3$. Since the spectrum is non-degenerate away from $\delta\mathbf{k}=0$ for all $\phi$ in this range, we focus on $\phi=\pi/6$ without loss of generality; this forces the first term in Eq.~(\ref{eq:199charp}) to vanish. We then see that the three energies in this case are given by $\epsilon_{\pm}=\pm|a||\delta \mathbf{k}|$ and $\epsilon_0=\mathcal{O}(|\delta\mathbf{k}|^2)$. Going to polar coordinates $(\delta k_x,\delta k_y,\delta k_z)=|\delta\mathbf{k}|(\sin\theta\cos\varphi,\sin\theta\sin\varphi,\cos\theta)$, the corresponding eigenfunctions are
\begin{equation}
\psi_{\pm}=\frac{1}{\sqrt{2}}\left(\begin{array}{c}
\pm \sin\theta \\
(\cos\varphi\pm i\cos\theta\sin\varphi)e^{-i\pi/6} \\
(\mp i\cos\theta\cos\varphi +\sin\varphi)e^{i\pi/6}
\end{array}
\right),
\psi_0=\left(\begin{array}{c}
i\cos\theta \\
e^{-i\pi/6}\sin\theta\sin\varphi  \\
- e^{i\pi/6}\sin\theta\cos\varphi\end{array}\right).
\end{equation}
To each of these eigenfunctions, the associated $U(1)$ Berry curvature is given by,
\begin{align}
\Omega_{\pm}&=i\nabla\times\langle\psi_{\pm}|\nabla\psi_{\pm}\rangle=\pm\frac{\delta\mathbf{k}}{|\delta\mathbf{k}|^3}, \nonumber \\
\Omega_0&=i\nabla\times\langle\psi_{\pm}|\nabla\psi_{\pm}\rangle=0. \label{eq:199berryc}
\end{align}
The integral of $\Omega_\pm$ over any surface enclosing the point $\mathbf{k}_0$ is
\begin{equation}
\nu_{\pm}=\frac{1}{2\pi}\oint \Omega_{\pm}\cdot d\mathbf{a}=\pm 2,
\end{equation}
and this will be true for all $\phi\in(0,\pi/3)$ by adiabatic continuity. This means that the threefold degeneracy point acts as a charge $+2$ monopole source of Berry flux, much in the same way as a double Weyl point (the sign is $+$ in our convention because the highest energy band carries positive Berry flux). Crucially, however, the presence of a third trivial band allows the dispersion near the degeneracy point to be linear, whereas for an ordinary double Weyl point the dispersion is necessarily quadratic\cite{Fang12}. Due to the Berry flux, a surface will have two Fermi arcs emanating from the projection of the point $\mathbf{k}_0$. There is, of course, a time-reversal conjugate source of Berry curvature at the point $-\mathbf{k}_0$. {Because time reversal both takes $\Omega(\mathbf{k})$ to $-\Omega(\mathbf{k})$ and reverses the direction of the unit normal $d\mathbf{a}$, it does not change the monopole charge. Hence the $3-$fold degeneracies at both points $\pm\mathbf{k}_0$ have the same monopole charge.} The surface projection of the point $-\mathbf{k}_0$ thus also harbors two Fermi arcs. The Nielsen-Ninomiya theorem then requires that there be compensating sinks of Berry curvature located somewhere in the Brillouin zone; their surface projections are the termination points for the four Fermi arcs. These provide the first example of Fermi arcs without the existence of Weyl points. 

A similar situation occurs when $\pi/3<\phi<2\pi/3$. In this case it is most convenient to perform the analysis with $\phi=\pi/2$. The spectrum in this case is identical to that above, except the Berry curvatures of the eigenfunctions are reversed. We have in this case that
\begin{equation}
\nu_{\pm}=\frac{1}{2\pi}\oint\Omega_{\pm}\cdot d\mathbf{a}=\mp2,
\end{equation}
and so in this phase the threefold degenerate point acts as a sink of Berry flux. This alternating pattern of source and sink continues as we rotate the Hamiltonian parameter $\phi$. At the transition points $\phi=n\pi/3$, the middle trivial band becomes degenerate with \emph{both} the $+$ and $-$ bands at different points of the Brillouin zone, and -- although itself trivial -- mediates the transfer of Berry flux between them.

{
\subsubsection{Magnetic Field -- Zeeman splitting of the spin-$1$ Weyl}
{ We now propose several experimentally accessible consequences of the spin-$1$ Weyl fermion. We first investigate the behavior of the degeneracy under externally applied symmetry-breaking perturbations.} However, since the spin-$1$ Weyl is not time-reversal symmetric, the following analysis generally applies to \emph{all} potential perturbations{ , such as strain or lattice distortion.}

Because the spin-$1$ fermion carries Chern number $\nu=\pm2$, an external potential will generically split the threefold degeneracy into an even number of Weyl points. When the magnetic field is aligned along a high-symmetry axis, however, a subgroup of the little group will be preserved, which places restrictions on the allowed configuration of Weyl points. Consider first a uniform external magnetic field aligned with the $[111]$ crystal direction. This breaks the symmetry group of the $P$ point down to the cyclic group generated by $\mathcal{G}_3=\{C_{31}^{-1}|101\}$. The most general Zeeman Hamiltonian $H_Z$ consistent with this symmetry group must satisfy
\begin{equation}
[H_Z,G_3]=0,
\end{equation}
where $G_3$ was defined in Eq. (\ref{eq:199irrep}). From this we deduce that $H_Z$ is diagonal in the basis of eigenvectors $\mathbf{v_i}$ satisfying $G_3\mathbf{v}_i=c_i\mathbf{v_i}$ which we denote by
\begin{equation}
\mathbf{v_1}=\frac{1}{\sqrt{3}}\left(\begin{array}{c} 1 \\ 1\\ 1 \end{array}\right),\;\;\mathbf{v_2}=\frac{1}{\sqrt{3}}\left(\begin{array}{c} e^{-2\pi i/3} \\ e^{2\pi i/3}\\ 1 \end{array}\right),\;\;\mathbf{v_1}=\frac{1}{\sqrt{3}}\left(\begin{array}{c}  e^{2\pi i/3} \\  e^{-2\pi i/3}\\ 1 \end{array}\right),
\end{equation}
with eigenvalues
\begin{equation}
c_1=1,\;\;c_2=e^{2\pi i/3},\;\;c_3=e^{-2\pi i/3}.	
\end{equation}
In this basis,
\begin{equation}
H_Z=B\left(\begin{array}{ccc} g_1 & 0 & 0\\0 & g_2 & 0\\0 & 0 &-g_1-g_2\end{array}\right)
\end{equation}
where, without loss of generality, we have kept the zero of energy fixed. Along the high symmetry { $(k,k,k)$} axis, the full Hamiltonian then takes the form,
\begin{align}
H(k)&=\left(\begin{array}{ccc} Bg_1+2|a|k\cos\phi & 0 & 0\\
0 & Bg_2+|a|k(\sqrt{3}\sin\phi-\cos\phi)& 0\\ 
0 & 0 & -B(g_1+g_2)-a|k|(\sqrt{3}\sin\phi+\cos\phi)\end{array}\right) \equiv \left(\begin{array}{ccc}E_1 & 0 & 0\\ 0 & E_2 & 0 \\ 0 & 0 & E_3\end{array}\right)
\end{align}
{ Let us index these energies by $a,b\in\{1,2,3\}$. We see that there are, in general, three Weyl points on this axis, when $E_a=E_b$ for each pair of energies, located at points $\delta \mathbf{k}_{ab}=k_{ab}(1,1,1)$. Note that these Weyl points do \emph{not} all occur between the same two bands. For instance, let $E_1>E_2>E_3$. Then the degeneracies at $\delta\mathbf{k}_{12}$ and $\delta\mathbf{k}_{13}$ occur between the highest energy band and the middle energy band, while the degeneracy at $\delta\mathbf{k}_{23}$ occurs between the middle energy band and the lowest energy band. In order for the Berry flux through a large Fermi surface to remain unaffected by small Zeeman splitting, there must exist additional Weyl points away from this high symmetry axis, and $C_3$ symmetry demands that they come in triplets. The total monopole charge of all Weyl points occuring between the upper two bands must be equal to the total monopole charge of all Weyl points occuring between the lower two bands; both of these must be equal to the original monopole charge of the underlying spin-$1$ fermion. Only in this case will the chern number of a large Fermi surface enclosing all of the degeneracies remain unchanged for all values of the Fermi energy. We will return to this issue shortly.

First, however, we compute the the linearized Hamiltonian around each of the Weyl points on the high symmetry axis. We find that the Hamiltonian expanded around each crossing $E_a=E_b$ can be written}
\begin{equation}
H_{ab}=(\mathbf{\delta k}-\delta\mathbf{k}_{ab})\cdot\mathbf{u}^{ab}\sigma_0+(\mathbf{\delta k}-\delta\mathbf{k}_{ab})_iA^{ab}_{ij}\sigma_j,
\end{equation}
with velocity matrices
\begin{align}
A^{12}&=|a|\left(\begin{array}{ccc}-\frac{1}{3}(\cos\phi+\sqrt{3}\sin\phi) & 0 & \frac{1}{6}(3\cos\phi-\sqrt{3}\sin\phi)\\
\frac{1}{6}(\cos\phi+\sqrt{3}\sin\phi) & \frac{1}{6}(\sqrt{3}\cos\phi+3\sin\phi) & \frac{1}{6}(3\cos\phi-\sqrt{3}\sin\phi) \\
\frac{1}{6}(\cos\phi+\sqrt{3}\sin\phi) & -\frac{1}{6}(\sqrt{3}\cos\phi+3\sin\phi) & \frac{1}{6}(3\cos\phi-\sqrt{3}\sin\phi)\end{array}\right),\nonumber \\
 A^{23}&=|a|\left(\begin{array}{ccc}\frac{2}{3}\cos\phi & 0 & \frac{1}{\sqrt{3}}\sin\phi \\
 -\frac{1}{3}\cos\phi & -\frac{1}{\sqrt{3}}\cos\phi & \frac{1}{\sqrt{3}}\sin\phi\\
 -\frac{1}{{3}}\cos\phi & \frac{1}{\sqrt{3}}\cos\phi & \frac{1}{\sqrt{3}}\sin\phi\end{array}\right),\nonumber \\
 A^{13}&=|a|\left(\begin{array}{ccc}-\frac{1}{3}(\cos\phi+\sqrt{3}\sin\phi) & 0 & \frac{1}{6}(3\cos\phi+\sqrt{3}\sin\phi)\\
\frac{1}{6}(\cos\phi-\sqrt{3}\sin\phi) & \frac{1}{6}(\sqrt{3}\cos\phi+3\sin\phi) & \frac{1}{6}(3\cos\phi+\sqrt{3}\sin\phi) \\
\frac{1}{6}(\cos\phi-\sqrt{3}\sin\phi) & \frac{1}{6}(\sqrt{3}\cos\phi-3\sin\phi) & \frac{1}{6}(3\cos\phi+\sqrt{3}\sin\phi)\end{array}\right)
\end{align}
and tiling vectors
\begin{equation}
\mathbf{u}^{12}=\frac{|a|}{6}(\cos\phi+\sqrt{3}\sin\phi)(1,1,1),\;\;\mathbf{u}^{23}=-\frac{|a|}{3}\cos\phi(1,1,1),\;\;\mathbf{u}^{13}=\frac{|a|}{6}(\cos\phi-\sqrt{3}\sin\phi)(1,1,1),
\end{equation}
{ where $a$ and $\phi$ are the parameters of the spin-$1$ Weyl Hamiltonian defined previously. Note that $g_1$ and $g_2$ do not affect the velocities or tilt vectors of the Weyl points due to the linear form of the unperturbed Hamiltonian; these g-factors only affect \emph{where} in the BZ the Weyls appear.} 

Recall that the Chern number of a Weyl point is given by $\nu^{ab}=\mathrm{sign}\left[\det A^{ab}\right]$, and its ``type'' is characterized by\cite{Soluyanov2015} $t^{ab}=\mathrm{sign}\left[\det\left(A^{ab}(A^{ab})^T-\mathbf{u}^{ab}(\mathbf{u}^{ab})^T\right)\right]$. Using the above expressions, we make some useful observations. First, we note that the Chern numbers of the three Weyl points are never all equal. { Without loss of generality, let us take the spin-$1$ Weyl to have monopole charge $+2$ (in the opposite case, the same argument will apply with all signs reversed). Then two of the Weyl points have positive Chern number, and one negative (and vice-versa); the two Weyls with positive Chern number appear between the highest energy pair of bands. This implies that there are exactly three Weyl points off of the high symmetry axis, and that they all have positive monopole charge as well, and occur between the lowest two energy bands. This ensures that the total monopole charge of degeneracies between each pair of bands is the same.} As a corollary, we note that at the exactly solvable points $\phi=(2n+1)\pi/6$, one of the matrices $A^{ab}$ becomes singular. This occurs when the three off-axis Weyl points coallesce with the opposite-chirality Weyl point on the symmetry axis, leading to a double-Weyl (whose Hamiltonian is degenerate to linear order). 

Finally, we note that the types of these Weyl points are given by
\begin{align}
t^{12}&=\mathrm{sign}\left(\frac{1}{54}\cos\phi(3\cos\phi-2\cos3\phi+3\sqrt{3}\sin\phi)\right)\\
t^{23}&=\mathrm{sign}\left(\frac{4}{27}\cos^4\phi(1-2\cos2\phi)\right),\\
t^{13}&=\mathrm{sign}\left(\frac{1}{54}\cos\phi(3\cos\phi-2\cos3\phi-3\sqrt{3}\sin\phi)\right),
\end{align}
which change sign as a function of $\phi$. These field-split Weyl points thus transition between type-I and type-II as a function of material parameters. In the linearized model, these transitions occur exactly at the points $\phi=n\pi/6$, although the location of these transitions can be moved as a function of higher-order terms in the Hamiltonian. Splitting of a spin-$1$ fermion with a magnetic field thus offers a possible experimental route to tunable type-II Weyl points.

We can carry out a similar analysis for a magnetic field aligned along the $[100]$ crystal axis. In this case the only remaining symmetry is $\mathcal{G}_1=\{C_{2x}|\bar{\frac{1}{2}}\frac{1}{2}0\}$. Because the representation matrix $G_1$ of this symmetry is already diagonal, it is easy to see that the most general Zeeman Hamiltonian (ignoring overall shifts in the zero of energy) can be written in terms of the Gell-Mann matrices as
\begin{equation}
H_Z=B(g_1\lambda_1+g_2\lambda_2+g_3\lambda_3+g_0\lambda_8)
\end{equation}
Along the high symmetry $\delta\mathbf{k}=k(1,0,0)$ we may write the full Hamiltonian as
\begin{equation}
H(k)=H_0+H_Z=Bg_0\lambda_8+Bg_3\lambda_3+(Bg_1+a_1k)\lambda_1+(Bg_2-a_2k)\lambda_2
\end{equation}
where we have introduced $a_1=|a|\cos\phi$ and $a_2=|a|\sin\phi$. Because the set $\{\lambda_1,\lambda_2\lambda_3,\lambda_8\}$ satisfy the same commutation algebra as $\{\sigma_1,\sigma_2,\sigma_3,\sigma_0\}$, we can trivially diagonalize the Hamiltonian along this axis, obtaining energies
\begin{equation}
E_{\pm}=\frac{1}{\sqrt{3}}Bg_0\pm\sqrt{(Bg_1+a_1k)^2+(Bg_2+a_2k)^2+B^2g_3^2},\;\; E_0=-\frac{2}{\sqrt{3}}g_0B
\end{equation}
There are two distinct parameter regions of interest. Let us define 
\begin{equation}
g_*=\frac{1}{|B|}\inf_{k}\left(E_+-E_-\right)=\sqrt{g_3^2+\frac{(a_2g_1-a_1g_2)^2}{|a|^2}}
\end{equation}
When $g_*<|\sqrt{3}g_0|$, there are two Weyl points on the $[100]$ axis, occurring between the $E_0$ band and either the $E_+$ or $E_-$ bands. Because the $E_0$ band is identically flat on this axis, these Weyls are at the transition between type-I and type-II; the inclusion of higher order terms will shift them into either of these two phases. Additionally, Chern number counting and symmetry requires that there exist two additional Weyl points between the $E_+$ and $E_-$ bands, that occur off of the high symmetry axis.

As $g_*$ increases towards $|\sqrt{3}g_0|$, the two Weyl points on the $[100]$ axis move together, until they coalesce in a double Weyl point at exact equality. For $g_*>|\sqrt{3}g_0|$, this double Weyl splits again into two single Weyl points, which move away from the $[100]$ axis. Hence in this parameter range, all four Weyl points exist away from the high-symmetry axis.
}

\subsubsection{Magnetic field -- Landau levels for SG 199}
Here, we consider adding an orbital magnetic field, $\mathbf{B}=B\hat{z}$, to the Hamiltonian (\ref{eq:H199}) for SG 199.
%\begin{equation}H_{199}(\phi,\mathbf{k})=\begin{pmatrix} 0 & e^{i\phi}k_x & e^{-i\phi}k_y \\ e^{-i\phi}k_x & 0 & e^{i\phi}k_z \\ e^{i\phi}k_y & e^{-i\phi}k_z & 0\end{pmatrix} \end{equation}
%and add a magnetic field $\mathbf{B}=B\hat{z}$. 
We make the canonical substitution $k_{x,y} \rightarrow \Pi_{x,y} \equiv k_{x,y}+eA_{x,y}$, where $ \left[ \Pi_x,\Pi_y\right]=-iB$, and define the raising and lowering operators:
\begin{equation} a=\frac{1}{\sqrt{2B}}\left(\Pi_x-i\Pi_y\right),\quad a^\dagger = \frac{1}{\sqrt{2 B}}\left(\Pi_x+i\Pi_y\right)\end{equation} which obey $ \left[ a,a^\dagger\right] = 1$. 
The Hamiltonian then takes the form,
\begin{equation} H(B,k_z) = \sqrt{\frac{B}{2}}\begin{pmatrix} 0 & e^{i\phi}(a+a^\dagger) & ie^{-i\phi}(a-a^\dagger) \\ e^{-i\phi}(a+a^\dagger) & 0 & e^{i\phi}\bar{k}_z \\ ie^{i\phi}(a-a^\dagger) & e^{-i\phi}\bar{k}_z & 0\end{pmatrix} \end{equation}
where $\bar{k}_z \equiv k_z/\sqrt{B/2}$.

An eigenstate, $\psi$, of $H(B,k_z)$, with energy $E\sqrt{B/2}$, can be written as a linear combination:
\begin{equation}\psi = \begin{pmatrix} \sum_{n\geq 0} a_n|n\rangle \\ \sum_{n\geq 0} b_n|n\rangle  \\\sum_{n\geq 0} c_n|n\rangle\end{pmatrix} 
\label{eq:psidef}
\end{equation}
Using $a|n\rangle = \sqrt{n}|n\rangle, a^\dagger|n\rangle = \sqrt{n+1}|n+1\rangle$, $H\psi = E\psi$ yields three equalities
%\begin{align}
%\sum_{n\geq 0} \left( \left( e^{i\phi}b_n + i e^{-i\phi}c_n \right)\sqrt{n}|n-1\rangle + \left( e^{i\phi}b_n - i e^{-i\phi}c_n \right)\sqrt{n+1}|n+1\rangle \right)&= E\sum_{n\geq 0}a_n|n\rangle \nonumber\\
%\sum_{n\geq 0} \left(  e^{-i\phi}a_n\left(\sqrt{n}|n-1\rangle + \sqrt{n+1}|n+1\rangle \right) + c_ne^{i\phi}\bar{k}_z|n\rangle \right) &=E \sum_{n\geq 0}b_n|n\rangle \nonumber\\
%\sum_{n\geq 0} \left( ie^{i\phi} a_n \left(\sqrt{n}|n-1\rangle - \sqrt{n+1}|n+1\rangle \right) + b_ne^{-i\phi}\bar{k}_z|n\rangle \right) &= E\sum_{n\geq 0}c_n|n\rangle
%\end{align}
%Shifting the index in the sums yields three equations 
that hold for all $n\geq 0$:
\begin{align}
\left( e^{i\phi}b_{n+1} + i e^{-i\phi}c_{n+1} \right)\sqrt{n+1} + \left( e^{i\phi}b_{n-1} - i e^{-i\phi}c_{n-1} \right)\sqrt{n} &= Ea_n\label{eq:a}\\
e^{-i\phi}\left(a_{n+1}\sqrt{n+1} + a_{n-1}\sqrt{n}\right) + c_ne^{i\phi}\bar{k}_z  &=E b_n \label{eq:b}\\
 ie^{i\phi} \left(a_{n+1}\sqrt{n+1} - a_{n-1}\sqrt{n} \right) + b_ne^{-i\phi}\bar{k}_z &= Ec_n\label{eq:c}
\end{align}
Combining equations (\ref{eq:b}) and (\ref{eq:c}),
\begin{align}
(E^2-\bar{k}_z^2)b_n &= a_{n+1}\sqrt{n+1}\left( Ee^{-i\phi} + i\bar{k}_z e^{2i\phi} \right) + a_{n-1}\sqrt{n}\left( Ee^{-i\phi} -i\bar{k}_z e^{2i\phi} \right) \label{eq:b2}\\
(E^2-\bar{k}_z^2)c_n &= a_{n+1}\sqrt{n+1}\left( iEe^{i\phi} + \bar{k}_z e^{-2i\phi} \right) + a_{n-1}\sqrt{n}\left( -iEe^{i\phi} + \bar{k}_z e^{-2i\phi} \right) \label{eq:c2}
\end{align}
Substituting into Eq~(\ref{eq:a}) yields an equation only in terms of the $a_n$, which must hold for all $n\geq 0$:
\begin{equation} a_n\left( E(E^2-\bar{k}_z^2-2(2n+1))-2\bar{k}_z\sin(3\phi) \right) = 2i\bar{k}_z\cos(3\phi)\left( a_{n+2}\sqrt{(n+1)(n+2)}- a_{n-2}\sqrt{n(n-1)}\right)\label{eq:a2}
\end{equation}

In the specific case when $\cos(3\phi) = 0$, we can solve the entire spectrum exactly. The right-hand side of Eq~(\ref{eq:a2}) disappears, yielding 
\begin{equation} a_n\left( E(E^2-\bar{k}_z^2-2(2n+1))-2\bar{k}_z\sin(3\phi) \right) =0 \,\, \forall \,\, n\geq 0\label{eq:a0}\end{equation}
There are two solutions to (\ref{eq:a0}): either all $a_n=0$ for all $n$ or $a_n=\delta_{nj}$ for some $j\geq 0$. 

First, when $a_n=0$ for all $n$, Eqs~(\ref{eq:b2}) and (\ref{eq:c2}) simplify to $(E^2-\bar{k}_z^2)b_n = (E^2-\bar{k}_z^2)c_n = 0$. 
Thus, $E=\pm \bar{k}_z$. 
Then Eqs~(\ref{eq:b}) and (\ref{eq:c}) yields the single equality, 
$c_ne^{i\phi} =\pm b_n$, which, when substituted into Eq~(\ref{eq:a}), yields, 
\begin{equation}
\left( 1 + s \right)b_{n+1}\sqrt{n+1} + \left( 1 - s \right)b_{n-1}\sqrt{n}=0
\end{equation}
where $s=(E/\bar{k}_z)(ie^{-3i\phi})$.
If $s=-1$, then all $b_n=0$ and there is no nontrivial solution for $\psi$, while if $s=+1$, $b_0$ does not enter the constraint and it follows that $b_{n>0}\neq 0$, while $b_0=0$. Thus, when $e^{-3i\phi} = \pm i$, there is a chiral (or anti-chiral) solution, 
\begin{equation} \psi = \begin{pmatrix} 0 \\ |0\rangle \\ \mp e^{-i\phi}|0\rangle \end{pmatrix}, E = \mp \bar{k}_z \label{eq:specialcasechiral}
\end{equation}

Second, when $a_n=\delta_{nj}$ for some $j$,
Eq~(\ref{eq:a0}) shows that the energy of the state $\psi$ is a solution to 
\begin{equation} E(E^2 - \bar{k}_z^2-2(2j+1))-2\bar{k}_z\sin(3\phi) = 0
\end{equation}
%If $j=0$, then $E(E^2-\bar{k}_z^2-2)=2\bar{k}_z\sin(3\phi)$.
Further, we can find the $b_n$ and $c_n$ by Eqs~(\ref{eq:b2}) and (\ref{eq:c2}); taking $e^{-3i\phi} = \pm i$, the only non-zero values are
\begin{align} 
%(E^2-\bar{k}_z^2)b_{j+1} &= \sqrt{j+1}e^{-i\phi} (E \mp \bar{k}_z) =(E^2-\bar{k}_z^2)\left( \mp e^{i\phi} c_{j+1}\right) \nonumber\\
b_{j+1} &= \frac{\sqrt{j+1}e^{-i\phi}}{E \pm \bar{k}_z} = \mp e^{i\phi} c_{j+1} \nonumber\\
%(E^2-\bar{k}_z^2)b_{j-1} &= \sqrt{j}e^{-i\phi} (E \pm \bar{k}_z)=(E^2-\bar{k}_z^2)\left(\pm e^{i\phi} c_{j-1}\right),
b_{j-1} &= \frac{\sqrt{j}e^{-i\phi}}{E \mp \bar{k}_z}=\pm e^{i\phi} c_{j-1},
\end{align}
where we have assumed $j\neq 0$ by assuming $|E|\neq | \bar{k}_z|$.
Eq~(\ref{eq:a2}) does not yield any additional constraints.

When $j=0$ and $E=s\bar{k}_z$, Eqs~(\ref{eq:a}), (\ref{eq:b}) and (\ref{eq:c}) yield two solutions, where $E=\frac{1}{2}\left( \sin(3\phi) \bar{k}_z \pm \sqrt{\bar{k}_z^2+8}\right)$ and $b_1=\frac{1}{2}Ee^{-i\phi}= ie^{-2i\phi}c_1$, $b_{n\neq 1 } = c_{n\neq 1} = 0$. Notice that these solutions are both chiral, in the sense that their slope never changes sign. Thus, it naively appears that we have found two more chiral solutions, in addition to that described by Eq~(\ref{eq:specialcasechiral}). However, since one solution has $E>0$ and the other $E<0$, these can be considered as two halves of a single chiral solution. The full spectrum is shown in Fig~\ref{fig:LL199}.

We can now compare the Landau levels of SG 199 when $\phi = \pi/2$ to those of single $(H=k_i\sigma_i)$ and double ($H=k_z\sigma_z + \left((k_x^2-k_y^2)\sigma_x + 2k_xk_y\sigma_y\right)/k_0$, where $k_0$ is some parameter with units of $k_{x,y,z}$) Weyl points:
\begin{align}
E_{\text{single Weyl}} &= \begin{cases} \pm \sqrt{k_z^2 + 2Bn}& n> 0 \\ -k_z & n=0\end{cases}\\
E_{\text{double Weyl}} &= \begin{cases} \pm \sqrt{k_z^2 + n(n-1)(2B)^2/k_0^2} & n>1\\ -k_z & n=0,1 \end{cases}
\end{align}
In SG 199, at $k_z=0$, $E\sim \sqrt{Bn}$, like a single Weyl point, but has two chiral modes, like a double Weyl point.
Furthermore, while the two chiral modes of the double Weyl point span continuously from 
$k_z\rightarrow -\infty$ to $k_z\rightarrow +\infty$, in SG 199, there are several chiral bands that, together, give the +2 Chern number. Last, there is a large number of bands near zero energy in SG 199 that do not exist for the double Weyl. 
The existence of these bands, with their large density of states, suggests that there is enough spectral weight for the system to undergo a superconducting transition, an impossible feat for the single or double Weyls. If this happens, the chiral modes cannot disappear. Depending on the structure of the gap between the $P$ and $-P$ points (FFLO vs $k=0$ gap), different superconducting scenarios, all interesting, can be envisioned. The possibility that spin-1 and spin-3/2 Weyls could be superconducting even at zero energy is fundamentally different from the spin-1/2 Weyls and will be investigated in future work.
%These differences should be observable via magnetic oscillations.

\begin{figure}
\includegraphics[width=.45\textwidth]{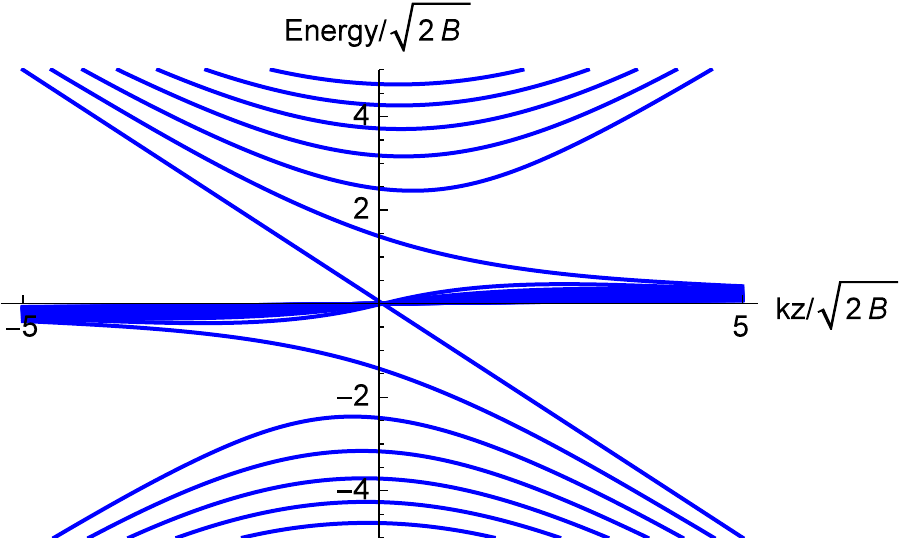}
\caption{Landau levels in SG 199 at $\phi = \pi/2$ when $B=B\hat{z}$. Higher order terms split the degeneracy at $k_z=0$, as shown in the main text. At this value of $\phi$, the spectrum does not depend on the direction of $B$ since, to linear order, the Hamiltonian is rotationally invariant.}
\label{fig:LL199}
\end{figure}

The general case with $\cos(3\phi) \neq 0$ is more complicated because Eq~(\ref{eq:a2}) mixes $a_n$ with $a_{n\pm 2}$. 
We find the spectrum exactly at $k_z=0$: the right-hand side of Eq~(\ref{eq:a2}) disappears, simplifying to:
\begin{equation} E_0(E_0^2 - 2(2n+1))a_n = 0 ,\,\ \forall \,\ n\geq 0,\end{equation}
where $E_0 \equiv E(k_z=0)$.

If $E_0 = 0$ then Eqs~(\ref{eq:b}) and (\ref{eq:c}) imply that $a_n=0$ for all $n$, while the $b_n$ and $c_n$ are constrained only by Eq~(\ref{eq:a}):
\begin{equation} \left(e^{i\phi}b_{n+1}+ie^{-i\phi} c_{n+1}\right)\sqrt{n+1} + \left(e^{i\phi}b_{n-1}-ie^{-i\phi} c_{n-1}\right)\sqrt{n}  = 0
\end{equation}
Define $d_{n\geq 0}=e^{i\phi}b_{n}+ie^{-i\phi} c_{n}$ and $e_{n\geq 2} = e^{i\phi}b_{n-2} -ie^{-i\phi}c_{n-2}$. The $n$ are now decoupled: 
\begin{equation} d_n\sqrt{n} + e_n\sqrt{n-1}, \text{ for } n\geq 2, \text{ and }d_1= 0  \end{equation}
%and\begin{equation} |b_n|^2 + |c_n|^2 = \frac{1}{2}\left( |d_n|^2 + |e_{n+2}|^2 \right)\end{equation}
Thus, there is a family of orthogonal states with $E_0=0$ defined by $d_n=\delta_{nj}=-\sqrt{1-1/j}e_n$, for $j>1$. Since $d_0$ is unconstrained, there is an additional state with $d_n=\delta_{n0}, e_n=0$.

If $E_0\neq 0$, then it must be that $a_n = \delta_{nj}$ for some $j\geq 0$ and $E_0 = \pm \sqrt{2(2j+1)}$, while the $b_n$ and $c_n$ are determined by Eqs~(\ref{eq:b2}) and (\ref{eq:c2}).

Thus, we have the complete spectrum at $\bar{k}_z = 0$: there is an infinite family of states with $E_0=0$ and a set of non-degenerate states with $E_0 = \pm \sqrt{2(2j+1)}$.

\subsection{SG 214}

We now to space group $214$. If we ignore time-reversal symmetry, the little group of the $P=(\pi,\pi,\pi)$ point here is identical to that of space group 199 from Sec.\ref{sec:199kdotp}, and consequently hosts the same representation and an identical Fermi arc structure originating from a non-Weyl point. To include time reversal, we look for an element of the full space group $\mathcal{G}_0$ such that $\mathcal{T}\mathcal{G}_0$ leaves the $P$ point invariant. Unlike SG $199$, SG $214$ has such an element, which we take to be
\begin{equation}
\mathcal{G}_0=\{C_{2,1\bar{1}0}|\frac{1}{2}\frac{1}{2}\frac{1}{2}\}
\end{equation}
As shown in Eqs.~(\ref{eq:rels214T}) and (\ref{eq:TR2-214}), we have $(\mathcal{T}\mathcal{G}_0^2)=1$ in this case, and hence the inclusion of time-reversal symmetry does not necessarily change the dimension of the $3d$ representation. We thus take $G_3,G_1$ to be given by Eq.~(\ref{eq:199irrep}). To find the antiunitary matrix representative $G_0$ of $\mathcal{T}\mathcal{G}_0$, we impose Eq.~(\ref{eq:tr1}) along with the constraint $G_0^2=1$. Using the relations of Eq.~(\ref{eq:rels214T}), along with
\begin{equation}
\mathcal{G}_0\mathcal{G}_3\mathcal{G}_0^{-1}=\mathcal{G}_3^{-1},
\end{equation}
 we find
\begin{equation}
G_0=\left(\begin{array}{ccc}1&0&0\\0&0&1\\0&1&0\end{array}\right)K. \label{eq:sg214TR}
\end{equation}
We can now repeat the same procedure as above, now with the additional constraint imposed by Eq.~(\ref{eq:tr2}). To do so we need the additional fact that, acting on $\mathbf{k}$ points, $\mathcal{T}\mathcal{G}_0$ takes $(\delta k_x,\delta k_y,\delta k_z)$ to $(\delta k_y, \delta k_x, \delta k_z)$.
 However, if we are content to consider topological properties, we need do no additional work -- the linearized Hamiltonian Eq.~(\ref{eq:H199}) is \emph{already} time-reversal covariant. We thus conclude that the threefold fermion in SG $214$ is topologically identical to that in SG $199$. 

{ Furthermore, because external perturbations (both magnetic and nonmagnetic) break the $G_0$ symmetry, the analysis of the splitting of the threefold degeneracy is nearly identical to that in SG 199 presented above. There are, however, two exceptions: both a time-reversal preserving (i.e. strain) perturbation aligned along the $[1\bar{1}0]$ axis and a magnetic field along the $[110]$ axis preserve the $G_0$ symmetry. The most general Hamiltonian $V$ consistent with the representation (\ref{eq:sg214TR}) is
\begin{equation}
V=V_1(\lambda_1+\lambda_4)+V_2(\lambda_2-\lambda_5)+V_3\lambda_6+V_4\left(\lambda_3-\frac{1}{\sqrt{3}}\lambda_8\right)
\end{equation}
However, because the energy bands do not have distinct eigenvalues under any symmetry operation along the $[110]$ axis preserved by $G_0$, band-crossings on this axis will be avoided; The spin-$1$ Weyl will split into four Weyls away from the high-symmetry axis. { As before, two these Weyls occurs between the highest energy pair of bands, and  two between the lowest energy pair of bands. Each have the same monopole charge, and are mapped into each other by the twofold $G_0$ symmetry.}
}
\subsection{SG 220}

Lastly, we examine the threefold degeneracy in SG $220$. Like SG $199$, the little group of the $G_3$ point here contains no antiunitary symmetries. As noted in Section \ref{sec:220}, there are two $3$d irreps of the little group $G^{\mathbf{k}_0}$. However since these representations differ only in the sign of the matrix representative $G_4$, and this overall sign does not affect the conjugation in Eq.~(\ref{eq:Htrans}), the two representation give rise to the same $\mathbf{k}\cdot\mathbf{p}$ Hamiltonian. It thus suffices to consider only one of them. We may take
\begin{align}
G_3&=\Delta(\{C_{3,111}|000\})=-i\left(\begin{array}{ccc}
					0 & 1 & 0\\
					0 & 0 &-1\\
					1 & 0 & 0
					\end{array}\right),\\
G_4&=\Delta(\{S_{4x}|\frac{1}{2}11\})=e^{i\pi/4}\left(\begin{array}{ccc}
					0 & 0 & 1\\
					0 & 1 & 0\\
					-1 & 0 & 0
					\end{array}\right)\\
\end{align}
as a minimal generating set for the 3d irrep of $G^{\mathbf{k}_0}$. This is unitarily equivalent to the representation presented in Section \ref{sec:220}, but is more convenient for our purposes. Before solving for the $\mathbf{k}\cdot\mathbf{p}$ Hamiltonian, let us make some observations. Assume we have a Hamiltonian $H_{220}$ that transforms covariantly under the little group $G^{\mathbf{k}_0}$. Consider then $H_{220}(k,k,k)$, the Hamiltonian evaluated along the line $\delta k_x=\delta k_y=\delta k_z=k$. The elements $\mathcal{G}_3$ and 
\begin{equation}
\mathcal{M}=\{\sigma_{\bar{x}y}|\frac{1}{2}\frac{1}{2}\frac{1}{2}\}=\mathcal{G}_4^2\mathcal{G}_3^{-1}\mathcal{G}_4
\end{equation}
of the little group $G^{\mathbf{k_0}}$ leave the entire line invariant, and hence their matrix representatives $M$ and $G_3$ commute with $H_{220}(k,k,k)$. These two matrices form a representation of the symmetry group of the line (in fact, this symmetry group is isomorphic to the group $S_3$ of permutations of three elements. $\mathcal{M}$ is a pairwise exchange, and $\mathcal{G}_3$ is a cyclic permutation.) The representation matrix for $\mathcal{M}$ is
\begin{align}
M&=\Delta(\mathcal{M})=e^{3i\pi/4}\left(\begin{array}{ccc}1&0&0\\0&0&1\\0&1&0\end{array}\right). \\
\end{align}
This representation, however, is reducible - the vector $\psi=(1,-1,-1)$ is an eigenvector of both $M$ and $G_3$. Thus, the group representation generated by $M$ and $G_3$ splits into a 1d irrep and a 2d irrep (these are, respectively, the ``sign representation'' and the ``rotation representation'' of $S_3$). Consequently, as we move along the line $\delta k_x=\delta k_y=\delta k_z$, the threefold degeneracy splits into one non-degenerate band, and two degenerate bands that disperse linearly away from this line. Thus, $H_{220}$ has a symmetry protected line node -- the two degenerate bands have differing eigenvalues under the mirror symmetry $\mathcal{M}$. A similar analysis holds for all the lines satisfying $|\delta k_x|=|\delta k_y|=|\delta k_z|$ by symmetry. The location of these line nodes is shown in Fig.~\ref{fig:nodes220-3}.

Anticipating these line nodes, we can now implement the constraints Eq.~(\ref{eq:Htrans}). To linear order, we find
\begin{equation}
H_{220}=a_1\left(\begin{array}{ccc}0&\delta k_y&\delta k_x\\\delta k_y & 0 &-\delta k_z\\ \delta k_x & -\delta k_z & 0 \end{array}\right),
\end{equation}
where $a_1\in\mathbb{R}$ is a free real parameter. From Eq.~(\ref{eq:H199}), we recognize this, after a rotation of the Brillouin zone, as $H_{199}$ at the transition point $\phi=0$. To analyze the line-nodes, we write
\begin{equation}
\delta\mathbf{k}_i=(k,k,k)+\delta\mathbf{q}
\end{equation}
and we perform degenerate perturbation theory in the small parameters $\delta\mathbf{q}$. We find that in the basis of degenerate states, the Hamiltonian can be written
\begin{align}
H_{220}&\approx a_1 k\sigma_0 +\delta q_i A_{ij}\sigma_j, \\
A_{ij}&=\left(\begin{array}{ccc}0 & 0 & \frac{2}{3}\\\frac{1}{\sqrt{3}} & 0 & \frac{1}{3} \\ -\frac{1}{\sqrt{3}} & 0 & \frac{1}{3}\end{array}\right).
\end{align}
This Hamiltonian describes an ordinary line node. As such, we know that the holonomy of the eigenstates $\psi_\pm$ along any loop encircling the line
 %(and respecting the $\mathcal{G}_4$ and $\mathcal{M}$ symmetries) 
 is given by
\begin{equation}
w_{\pm}=\mathcal{P}\exp\left(\oint\langle\psi_\pm|\nabla\psi_\pm\rangle\cdot d\ell\right)=-1.
\end{equation}
{ Using the relationship between the Berry phase and the Wannier-center polarization\cite{polarization}, we conclude that this nontrivial holonomy implies the existence of ``drumhead'' surface states\cite{Chan15}.

Finally, we note that due to the equivalence between $H_{199}$ and $H_{220}$, the analysis of the splitting of the threefold degeneracy can be deduced by setting $\phi=0$ in the analysis of SG 199 in Section \ref{sec:199kdotp} above.}

\subsection{$\mathbf{k}\cdot\mathbf{p}$ Hamiltonians for Sixfold degeneracies}
\subsection{SG $206$}\label{sec:sg206}
The simplest of the sixfold degeneracy points is that in SG $206$ at the $P$ point. If we ignore time-reversal symmetry, the little group of SG $206$ is isomorphic to that of SG $199$. However, the full space group contains the element $\mathcal{G}_0=\{I|000\}$, and hence $\mathcal{T}\mathcal{G}_0$ is an antiunitary element in $G^{\mathbf{k}_0}$. Furthermore, $(\mathcal{T}\mathcal{G}_0)^2=-1$, and it fixes \emph{every} point in the Brillouin zone. Thus, not only does the threefold irrep without time-reversal double to a 6d irrep (as shown in Section \ref{sec:TR3}, but, in addition, all energy bands of the $\mathbf{k}\cdot\mathbf{p}$ Hamiltonian are doubly degenerate. From Eq.~(\ref{eq:rels206T}), we see that $\mathcal{G}_0$ commutes with the generators of the little group up to the sum of a multiple of $4$ full lattice translations. Since at the $P$ point such translations must be represented by the identity, we have
\begin{align}
(G_3)_{206}&=(G_3)_{199}\oplus (G_3)_{199},\\
(G_1)_{206}&=(G_1)_{199}\oplus (G_1)_{199}, \\
G_0&=\left(\begin{array}{cc} 0 & \lambda_0 \\ -\lambda_0 & 0\end{array}\right)K.
\end{align}
Directly imposing these symmetries yields, to linear order,
\begin{equation}
H_{206}=E_0\lambda_0\sigma_0+\left(\begin{array}{cccccc}
0 & a\delta k_x & a^*\delta k_y&0&b\delta k_x &-b\delta k_y\\
a^*\delta k_x & 0 & a\delta k_z & -b\delta k_x&0&b\delta k_z\\
a  \delta k_y & a^*\delta k_z & 0 & b\delta k_y & -b\delta k_z & 0\\
0&-b^*\delta k_x &b^*\delta k_y & 0 & a^*\delta k_x & a\delta k_y \\
b^*\delta k_x & 0 & -b^*\delta k_z & a\delta k_x & 0 & a^*\delta k_z \\
-b^*\delta k_y & b^*\delta k_z & 0 & a^*\delta k_y & a\delta k_z & 0
\end{array}\right),
\end{equation}
where $a=a_1+ia_2$ and $b=b_1+ib_2$ are two complex parameters. We can gain more insight by examining the characteristic polynomial of $H_{206}$, given by,
\begin{equation}
p_{206}(a,b)=\det\left(H_{206}-\epsilon\lambda_0\otimes\sigma_0\right)=\left(\epsilon^3-(|a|^2+|b|^2)|\mathbf{\delta k}|^2-2a_1(a_1^2-3a_2^2-3|b|^2)\right)^2.
\end{equation}
after setting $E_0=0$.
Defining a new family of complex parameters,
\begin{align}
c(\alpha)&=a_1+i\mathrm{sgn}(a_2)\sqrt{a_2^2+\alpha^2|b|^2},\\
d(\alpha)&=\sqrt{1-\alpha^2}b,\end{align} 
we have,
\begin{equation}
p_{206}(a,b)=p_{206}(c(\alpha),d(\alpha))
\end{equation}
for all $\alpha\in[0,1]$. This means $H_{206}(\alpha)=H_{206}(c(\alpha),d(\alpha))$ are unitarily equivalent for all $\alpha$. We see that
\begin{align}
H_{206}(\alpha=0)&=H_{206}(a,b), \\
H_{206}(\alpha=1)&=\left.H_{199}\oplus H_{199}^*\right|_{a=c(1)}, \label{eq:206homotopy}
\end{align}
where $H_{199}$ was given in Eq.~(\ref{eq:H199}). This homotopy shows that $H_{206}$ and $H_{199}\oplus H_{199}^*$ are isospectral and topologically equivalent. There are distinct phases separated by gap-closing transitions exactly as in SG $199$. Thus, the six bands of eigenstates of $H_{206}$ organize themselves into three degenerate pairs; although there is no net Berry flux, two of these pairs consist of $\nu=\pm2$ bands. This can be seen most clearly by looking at the eigenvalues of the $SU(2)$ Wilson loop operators in each degenerate pair. Considering a sphere { centered on} the degeneracy point, we compute the $SU(2)$ Berry connection along a circle at polar angle $\theta$, in each of the fully gapped phases. Making use of Eq.~(\ref{eq:206homotopy}), we have, in any of the distinct phases,
\begin{align}
w_{\pm}(\theta)&=\mathcal{P}\exp(\oint\langle\psi_{\pm}^i|\nabla\psi_{\pm}^j\rangle)=\left(\begin{array}{cc}e^{i\Omega(\theta)} & 0\\0 &e^{-i\Omega(\theta)}\end{array}\right),\\
w_0(\theta)&=\mathcal{P}\exp(\oint\langle\psi_{0}^i|\nabla\psi_{0}^j\rangle)=\sigma_0,
\end{align}
where 
\begin{equation}
\Omega(\theta)=2\pi(1-\cos\theta)
\end{equation}
is the solid angle subtended by the spherical cap with opening angle $\theta$. We thus see that in the topologically nontrivial pair of bands, the Wilson loop eigenvalues wind twice, one chirally and one antichirally, as the loop is moved through the Brillouin zone. { When higher-order symmetry-allowed terms are included, the precise form of the Wilson loop matrices $w_{\pm}(\theta)$ will change. However, the degeneracies in the Wilson loop spectrum  -- and hence the windings of the eigenvalues -- are robust. This is due to the fact that we have chosen loops which respect the $C_2$ and antiunitary little group symmetries\cite{Aris-new}.}

{ We can also consider the effect of uniaxial uniform strain and magnetic field perturbations on the sixfold degeneracy. Due to the direct sum structure in Eq.~\ref{eq:206homotopy}, we know that any perturbation that breaks \emph{either} inversion or time-reversal symmetry will split the spin-$1$ Dirac point into two spin-$1$ Weyl points of opposite chirality, each of which will then split in accord with the analysis of Section \ref{sec:199kdotp}. This will be the case for almost every field direction. The exceptions are for strain and magnetic field aligned along the $x$-axis (and $y,z$ by symmetry). For this case, $G_1G_0$ remains a good antiunitary symmetry along the high-symmetry axis, and so we find three sets of line nodes by Kramers' theorem, rather than isolated point degenercies.}

\subsection{SG $230$}
The analysis of the sixfold degeneracy at the $P$ point in  SG $230$ is similar to that for SG $206$. If we ignore time reversal, the little groups of SG $220$ and SG $230$ at the $P$ point are identical. Because the full SG $230$ also contains inversion, however, in SG $230$ we have that $I\mathcal{T}$ is an antiunitary element of the little group. Taking into account the effect of inversion on non-symmorphic generators given in Eq.~(\ref{eq:rels230T}), in conjunction with Eq.~(\ref{eq:tr1}), we can take
\begin{align}(G_3)_{230}&=(G_3)_{220}\oplus (-(G_3)_{220}^*), \\
(G_4)_{230}&=(G_4)_{220}\oplus (-i(G_4)_{220}^*), \\
G_0&=\left(\begin{array}{cc} 0 & \lambda_0 \\ -\lambda_0 & 0\end{array}\right)K,
\end{align}
where the extra signs in the direct sum are needed to ensure that $G_0$ commutes (up to complex conjugation) with all unitary little group elements. Imposing the constraints from Eq.~(\ref{eq:Htrans}) yields, to linear order
\begin{equation}
H_{230}=E_0\lambda_0\sigma_0\left(\begin{array}{cccccc}
0 & a_1\delta k_y & a_1\delta k_x&0&b\delta k_y &-b\delta k_x\\
a_1\delta k_y & 0 & -a_1\delta k_z & -b\delta k_y&0&-b\delta k_z\\
a_1  \delta kx & -a_1\delta k_z & 0 & b\delta k_x & b\delta k_z & 0\\
0&-b^*\delta k_y &b^*\delta k_x & 0 & a_1\delta k_y & a_1\delta k_x \\
b^*\delta k_y & 0 & b^*\delta k_z & a_1\delta k_y & 0 & -a_1\delta k_z \\
-b^*\delta k_x & -b^*\delta k_z & 0 & a_1\delta k_x & -a_1\delta k_z & 0
\end{array}\right).
\end{equation}
We note that this is, up to a rotation, equivalent to $H_{206}$ with $a_2=0$. Thus, the homotopy constructed above is unchanged, and
\begin{equation}
H_{230}\sim \left.H_{199}\oplus H_{199}^*\right|_{a=a_1+i|b|}.
\end{equation}  
Naturally, the same Wilson loop { and perturbation} analysis as above applies.

\subsection{SG $205$}
Unlike the cases considered up to now, the $R$ point is time reversal invariant, and in SG $205$ it is also inversion symmetric. There are two different 6d representations of the little group which differ only in the sign of the inversion matrix, which is proportional to the identity in this case. Thus, these two mathematically distinct representations are physically indistinguishable in isolation, and both give the same structure for the $\mathbf{k}\cdot\mathbf{p}$ Hamiltonian; we will therefore analyze only one of them in detail. However, the existence of these two representations which differ in their inversion eigenvalues opens up the possibility of creating a \emph{twelve-fold} degeneracy, by tuning a single parameter, namely the energy splitting between two distinct 6$-$fold degeneracies.

For the representation of the little group we may take
\begin{align}
(G_3)_{205}&=\Delta(\{C_{3,111}^{-1}|010\})=(G_3)_{199}\oplus (G_3)_{199},\\
(G_1)_{205}&=\Delta(\{C_{2x}|\frac{1}{2}\frac{3}{2} 0\})=(G_1)_{199}\oplus (G_1)_{199}, \\
(G_4)_{205}&=\Delta(\{I|000\})=\lambda_0\otimes\sigma_0,\\					
T&=\left(\begin{array}{cc}
					0 & \lambda_0 \\
					-\lambda_0 & 0
					\end{array}\right)K.
\end{align}
Since the little group contains both inversion and time reversal, we expect not only that bands will come in degenerate pairs, but also that to lowest non-vanishing order the Hamiltonian will be quadratic. Indeed, imposing the constraints Eq.~(\ref{eq:Htrans}) we find
\begin{equation}
H_{205}=\left(\begin{array}{cccccc}
f_1(\delta\mathbf{k}) & a\delta k_y\delta k_z & a^*\delta k_x\delta k_z&0&b\delta k_y\delta k_z &-b\delta k_x\delta k_z\\
a^*\delta k_y\delta k_z & f_2(\delta\mathbf{k}) & a\delta k_x\delta k_y & -b\delta k_y\delta k_z&0&b\delta k_x\delta k_y\\
a  \delta k_x\delta k_z & a^*\delta k_x \delta k_y & f_3(\delta\mathbf{k}) & b\delta k_x\delta k_z & -b\delta k_x\delta k_y & 0\\
0&-b^*\delta k_y\delta k_z &b^*\delta k_x\delta k_z & f_1(\delta\mathbf{k}) & a^*\delta k_y\delta k_z & a\delta k_x\delta k_z \\
b^*\delta k_y\delta k_z & 0 & -b^*\delta k_x\delta k_y & a\delta k_y\delta k_z & f_2(\delta\mathbf{k}) & a^*\delta k_x\delta k_y \\
-b^*\delta k_x\delta k_z & b^*\delta k_x\delta k_y & 0 & a^*\delta k_x\delta k_z & a\delta k_x\delta k_y & f_3(\delta\mathbf{k})
\end{array}\right),
\end{equation}
with
\begin{align}
f_1(\mathbf{\delta k})=E_0+E_1\delta k_x^2+E_2\delta k_y^2&+E_3\delta k_z^2,\;\; f_2(\mathbf{\delta k})=E_0+E_1\delta k_z^2+E_2\delta k_x^2+E_3\delta k_y^2,\nonumber \\
f_3(\mathbf{\delta k}) &=E_0+E_1\delta k_y^2+E_2\delta k_z^2+E_3\delta k_x^2.
\end{align}
% We recognize this as deriving from $H_{206}$ under the substitution
% \begin{equation}
% \delta k_\mu\rightarrow \frac{1}{2}|\epsilon^{\mu\nu\lambda}|\delta k_\nu\delta k_\lambda \label{eq:205coordtrans}
% \end{equation}
% As such, we know that it is unitarily equivalent to 
% \begin{equation}
% H_{205}\sim\left.H_{199}\oplus H_{199}^*\right|_{a=a_1+i\sqrt{a_2^2+|b|^2},\delta k_\mu\rightarrow \frac{1}{2}|\epsilon^{\mu\nu\lambda}|\delta k_\nu\delta k_\lambda}
% \end{equation}
% Unlike the previous cases, however, all the bands of 
% \begin{equation}
% H_{199}(\delta k_\mu\rightarrow \frac{1}{2}|\epsilon^{\mu\nu\lambda}|\delta k_\nu\delta k_\lambda)
% \end{equation}
% carry vanishing Berry curvature, which can be seen by applying the coordinate transformation (\ref{eq:205coordtrans}) to Eqs. (\ref{eq:199berryc}). As such, we find that there exists a gauge where the {\red $SU(2)$ Wilson loop matrices $w_{\pm}(\theta)$ and $w_0(\theta)$ in any of the degenerate pairs of bands} are independent of $\theta$ and proportional to the identity matrix{\red , and so there is no non-abelian Berry curvature.}
A similar argument to that for SG $206$ shows that this isospectral with and homotopic to 
\begin{equation}
H_{205}\sim H'_{205}\oplus (H'_{205})^*, H'_{205}=\left(\begin{array}{ccc}f_1(\delta\mathbf{k}) & c\delta k_y\delta k_z & c^*\delta k_x\delta k_z\\c^*\delta k_y\delta k_z & f_2(\delta\mathbf{k}) & c\delta k_x\delta k_y\\c  \delta k_x\delta k_z & c^*\delta k_x \delta k_y & f_3(\delta\mathbf{k})\end{array}\right), \label{eq:205blocks}
\end{equation}
with $c=a_1+i\sqrt{a_2^2+|b|^2}$. Comparing with the analysis of Eq.~(\ref{eq:199charp}) from Section \ref{sec:199kdotp}, we note that for generic $E_1\neq E_2\neq E_3\neq 0$ this Hamiltonian is non-degenerate except when $\phi=\mathrm{arg}(c)=n\pi/3$, for $n$ integer, in which case line nodes exist when $|\delta k_x|=|\delta k_y|=|\delta k_z|$.

To analyze the topological properties of this Hamiltonian, we note first that in the basis where $H_{205}$ takes the form of Eq.~(\ref{eq:205blocks}), the non-Abelian Berry curvature between degenerate $I\mathcal{T}$ partner bands is diagonal, and the entries are given by the \emph{Abelian} Berry curvatures of $H_{205}'$. Furthermore, because $H_{205}'$ is inversion symmetric, we know that these Abelian Berry curvatures vanish. Stokes theorem then tells us that the non-Abelian Wilson loop eigenvalues between the degenerate pairs of bands do not wind as the loops are moved throughout the Brillouin zone.

\subsection{SGs $198,212$ and $213$}
Finally, we look at the sixfold degeneracy in SGs $198,212$ and $213$ at the $R$ point. Unlike the previous sixfold fermions, the little group in these SGs does not contain $I\mathcal{T}$, and so bands are not forced to come in degenerate pairs. However, we note that the planes $\delta k_i=0, i=x,y,$ or $z$ contain in their little group the antiunitaries
\begin{align}
\mathcal{T}\mathcal{G}_{0x}&=\mathcal{T}\{C_{2x}|\frac{1}{2}\frac{1}{2}0\}, \nonumber \\
\mathcal{T}\mathcal{G}_{0y}&=\mathcal{T}\{C_{2y}|0\frac{1}{2}\frac{1}{2}\}, \nonumber \\
\mathcal{T}\mathcal{G}_{0z}&=\mathcal{T}\{C_{2z}|\frac{1}{2}0\frac{1}{2}\}, \label{eq:surfacekramers}
\end{align}
which square to $-1$, giving a Kramers type degeneracy. Thus, on these three planes, bands \emph{are} forced to stick together in pairs. Thus, these space groups will have \emph{surface nodes} along the faces of the Brillouin zone, which we have indicated schematically in Fig.~\ref{fig:nodes198}. Away from the zone boundaries, the six bands will in general be non-degenerate. To construct the 6d representation for SG $198$, we note that the unitary subgroup of the little group is isomorphic to the little group of SG $199$ (c.f. Section \ref{sec:condition}). As such, we can make use of the representation matrices Eq.~(\ref{eq:199irrep}) and the commutation relations Eq.~(\ref{eq:tr1}) for time-reversal to arrive at the representation
\begin{align}
(G_3)_{198}&=\Delta(\{C_{3,111}^{-1}|010\})=(G_3)_{199}\oplus (G_3)_{199}, \\
(G_1)_{198}&=\Delta(\{C_{2x}|\frac{1}{2}\frac{3}{2}0\})=(G_1)_{199}\oplus (G_1)_{199}, \\
T_{198}&=\left(\begin{array}{cc} 0 & \lambda_0\\ -\lambda_0 & 0\end{array}\right)\mathcal{K},
\end{align}
for the little group of SG $198$. 

For SGs $212$ and $213$ we can perform a unitary transformation on the representation Eq.~(\ref{eq:212rep}) to obtain (n=0 for SG212, n=1 for SG213)
\begin{align}
(G_1)_{212}&=\Delta(\{C_{2x}|\frac{1}{2}\frac{1}{2}0\})=\left(\begin{array}{ccc}
					1 & 0 & 0\\
					0 & -1 & 0\\
					0 & 0 & 1
					\end{array}\right)\otimes\sigma_0, \\
(G_3)_{212}&=\Delta(\{C_{3,111}^{-1}|000\})=\left(\begin{array}{ccc}
					0 & 0 & 1\\
					1 & 0 & 0\\
					0 & { -1} & 0
					\end{array}\right)\otimes\sigma_0, \\
(G_4)_{212}&=\Delta(\{C_{2\bar{1}10}|\frac{2n+1}{4}\frac{2n+1}{4}\frac{2n+1}{4}\})=\left(\begin{array}{ccc}
					0 & i & 0\\
					i & 0 & 0\\
					0 & 0 & -i
					\end{array}\right)\otimes\sigma_3, \\
T_{212}&=\left(\begin{array}{cc} 0 & \lambda_0\\ -\lambda_0 & 0\end{array}\right)\mathcal{K},
\end{align}
for the little group of SG $212$ and $213$. 
%Note that, like with SG 205, the existence of two distinct 6d representations allows for the possibility of an accidental $12-$fold degeneracy.
%Here we have used $G_0$ to denote the antiunitary matrix representation of time reversal $\mathcal{T}$. 
Finally, after imposing the symmetry constraints, we find for the respective Hamiltonians
\begin{align}
H_{198}(\delta\mathbf{k})&=\left(\begin{array}{cc}
H_{199}(a,\delta\mathbf{k}) &  bH_{199}(1,\delta\mathbf{k}) \\
b^* H_{199}(1,\delta\mathbf{k}) & -H^*_{199}(a,\delta\mathbf{k})
\end{array}\right), \nonumber \\
H_{212}(\delta\mathbf{k})&=H_{213}(\delta\mathbf{k})=\left(\begin{array}{cc}
H_{199}(ia_2,\delta\mathbf{k'}) & b H_{199}(1,\delta\mathbf{k'}) \\
b^* H_{199}(1,\delta\mathbf{k'}) & -H^*_{199}(ia_2,\delta\mathbf{k'})
\end{array}\right),
\end{align}
to linear order, where
\begin{equation}
\delta\mathbf{k}'=(\delta k_z,\delta k_x,-\delta k_y).
\end{equation}

{Let us examine how these sixfold fermions -- and in particular the surface degeneracies -- split in the presence of strain or magnetic field perturbations. First, note that a uniaxial strain field parallel to the $x,y,$ or $z$ axis preserves one of the symmetries Eq.~\ref{eq:surfacekramers}, while a magnetic field \emph{perpendicular} to any of the $x,y,$ or $z$ axes preserves two of the symmetries. Hence for these choices of external field, one or more surface degeneracies remain.

Hence, we consider the effect of a magnetic field aligned along the $[111]$ high-symmetry direction. Not only does this break all the surface degeneracies, but the reduced symmetry group will be isomorphic to the case of $SG 206$ in the presence of a magnetic field considered in Section \ref{sec:sg206}. { The most convenient way to analyze this splitting is to imagine that it occurs as a two step process, where we first remove time reversal symmetry, and then remove the spatial symmetries broken by the field direction. Breaking time reversal symmetry alone will cause the sixfold degeneracy to split into a pair of spin-$1$ Weyls. Next, including crystal symmetry shows that these each split into a set of six ordinary Weyl points. The net effect of the magnetic field then is to create twelve total Weyl points.}

We can also consider the effect of uniaxial strain fields aligned along the $[111]$ direction, and in the case of SG $212$ and $213$ also along the $[\bar{1}10]$ direction. Since these perturbations preserve time-reversal symmetry, the sixfold degeneracy will split into a Weyl semimetal phase with three Kramers doublets of Weyl points at the $R$ point. 
 }

\subsection{$\mathbf{k}\cdot\mathbf{p}$ Hamiltonians for Eightfold degeneracies}
We now analyze the eightfold degenerate fermions. For most of this section, we will use the tensor product basis 
\begin{equation}
g_{ijk}=\sigma_i\otimes\sigma_j\otimes\sigma_k
\end{equation}
as our $8$-dimensional matrix basis. Each of these square to the $8\times8$ identity matrix $(g_{ijk})^2=g_{000}\equiv E_8$,
and satisfy
\begin{equation}
\frac{1}{8}\mathrm{tr}(g_{ijk}^\dag g_{\ell mn})=\delta_{i\ell}\delta_{jm}\delta_{kn}.
\end{equation}
\subsection{SGs $130$ and $135$}
Let us start with the $8-$fold degeneracy at the $A$ point in SGs $130$ and $135$. Unlike all our other examples, these have a tetragonal rather than a cubic Bravais lattice. The little groups $G^{\mathbf{k}_0}$ of the $A$ point of SGs $130$ and $135$ are isomorphic, and both contain inversion. Thus, in the presence of time reversal symmetry, we expect all bands to be doubly degenerate. For concreteness, we focus on SG $130$. The analysis for SG $135$ is identical. For the $8$-dimensional representation matrices of the little group we take
\begin{align}
G_1&=\Delta(\{C_{4z}|000\})=\left(\begin{array}{cc}P_4 & 0 \\ 0 & P_4\end{array}\right),\label{eq:reps130-1} \\
G_2&=\Delta(\{C_{2x}|\frac{1}{2}\frac{1}{2}0\})=\left(\begin{array}{cc}Q_4 & 0 \\ 0 & Q_4\end{array}\right),\\
G_3&=\Delta(\{I|\frac{1}{2}\frac{1}{2}\frac{1}{2}\})=\left(\begin{array}{cc}S_4 & 0 \\ 0 & S_4\end{array}\right),\\
T&=\left(\begin{array}{cc}0 & S_4 \\ -S_4 & 0\end{array}\right)K,\label{eq:reps130-2}
\end{align}
where we have defined
\begin{align}
P_4&=\frac{i}{\sqrt{2}}\left(\begin{array}{cccc}-1&-1&0&0\\1&-1&0&0\\0&0&1&1\\0&0&-1&1\end{array}\right), \;\;
Q_4=\frac{1}{\sqrt{2}}\left(\begin{array}{cccc}1&-1&0&0\\-1&-1&0&0\\0&0&1&-1\\0&0&-1&-1\end{array}\right),\;\;
S_4=\left(\begin{array}{cccc}0&0&1&0\\0&0&0&1\\1&0&0&0\\0&1&0&0\end{array}\right). \\
\end{align}
Before constructing the Hamiltonian explicitly, we make the following general remarks. First, consider the high symmetry line $\delta k_x=\delta k_y=0$. The little group of this line is generated by $\mathcal{G}_1,\mathcal{G}_3\mathcal{G}_2$ and the antiunitary element $\mathcal{G}_3\mathcal{T}$; $\mathcal{G}_3$ alone is not a symmetry of the line. Due to the block diagonal structure of the symmetry elements, it is obvious that along this line the $8-$dimensional representation splits into two fourfold degenerate representations, and so we expect to find two ``Dirac line nodes'' along this line, which we show in Fig.~\ref{fig:Diracnodes}.
This is a new type of line node not mentioned in prior studies of nodal lines. 
Similarly, along the line $\delta k_y=\delta k_z=0$, we have that the little group is generated by $G_2$ and $G_3G_1^2$, which also leads to two four-fold degenerate representations. Finally, since $C_{4z}$ relates this line to one with $\delta k_x=\delta k_z=0$, we have that along all three edges of the Brillouin zone which meet at the $A$ point, there are two fourfold degenerate bands, i.e. two Dirac line nodes. Fig.~\ref{fig:nodes130} shows the location of these line nodes in the Brillouin zone.
\begin{figure}
\includegraphics[width=.5\textwidth]{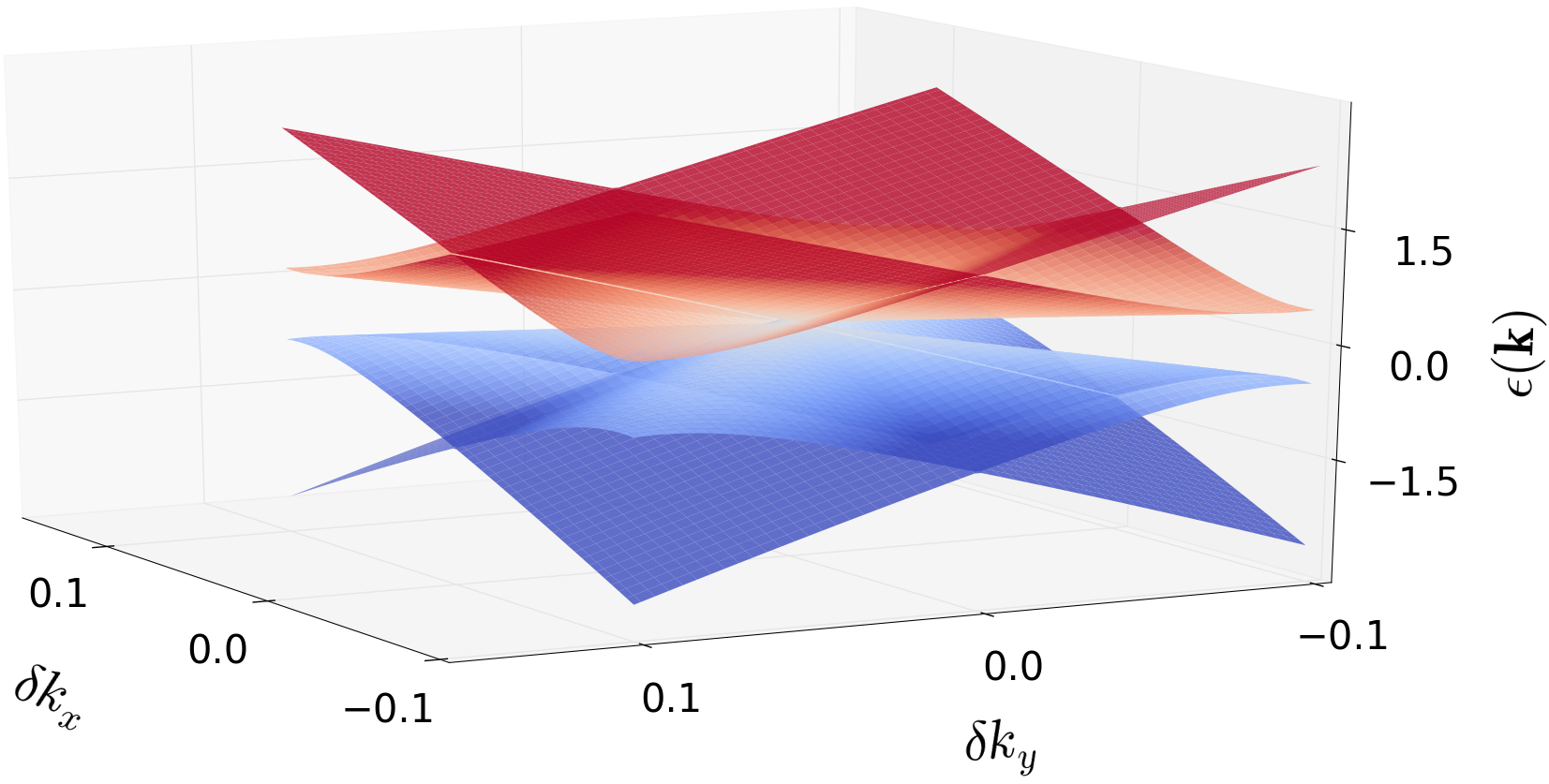}
\caption{Low energy dispersion relation near the $A$ point of SGs $130$ and $135$ in the plane $\delta k_z=0$. Each band is doubly degenerate due to $I\mathcal{T}$ symmetry. Fourfold Dirac line nodes can be seen along the lines $\delta k_x=0$ and $\delta k_y=0$.}\label{fig:Diracnodes}
\end{figure}

Keeping this in mind, we solve the constraints Eq.~(\ref{eq:tr2}) to find to linear order
\begin{align}
H_{130}&=H_{135}=\delta k_z(ag_{233}+bg_{232}+cg_{231})+\delta k_x(dg_{123}-eg_{130}+fg_{122}+hg_{121})+\delta k_y(dg_{323}+eg_{330}+fg_{322}+hg_{321}),
\end{align}
where $a,b,\dots,h$ are real parameters. This form of the Hamiltonian makes manifest the degeneracies discussed above: first, note that the matrices multiplying any one of the $\delta k_i$ all anticommute with each other, and all have only two distinct eigenvalues $\pm 1$. Each matrix then gives an subalgebra of the algebra $Cl_{6}$ of $8\times8$ Clifford algebra matrices. By a basis rotation, it then follows that there are only two distinct eigenvalues, and hence two fourfold degenerate line nodes on each of the high symmetry lines. 

Let us now look at the Wilson loop eigenvalues associated to these fourfold line nodes. For simplicity, we focus on the line nodes along $\delta k_x=\delta k_y=0$; the analysis of the others follows identically. We start by diagonalizing $H_0=H_{130}(\delta k_x=\delta k_y=0)$, which has eigenvalues $\epsilon_\pm=\pm |\delta k_z|\sqrt{a^2+b^2+c^2}$. Denote the corresponding eigenvectors as $\phi_\pm^{i}(\delta k_z)$, where $i=1,2,3,4$. We now project the eigenstates $\psi$ of the full Hamiltonian into the basis of $\phi_-^i$, i.e. we perform degenerate perturbation theory. Without loss of generality, we may take $b=c=f=h=0$, since this still allows the coefficients on each of the $\delta k_i$ to vary independently. In this case, the Hamiltonian is block diagonal, consisting of two $4\times 4$ blocks. We find that to lowest order in perturbation theory we can write the eigenstates of the Hamiltonian as
\begin{align}
\phi_{-}^1=\left(\begin{array}{c}\mathbf{e}\\0\\0\\0\end{array}\right), \phi_{-}^2=\left(\begin{array}{c}0\\ \mathbf{e}^*\\0\\0\end{array}\right), \phi_{-}^3=\left(\begin{array}{c}0\\0\\ \mathbf{e}^*\\0\end{array}\right), \phi_{-}^4=\left(\begin{array}{c}0\\0\\0\\ \mathbf{e}\end{array}\right),
\end{align}
where $\mathbf{e}=\frac{1}{\sqrt{2}}\left(1,-i\right)$.

In this basis, the remainder of the Hamiltonian takes the form
\begin{equation}
V_{ij}=\langle \phi_-^i|H_{130}-H_0|\phi_-^j\rangle=\left(\begin{array}{cccc}0&d\delta k_- & 0 &0\\ d\delta k_+ &0 &0 &0\\0&0&0&d\delta k_+\\0&0&d\delta k_-&0\end{array}\right),
\end{equation}
where $\delta k_+=\delta k_x+i\delta k_y$ and $\delta k_-=(\delta k_+)^*$. Because of $I\mathcal{T}$ symmetry, the eigenvectors of $V$ come in degenerate pairs, as reflected in the block diagonal structure of $V_{ij}$. Examining the two pairs emerging from the Dirac line node, which we denote by $\pm$, we see immediately that the Wilson loop matrix within each degenerate pair takes the form
\begin{equation}
w_{\pm}=\left(\begin{array}{cc}-1&0\\0&-1\end{array}\right).
\end{equation}
A similar analysis holds had we instead started with the positive energy Dirac line node. These eigenvalues are gauge invariant and insensitive to the particular choice of parameters we used above. These Wilson loop eigenvalues are identical to those of two decoupled Weyl line nodes. {Similar to SG 220, this nontrivial Berry phase implies the existence of \emph{pairs} of ``drumhead'' surface states associated to each Dirac line.

The Dirac lines serve as a ``parent phase'' for all known topological semimetals: external perturbations may split the eightfold intersection of Dirac lines into Dirac points, twofold degenerate Weyl line nodes, and isolated Weyl points. Splitting in the presence of time-reversal preserving (i.e. strain) perturbations was considered in Ref.~\onlinecite{Wieder2015};  { it is trivial to extend the analysis to time-reversal breaking (magnetic) perturbations, yielding similar conclusions.} Here we emphasize that it is the presence of Dirac \emph{lines}, rather than just the eightfold degeneracy, that is crucial for { the emergence of a Dirac semimetal phase:} external perturbations split the fourfold Dirac lines into twofold degenerate bands along the Brillouin zone edge, and it is these twofold lines that intersect to form topological Dirac points.

{ More concretely, consider an external magnetic field in the $z$ direction. This breaks time-reversal symmetry, as well as the $C_{2x}$ little group symmetry of the $A$ point. However, $G_1$ ($C_{4z}$), $G_3$ (inversion), and $G_2T$ (the product of $C_{2x}$ and time reversal) remain good symmetries. The eight-dimensional representation of these symmetries now splits at the $A$ point. Noting from Eqs.~(\ref{eq:reps130-1}--\ref{eq:reps130-2}) that the matrix representatives of these symmetries are block diagonal and anticommuting, we deduce that the representation splits into two fourfold representations at the $A$ point, which we may write as
\begin{equation}
G_1=e^{i\pi/4}\left(\begin{array}{cccc}0&0&1&0\\0&0&0&1\\1&0&0&0\\0&1&0&0\end{array}\right),\;\;
G_3=\left(\begin{array}{cccc}1&0&0&0\\0&1&0&0\\0&0&-1&0\\0&0&0&-1\end{array}\right),\;\;
G_2T=e^{-i\pi/4}\left(\begin{array}{cccc}0&1&0&0\\-1&0&0&0\\0&0&0&-1\\0&0&1&0\end{array}\right)K,
\end{equation}
along with its conjugate representation. These two fourfold degeneracies will move away from the Fermi level.  Furthermore, following our discussion of the Dirac lines, bands will be twofold degenerate the entire line $\delta k_x=\delta k_y=0$, where only $G_1$ and $G_2T$ remain good symmetries. As these degenerate pairs differ in their $C_{2z}$ eigenvalues, they are allowed to cross along this line, forming a Dirac semimetal phase. Inversion symmetry demands that a Dirac point at $\mathbf{k}^*=(\pi,\pi,k_z)$ have a partner at $-\mathbf{k}^*\equiv(\pi,\pi,-k_z)$. Finally, note that this pair of Dirac points is separated by the high symmetry $(k_x,k_y,0)$ and $(k_x,k_y,\pi)$ mirror planes. Whether or not these Dirac points form is a question of energetics, as shown in Fig.~\ref{fig:diracsplit}. For the remainder of this discussion, we will assume the Dirac points form, and analyze the resulting phase.

\begin{figure}
\includegraphics[width=\textwidth]{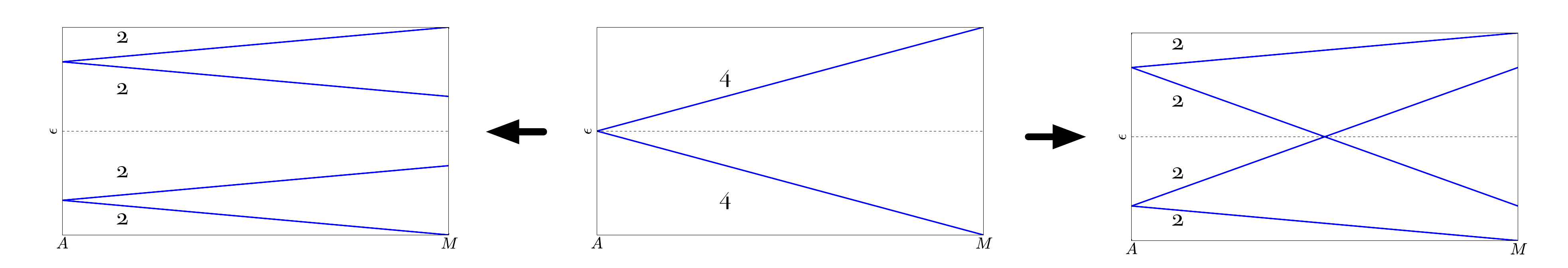}\label{fig:diracsplit}
\caption{Schematic diagram of the splitting of fourfould Dirac lines. The center shows the spectrum along the line $A\rightarrow M$ before the magnetic field is applied. Left shows the trivial splitting, and right shows the Dirac semimetal phase. The number of degenerate bands in each multiplet is labeled.}
\end{figure}

To proceed further, we must distinguish between space groups $130$ and $135$. 
SG $130$ contains the non-symmorphic glide reflection $\{\sigma_z|\frac{1}{2}\frac{1}{2}0\}$, while SG $130$ contains the symmorphic mirror $\{\sigma_z|000\}$, as can be seen by considering the products of inversion and $C_{4z}$ from Eq.~(\ref{eq:rels130}).
Hence, in SG $130$ the $k_z=0$ and $k_z=\pi$ planes are glide planes, while in SG $135$ they are mirror planes. Thus, in SG $135$ the Dirac semimetal phase can be characterized by a mirror Chern number\cite{Teo2008}.
To calculate this invariant (up to parity), we will focus on the $k_z=\pi$ plane. We can characterize the topological state of the system by the product of inversion eigenvalues in each $\sigma_z=\pm i$ eigenspace at the two inversion invariant momenta $A$ and $Z$ (c.f. Fig. \ref{fig:tetrabz}) in this plane ($C_{4z}$ symmetry fixes the product of inversion eigenvalues at $(\pi,0,\pi)$ and $(0,\pi,0)$ to be $1$). At the $A$ point, we found above that we will have four occupied bands: two in the $+i$ mirror eigenspace, and two in the $-i$ mirror eigenspace. Noting that $\sigma_z=G_1^2G_3=iG_3,$ we see that in each mirror subspace, pairs of bands have the same inversion eigenvalues, and so the product will be $+1$.

To complete the argument, we compute the product of inversion eigenvalues at the $Z$ point. Using Ref.~\cite{Cracknell}, we note that there is only one allowed representation at the $Z$ point, and -- crucially for our purposes -- bands with the same mirror eigenvalue have \emph{opposite} inversion eigenvalues in this representation. Since the magnetic field does not break inversion symmetry, this is sufficient to conclude that the product of inversion eigenvalues in each mirror subspace will be $-1$. 
Thus, we conclude that in SG $135$ the mirror Chern number will be nonzero when the Dirac points form. It follows that the Dirac points are connected by a topologically protected pair of Fermi arcs\cite{Yang2014}; these Fermi arcs are expected to emerge from the line-node surface states. 
}

\subsection{SGs $218$ and $220$}

We look next at the $8-$fold degeneracy in SG $218$ at the $R$ point, as well as SG $220$ at the $H$ point. Although these groups have different Bravais lattices, it was shown in Section \ref{sec:218} that the little groups of these respective points are isomorphic, and so their respective $\mathbf{k}\cdot\mathbf{p}$ Hamiltonians are identical. Neither of these groups contains $I\mathcal{T}$, since neither of these space groups is inversion symmetric\cite{Cracknell}, and so we in general expect to find $8$ non-degenerate bands. Using the same notation as Eq.~(\ref{eq:rels218}), we can take for our $8-$dimensional representation
\begin{align}
G_1&=\Delta(\mathcal{G}_1)=\left(\begin{array}{cc}Q_4 & 0 \\ 0 & Q_4\end{array}\right),\;\;
G_2=\Delta(\mathcal{G}_2)=\left(\begin{array}{cc}R_4 & 0 \\ 0 & R_4\end{array}\right),\label{eq:218rep1}\\
G_3&=\Delta(\mathcal{G}_3)=\left(\begin{array}{cc}P_4 & 0 \\ 0 & P_4\end{array}\right),\;\;
G_4=\Delta(\mathcal{G}_4)=\left(\begin{array}{cc}S_4 & 0 \\ 0 & S_4\end{array}\right),\label{eq:218rep2}\\
T&=\left(\begin{array}{cc}0 & -N_4 \\ N_4 & 0\end{array}\right)K,\label{eq:218rep3}
\end{align}
where we define
\begin{align}
P_4&=\frac{1}{\sqrt{2}}\left(\begin{array}{cccc}-i&-1&0&0\\1&i&0&0\\0&0&i&-1\\0&0&1&-i\end{array}\right),\;\;
Q_4=\frac{1}{\sqrt{2}}\left(\begin{array}{cccc}i&-1&0&0\\1&-i&0&0\\0&0&-i&-1\\0&0&1&-i\end{array}\right),\\
S_4&=\left(\begin{array}{cccc}1&0&0&0\\0&-1&0&0\\0&0&1&0\\0&0&0&-1\end{array}\right),\;\;
N_4=\left(\begin{array}{cccc}0&0&1&0\\0&0&0&1\\1&0&0&0\\0&1&0&0\end{array}\right), \\
R_4&=\frac{1}{4}\left(\begin{array}{cccc}1& -i-\sqrt{2}&\sqrt{3}(i+\sqrt{2})&\sqrt{3}\\-i+\sqrt{2}&1&-\sqrt{3}&\sqrt{3}(-i+\sqrt{2})\\ \sqrt{3}(-i+\sqrt{2})&\sqrt{3}&1&i-\sqrt{2}\\-\sqrt{3}&\sqrt{3}(i+\sqrt{2})&i+\sqrt{2}&1\end{array}\right).\\
\end{align}
These matrices satisfy the relations Eq.~(\ref{eq:matrixrels218}), as needed for a representation. Although this $8-$fold representation is quite cumbersome, we can make some general comments about the degeneracy structure of this fermion. First, let us consider the high symmetry line $\delta k_x=\delta k_y=\delta k_z$. The little group of this line is generated by $G_2$ and $G_4$ alone. This $8$ dimensional representation of the little group of the line is reducible. To see this most clearly, we let $U$ denote the unitary transformation that diagonalizes $G_2$. Then we have
\begin{equation}
UG_2U^\dag=\left(\begin{array}{cccc}\sigma_0&0&0&0\\0&\sigma_0&0&0\\0&0&e^{2i\pi/3}\sigma_0&0\\0&0&0&e^{4i\pi/3}\sigma_0\end{array}\right),\;\; \mathrm{and}\;\; UG_4U^\dag=\left(\begin{array}{cccc}\sigma_3&0&0&0\\ 0&\sigma_3&0&0\\0&0&0&-\sigma_0\\0&0&-\sigma_0&0\end{array}\right),
\end{equation}
from which we can directly see that the $8d$ representation reduces to four $1d$ representations and two $2d$ representations. Along this line we thus expect to find four singly degenerate bands, and two doubly degenerate bands.

Next, we consider the line $\delta k_x=\delta k_y=0$. The little group of this line is generated by $G_3T,G_3G_1$, and $G_4$. Although $G_3T$ is antiunitary, it squares to $+1$, and so it need not affect the degeneracy of the representation. The matrix representation of these elements is, from Eqs.~(\ref{eq:218rep1}-\ref{eq:218rep3}) 
\begin{equation}
G_3T=\left(\begin{array}{cccc}0&0&0&B\\0&0&B^*&0\\0&-B&0&0\\-B^*&0&0&0\end{array}\right)\mathcal{K},\;\;G_3G_1=i\left(\begin{array}{cccc}\sigma_1&0&0&0\\0&-\sigma_1&0&0\\0&0&\sigma_1&0\\0&0&0&-\sigma_1\end{array}\right), \mathrm{and} \;\;G_4=\left(\begin{array}{cccc}\sigma_3&0&0&0\\0&\sigma_3&0&0\\0&0&\sigma_3&0\\0&0&0&\sigma_3\end{array}\right),
\end{equation}
where
\begin{equation}
B=\frac{i}{\sqrt{2}}\left(\sigma_z+\sigma_y\right).
\end{equation}
Examining these matrices, we see that the $8d$ representation splits into four irreducible $2d$ representations along this line; the antiunitary $G_3T$ does not in fact double the degeneracy of the representations, but simply enacts the \emph{unitary} transformation $\sigma_1\rightarrow\sigma_1,\sigma_3\rightarrow\pm\sigma_2$ within each representation.

Finally, we examine the line $\delta k_x=\delta k_y=k, \delta k_z=0$. The little group of this line contains the element $G_1G_4G_2T$, which is antiunitary and squares to $-1$. Kramers's theorem then tells us that along this line too, there are four sets of twofold degenerate bands. The location of these degeneracies, and the symmetries which protect them, has been indicated in Fig.~\ref{fig:nodes218} for SG 218 (The lines and symmetries for SG 220 are identical, but their placement in the BCC Brillouin zone does not easily translate to a visualization).

These are the only additional symmetry-enforced degeneracies. We find for the linear $\mathbf{k}\cdot\mathbf{p}$ Hamiltonian
\begin{align}
% H_{218}=H_{220}&=a\left((\delta k_x+\delta k_y)(\sqrt{3}g_{303}-g_{023})+\sqrt{3}(\delta k_x-\delta k_y)g_{233}+(\delta k_x-\delta k_y)g_{113}+2\sqrt{2}\delta k_zg_{313}\right)\nonumber \\
% &+b\left((\delta k_x+\delta k_y)g_{013}+(\delta k_x-\delta k_y)g_{123}-\sqrt{2}\delta k_zg_{323}\right)\nonumber \\
% &+c\left((\delta k_x +\delta k_y)g_{301}\right)
H_{218}=H_{220}=\left(\begin{array}{cc}U & V \\ V^\dag & -U\end{array}\right),
\end{align}
with
\begin{align}
U&=\left(
\scalemath{0.4}{
\begin{array}{cccc}
 \sqrt{3} a (\delta k_x+\delta k_y) & -i \sqrt{3} a (\delta k_x-\delta k_y) & b \left(\delta k_x+\delta k_y+i
   \sqrt{2} \delta k_z\right)+a \left(i \delta k_x+i \delta k_y+2 \sqrt{2} \delta k_z\right) & (a-i b)
   (\delta k_x-\delta k_y) \\
 i \sqrt{3} a (\delta k_x-\delta k_y) & -\sqrt{3} a (\delta k_x+\delta k_y) & (a-i b) (\delta k_x-\delta k_y) & b
   \left(\delta k_x+\delta k_y-i \sqrt{2} \delta k_z\right)+i a \left(\delta k_x+\delta k_y+2 i \sqrt{2}
   \delta k_z\right) \\
 b \left(\delta k_x+\delta k_y-i \sqrt{2} \delta k_z\right)+a \left(-i \delta k_x-i \delta k_y+2 \sqrt{2}
   \delta k_z\right) & (a+i b) (\delta k_x-\delta k_y) & \sqrt{3} a (\delta k_x+\delta k_y) & i \sqrt{3} a
   (\delta k_x-\delta k_y) \\
 (a+i b) (\delta k_x-\delta k_y) & b \left(\delta k_x+\delta k_y+i \sqrt{2} \delta k_z\right)-i a
   \left(\delta k_x+\delta k_y-2 i \sqrt{2} \delta k_z\right) & -i \sqrt{3} a (\delta k_x-\delta k_y) & -\sqrt{3}
   a (\delta k_x+\delta k_y) \\
\end{array}
}
\right),\\
V&=\left(
\scalemath{0.4}{
\begin{array}{cccc}
 (c+i d) (\delta k_x+\delta k_y) & (d-i c) (\delta k_x-\delta k_y) & \sqrt{6} (c+i d) \delta k_z+(e+i f)
   \left(\delta k_x+\delta k_y+i \sqrt{2} \delta k_z\right) & (f-i e) (\delta k_x-\delta k_y) \\
 i (c+i d) (\delta k_x-\delta k_y) & -(c+i d) (\delta k_x+\delta k_y) & (f-i e) (\delta k_x-\delta k_y) & (e+i f)
   \left(\delta k_x+\delta k_y-i \sqrt{2} \delta k_z\right)-\sqrt{6} (c+i d) \delta k_z \\
 \frac{1}{3} \left(\left(-2 i \sqrt{3} c+2 \sqrt{3} d+3 e+3 i f\right) (\delta k_x+\delta k_y)+\sqrt{2}
   \left(\sqrt{3} c+i \sqrt{3} d-3 i e+3 f\right) \delta k_z\right) & \frac{1}{3} \left(2 \sqrt{3} c+2 i
   \sqrt{3} d+3 i e-3 f\right) (\delta k_x-\delta k_y) & (c+i d) (\delta k_x+\delta k_y) & i (c+i d)
   (\delta k_x-\delta k_y) \\
 \frac{1}{3} \left(2 \sqrt{3} c+2 i \sqrt{3} d+3 i e-3 f\right) (\delta k_x-\delta k_y) & \frac{1}{3}
   \left(\left(-2 i \sqrt{3} c+2 \sqrt{3} d+3 e+3 i f\right) (\delta k_x+\delta k_y)-\sqrt{2} \left(\sqrt{3}
   c+i \sqrt{3} d-3 i e+3 f\right) \delta k_z\right) & (d-i c) (\delta k_x-\delta k_y) & -(c+i d)
   (\delta k_x+\delta k_y) \\
\end{array}
}
\right),
\end{align}
which we checked respects the symmetries we have enumerated.
\begin{figure}
\subfloat[Line Nodes in SG 220\label{fig:nodes220-3}]{\includegraphics[width=0.3\textwidth]{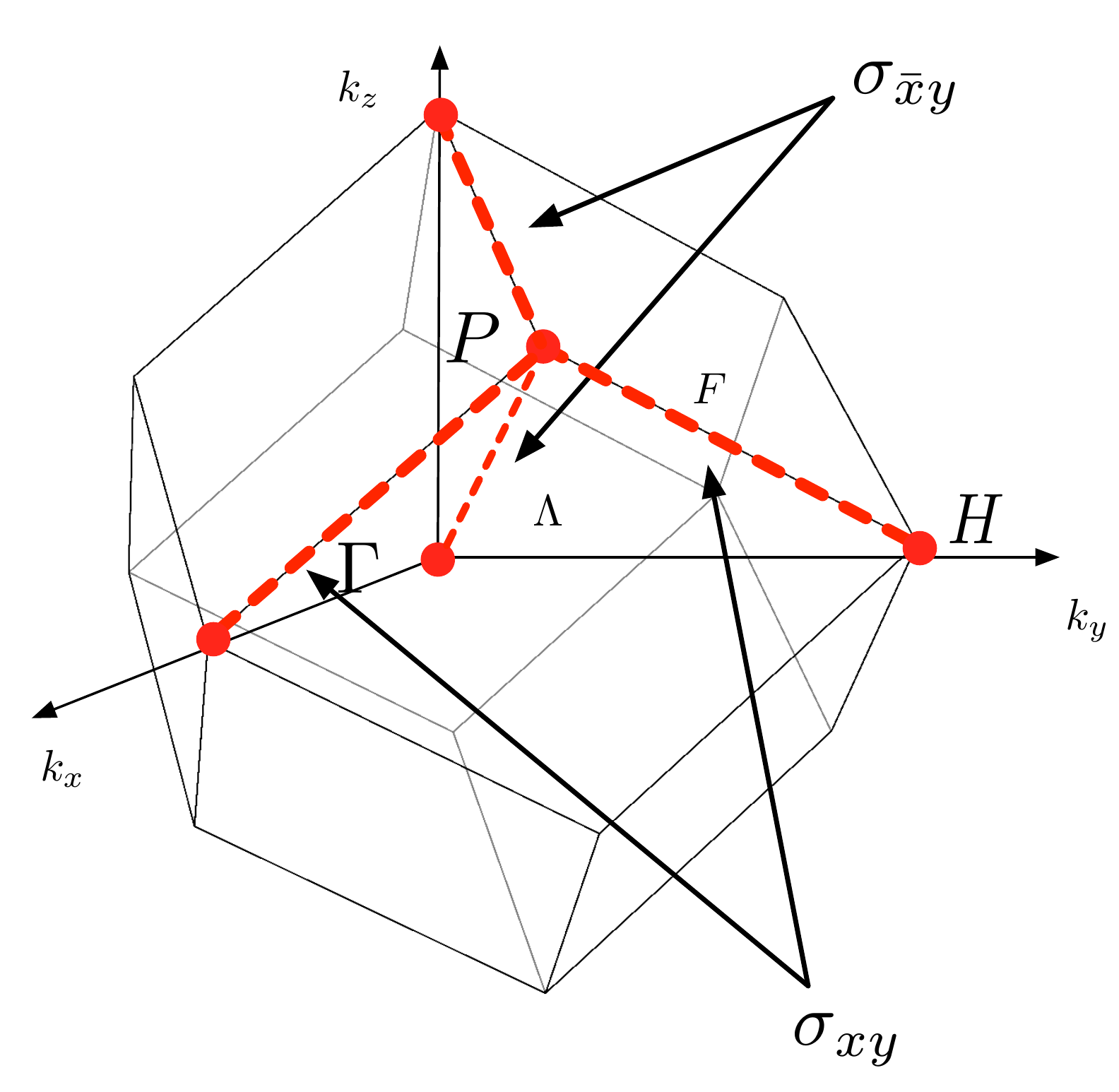}}
\subfloat[Surface Nodes in SGs 198,212, and 213\label{fig:nodes198}]{\includegraphics[width=0.3\textwidth]{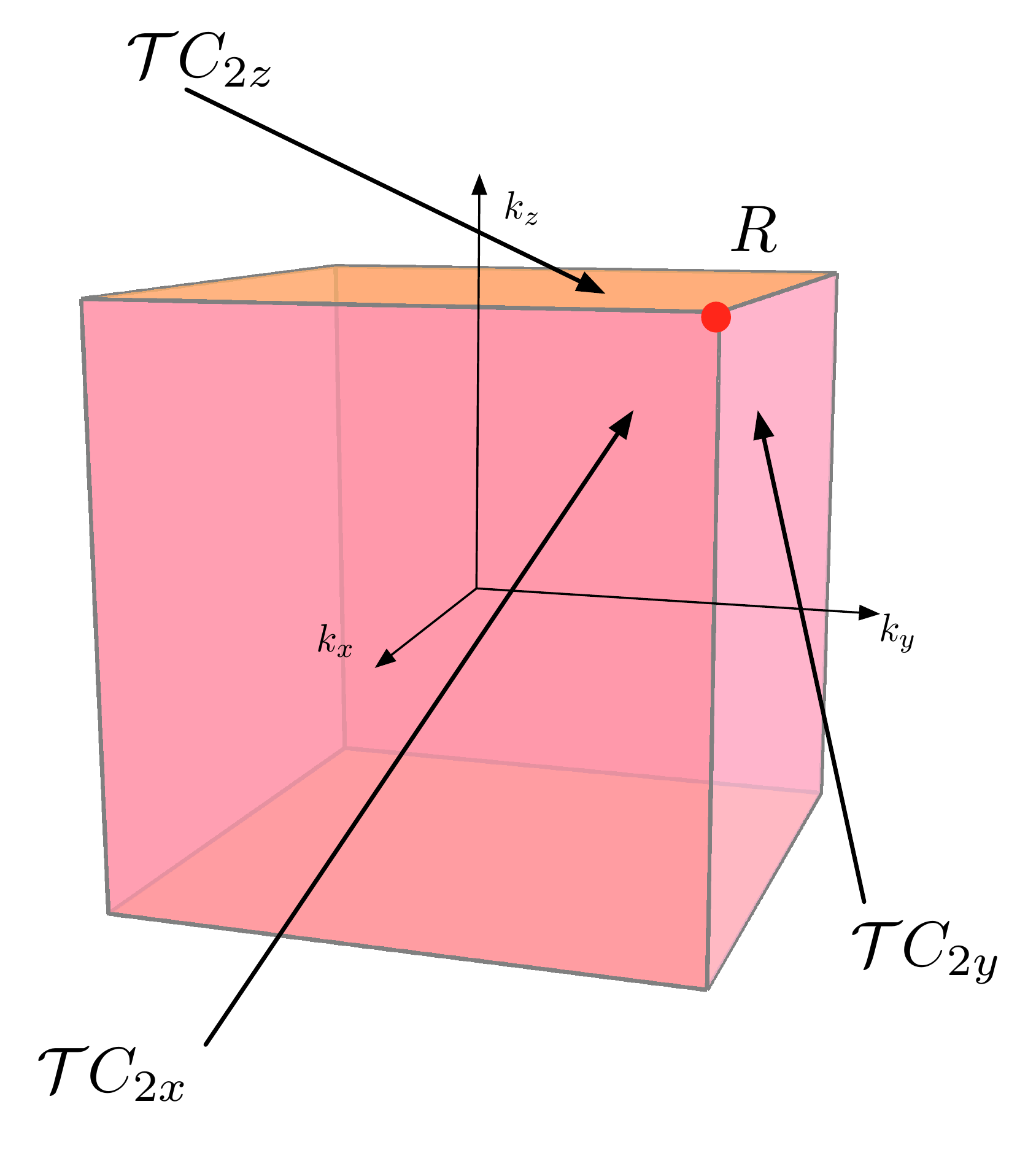}}\quad
\subfloat[Dirac line nodes in SGs 130 and 135\label{fig:nodes130}]{\includegraphics[width=0.3\textwidth]{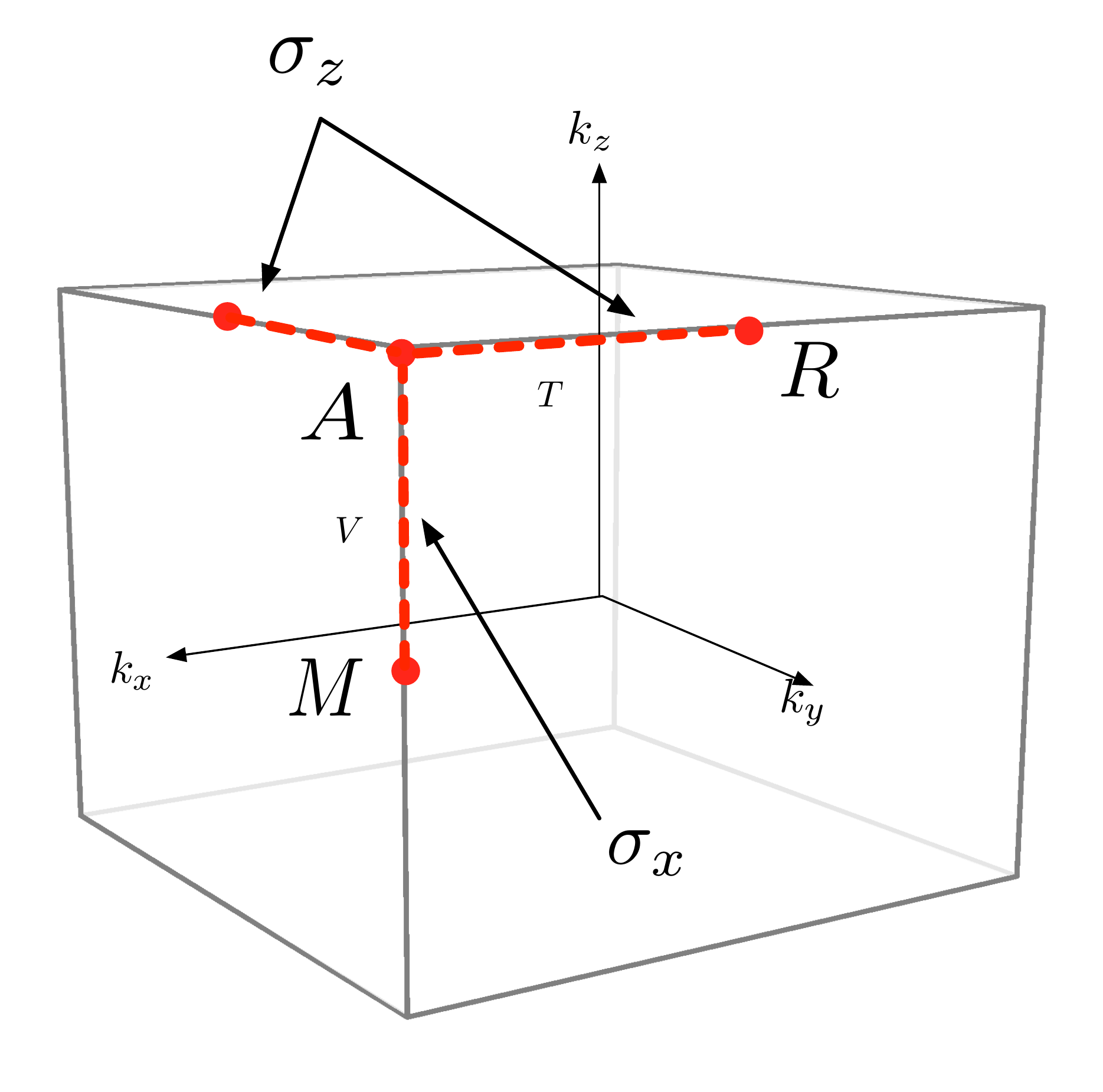}}
\subfloat[Line nodes in SG 218\label{fig:nodes218}]{\includegraphics[width=0.3\textwidth]{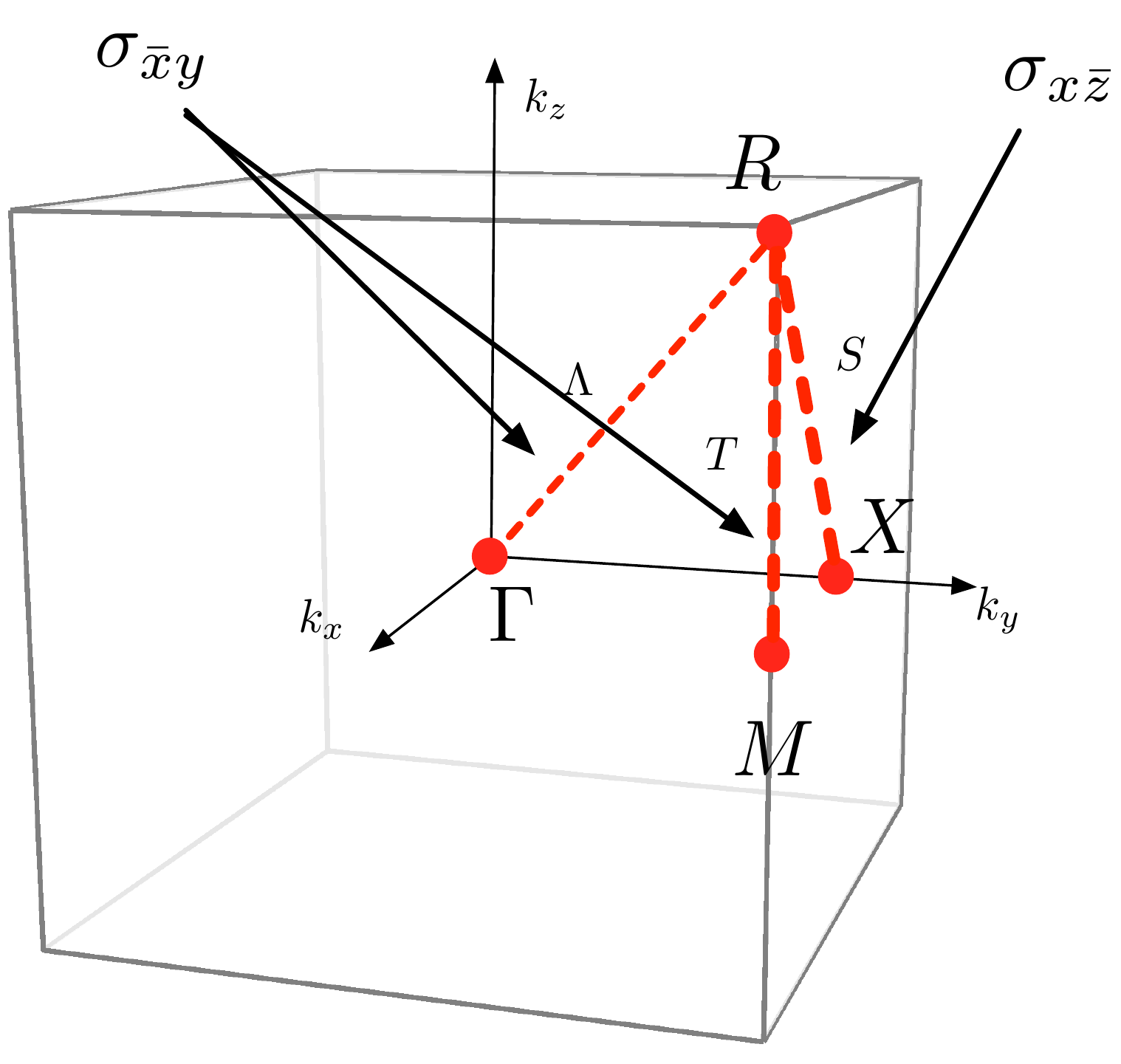}}
\caption{Highlighted are the line and surface degeneracies in (a) SG 220, (b) SG 198,212, and 213, (c) SGs 130 and 135, and (d) SG 218. We have indicated the mirror symmetries which protect the line nodes, as well as the antiunitary symmetries that protect the Kramers degenerate surface nodes.}
\end{figure}
\subsection{SGs $222$ and $223$, and $230$}
Lastly, we examine the $8$-fold degeneracies at the $R$ point in SGs $222$ and $223$, as well as the $H$ point in SG $230$. As noted in Section \ref{sec:222}, all of these points have isomorphic little groups, as enumerated in Eq.~(\ref{eq:rels222}). Unlike the previous case, however, these little groups all contain inversion, and so in the presence of time reversal all bands are twofold degenerate. 
%There are two different $8-$dimensional representations of the little group, but since the representations differ only by the sign of the inversion matrix $G_4$, they lead to the same $\mathbf{k}\cdot\mathbf{p}$ Hamiltonian. As such, we focus on only one of them, without loss of generality.
Using the notation of Eq.~(\ref{eq:rels222}), we can take as our generating set the matrices
\begin{align}
G_1&=\Delta(\mathcal{G}_1)=\left(\begin{array}{cc}P_4 & 0 \\ 0 & P_4^*\end{array}\right),\\
G_3&=\Delta(\mathcal{G}_3)=\left(\begin{array}{cccc}\sigma_0 & 0 & 0 &0\\0 & -\sigma_0 & 0 &0\\0 & 0 & \sigma_0 & 0\\ 0&0&0&-\sigma_0\end{array}\right),\\
G_4&=\Delta(\mathcal{G}_4)=\left(\begin{array}{cc}R_4 & 0 \\ 0 & R_4^*\end{array}\right),\\
T&=\left(\begin{array}{cc}0 &-\sigma_0\otimes\sigma_0 \\ \sigma_0\otimes\sigma_0 & 0\end{array}\right)\mathcal{K},
\end{align}
where we have defined
\begin{equation}
P_4=\left(\begin{array}{cccc}0&0&e^{-i\pi/4}&0\\0&0&0&e^{i\pi/4}\\e^{-i\pi/4}&0&0&0\\0&e^{i\pi/4}&0&0\end{array}\right), \mathrm{and}\;\; R_4=\frac{1}{\sqrt{2}}\left(\begin{array}{cccc}e^{5i\pi/12}&e^{5i\pi/12}&0&0\\e^{-i\pi/12}&-e^{-i\pi/12}&0&0\\0&0&-e^{i\pi/12}&-e^{i\pi/12}\\0&0&e^{7i\pi/12}&e^{-5i\pi/12}\end{array}\right).
\end{equation}
Note that in order to make the nonvanishing blocks of $G_3T$ diagonal, this representation differs by a unitary transformation from that constructed in Section \ref{sec:222}. 
%Also note that, analogous to SG $205,212$ and $213$, there exists the possibility for a $16-$fold degeneracy, by tuning the energy splitting between two inequivalent $8d$ representations. 
Imposing these symmetry constraints, we find
\begin{align}
H_{222}=H_{223}&=\delta k_z\left(ag_{313}+bg_{111}+cg_{112}\right)
-\delta k_x\left(\frac{a}{2}g_{113}+\frac{\sqrt{3}a}{2}g_{120}+bg_{311}+cg_{312}\right)\nonumber \\
&+\delta k_y\left(\frac{a}{2}g_{210}-\frac{\sqrt{3}a}{2}g_{223}+bg_{012}-cg_{011}\right),
\end{align}
and similarly
\begin{align}
H_{230}&=\delta k_z\left(ag_{313}+bg_{112}-cg_{111}\right)-\delta k_x\left(-\frac{a}{2}g_{210}+\frac{\sqrt{3}a}{2}g_{223}+bg_{011}+cg_{012}\right)\nonumber \\
&-\delta k_y\left(\frac{a}{2}g_{113}+\frac{\sqrt{3}a}{2}g_{120}-cg_{311}+bg_{312}\right).
\end{align}
Unlike the case with SGs $130$ and $135$, there are no additional line nodes here, as can be seen from examining the algebra of matrices appearing in the Hamiltonians. There is generically no degeneracy away from the $8-$fold point beyond the doubling required by $I\mathcal{T}$.

{ \section{Symmetry-protected Spin-3/2}
As noted in the main text, our $3$-fold degenerate fermions exist in the same topological phase as the Hamiltonian
\begin{equation}
H=\mathbf{k}\cdot\mathbf{S},
\end{equation}
where $\mathbf{S}$ is the vector of spin-$1$ matrices. A natural question to ask is whether there exist fermions in the topological phase of $\mathbf{k}\cdot\mathbf{S}$ for higher-spin representations. { For the spin $j$ representation, this phase consists of a $2j+1$ fold degeneracy, with bands having Berry flux $2n$ for every $n\in\{-j,-j+1,\dots,j-1,j\}$. Given that our exhaustive classification of $6$ and $8$-fold degeneracies does not include any band crossings with non-zero Chern number, we know that such fermions cannot exist at high symmetry points for $j\geq 2$.} To answer this question for $j=3/2$, we must examine symmetry-protected four-band crossings.

In our classification of new fermions, we did not { focus} on four-fold band crossings, { which will be investigated in a future publication and do not require non-symmorphic symmetries to be stabilized.} 
We, and others\cite{Young12,Yang2014}, implicitly assumed that linear four band crossings in spin-orbit coupled (SOC) materials with time-reversal (TR) symmetry are essentially Dirac-like, i.e., consisting of doubly-degenerate bands with zero Chern number. Here we will show that this is not the case -- crystal symmetry can stabilize linearly dispersing fermions in the topological class of
\begin{equation}
H_{3/2}=\mathbf{k}\cdot\mathbf{S} \label{eq:Ham3/2}
\end{equation}
where
\begin{equation}
S_x=\frac{1}{2}\left(\begin{array}{cccc}
					0 & \sqrt{3} & 0 & 0\\ 
					\sqrt{3} & 0 & 2 & 0 \\ 
					0 & 2 & 0 & \sqrt{3} \\ 
					0 & 0 & \sqrt{3} & 0\end{array}\right),\;\;
S_y=\frac{1}{2i}\left(\begin{array}{cccc}
					0 & \sqrt{3} & 0 & 0\\ 
					-\sqrt{3} & 0 & 2 & 0 \\ 
					0 & -2 & 0 & \sqrt{3} \\ 
					0 & 0 & -\sqrt{3} & 0\end{array}\right),\;\; \mathrm{and}\;\;
S_z=\frac{1}{2}\left(\begin{array}{cccc}
					3 & 0 & 0 & 0\\ 
					0 & 1 & 0 & 0 \\ 
					0 & 0 & -1 & 0 \\ 
					0 & 0 & 0 & -3\end{array}\right) \label{eq:S32def}
\end{equation}
are the spin-$3/2$ matrices. Hamiltonians in this class have four energy bands non-degenerate away from $\mathbf{k}=0$, with Chern numbers $\pm 3,\pm 1$;\cite{Berry1984} unlike Dirac fermions, these are chiral topological quasiparticles -- the total monopole charge of these fermions is \emph{four} ($=3+1$).

We now find the space groups which can protect spin-$3/2$ fermions at high-symmetry points in TR invariant systems with SOC (c.f. Ref.~\onlinecite{Liang2016}, where the issue of symmetry protection and physical realizability were overlooked). We will then construct the low-energy Hamiltonian for the particle and examine its phase diagram. Finally, we remark on experimental consequences. 

\subsection{Where can topological fourfold crossings exist?}
To find crystals which support spin-$3/2$ fermions, we follow a similar strategy as in the main text: we look for high symmetry points in the Brillouin zone (BZ) of the $230$ space groups where the double-valued representation of the little group has a $4d$ irrep. We first note that not all $4d$ irreps can be topological. In particular, we need not consider space groups with inversion symmetry: these have the product $\mathcal{T}I$ of TR and inversion as a symmetry and hence all Chern numbers vanish. The non-centrosymmetric space groups are: $1,3-9,16-46,75-82,89-122,143-146,149-161,168-174,177-190,195-199$ and $207-220$. Furthermore, we note that because TR symmetry preserves the Chern number { of Fermi surfaces (monopole charge)}, a spin-$3/2$ fermion may only occur in irreps that are four-dimensional \emph{before} imposing TR symmetry.

With these constraints in mind, we find that only SGs $207-220$ have the possibility of hosting spin-$3/2$ fermions. The relevant fourfold degeneracies occur at the $\Gamma$ point in all fourteen cases, as well as at the $R$ point in SGs $207,208$ and $215$, the $H$ point in $211,214$ and $217$, and the $P$ point in SG $217$. { Even at the Brillouin zone edge, non-symmorphicity is not essential to stabilize these fourfold degeneracies. In fact, SGs $207,209,211,215,216$ and $217$ are \emph{symmorphic} } In Table \ref{tab:littlegrps} below we enumerate the generators of the little group of all of these points.

%%%%%%%%%%%%%%%%%%%%%%%%%%%%%%%%%%%%%%%%%%%%%%%%%%%%%%%%%%%%%%%%%%%%%%%%%%%%%%%%%%%%%%%%%
%%NOTE TO SELF - IVE SWITCHED THE ORDER OF GENERATORS 2 AND 3 AS COMPARED WITH THE BOOK%%
%%%%%%%%%%%%%%%%%%%%%%%%%%%%%%%%%%%%%%%%%%%%%%%%%%%%%%%%%%%%%%%%%%%%%%%%%%%%%%%%%%%%%%%%%
\begin{table}[t]
\begin{tabular}{l | c | c | l  }
SG& La & $k$ & Point group Generators \\
  \hline		
207 & cP & $\Gamma$, R & $\mathcal{G}_1=C_{4x}, \mathcal{G}_2=C_{2,1\bar{1}0},\mathcal{G}_3=C_{3,111}^{-1}$ \\
208 & cP & $\Gamma, R$ & \\
209 & cF & $\Gamma$ & \\
210 & cF & $\Gamma$ &  \\
211 & cI & $\Gamma, H$ & \\
212 & cP & $\Gamma$ & \\
213 & cP & $\Gamma$ & \\
214 & cI & $\Gamma, H$ & \\
215 & cP & $\Gamma, R$ & $\mathcal{G}_1=IC_{4x}, \mathcal{G}_2=\sigma_{1\bar{1}0},\mathcal{G}_3=C_{3.110}^{-1}$ \\
216 & cF & $\Gamma$ & \\
217 & cI & $\Gamma, H, P$ & \\
218 & cP & $\Gamma$ & \\
219 & cF & $\Gamma$ & \\
220 & cI & $\Gamma$ &
\end{tabular}
\caption{High symmetry points that can potentially host topological four-fold degeneracies, and the generators of the little group of each point.}\label{tab:littlegrps}
\end{table}
Conveniently, all of these groups are isomorphic. The representation matrices \emph{in all cases} are given by,
\begin{align}
G_1=\Delta(\mathcal{G}_1)=\left(\begin{array}{cc}M & -iM \\ -iM & M\end{array}\right),
G_2=\Delta(\mathcal{G}_2)=\left(\begin{array}{cc}0 & e^{5i\pi/4}\sigma_z \\ e^{-i\pi/4}\sigma_z & 0\end{array}\right),
G_3=-G_1^{-1}G_2G_1^{-2},\label{eq:4drep}
\end{align}
where we have introduced
\begin{equation}
M=-\frac{1}{2}(\sigma_z+\sqrt{3}\sigma_x)
\end{equation}

We can now make one final observation: to recover the Hamiltonian (\ref{eq:Ham3/2}) as the low-energy $\mathbf{k}\cdot\mathbf{p}$ Hamiltonian, there cannot be any symmetry-required degeneracies along lines emanating from the fourfold point. This condition is violated in SGs $215-220$, as we will now argue. Consider first the line $(0,k,0)$ originating from a fourfold degeneracy at the $\Gamma$ (and $H$, where relevant) point. The little group of this line is represented as
\begin{align}
\Delta(C_{2y})=G_3^{-1}G_1^2G_3,\;\mathrm{and}\;\;
\Delta(\sigma_{10\bar{1}})=G_2G_3	
\end{align}
This representation is reducible. { To see this, let us work in the basis where $\Delta(C_{2y})$ is diagonal. In this basis we have
\begin{equation}
\Delta(C_{2y})=\left(\begin{array}{cccc}i&0&0&0\\0&-i&0&0\\0&0&i&0\\0&0&0&-i\end{array}\right), \Delta(\sigma_{10\bar{1}})=\left(\begin{array}{cccc}0&e^{i\pi/4}&0&0\\e^{3i\pi/4}&0&0&0\\0&0&0&e^{5i\pi/4}\\0&0&e^{7i\pi/4}&0\end{array}\right).
\end{equation}
 It is easy to see that this representation splits into two two-dimensional irreps.} Hence, the low-energy Hamiltonian for these space groups have twofold degeneracies along this line, and hence these fourfold degeneracies cannot support a monopole charge. A similar story holds for the $P$ point in SG $217$ and the $R$ point in SG $215$ (there the relevant line there is $(0,0,k)$ and same argument goes through after permuting the components of $\mathbf{k}$.)

We thus see that the eight space groups SG $207-214$ can host fourfold degeneracies, which may be topological. These may occur at the $\Gamma$ point in all eight SGs, as well as at the $R$ point in SGs $207$ and $208$, and the $H$ point in SGs $211$ and $214$. We will now proceed to build the Hamiltonian.

\subsection{$\mathbf{k}\cdot\mathbf{p}$ near the Spin-$3/2$ point}
We can now find the most general $4\times 4$ linear Hamiltonian around the non-degenerate spin-3/2 point(s) in SGs $207-214$. To do so, we enforce the constraints
\begin{equation}
G_1H(\delta k_x,\delta k_y,\delta k_z)G_1^{-1}=H(\delta k_x,-\delta k_z,\delta k_y),\mathrm{and}\;G_2H(\delta k_x,\delta k_y,\delta k_z)G_2^{-1}=H(-\delta k_y,-\delta k_x,-\delta k_z).
\end{equation}
Since in these SGs we are always considering TR invariant momenta, we must also construct the matrix representation { $G_0=\Delta(\mathcal{T})K$ of time reversal, where $K$ is complex conjugation. Demanding that 
\begin{equation}
G_0^2=-1,\;\;G_0G_1G_0^{-1}=G_1,\;\;G_0G_2G_0^{-1}=G_2,
\end{equation}}
we find we can take
\begin{equation}
T=\left(\begin{array}{cc} 0 & -\mathbb{I}_2 \\ \mathbb{I}_2 & 0 \end{array}\right) K
\end{equation}
where $\mathbb{I}_2$ is the $2\times2$ identity matrix. It turns out, however, that to linear order time reversal imposes no additional constraints on the Hamiltonian. We have, after setting the zero of energy to $0$,
\begin{equation}
H=\left(\begin{array}{cccc}
a\delta k_z & 0 &-\frac{a+3b}{4}\delta k_+ & \frac{\sqrt{3}}{4}(a-b)\delta k_- \\
0 & b\delta k_z & \frac{\sqrt{3}}{4}(a-b)\delta k_- & -\frac{3a+b}{4}\delta k_+ \\
-\frac{a+3b}{4}\delta k_- & \frac{\sqrt{3}}{4}(a-b)\delta k_+ & -a \delta k_z & 0 \\
\frac{\sqrt{3}}{4}(a-b)\delta k_+ & -\frac{`3a+b}{4}\delta k_- & 0 & -b\delta k_z
\end{array}\right),
\end{equation}
where  $a$ and $b$ are two real-valued parameters. The parameter space of this Hamiltonian is very rich. First, note that when $b=-a/3$ we have, afer permuting rows and columns
\begin{equation}
H_{b=-a/3}\approx \frac{a}{3} \delta \mathbf{k}\cdot \mathbf{S} \label{eq:kdotS}
\end{equation}
{ with $\mathbf{S}$ defined in Eq.~(\ref{eq:S32def})} as desired. Similarly, when $b=-3a$ we have 
\begin{equation}
H_{b=-3a}=a\delta\mathbf{k}\cdot\mathbf{S}^* \label{eq:mkdotS}
\end{equation}
Another interesting point occurs when $a=b$, where we have
\begin{equation}
H_{a=b}\approx a (H_1\oplus H_1) \label{eq:twoweyl}
\end{equation}
where
\begin{equation}
H_1=\delta k_z\sigma_z-\delta k_x\sigma_x+\delta k_y\sigma_y
\end{equation}
which corresponds to two degenerate negative chirality Weyl points. { This is distinct from the usual Dirac point\cite{Young12,Yang15}, which can be viewed as a superposition of two \emph{opposite} chirality Weyl points.} We now map the full phase diagram.

\subsection{Full phase diagram of the spin-3/2 fermion}

Let us now systematically analyze the different possible phases of the Hamiltonian Eq.~(\ref{eq:Ham3/2}). First, let us define
{ \begin{equation}
 a=C\cos\phi,\;\mathrm{and}\;b=C\sin\phi,
 \end{equation}
 where $\phi\in[-\pi,\pi)$ and $C>0$.} We would like to know the different possible Chern numbers of bands as $\phi$ is varied. We start by looking for values of $\phi$ at which the Chern numbers of bands change, i.e., when bands become degenerate. As in Section \ref{sec:199kdotp}, we do this by looking at the characteristic polynomial,
\begin{equation}
p(\mathbf{k},\phi,\epsilon)=\mathrm\det(H-\epsilon\mathbb{I}),
\end{equation}
and ask for what values of $\mathbf{k},\phi$ it can factor as
\begin{equation}
p\stackrel{?}{=}(\epsilon-\alpha)^2(\epsilon-\beta)(\epsilon-\gamma). \label{eq:polyfactor}
\end{equation}
Proceeding, we have
\begin{equation}
p(\mathbf{k},\phi,\epsilon)=\epsilon^4-\epsilon^2|\mathbf{k}|^2+\frac{1}{8}(1-\cos4\phi)(k_x^4+k_y^4+k_z^4)+\frac{1}{8}(\frac{11}{4}+\frac{7}{4}\cos 4\phi+3\sin 2\phi)(k_x^2k_y^2+k_x^2k_z^2+k_y^2k_z^2)\label{eq:charpoly}
\end{equation}
The fact that this contains no linear or cubic terms imposes the following constraints on the $\{\alpha,\beta,\gamma\}$ appearing in the factorization (\ref{eq:polyfactor}):
\begin{equation}
2\alpha+\beta+\gamma=\alpha^2(\beta+\gamma)+2\alpha\beta\gamma=0
\end{equation}
These constraints have two distinct types of solution: either
\begin{equation}
\beta=\gamma=-\alpha
\end{equation}
or
\begin{equation}
\alpha=0,\;\;\beta=-\gamma
\end{equation}
Let us examine the type-I solutions first. These require
\begin{equation}
p=(\epsilon+\alpha)^2(\epsilon-\alpha)^2=\epsilon^4-2\alpha^2\epsilon^2+\alpha^4
\end{equation}
Comparing with Eq.~(\ref{eq:charpoly}), we see this requires
\begin{equation}
2|\mathbf{k}|^4=(1-\cos4\phi)(k_x^4+k_y^4+k_z^4)+(\frac{11}{4}+\frac{7}{4}\cos 4\phi+3\sin 2\phi)(k_x^2k_y^2+k_x^2k_z^2+k_y^2k_z^2)
\end{equation}
This is satisfied for all $k$ at
\begin{equation}
\phi=\frac{(4n+1)\pi}{4},\;\; n\in\mathbb{Z}
\end{equation}
and along the lines $k_i=k_j=0$ with $i,j\in\{x,y,z\}$, $i\neq j$ and
\begin{equation}
\phi=\frac{(4n+3)\pi}{4}
\end{equation}
Examining solutions of type 2, we find that there are degeneracies along the lines $k_i=k_j=0$ when
\begin{equation}
\phi=\frac{n\pi}{2},
\end{equation}
and that there is a second set of degeneracies when $\phi=\phi_2$, along the lines $|k_x|=|k_y|=|k_z|$.

Having identified all parameter values where the bands become degenerate (Lifschitz topological phase transitions), we can now move on to identify the distinct phases of the Hamiltonian. We focus only on the parameter range $\phi\in[-\pi/2,\pi/2]$; the rest of the phase diagram can be obtained by taking $H\rightarrow-H$ Using the identifications from Eq. (\ref{eq:kdotS}) and Eq. (\ref{eq:mkdotS}), we note that for $-\pi/2<\phi<-\pi/4$ the Hamiltonian is in the same topological phase as $-\mathbf{k}\cdot\mathbf{S}$, and the bands have Chern numbers $\nu=3,1,-1,-3$ ordered in decreasing energy. For $-\pi/4<\phi<0$, we have that the Hamiltonian is in the same topological phase as $\mathbf{k}\cdot\mathbf{S}$, with bands having Chern numbers $\nu=-3,-1,1,3$ ordered in decreasing energy. As with the spin-$1$ Weyl, the two sets of band degeneracies for $\phi=-\pi/4$ allow for the Chern numbers of the bands to reverse ordering in energy at this transition.

Next, note that for $\phi=0$ and $\phi=\pi/2$, the only degeneracies are between the two middle bands, at zero energy. This, along with Eq. (\ref{eq:twoweyl}) allows us to deduce that the degeneracy at $\phi=\pi/4$ { need not be stable when higher-order terms in the $\mathbf{k}\cdot\mathbf{p}$ Hamiltonian are considered (we cannot exclude the possibility that it broadens into a distinct phase, however). Finally, we conclude that for $0<\phi<\pi/2, \phi\neq\pi/4$ bands have} Chern numbers $\nu=-3,1,-1,3$ ordered in decreasing energy. As far as we are aware, this phase has no precedent in prior experimental or theoretical work.
}

%%%%%%%%%%%%%%%%%%%%%%%%%%%%%%%%%%%%%%%%%%%%%%%%%%%%%%%%
%%%%%%%%%%%%%%%%%%%%%%%%%%%%%%%%%%%%%%%%%%%%%%%%%%%%%%%%
%%%%%%%%%%%%%%%%%%%%%%%%%%%%%%%%%%%%%%%%%%%%%%%%%%%%%%%%

\section{Additional candidate materials}

Here we show additional candidate materials that realize our new fermions. In Table~\ref{tab:materials}, we provide a complete list of materials (from the main text and this appendix) and indicate whether the materials exist in single crystal or powder form, or have not been reported but are predicted to be stable.
\begin{table}
\begin{center}
\subfloat[ ]{
\begin{tabular}{l|l}
SG & Compound \\ \hline
130 & PdBi$_2$O$_4$ \\
130 & WO$_3$ \\
135 & PdSe \\
198 & AsPdS \\
198 & FeSi \\
198 & K$_3$BiTe$_3$ \\
198 & SiCo \\
199 & Pd$_3$Bi$_2$S$_2$ \\
205 & PdSb$_2$ \\
212 & Li$_2$Pd$_3$B \\
212 & SrSi$_2$ \\
213 & Mg$_3$Ru$_2$ \\
214 & Ag$_3$Se$_2$Au \\
214 & La$_3$PbI$_3$ \\
218 & CsSn \\
220 & Ba$_4$Bi$_3$ \\
220 & La$_4$Bi$_3$ \\
\end{tabular}} \quad
\subfloat[ ]{
\begin{tabular}{l|l}
SG & Compound \\ \hline
130 & CuBi$_2$O$_4$  \\
135 & PdS \\
198 & MgPt  \\
198 & AlPt  \\
198 & AlPd  \\
198 & RuSi  \\
198 & OsSi  \\
206 & KBiF$_6$  \\
213 & Li$_2$Pt$_3$B  \\
213 & Re$_2$W$_3$C  \\
223 & LaPd$_3$S$_4$ \\
223 & Nb$_3$Bi \\
223 & Ta$_3$Sb \\
230 & RhBi \\ 
\end{tabular}}
\quad
\subfloat[ ]{
\begin{tabular}{l|l}
SG & Compound\\
\hline
130 & AgBi$_2$O$_4$ \\
130 & AuBi$_2$O$_4$ \\
135 & PdTe \\
199 & Ni$_3$(BiS)$_2$ \\
199 & Pd$_3$Bi$_2$Se$_2$ \\
199 & Pd$_3$Bi$_2$Te$_2$ \\
230 & SiO$_2$ \\
\end{tabular}}
\end{center}
\label{tab:materials}\caption{Candidate materials that (a) exist in single crystal form, (b) exist in powder form, or (c) are predicted.}
\end{table}

We begin with space group 198, of which MgPt was an example shown in the main text. In Fig~\ref{fig:dft198}, we show seven other binary compounds (AlPt\cite{Ferro63}, AlPd\cite{Ferro64}, SiCo\cite{Demchenko08}, RuSi\cite{Weitzer97}, ReSi\cite{Skolozdra69}, OsSi\cite{Korst57} and FeSi\cite{Neef06}) in the space group that also shown band crossings within 1eV of the Fermi level and exist as either single crystals or in powder form.

In the family of Pd$_3$Bi$_2$S$_2$, shown in the main text, is Ni$_3$(BiS)$_2$; numerical calculations show that the electronic configuration for Ni$_3$(BiS)$_2$ in SG 12, which is known to exist in crystal form, is only slightly lower in energy than in SG 199.\cite{Weihrich07}
We show its band structure in Fig. \ref{fig:dft199}, which shows a three-band crossing at the $P$ point only .1eV above the Fermi level. 
In Fig. \ref{fig:dft199-2} and \ref{fig:dft199-3} we show that the predicted compounds Pd$_3$Bi$_2$Se$_2$ and Pd$_3$Bi$_2$Te$_2$, which exhibit 3-fold band crossings almost exactly at the Fermi level.
Analogous band crossings in SG 214 can be found in the family of La$_3$PbI$_3$\cite{Mattausch04}; Fig.~\ref{fig:dft214} shows that the Fermi level is almost exactly at the band crossing.

In the main text, we showed materials in SG 220 that exhibit 6-fold and 8-fold fermions near the Fermi level. Here, we show their counter parts in SG 230. One example is RhBi$_4$, which exists in powder form.\cite{Ross59} Fig~\ref{fig:dft230a} shows its 6-fold and 8-fold band crossings within .1eV of the Fermi level. SiO$_2$ can also feasibly exist in SG 230\cite{Foster04}. Fig~\ref{fig:dft230b} shows that it exhibits both types of new fermions, although well above the Fermi level.

Fig.~\ref{fig:dft212-1} shows a 6-fold fermion in SG 212 in SrSi$_2$, which is known to exist in single crystal form\cite{Palenzona04}, .7eV above the Fermi level.
Fig.~\ref{fig:dft213-2} and \ref{fig:dft213-3} show its  counterparts in SG 213, in Re$_2$W$_3$\cite{Lawson71} and Li$_2$Pt$_3$B\cite{Mochiku06}, which both exist in powder form.
Similarly, in SG 206, Fig.~\ref{fig:dft206} shows the 6-band crossing in KBiF$_6$ at the $P$ point, which is the counterpart of that in SG 205 shown in the main text. KBi$F_6$ is known to exist in powder form\cite{Hebecker71}.

8-fold fermions are required to exist in SG 130. In the main text, we showed two bismuth oxides in this space group. Figs~\ref{fig:dft130-3} and \ref{fig:dft130-4} show the band structures for the hypothetical compounds AuBi$_2$O$_3$ and AgBi$_2$O$_3$; the latter is slightly magnetic, which explains small band splittings that are protected in the former. Both show the band crossing at the $A$ point about .1eV below the Fermi level and isolated from all other bands. 
Another example in this space group is WO$_3$\cite{Charushnikova97}, shown in Fig.~\ref{fig:dft130-2}. This material exists in single crystal form.

8-fold fermions are also required to exist in SG 135. The main text showed PdS as an example. Figs~\ref{fig:dft135} and \ref{fig:dft135-3} show 8-band crossings in the related materials PdSe, which has been observed in single crystal form\cite{Ijjaali01}  and PdTe, which is predicted to be stable. In both cases, the band crossing is between .25eV  and .5eV above the Fermi level.

\begin{figure}[H]\centering
\subfloat[AlPt (SG 198)\label{fig:AlPt}]{%
  \includegraphics[width=.3\textwidth]{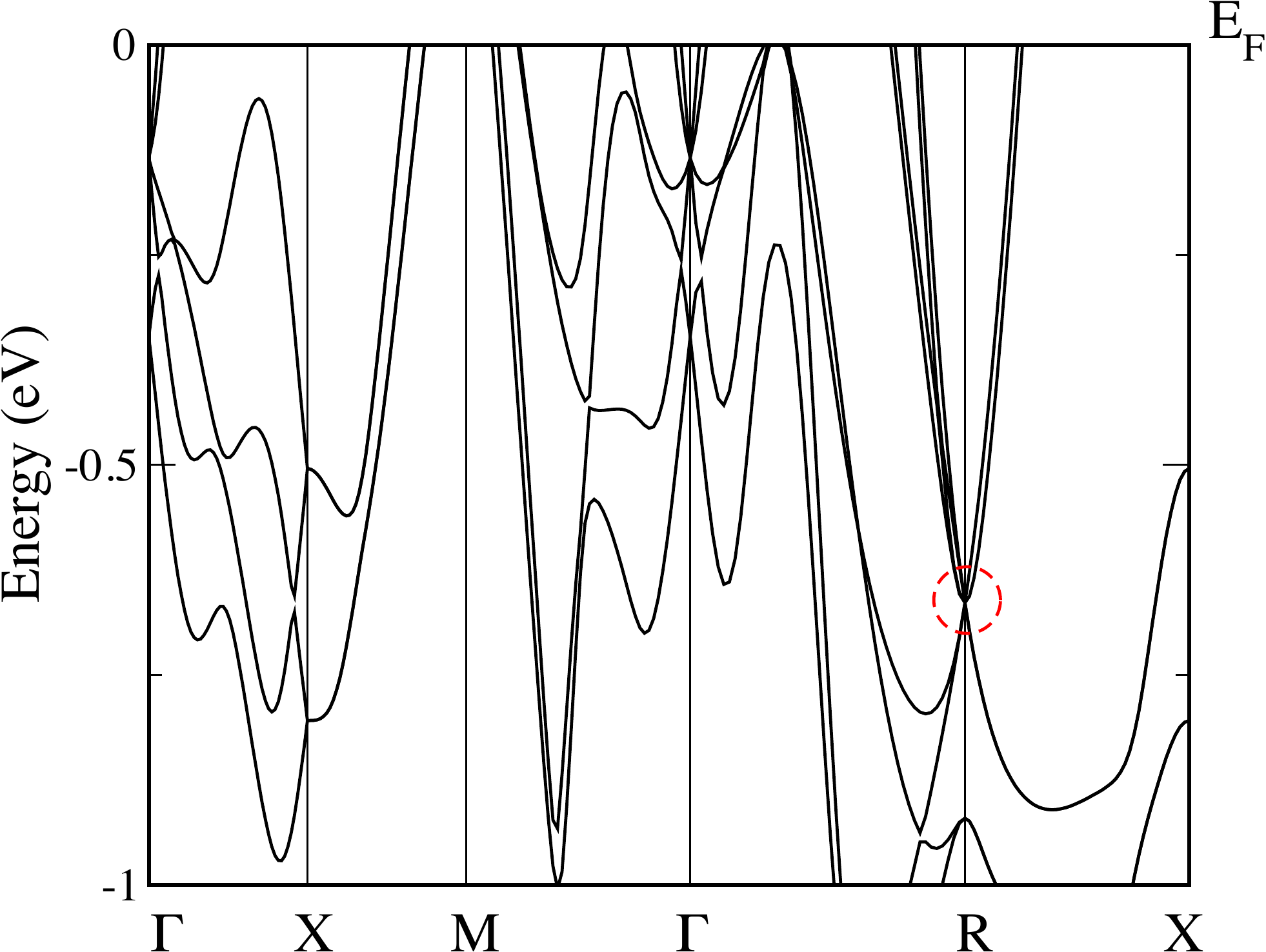}%
}\quad
\subfloat[AlPd (SG 198)\label{fig:AlPd}]{%
  \includegraphics[width=.3\textwidth]{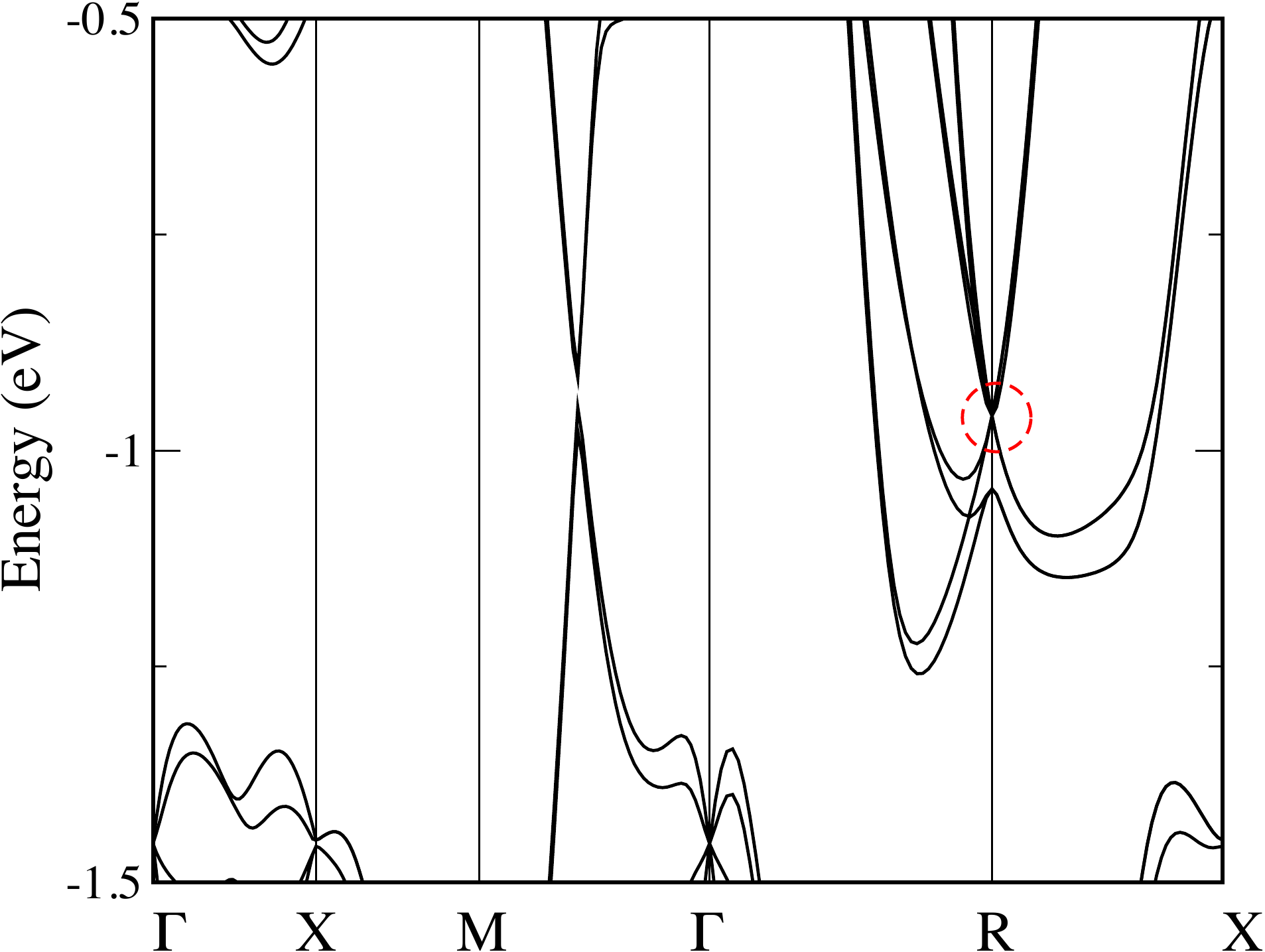}%
}\quad \\ 
\subfloat[SiCo (SG 198)\label{fig:SiCo}]{%
  \includegraphics[width=.3\textwidth]{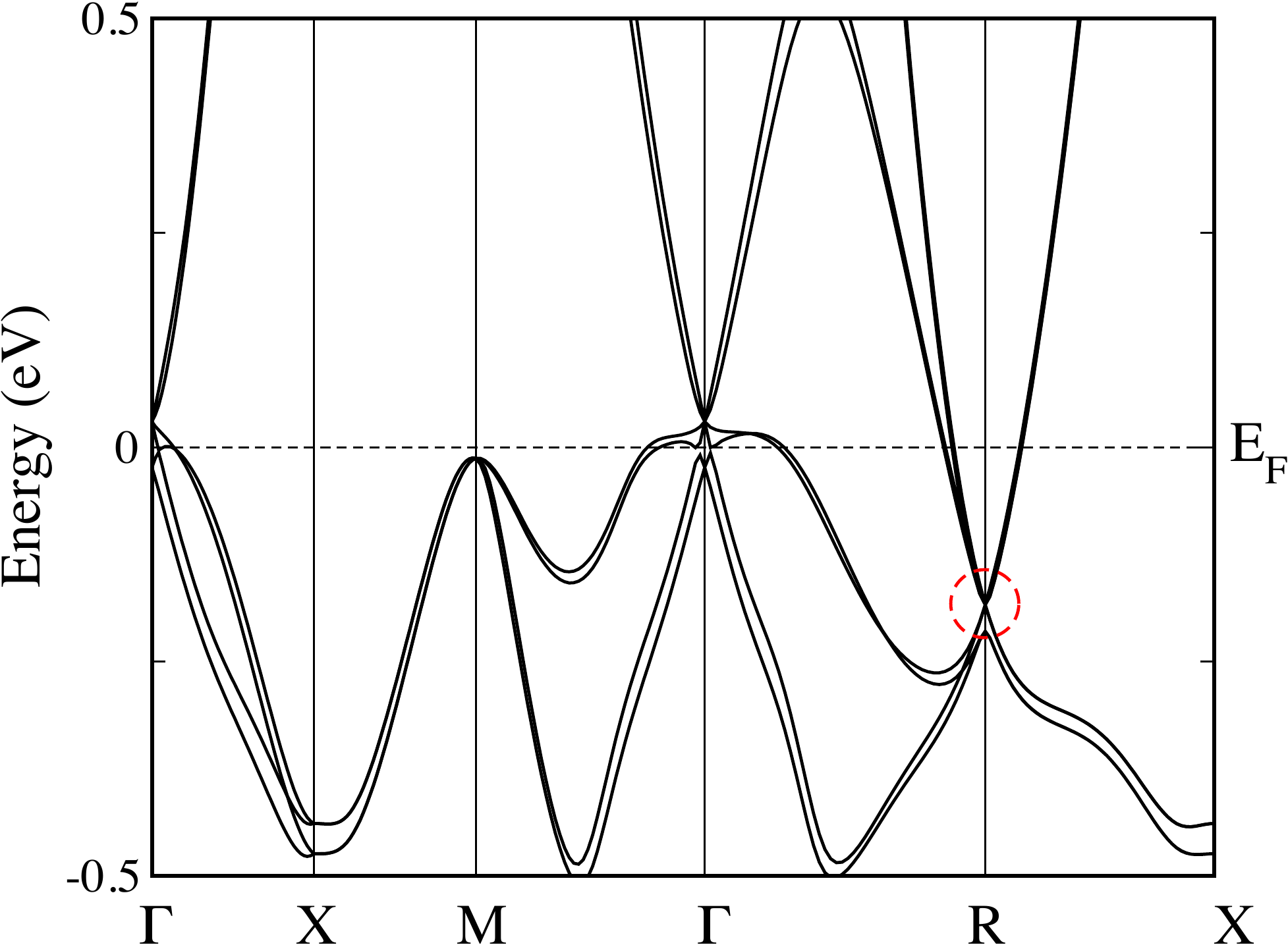}%
}\quad
\subfloat[RuSi (SG 198)\label{fig:RuSi}]{%
  \includegraphics[width=.3\textwidth]{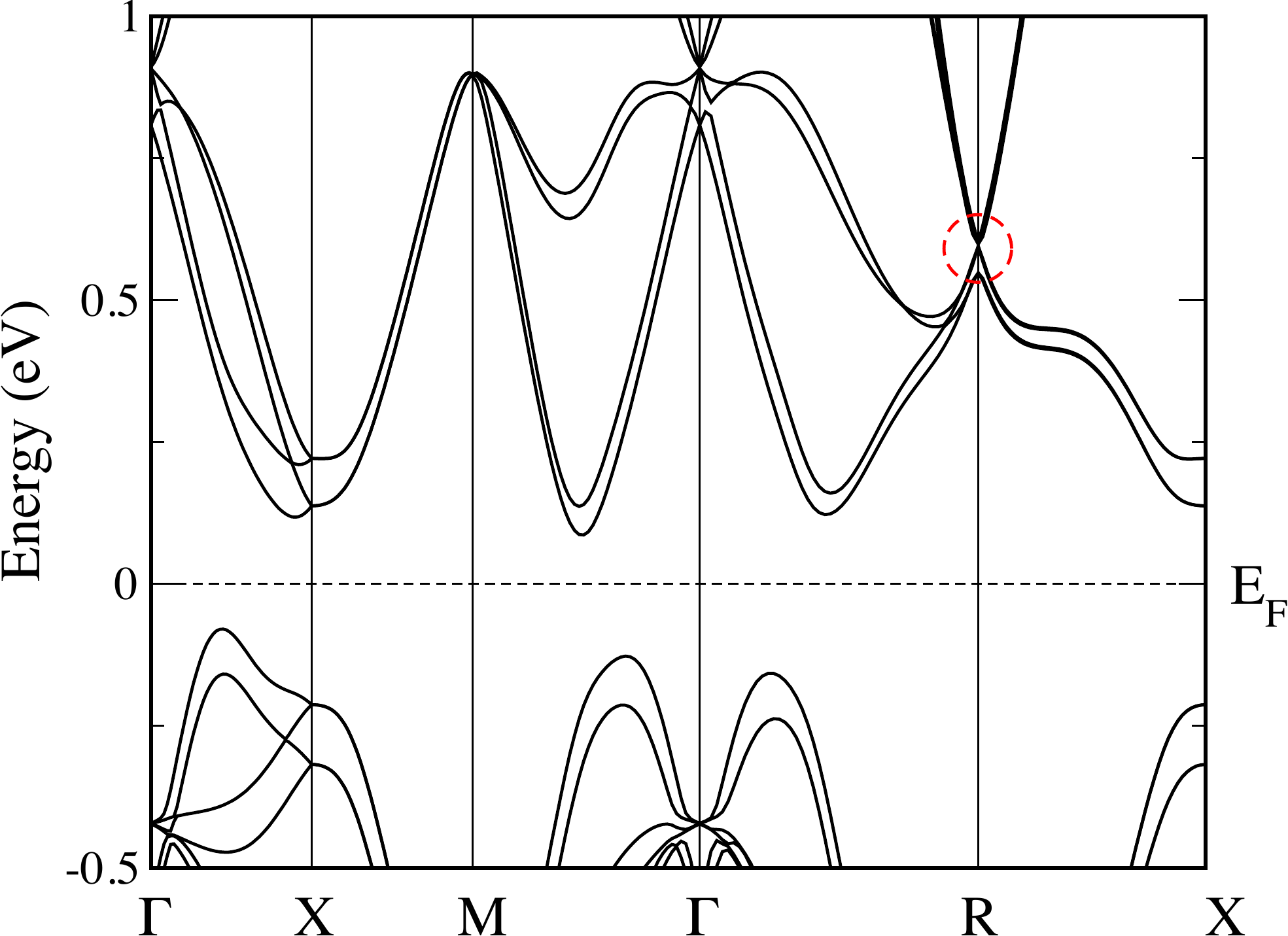}%
}\quad
\subfloat[ReSi (SG 198)\label{fig:ReSi}]{%
  \includegraphics[width=.3\textwidth]{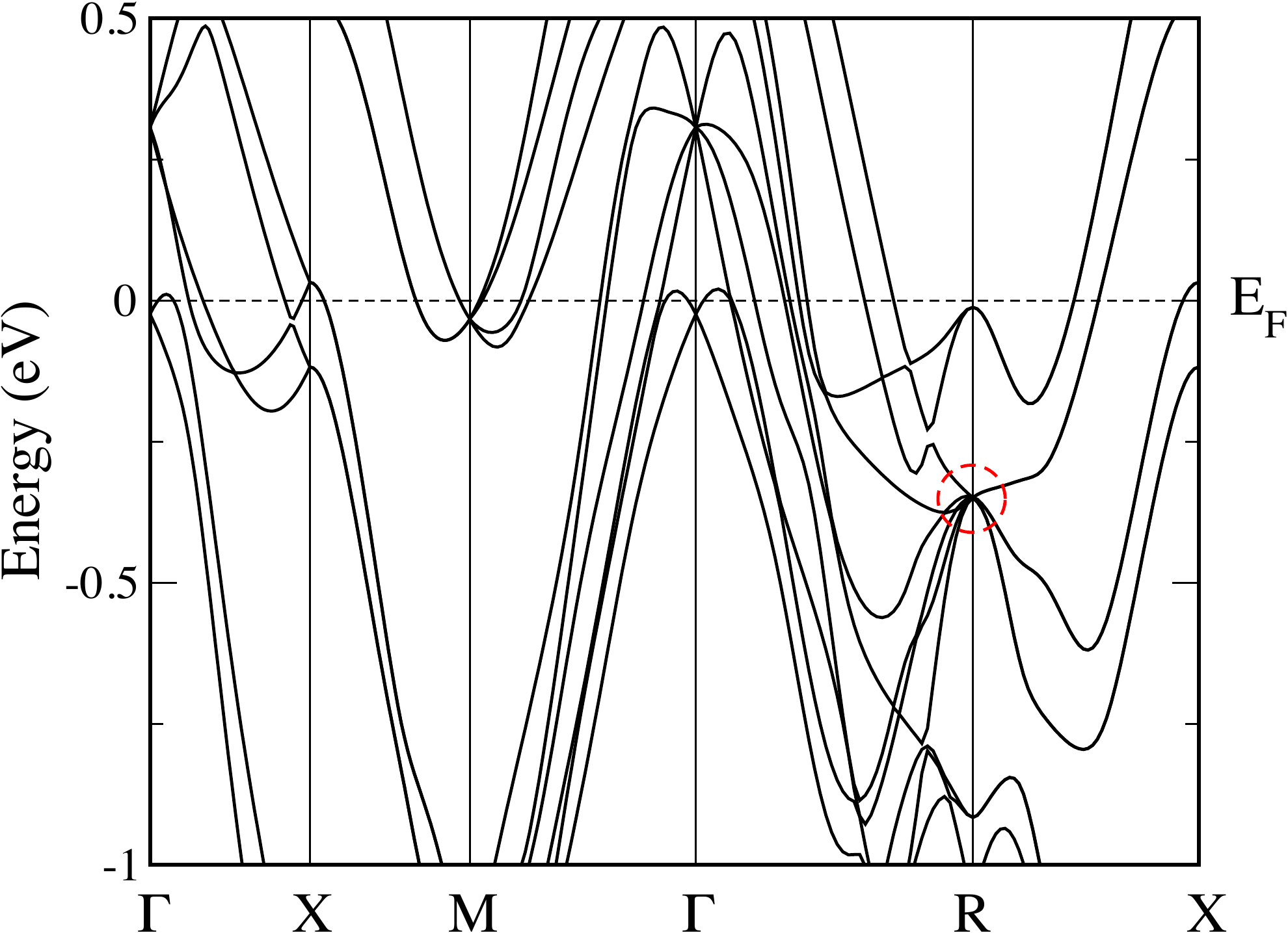}%
}\quad
\subfloat[OsSi (SG 198)\label{fig:OsSi}]{%
  \includegraphics[width=.3\textwidth]{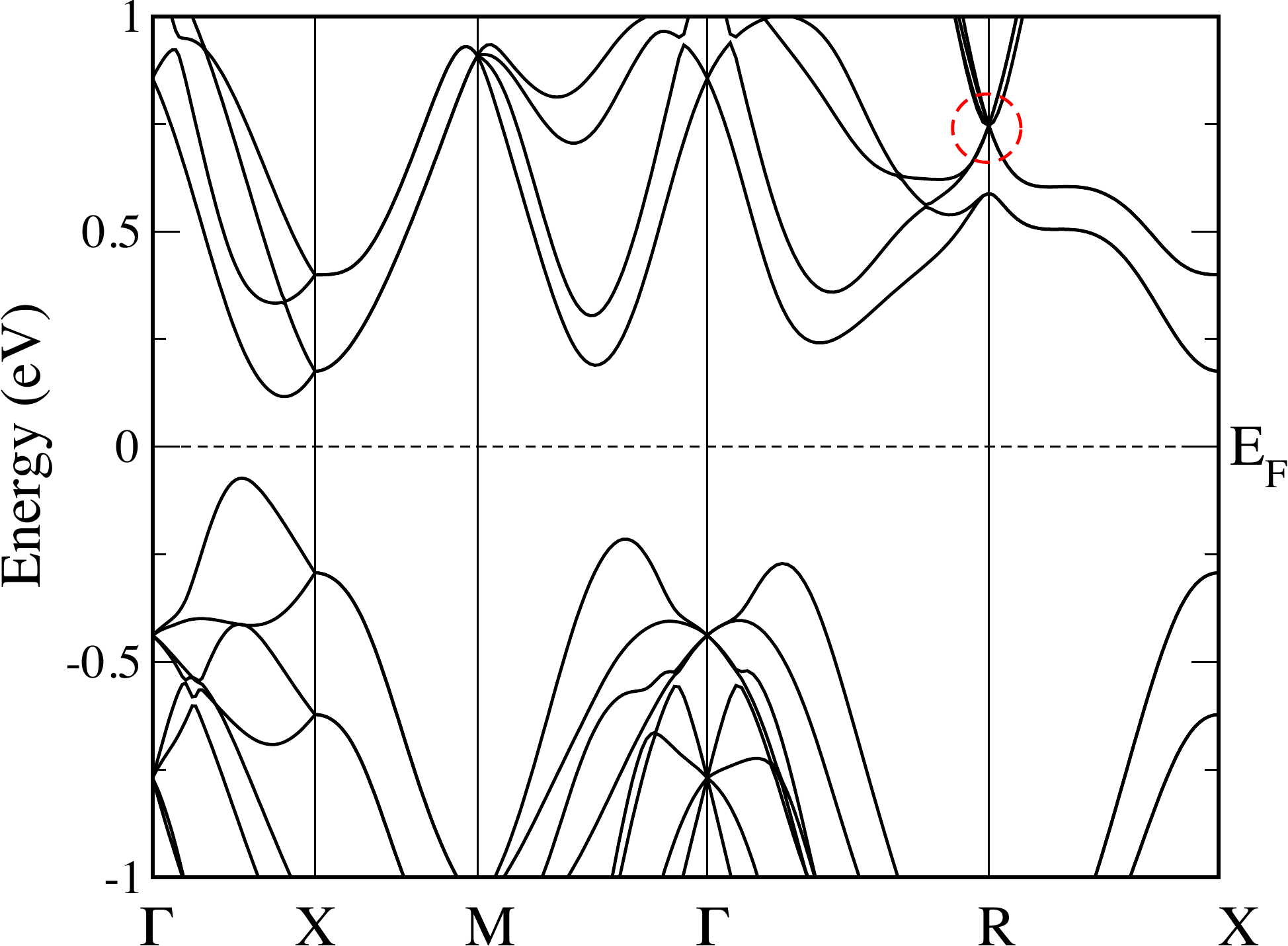}%
}\quad
\subfloat[FeSi (SG 198)\label{fig:FeSi}]{%
  \includegraphics[width=.3\textwidth]{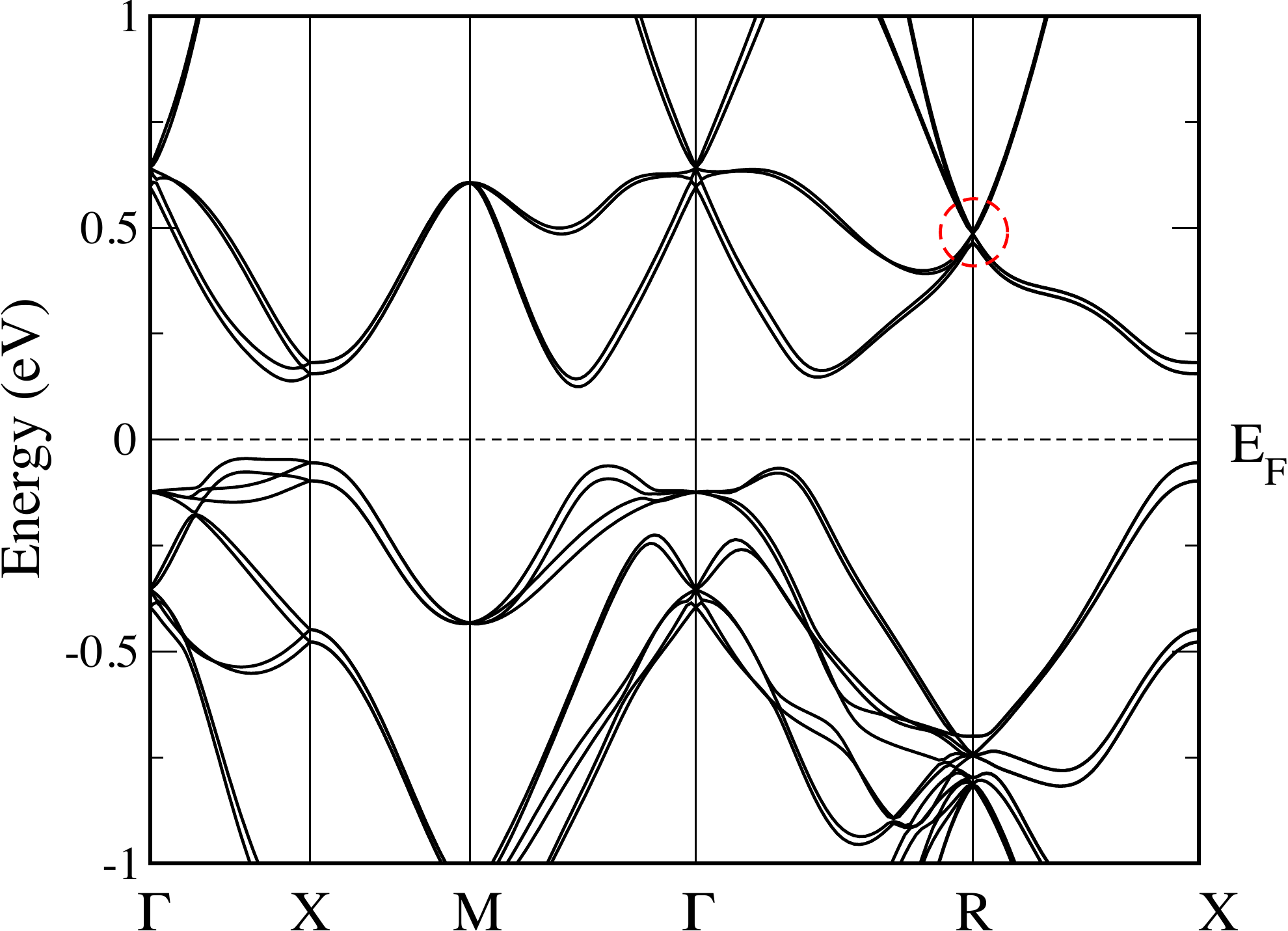}%
}
\caption{6-fold fermions at the $R$ point in SG 198}
\label{fig:dft198}
\end{figure}

\begin{figure}[H]\centering
\subfloat[Ni$_3$(BiS)$_2$ (SG 199)\label{fig:dft199}]{%
  \includegraphics[width=.3\textwidth]{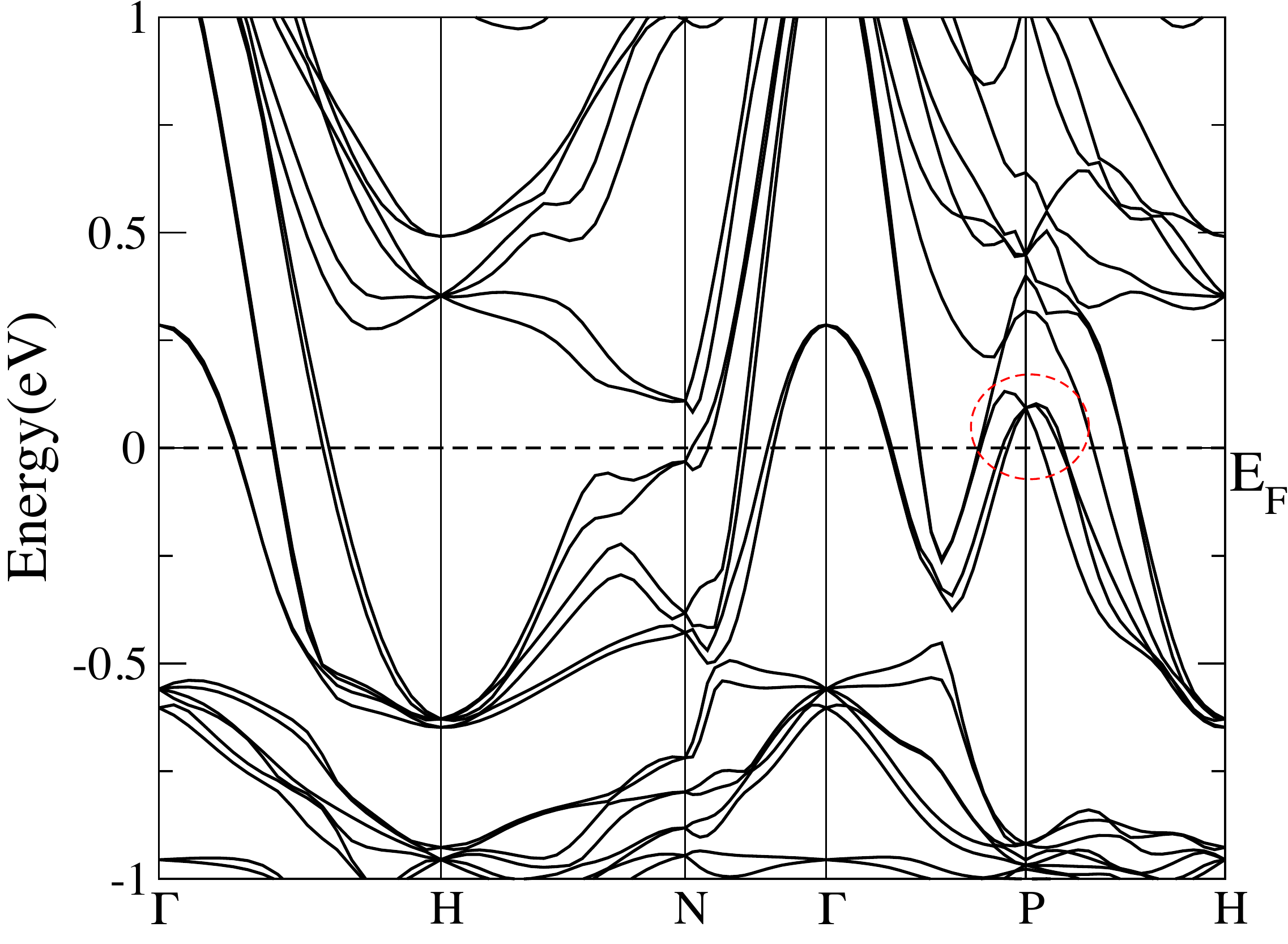}%
}\quad
\subfloat[Pd$_3$Bi$_2$Se (SG 199)\label{fig:dft199-2}]{%
  \includegraphics[width=.3\textwidth]{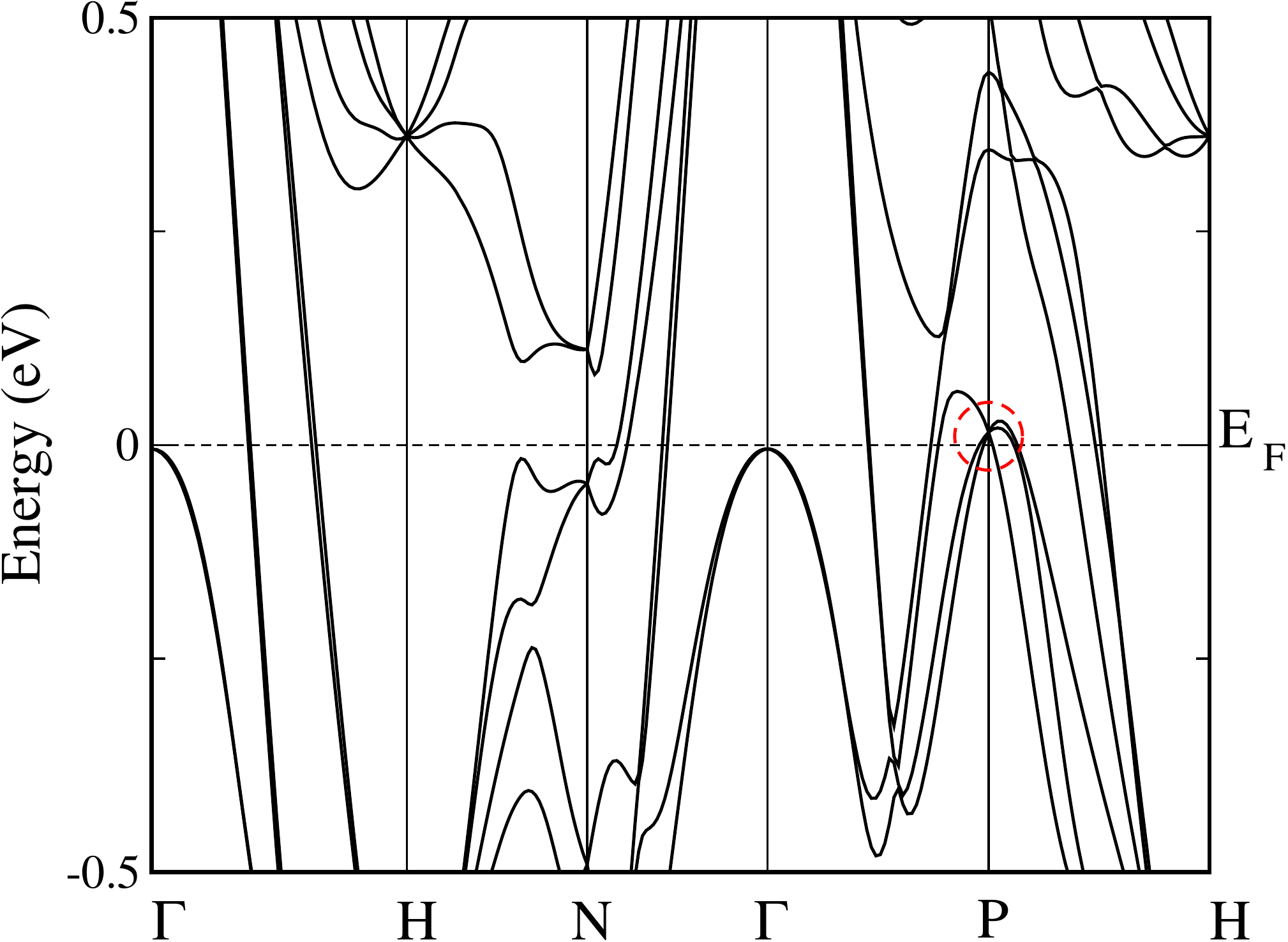}%
}\quad
\subfloat[Pd$_3$Bi$_2$Te (SG 199)\label{fig:dft199-3}]{%
  \includegraphics[width=.3\textwidth]{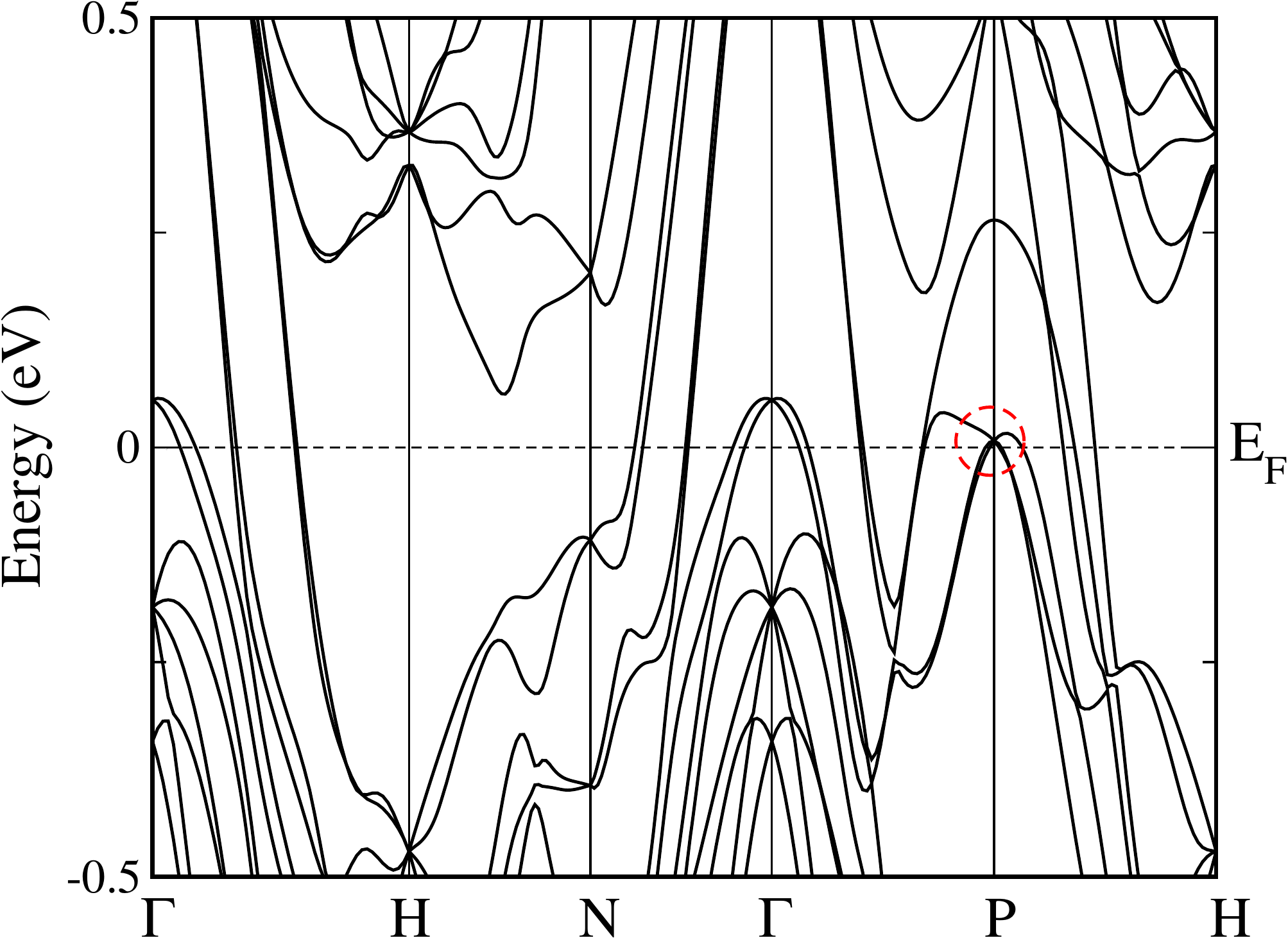}%
}\quad
\subfloat[La$_3$PbI$_3$ (SG 214)\label{fig:dft214}]{%
  \includegraphics[width=.3\textwidth]{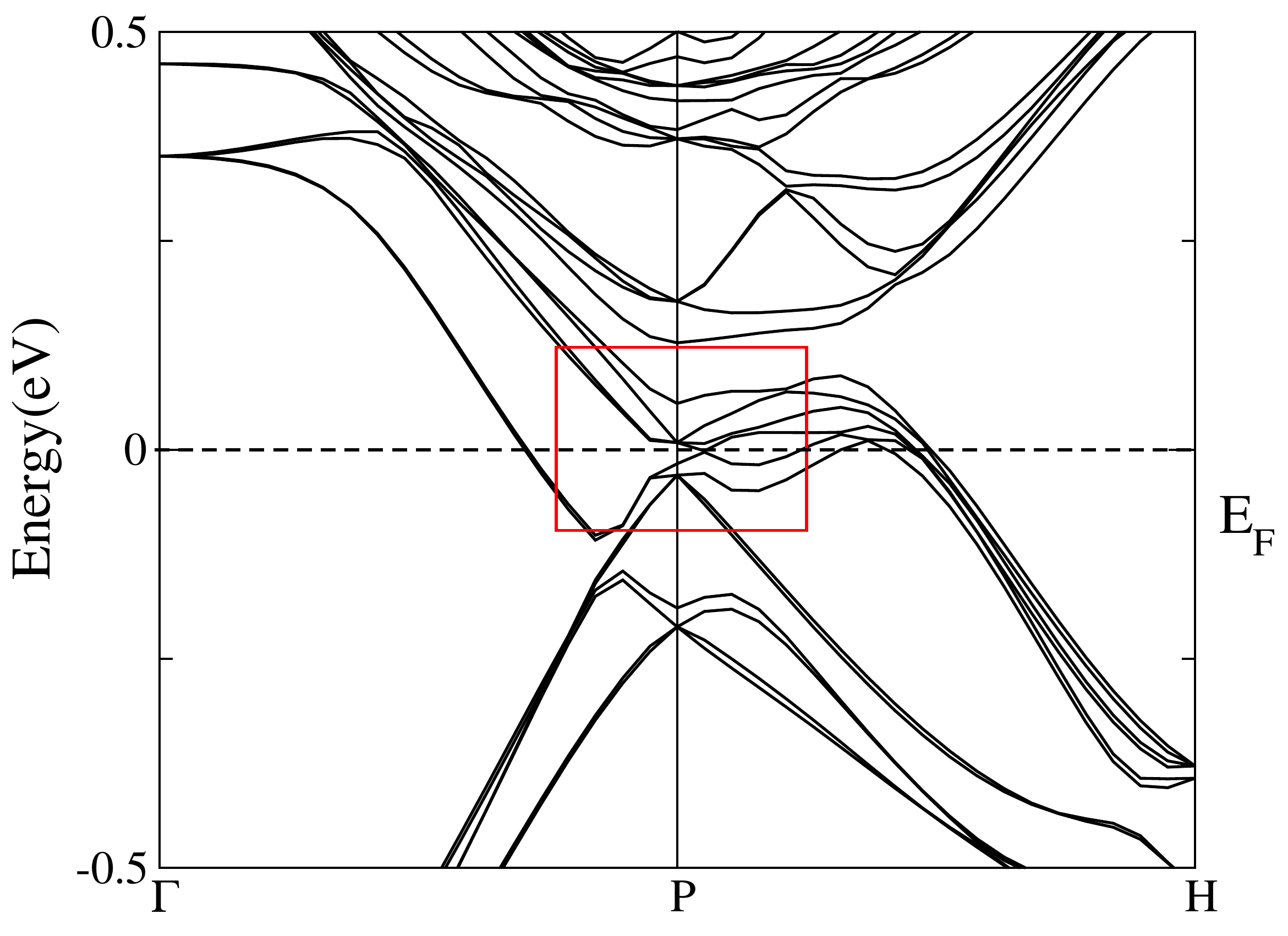}%
	}\quad
\subfloat[SrSi$_2$ (SG 212)\label{fig:dft212-1}]{
	\includegraphics[width=.3\textwidth]{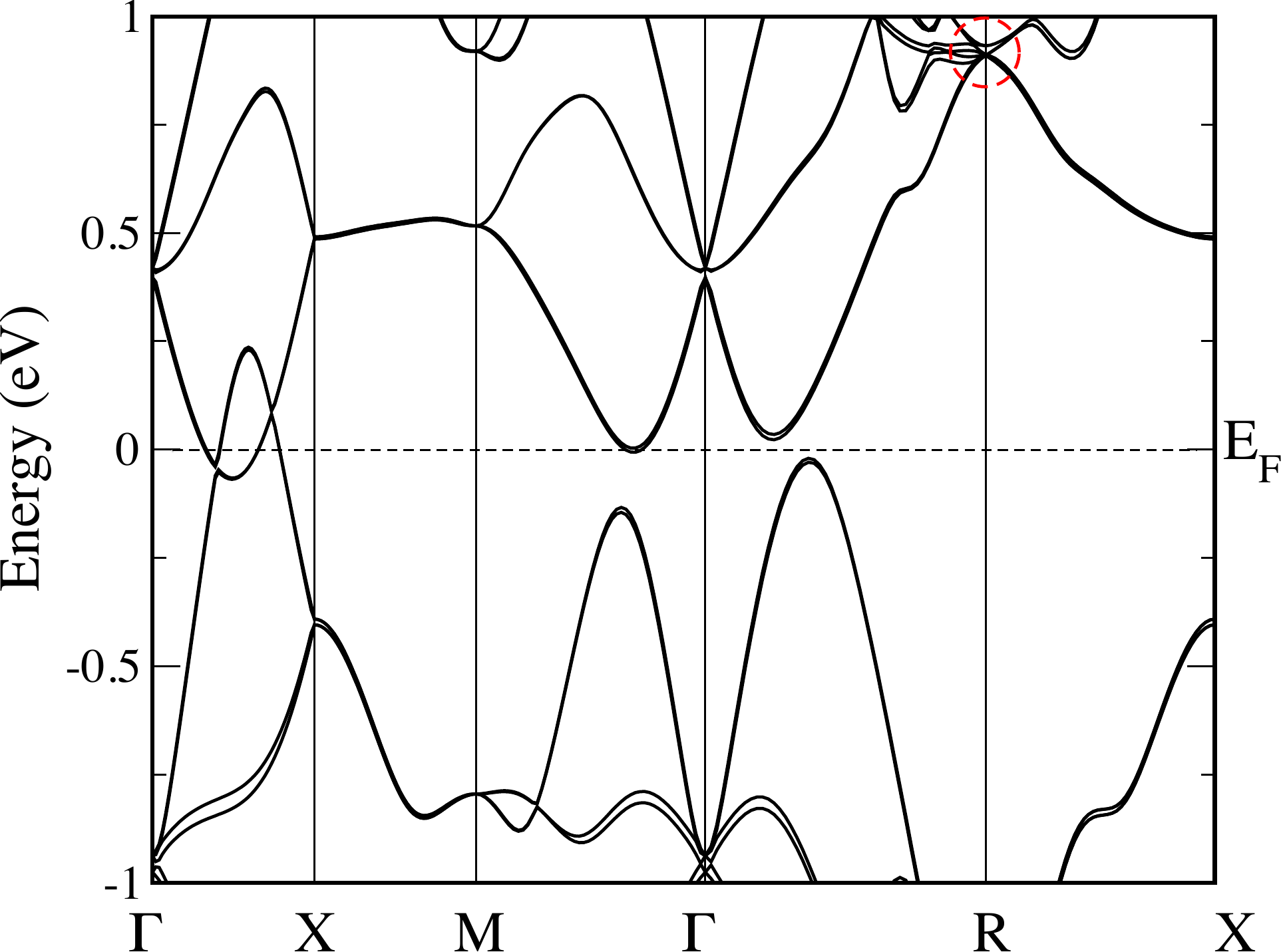}
}\quad
\subfloat[Re$_2$W$_3$C (SG 213)\label{fig:dft213-2}]{
	\includegraphics[width=.3\textwidth]{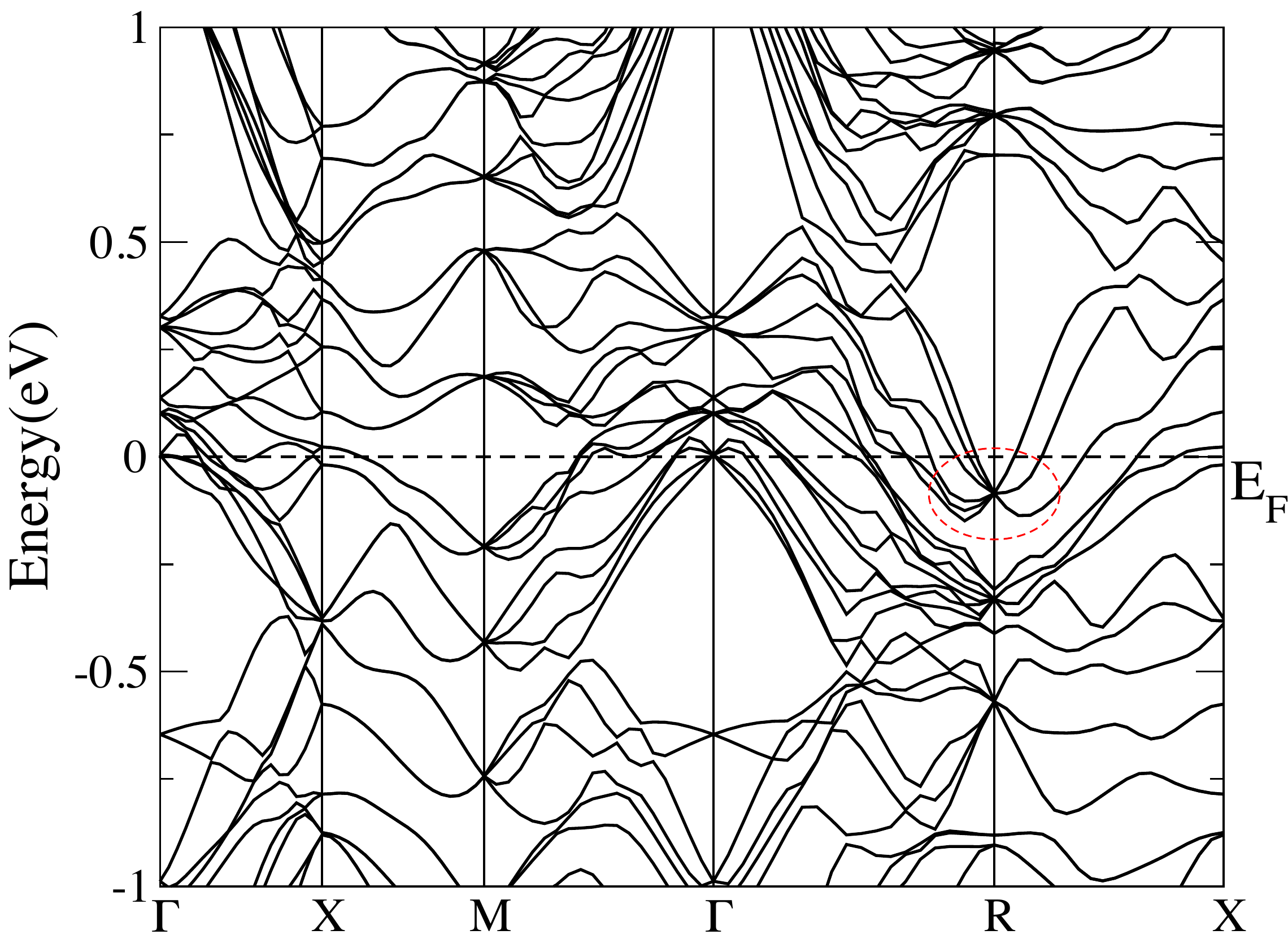}
}\quad 
\subfloat[Li$_2$Pt$_3$B (SG 213)\label{fig:dft213-3}]{
	\includegraphics[width=.3\textwidth]{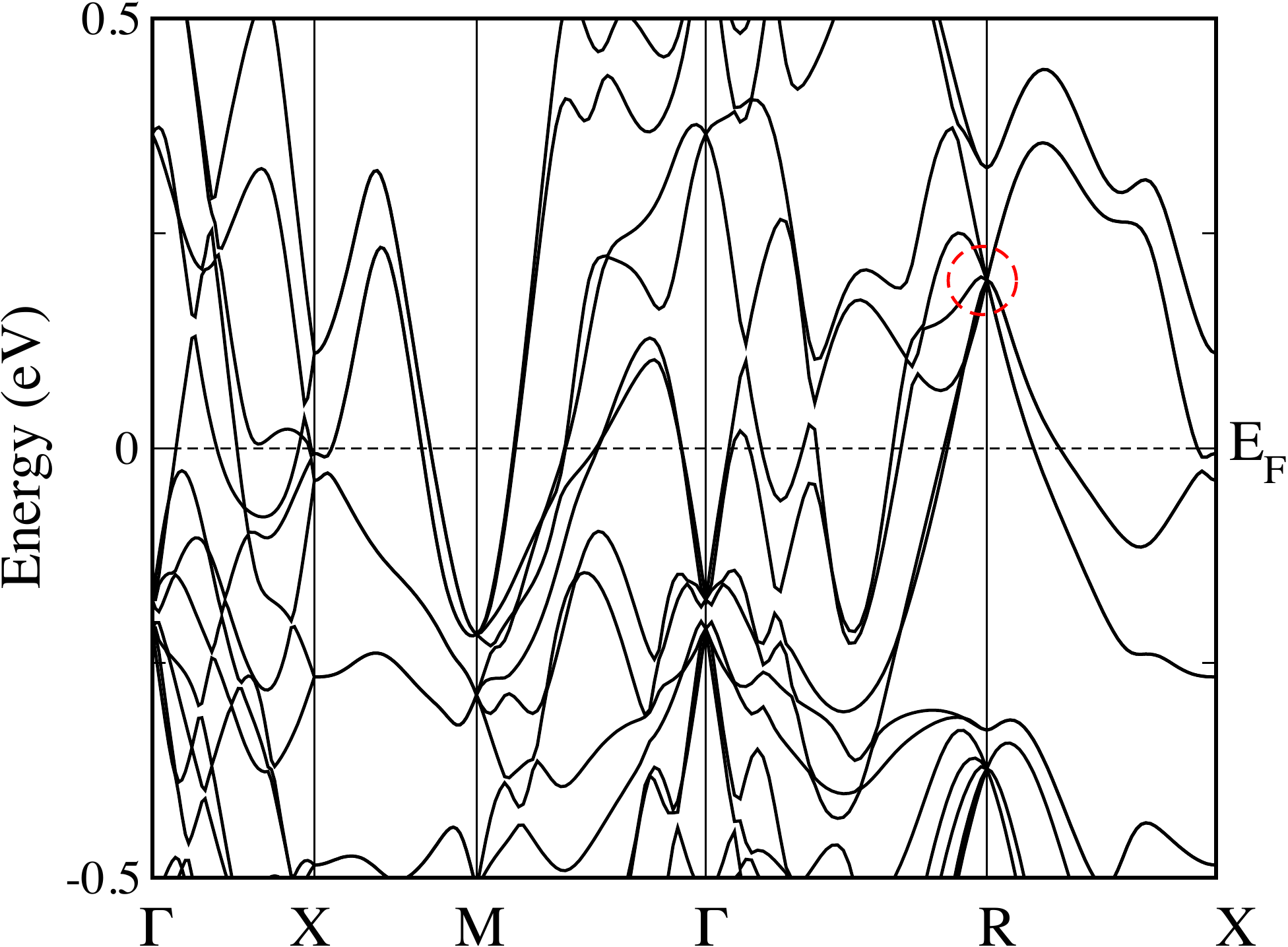}
}\quad 
\subfloat[KBiF$_6$ (SG 206)\label{fig:dft206}]{%
  \includegraphics[width=.3\textwidth]{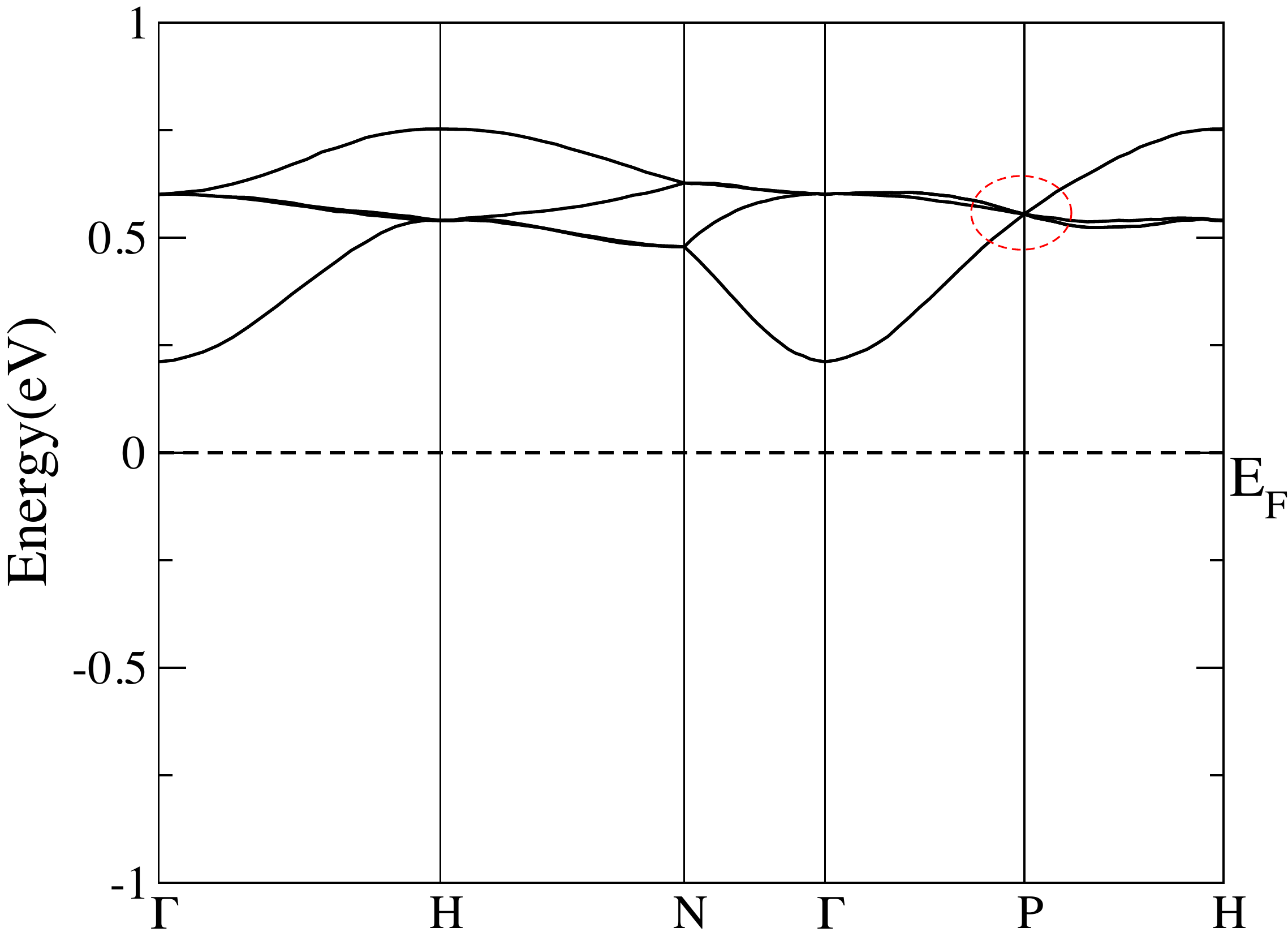}%
	}
\caption{(a),(b),(c) 3-fold fermions at the $P$ point in SG 199 and (d) SG 214. 6-fold crossings in (e) SG 212, (f),(g) SG 213, and (h) SG 206.}
\label{fig:dft-app1}
\end{figure}

\begin{figure}[H]\centering
	\subfloat[RhBi$_4$ (SG 230)\label{fig:dft230a}]{
	\includegraphics[width=.3\textwidth]{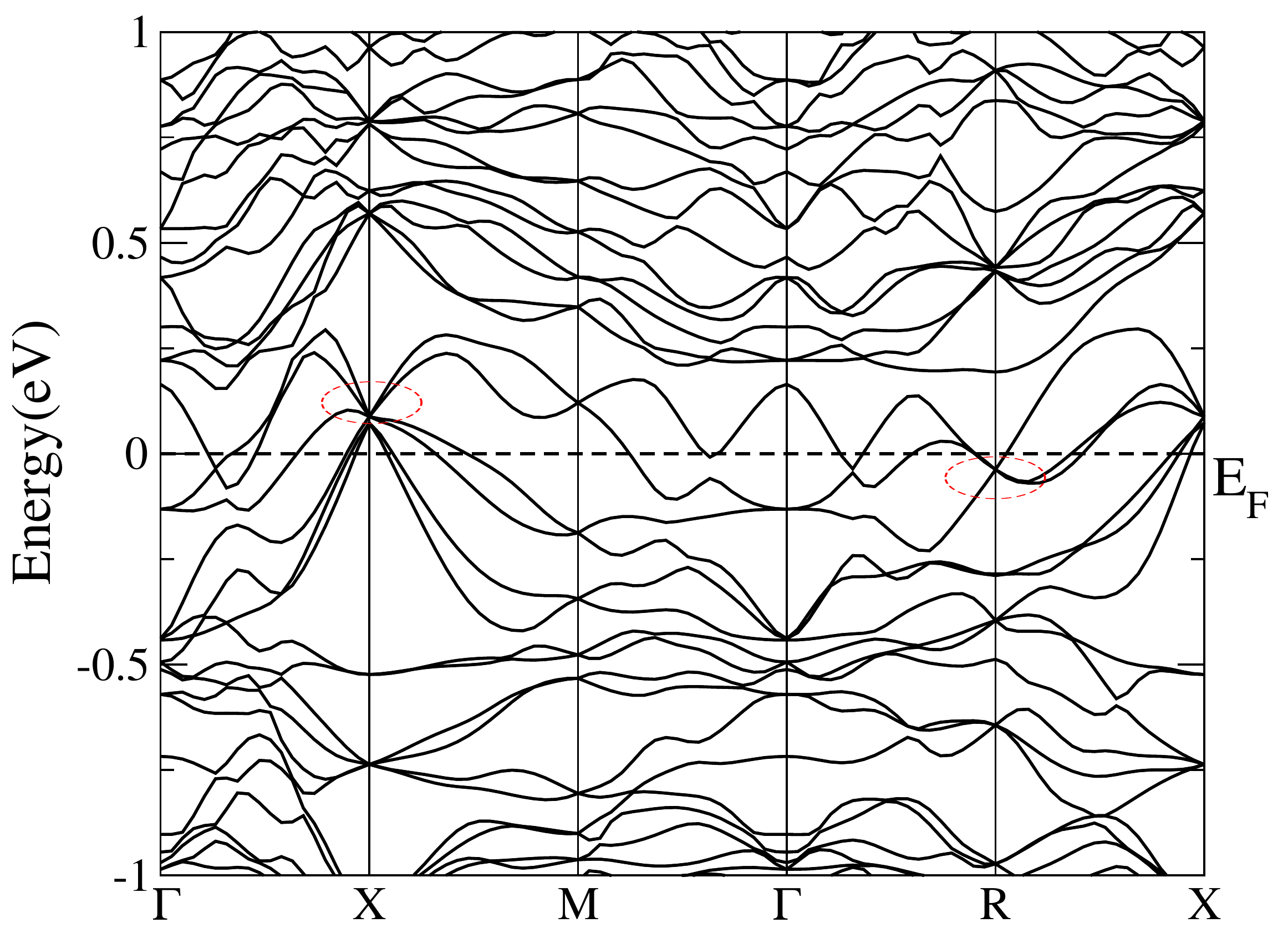}
	}\quad
	\subfloat[SiO$_2$ (SG 230)\label{fig:dft230b}]{
	\includegraphics[width=.3\textwidth]{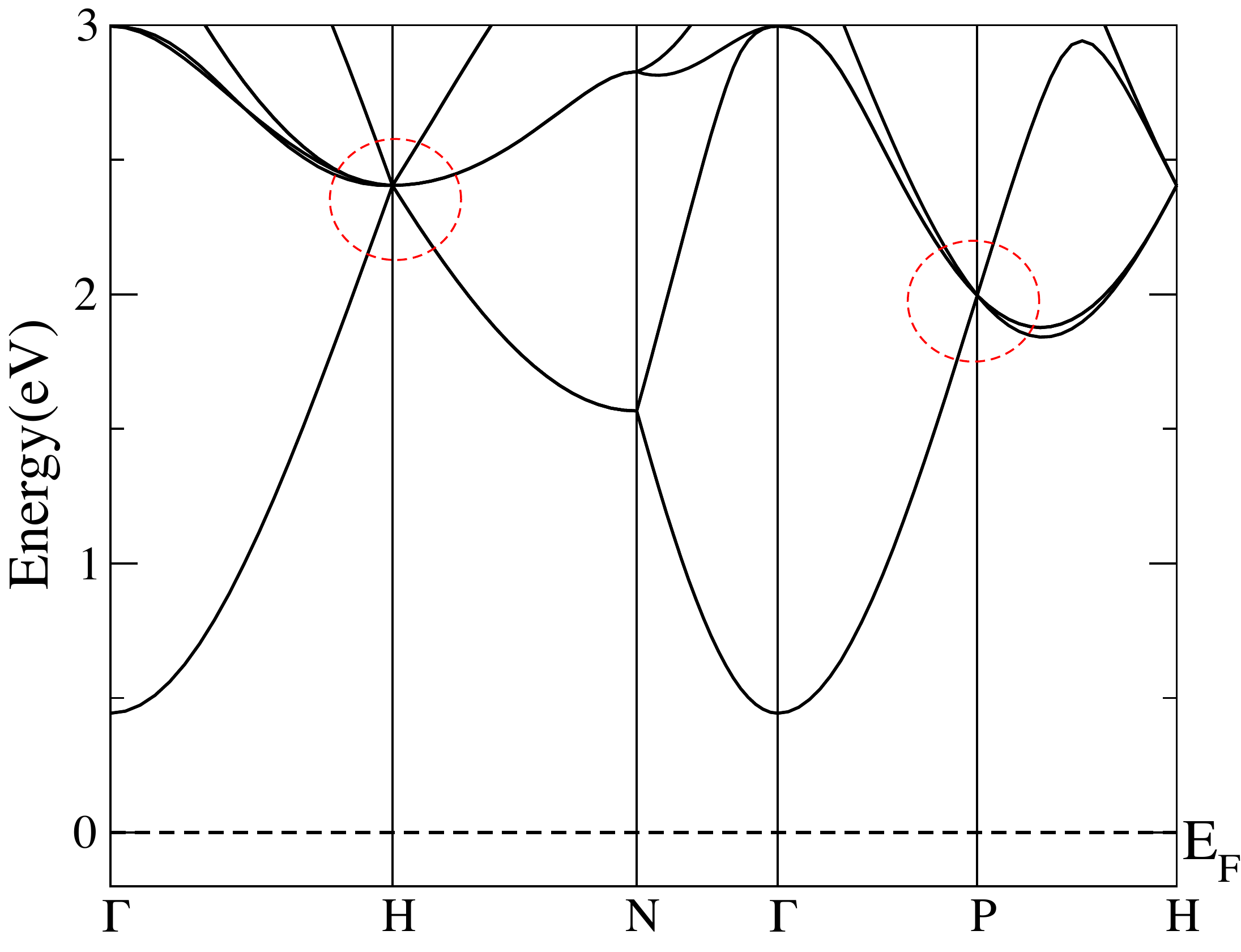}
	}
\caption{Compounds 230 display 3- and 8-fold fermions near the Fermi level at the $P$ and $H$ points, respectively.}
\label{fig:dft230}
\end{figure}

\begin{figure}[H]\centering
\subfloat[WO$_3$ (SG 130)\label{fig:dft130-2}]{
	\includegraphics[width=.3\textwidth]{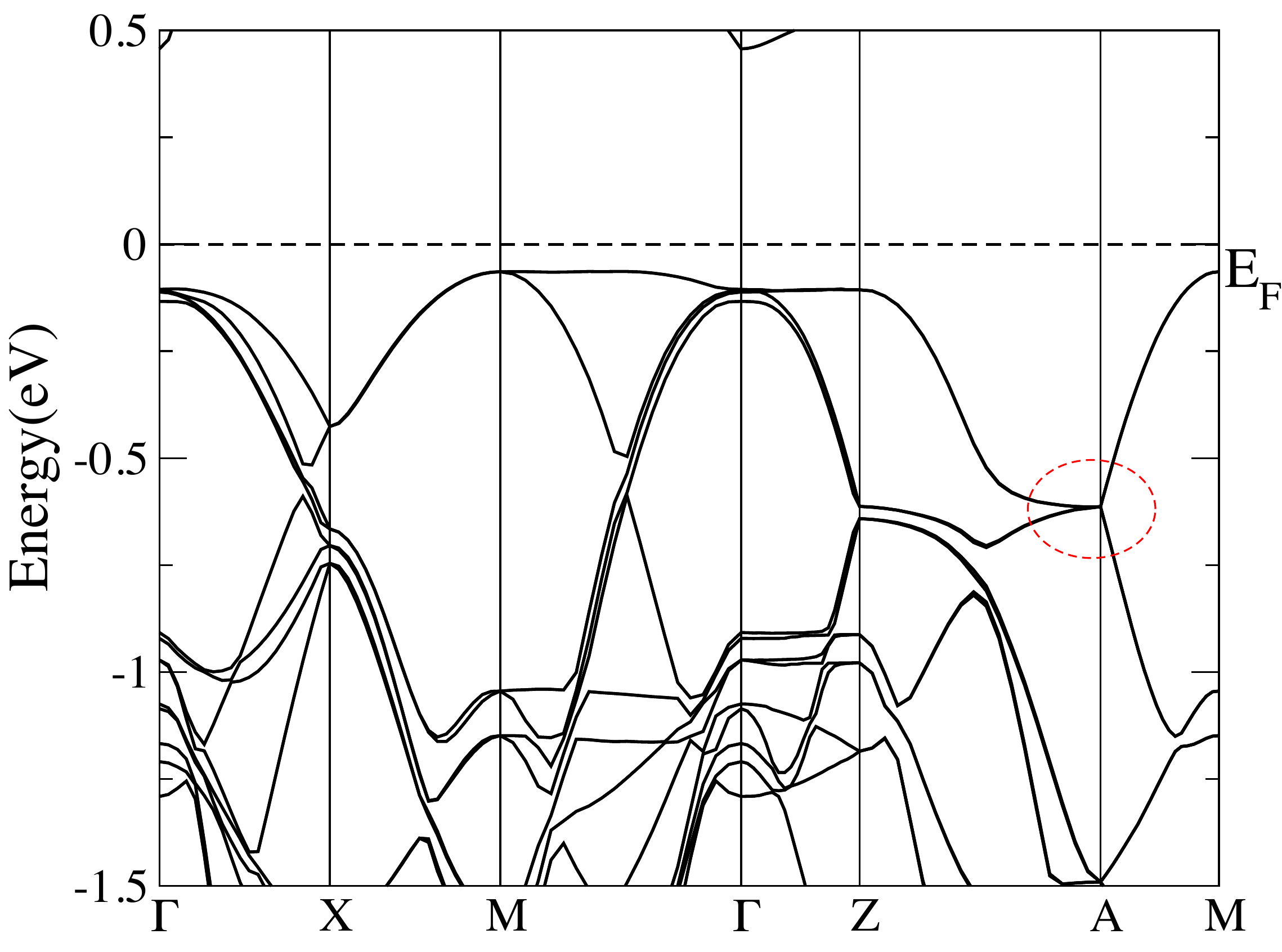}
}\quad 
\subfloat[AuBi$_2$O$_4$ (SG 130)\label{fig:dft130-3}]{
	\includegraphics[width=.3\textwidth]{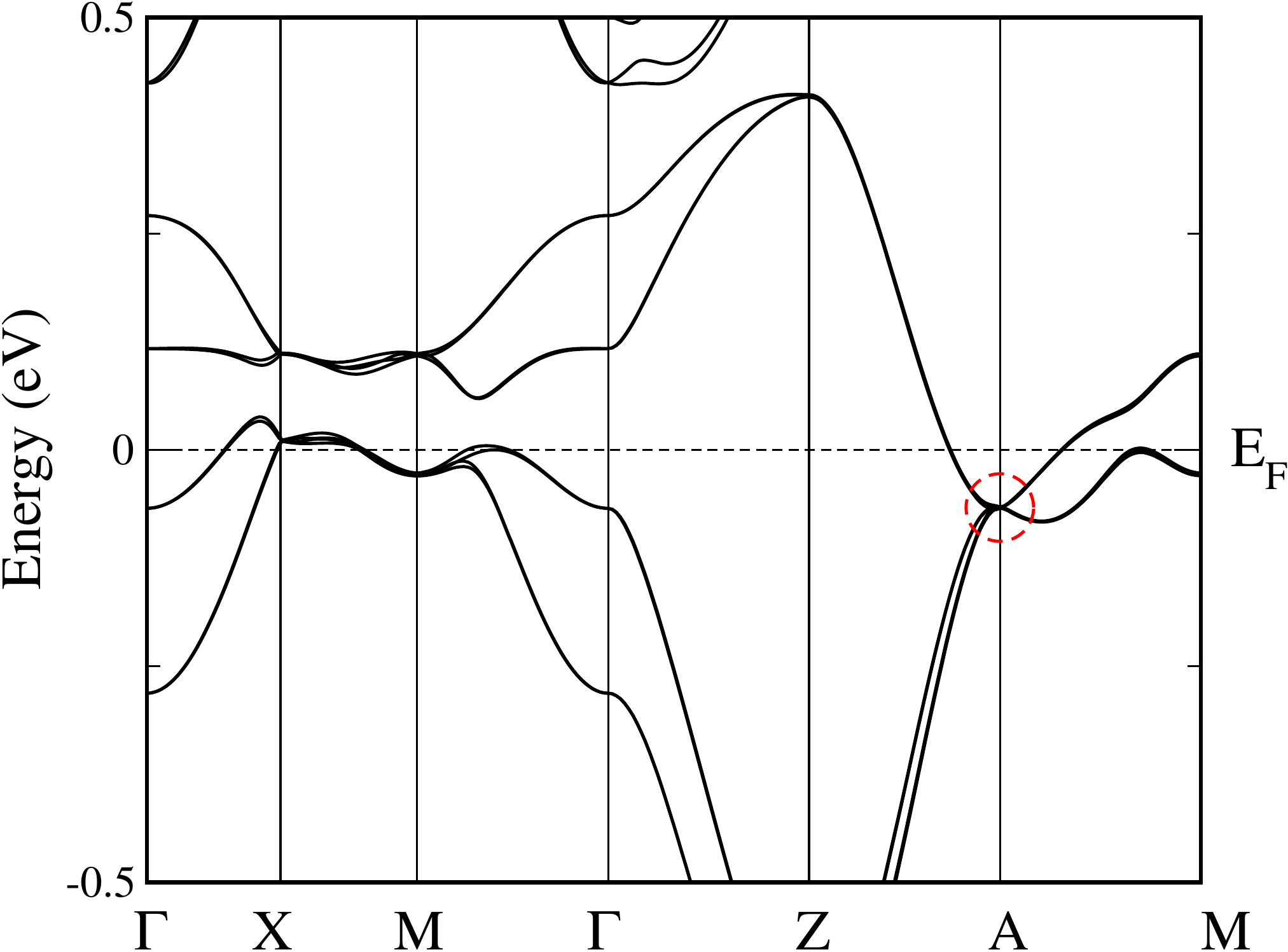}
}\quad
\subfloat[AgBi$_2$O$_4$ (SG 130)\label{fig:dft130-4}]{
	\includegraphics[width=.3\textwidth]{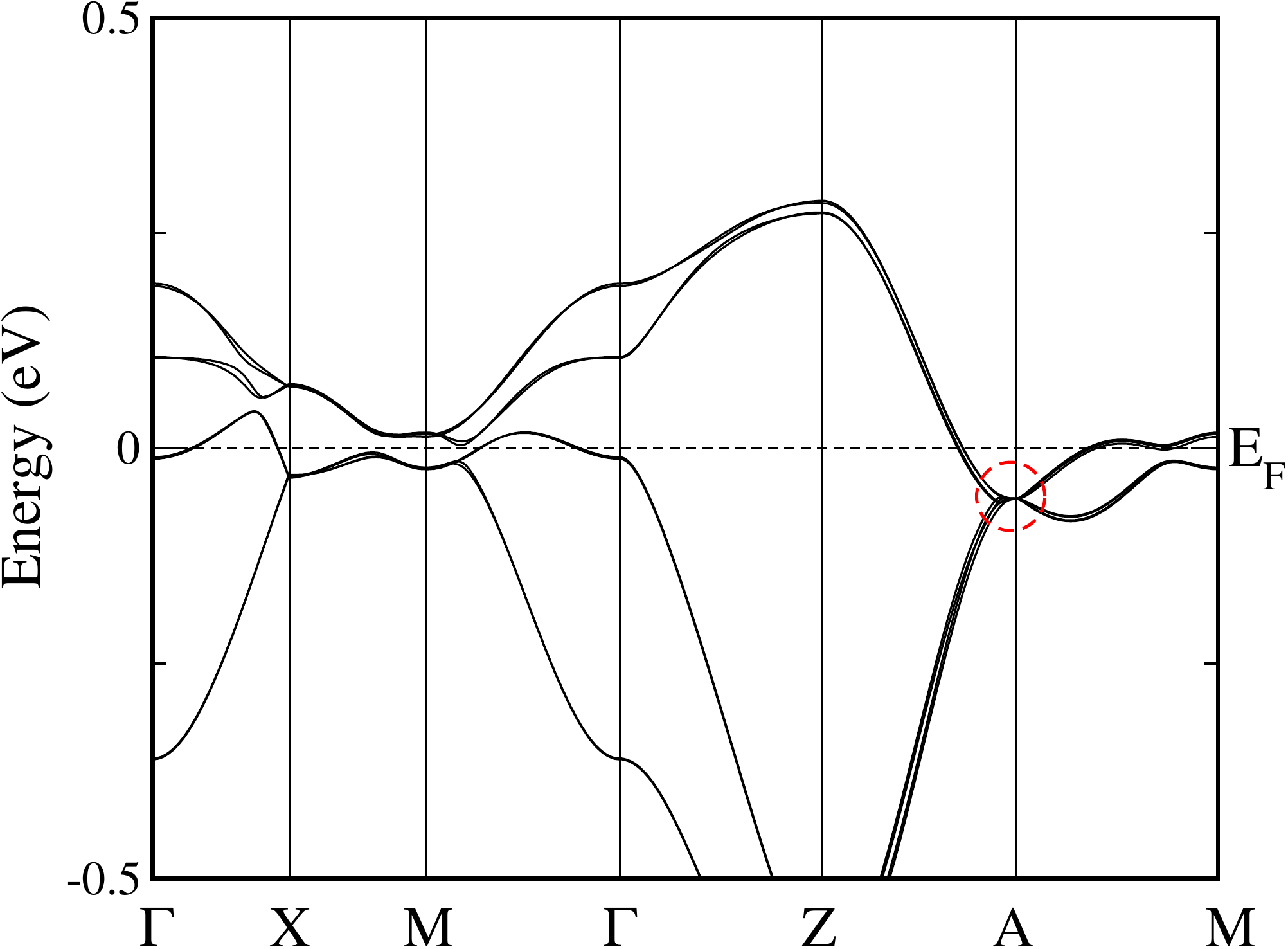}
}\quad
\subfloat[PdSe (SG 135)\label{fig:dft135}]{
	\includegraphics[width=.3\textwidth]{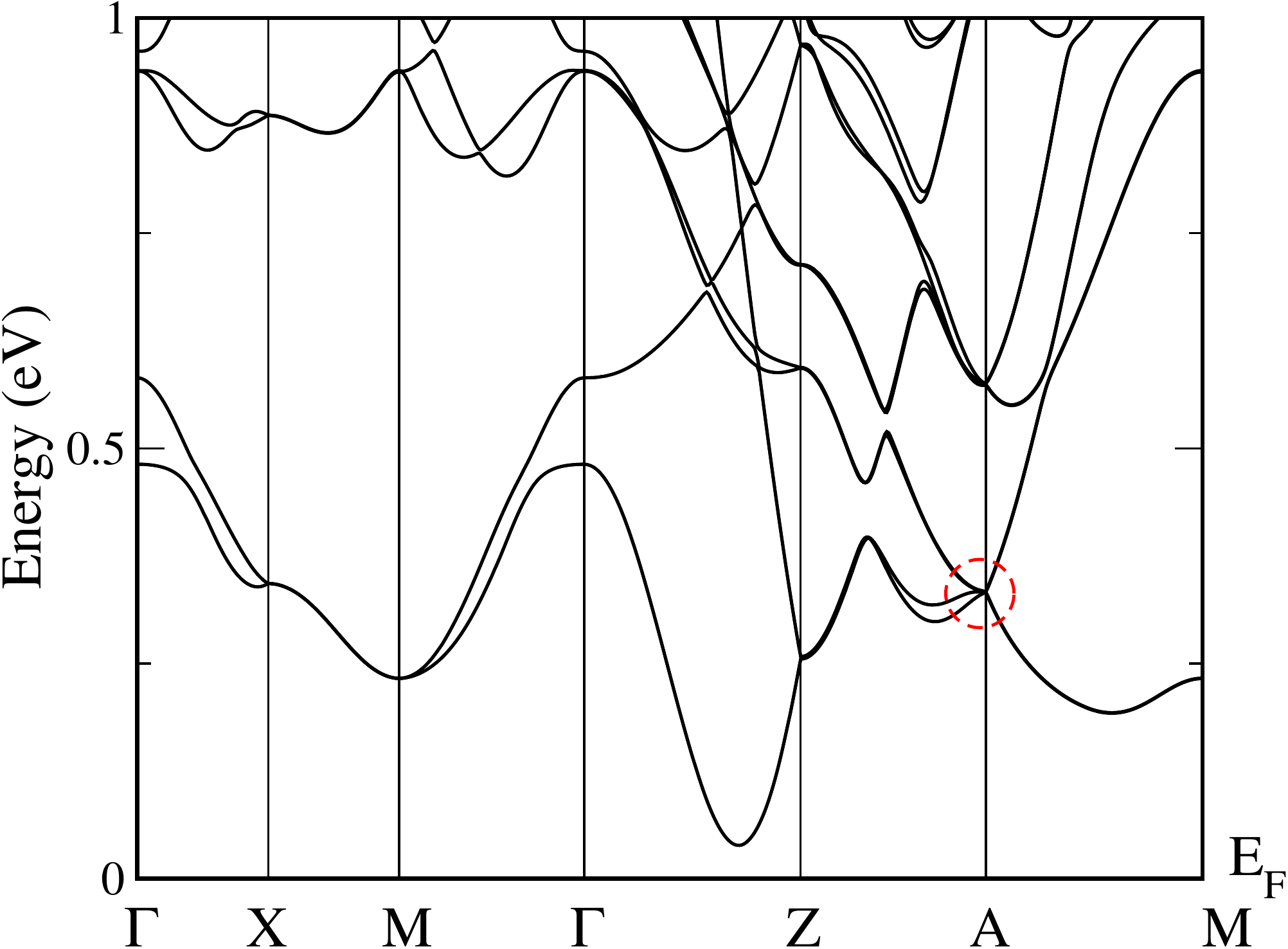}
}\quad
\subfloat[PdTe (SG 135)\label{fig:dft135-3}]{
	\includegraphics[width=.3\textwidth]{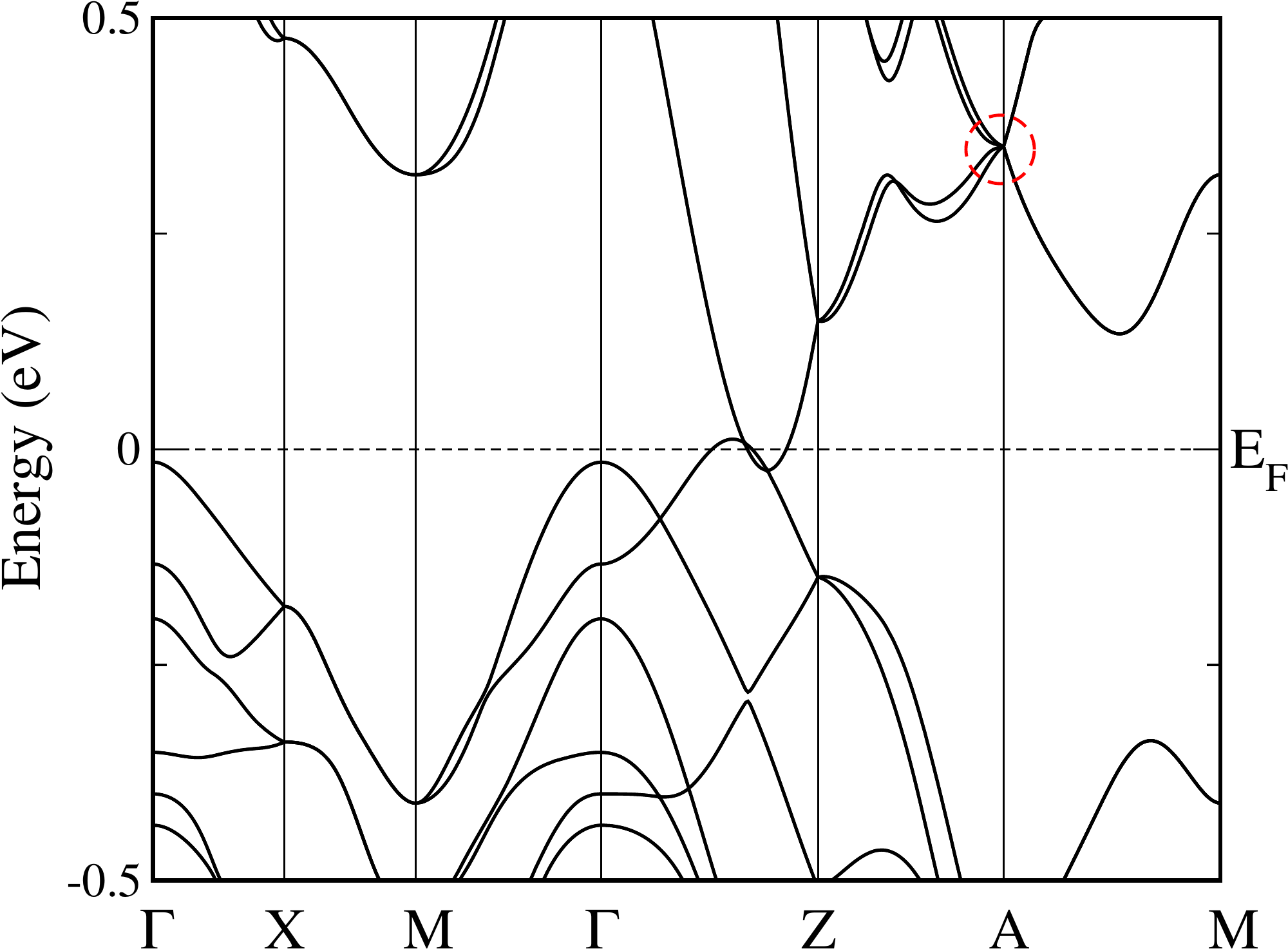}
}
\caption{Required 8-band crossings at the $A$ point in (a), (b), (c) SG 130 and (d), (e) 135.}
\label{fig:dft-app2}
\end{figure}

\bibliography{newfermions}